%% file: thesis.tex
\documentclass[a4paper,11pt,twoside,extrafontsizes,oldfontcommands]{memoir}
\usepackage{thesis}
\begin{document}
\frontmatter
\pagestyle{empty}
%
%
\include{OTHER/titlepage}

\pagestyle{plain}
%
%
\include{OTHER/declaration} 
%
%
\include{OTHER/summary} 
%
\include{OTHER/thanks} 
%
%
\include{OTHER/contents}

%
\mainmatter
%
\pagestyle{headings}
%
%
\graphicspath{{CHAP-1/FIGS/}{.}}
\include{CHAP-1/chapter1}

%
\graphicspath{{CHAP-2/FIGS/}{.}}
\include{CHAP-2/chapter2}

%
\graphicspath{{CHAP-3/FIGS/}{.}}
\include{CHAP-3/chapter3}
\graphicspath{{CHAP-4/FIGS/}{.}}
\include{CHAP-4/chapter4}

%
\graphicspath{{CHAP-5/FIGS/}{.}}
\include{CHAP-5/chapter5}

\graphicspath{{CHAP-6/FIGS/}{.}}
\include{CHAP-6/chapter6}

%
\graphicspath{{CHAP-7/FIGS/}{.}}
\include{CHAP-7/chapter7}

%
\graphicspath{{CHAP-8/FIGS/}{.}}
\include{CHAP-8/chapter8}

%
\graphicspath{{CHAP-9/FIGS/}{.}}
\include{CHAP-9/chapter9}

%
\graphicspath{{CHAP-10/FIGS/}{.}}
\include{CHAP-10/chapter10}

%
%
\appendix
\graphicspath{{APP-A/FIGS/}{.}}
\include{APP-A/appendixa}

\graphicspath{{APP-B/FIGS/}{.}}
\include{APP-B/appendixb}

\graphicspath{{APP-C/FIGS/}{.}}
\include{APP-C/appendixc}

\graphicspath{{APP-D/FIGS/}{.}}
\include{APP-D/appendixd}

%
\backmatter
%
%
%
%
%
\include{OTHER/references}

%
%
\end{document}

%% file: OTHER/titlepage.tex
\ifpdf
  \pdfbookmark[0]{Titlepage}{title}{}
\fi
%
%
\begin{center}
 \qquad\\[10mm]
 {\renewcommand\baselinestretch{1.2}\Huge\textbf{
%
%
   Physical modelling of galaxy clusters and Bayesian inference in astrophysics
 }\par}
 \qquad\\[50mm]
 {\LARGE Kamran Javid}\\[10mm]
 {\large Fitzwilliam College}\\[10mm]
 \includegraphics[width=50mm]{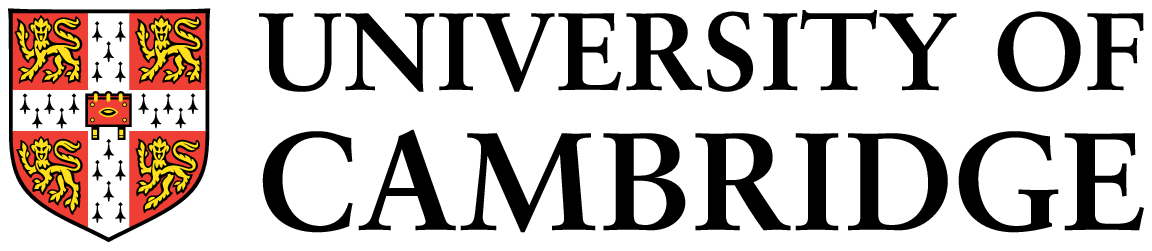}\\[60mm]
 {\Large May 2018}\\[10mm]
 {\large A dissertation submitted for the degree of Doctor of Philosophy}
\end{center}

\cleardoublepage

%% file: OTHER/declaration.tex
\chapter*{Declaration}


This dissertation is the result of my own work and includes nothing which is
the outcome of work done in collaboration except as declared in the Preface
and specified in the text.
It is not substantially the same as any that I have submitted, or, is being
concurrently submitted for a degree or diploma or other qualification at the
University of Cambridge or any other University or similar institution except
as declared in the Preface and specified in the text. I further state that no substantial part of my dissertation has already been submitted, or, is being
concurrently submitted for any such degree, diploma or other qualification
at the University of Cambridge or any other University of similar institution
except as declared in the Preface and specified in the text.
It does not exceed the prescribed word limit for the relevant Degree Committee.

\cleardoublepage

%% file: OTHER/summary.tex

\chapter*{Summary}



This thesis is concerned with the modelling of galaxy clusters, applying these models to real and simulated data using Bayesian inference, and the development of Bayesian inference algorithms applicable to a wide range of astrophysical problems.

I present a comparison of mass estimates for $54$ galaxy cluster candidates from the second Planck catalogue (PSZ2) of Sunyaev--Zel'dovich sources. I compare the mass values obtained with data taken from the Arcminute Microkelvin Imager (AMI) radio interferometer system and from the Planck satellite. The former of these uses a Bayesian analysis pipeline that parameterises a cluster in terms of its physical quantities, and models the dark matter \& baryonic components of a cluster using Navarro-Frenk-White (NFW) and generalised-NFW profiles respectively. The mass estimates derived from Planck data are obtained from the results of the Bayesian detection algorithm PowellSnakes (PwS). I also analyse simulated AMI data with input values based on PwS mass estimates.

I then compare three cluster models using AMI data for the 54 cluster sample. The two observational models considered only model the gas content of the cluster. To compare the physical and observational models I consider their posterior parameter estimates, including the calculation of a metric defined between two probability distributions. The models' fit to the cluster data is evaluated by looking at the Bayesian evidence values.

Improvements to the physical modelling of galaxy clusters are then considered, either by relaxing some of the assumptions underlying the physical model, or by introducing a new profile for the dark matter component of clusters. The resultant models are compared with the physical model introduced previously.

The final part of the cluster analysis work focuses on Bayesian analysis using a joint likelihood function of data from both AMI and the Planck satellite simultaneously. The results of this joint analysis are compared with those obtained from the individual likelihood analyses using simulated data and with real data taken from the $54$ cluster sample.

Finally, a new Bayesian inference algorithm based on nested sampling is presented. The algorithm, named the "geometric nested sampler", is an adaption of the Metropolis-Hastings nested sampler and makes use of the geometrical interpretation of sets of parameters to sample from their domains efficiently. The geometric nested sampler is tested on several toy models as well as a model representing the emission of gravitational waves from binary black hole mergers. The results obtained using the geometric nested sampler are compared with those from popular nested sampling algorithms.

\cleardoublepage

%% file: OTHER/thanks.tex
\chapter*{Acknowledgements}


Completing my PhD has been a bit of a roller coaster ride, and a journey that has been full of helpful and supportive people.

First of all I would like to thank my primary supervisor, Richard Saunders. Your support and attention through a number of issues including ones I couldn't have imagined experiencing before starting my PhD was wonderful. Your thoroughness and attention to detail have changed the way I approach problems and I am eternally grateful for it. You have been family-like figure during my time at Cambridge as well as my work colleague, and I know that our friendship will continue beyond Cambridge.

A massive thank you to my secondary supervisor Yvette Perrott, who has done a fabulous job of providing me support, and giving direction to my research. Your understanding of the underlying astrophysics associated with my research, ability to give recommendations and ways to improve my work will be very much missed post-Cambridge.

Another massive thank you to Will Handley for his support while we collaborated on the nested sampling project. You are a fantastic supervisor of research, and I know your career will go very far. You are a pleasure to work with and have helped me confirm that statistics is one of my true passions.

I would also like to thank everyone else who I have collaborated with during my PhD: Clare Rumsey, Pedro Carvalho, Keith Grainge, Mike Hobson, Anthony Lasenby, Farhan Feroz, and others. A special thanks goes to Malak Olamaie who was incredibly helpful as a stand-in supervisor during the early stages of my PhD while Yvette was on maternity leave.

Thanks to Dave Green for his invaluable \LaTeX{} support while writing my first year report, research papers, and my thesis. I would also like to thank Dave Titterington and Greg Willatt from the Cavendish Astrophysics IT team, and Stuart Rankin from the Cambridge High Performance Computing team. Your impeccable services and support have kept the `experimental equipment' of my PhD in tip-top shape, and so I am truly indebted to you.

A special shout out to the University food halls across the city, who have done a fantastic job of the not so easy task of keeping my belly full the past few years. In particular, Hughes Hall did an exquisite job of appreciating my love for food by giving me over-sized, nutritious and tasty portions.

I would also like to thank the students I have supervised at Trinity College during my PhD, you have provided a mental workout like no other. You are undoubtedly some of the smartest people I will ever meet, and it was a pleasure to get an insight into your way of thinking.

To my friends in the astrophysics office: Terry Jin, Maximilian G\"{u}nther, Marion Neveu, Richard Hall, Zoe Ye and Julia Riley. Thank you for providing fun and support in one of the places I needed it the most!

To my friends in Sheffield, Cambridge, and London, a big thank you for helping me keep sane during the last few years with our fun adventures together. A special thanks to Rob Hull for teaching me how to electronically draw images of three-dimensional shapes neatly for the nested sampling project!

The Mathematics department at Wales High School deserves a lot of credit for getting me interested in mathematics in the first place. Without your passion and enthusiasm for the subject, I doubt I would have gone to university at all. I am eternally grateful and I hope to re-pay you one day by helping to inspire our next generation of students to love the subject.

Saving the best until last, I would like to thank my family. Without your love there's no way I would have finished this PhD. Everything I do, I do it to make you proud of me. You are my drive and I will strive to make you proud for the rest of my days. To my mother, given we're a single-parent, only-child family, I'm sure no one would have expected us getting this far. Putting a smile on your face and making you proud is worth a lifetime of my blood, sweat and tears.

\cleardoublepage

%% file: OTHER/contents.tex
%
%
\bgroup
\renewcommand{\baselinestretch}{1.0}\normalsize

\tableofcontents

%
%

\egroup

\cleardoublepage

%% file: CHAP-1/chapter1.tex
\chapter{Introduction}\label{c:first}


\section{Galaxy clusters}\label{s:galaxy_clusters}

In the local Universe and out to redshifts of around two, clusters of galaxies are observed as massive gravitationally bound structures, often roughly spherical and with very dense central cores (see reviews by e.g. \citealt{2002ARA&A..40..539R}, \citealt{2005RvMP...77..207V}, \citealt{2011ARA&A..49..409A}, and \citealt{2013SSRv..177..247G}). 
It is over eighty years ago that it was first postulated that a galaxy cluster's mass is dominated by dark matter (\citealt{1933AcHPh...6..110Z} and \citealt{1937ApJ....86..217Z}). More recently it has been shown that dark matter contributes $\approx90\%$ of the cluster mass (see e.g. \citealt{2006ApJ...640..691V} and \citealt{2011ApJS..192...18K}). Stars, gas and dust in galaxies, as well as a hot ionised intra-cluster medium (ICM) make up the rest of the mass in a cluster, with the latter being the most massive baryonic component. The galaxies emit in the optical and infrared wavebands, whilst the ICM emits in X-ray via thermal Bremsstrahlung and also interacts with cosmic microwave background (CMB) photons via inverse Compton scattering. This last effect is what is known as the Sunyaev--Zel'dovich (SZ) effect \citep{1970CoASP...2...66S}.


\section{The Sunyaev--Zel'dovich effect}\label{s:sz_effect} 

\begin{figure}
\centerline{\includegraphics[width=\linewidth]{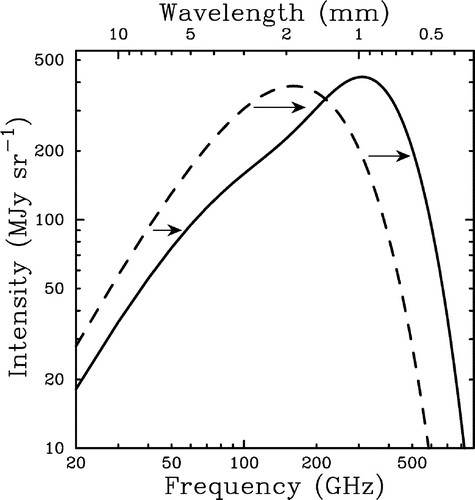}}
\caption{Radiation intensity as a function of frequency. Note the dashed line represents the incident radiation, whilst the solid line represents the energy-boosted inverse Compton scattered radiation. Taken from \citet{2002ARA&A..40..643C}.}\label{f:sz_intensity}
\end{figure}

The SZ effect is particularly strong in the cluster ICM, where temperatures range between $10^{7} - 10^{8}$~K. The nature of the CMB spectrum means that the effect leads to an increase in intensity at frequencies above $217$~GHz and a decrease for frequencies below (Figure~\ref{f:sz_intensity}). The measurement of the SZ surface brightness increment / decrement has the crucial characteristic that it is redshift independent (see Section~\ref{s:sz_theory}). 
The SZ effect has the additional advantage over X-ray analysis, that it only depends on the electron number density linearly (see Section~\ref{s:sz_theory}), whereas X-ray Bremsstrahlung emission is proportional to electron number density squared. This means that SZ can in practice be used to analyse a cluster at higher radius.
The Planck telescope (Section~\ref{s:pl_mission}) and the Arcminute Microkelvin Imager radio interferometer system (AMI, see Section~\ref{s:ami}) both observe galaxy clusters by measuring the SZ effect. 

\section{Planck mission}\label{s:pl_mission}

The Planck mission\footnote{\url{http://www.esa.int/Planck/}.} is a European Space Agency (ESA) mission, whose primary objective is to investigate the CMB. The Planck telescope was a space telescope which was launched in May 2009 and deactivated in October 2013. The combination of Planck's low-frequency and high-frequency instruments (LFI and HFI) provides nine frequency channels in the range $37$~GHz -- $857$~GHz. The LFI has angular resolutions of $33$, $24$, and $14$~arcminutes at respective frequencies of $30$, $44$, and $70$~GHz. The HFI has angular resolutions of $10$, $7.1$, and $5.5$~arcminutes at $100$, $143$, and $217$~GHz and $5.0$~arcminutes at each of $353$, $545$, and $857$~GHz. For more information on the Planck telescope I refer the reader to the Scientific Programme of Planck \citep{2006astro.ph..4069T}. 
In addition to all-sky coverage, Planck has its own advantages for SZ work: a very wide range of frequency channels, polarisation capability, and a channel at the 217-GHz null frequency of SZ all help to remove contamination from synchrotron, Bremsstrahlung and dust emissions.
Of particular importance for the work described here are the Planck cluster-catalogues (see \citealt{2014A&A...571A..29P}, \citealt{2015A&A...581A..14P} and \citealt{2016A&A...594A..27P} for papers relating to catalogues PSZ1, PSZ1.2 and PSZ2 respectively, where `PSZX' refers to the X\textsuperscript{th} Planck SZ catalogue). These provide e.g. cluster candidate positions, redshift ($z$) values (see Section~\ref{s:pl_z}), integrated Comptonisation parameter ($Y$) values and mass ($M$) estimates. PSZ2 is the most recent all-sky Planck cluster catalogue, and is the one which I refer to unless stated otherwise.


\section{AMI}\label{s:ami} 

AMI is an interferometer system near Cambridge, designed for SZ studies (see e.g. \citealt{2008MNRAS.391.1545Z}). It consists of two arrays: the Small Array (SA), optimised to couple to SZ signal, with an angular resolution of $\approx 3$~arcmin and sensitivity to structures up to $\approx 10$~arcmin in scale; and the Large Array (LA), with angular resolution of $\approx 30$~arcsec, which is largely insensitive to SZ, and is used to characterise and subtract confusing radio-sources (see Section~\ref{s:identified_sources}).  Both arrays operate at a central frequency of $\approx 15.7$~GHz and, at the time the AMI data for this paper were taken, with a bandwidth of $\approx 4.3\,$~GHz, divided into six channels. Both arrays actually operate over the wide frequency range of $\approx$ $12.0$ -- $18.0$~GHz for sensitivity, and the correlator splits this range into eight separate channels each approximately $0.72$~GHz wide to reduce chromatic aberration over the fields of view to manageable levels. However, due to satellite interference at the lower end of the spectrum, data from the bottom two channels are excluded, giving the effective bandwidth of $4.3$~GHz across six channels mentioned above). A summary of AMI's characteristics is given in Table~\ref{t:ami}. More detail on AMI is given in Section~\ref{s:ami_int}. Note that AMI has recently received a new digital correlator 
\citep{2018MNRAS.475.5677H}, but all data used in this thesis were obtained by the system with its analogue correlator.


\begin{table}
\centering
\begin{tabular}{{l}{c}{c}}
\hline
 & SA & LA \\
\hline 
Antenna diameter & $3.7~\rm{m}$ & $12.8~\rm{m}$ \\
Number of antennas & $10$ & $8$ \\
Baseline lengths (current) & $5-20~\rm{m}$ & $18-110~\rm{m}$ \\
Primary beam FWHM (at $15.7~\rm{GHz}$) & $20.1~\rm{arcmin}$ & $5.5~\rm{arcmin}$ \\
Typical synthesised beam FWHM & $3~\rm{arcmin}$ & $30~\rm{arcsec}$ \\
Flux sensitivity & $30~\rm{mJy}~\rm{s}^{1/2}$ & $3~\rm{mJy}~\rm{s}^{1/2}$ \\
\hline
\end{tabular}
\caption{Summary of AMI characteristics. Both arrays measure the same linear polarisation.}\label{t:ami} 
\end{table}

\section{Remainder of this thesis}

In Chapter~\ref{c:second} I give an overview of the theory underlying various topics which are heavily relied upon throughout the thesis: interferometry, measuring the SZ effect, galaxy cluster modelling, and Bayesian inference.\\
In Chapter~\ref{c:third} I apply a cluster model to data from AMI of clusters detected by Planck, and compare the results with those obtained directly from Planck data. I also analyse simulated cluster data whose inputs are based on the mass estimates obtained from Planck data, to see if AMI simulations \& the cluster model are capable of inferring the correct cluster masses.\\  
Chapter~\ref{c:fourth} presents the results of a cluster model comparison for the sample of 54 clusters considered in the previous Chapter; for the three models I compare the parameter estimates and Bayesian evidence values obtained for each cluster.\\
A new cluster model is presented in Chapter~\ref{c:fifth} which uses an Einasto profile to model the dark matter component of a cluster. By looking at cluster parameter profiles, and performing Bayesian analysis on simulated \& real data, I compare the new model with the one presented in Chapter~\ref{c:second}.\\
Chapters~\ref{c:sixth} and~\ref{c:seventh} detail further attempts to enhance galaxy cluster modelling. I first try to relax the mass assumption associated with the models detailed in Chapters~\ref{c:second} and~\ref{c:fifth}, and plot the resulting mass profiles for a range of clusters (Chapter~\ref{c:sixth}). I then try to incorporate non-thermal pressure into the cluster models in Chapter~\ref{c:seventh}, and plot the resultant parameter profiles. \\
In Chapter~\ref{c:eighth} I introduce a joint AMI-Planck analysis method, which revolves around evaluating the likelihood functions associated with each instrument simultaneously. I then present the results of this method applied to both simulated and real datasets, and compare with the results obtained from conducting the individual instrument analyses separately. \\
An overview of Monte Carlo sampling methods is given in Chapter~\ref{c:ninth}. This includes an introduction to nested sampling, the method upon which the algorithm presented in Chapter~\ref{c:tenth} is based on. I also explain briefly how samples can be used to approximate the distribution from which they originate. \\ 
In Chapter~\ref{c:tenth} I provide the motivation \& technical details of the nested sampling algorithm I have created and refer to as the "geometric nested sampler". I apply the algorithm to several toy models \& to an astrophysical application (detecting gravitational waves from a black hole binary merger system), and compare its performance with pre-existing nested sampling algorithms.

\section{Conventions}

A `concordance' flat $\Lambda$CDM cosmology is assumed: $\Omega_{\rm M} = 0.3$, $\Omega_{\Lambda} = 0.7$, $\Omega_{\rm R} = 0$, $\Omega_{\rm K} = 0$, $h=0.7$, $H_{0} = 100~h~\rm km~s^{-1}$~Mpc$^{-1}$, $\sigma_{8} = 0.8$, $w_{0} = -1$, and $w_{\rm a} = 0$. The first four parameters correspond to the (dark + baryonic) matter, the cosmological constant, the radiation, and the curvature densities respectively. $h$ is the dimensionless Hubble parameter, while $H_{0}$ is the Hubble parameter now and $\sigma_{8}$ is the power spectrum normalisation on the scale of $8$~$h^{-1}$~Mpc now. $w_{\rm 0}$ and $w_{\rm a}$ are the equation of state parameters of the Chevallier-Polarski-Linder parameterisation \citep{2001IJMPD..10..213C}.



%% file: CHAP-2/chapter2.tex
\chapter{Introductory theory}\label{c:second}


\section{Interferometry}\label{s:interferometry} 


In addition to the advantage of high angular resolution from long baselines, interferometers possess a number of advantages over single-dish telescopes, particularly for CMB work. Among these are their relative insensitivity to atmospheric emission (see e.g. \citealt{2003MNRAS.341.1057W}), the ease with which systematic errors such as ground spill (\citealt{2000ApJ...543..787L}) can be dealt with; and radio-source contamination (see e.g. \citealt{2002MNRAS.337.1207G}) can be kept to a minimum. Furthermore, the angular sensitivity of an interferometer can be fine-tuned by adjusting baseline lengths.

\begin{figure}
\centerline{\includegraphics[width=\linewidth]{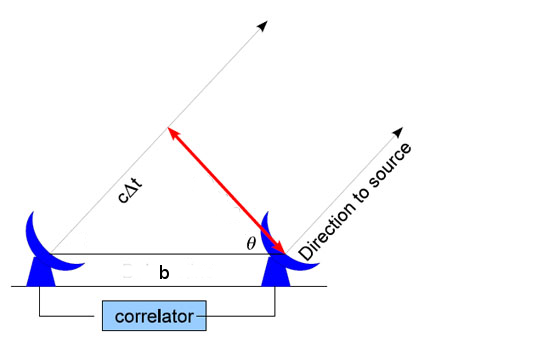}}
\caption{Simple east-west single baseline interferometer tracking a patch of sky containing a single radio-source. For a baseline $b$ and a source at angle $\theta$ from the vertical axis, the wavefront has to travel an additional distance $b\sin\theta$ to the further antenna. Image taken from \citet{2013ASSL..396...45Z}.}\label{f:interferometer}
\end{figure}

To understand how an interferometer works, consider a two-antenna system similar to the one constructed by Ryle and Vonberg \citep{1948RSPSA.193...98R}. Figure~\ref{f:interferometer} shows two antennas on an east-west baseline of length $b$ tracking a visible patch of sky which, initially, meets three conditions: (i) contains only one radio-source; (ii) this source is at the centre of the tracked patch; and (iii) this source is unresolved by the interferometer. At hour angle $\theta$ (as defined in Figure~\ref{f:interferometer}), the voltages $V_{1}$ and $V_{2}$ measured by each antenna at time $t$ are
\begin{equation}
\begin{aligned}\label{e:interferometer1}
V_{1} &= V_{0}e^{i(\omega t + kb \sin\theta)}, \\
V_{2} &= V_{0}e^{i\omega t},
\end{aligned}
\end{equation}
where $V_{0}$ is the signal voltage amplitude, $\omega$ is the angular frequency of the source radiation being observed and $k$ is the corresponding wavenumber. The $\omega t$-dependent parts are removed and the correlator multiplies the remaining components of \ref{e:interferometer1} together to give a response, termed visibility, proportional to
\begin{equation}\label{e:interferometer2}
e^{ikb\sin\theta},
\end{equation}
in which the constant of proportionality (including $V_0^2$, the effects of integration time, dish area and so on) which in practice is evaluated by observation of a bright, unresolved radio-source with well known properties. Unless the observing bandwidth $\Delta \omega$ is very low (and thus the coherence length $2 \pi c / \omega$ is very long), the baseline must be `phased up' by inserting an additional path equivalent to $b\sin\theta$ into the interferometer arm which the radiation hits first. This compensates for the extra path $c \Delta t =  b\sin\theta$ involved in the other arm. 

We now relax condition (ii). If the source is offset from the pointing centre by an angle $\alpha$, the extra path becomes $b\sin(\theta + \alpha)$. The path compensation is set for the pointing centre so that multiplying the equivalent expressions of \ref{e:interferometer1} now gives a visibility proportional to  
\begin{equation}\label{e:interferometer3}
\begin{aligned}
& e^{ikb(\sin(\theta + \alpha) - \sin\theta)} \\ 
&= e^{ikb\alpha \cos\theta},
\end{aligned}
\end{equation}
using the small angle approximation for $\alpha$. 

We now relax condition (iii). The response to a source, which has a top-hat surface-brightness distribution in $\alpha$-space of width $\Delta \alpha$ and centred on $\alpha$, is averaged over $\Delta \alpha$, giving a response proportional to
\begin{equation}\label{e:interferometer4}
\begin{aligned}
&\frac{1}{\Delta\alpha}\int_{-\frac{\Delta\alpha}{2}}^{\frac{\Delta\alpha}{2}}e^{ikb\alpha' \cos\theta}\,\mathrm{d}\alpha' \\
&= \mathrm{sinc}\left( \frac{kb\Delta\alpha\cos\theta}{2} \right)
\end{aligned}
\end{equation}
Thus sources with a large angular size on the sky ($\Delta\alpha \gg k b \cos \theta$) are resolved out by the interferometer since $\mathrm{sinc}(x) \rightarrow 0$ as $x \rightarrow \infty$. 

One can similarly examine the effect of the observing bandwidth. Repeating the above analysis for $k$ gives
\begin{equation}\label{e:interferometer5}
\begin{aligned}
&\frac{1}{\Delta k}\int_{-\frac{\Delta k}{2}}^{\frac{\Delta k}{2}}e^{ik'b\alpha \cos\theta}\,\mathrm{d}k' \\
&=  \mathrm{sinc}\left( \frac{\alpha b\Delta k\cos\theta}{2} \right).
\end{aligned}
\end{equation}
So a large enough bandwidth also causes the signal to fall, this time due to chromatic aberration. This explains the need for independent frequency channels which are a feature of AMI.

Finally, replacing condition (i) by a surface-brightness distribution $I(\theta,\alpha)$, and incorporating the primary beam function $A(\alpha)$, gives a visibility proportional to
\begin{equation}\label{e:interferometer6}
\iiint A(\alpha')\,I(\theta',\alpha')\,e^{ik'b\alpha' \cos\theta'}\,\mathrm{d}\theta'\mathrm{d}\alpha'\mathrm{d}k'.
\end{equation}

\subsection{AMI interferometry}
\label{s:ami_int}

The compensation for the path-length differences between each antenna and the cluster having been done in cables, the analogue correlator multiples the signal from each antenna at time $t$ by the signal at times $t - 7 \Delta t, t - 6 \Delta t,..., t,..., t + 7 \Delta t, t + 8 \Delta t$; Fourier transforming these lag products gives the amplitude and phase values of each of eight frequency channels. A problem with the analogue correlator is that each timelag $\Delta t = \Delta l / v_{\mathrm{group}}$ is not the same because each $\Delta l$, nominally $25$~mm, varies by some $5$--$10$\% because the circuit boards providing the $\Delta l$s have non-uniform relative permittivities and thicknesses. 

\citet{thompson} discuss cross-correlator performance in terms of cross-correlation correlation coefficient $\rho$,
\begin{equation}
\label{e:correlation}
\frac{\int (a_i - \langle a_i \rangle) (a_j - \langle a_j \rangle) \mathrm{d} t}{\sqrt{\int (a_i - \langle a_i \rangle)^2 \mathrm{d} t \int (a_j - \langle a_j \rangle)^2 \mathrm{d} t }},
\end{equation} 
where $a_i$ is the instantaneous voltage from antenna $i$, and $\langle \rangle$ denotes average over the few-second integration time $\tau$, and the integrals are over $\tau$. However, radio astronomy cross-correlators do not measure the denominator of equation~\ref{e:correlation}; what they do measure, for each lag, is effectively the numerator. The signal power is described as $A\exp (i \phi)$ where $A$ is "amplitude" and $\phi$ is "phase". The noise power is that from the front-end amplifiers, the atmosphere, and the CMB. The signal in the cross-correlation increases coherently over time, so the signal energy increases as time, while the noise increases incoherently so the noise energy increases as time$^{1/2}$. $\tau$ is chosen such that over it, signal energy $ \ll$ noise energy. For the measurements over $\tau$ to be meaningful, you want each receiver chain, from each front-end amplifier to correlator input, to produce a power that is \textit{stable} over the whole (typically $6$--$8$~hour) observation run. This is achieved with automatic gain controls designed to keep the power going into each correlator input constant. (Note that neither the gains of the receiver chains nor the output powers have to be the same -- astronomical calibration deals with this). 

However, ensuring the power at a correlator input is maintained at a constant level will bias measurements if, for example, the weather changes: cloud, rain, and raindrops on the receiver cover all emit at GHz-frequencies, thus raising (compared with fine weather) the noise power and so lowering the signal. 
This effect is (ideally) removed by the noise injection system (at Cambridge misleadingly called the `rain gauge') which works as follows. Low-level noise (of power $\approx 1$\% of the power due to front-end amplifier, CMB and atmosphere), of constant mean power and known signature $P(t)$ is injected into the waveguide that feeds the astronomical signal into the front-end amplifier. At each correlator input, the noise power due to $P(t)$ is extracted by synchronous detection and compared with the total noise power so that the noise power due to front-end amplifier, CMB and atmosphere, which determines the system temperature, is measured.  


\section{Measuring the SZ effect with an interferometer}\label{s:sz_theory} 

For a small field size, an interferometer samples from the two-dimensional complex visibility plane $\vec{u}$, also known as the $u$-$v$ plane, where $u$ and $v$ are orthogonal projected baselines in units of observing wavelength. For a given frequency $\nu$ the quantity measured by an interferometer corresponds (see equation~\ref{e:interferometer6}) to the Fourier components of the sky brightness distribution $\tilde{I}_{\nu}(\vec{u})$. $\tilde{I}_{\nu}(\vec{u})$ is given by the weighted Fourier transform of the surface brightness $I_{\nu}(\vec{x})$, 
\begin{equation}\label{e:sz1}
\tilde{I}_{\nu}(\vec{u}) = \int_{-\infty}^{\infty} A_{\nu}(\vec{x})I_{\nu}(\vec{x})e^{2\pi i\vec{u}\cdot\vec{x}}\,\mathrm{d}^{2}\vec{x},
\end{equation}
where $\vec{x}$ is the position in the sky relative to the phase centre and $A_{\nu}(\vec{x})$ is the primary beam of the (identical) antennas for a given frequency; note that $I$ and $A$ here are parameterised in terms of spatial coordinates rather than angular. The positions at which $\tilde{I}_{\nu}(\vec{u})$ are sampled from is therefore determined by the physical orientation of the antennas. 

The change in CMB surface brightness due to the thermal SZ effect in a galaxy cluster is given by (see e.g. \citealt{1999PhR...310...97B})
\begin{equation}\label{e:sz2}
\delta I_{\mathrm{cl}, \nu} = T_{\rm CMB}yf_{\nu}\frac{\partial B_{\nu}}{\partial T}\bigg|_{ T = T_{\rm CMB}}
\end{equation}
where the last factor is the derivative of the blackbody spectrum with respect to temperature evaluated at the temperature of the CMB, which at present is $T_{\rm CMB} = 2.728$~K \citep{1996ApJ...473..576F}. The surface brightness per unit frequency of blackbody radiation is given (see e.g. \citealt{2007Spr...41}) by
\begin{equation}\label{e:sz3}
B_{\nu}(T) = \frac{2h_{\rm p}\nu^{3}}{c^{2}}\frac{1}{e^{h_{\rm p}\nu/k_{\rm B}T}-1},
\end{equation}
where $h_{\rm p}$ is the Planck constant and $k_{\rm B}$ is the Boltzmann constant. Hence the derivative is given by
\begin{equation}\label{e:sz4}
\frac{\partial B_{\nu}}{\partial T}\bigg|_{T = T_{\rm CMB}} = \frac{2h_{\rm p}^{2}\nu^{4}}{c^{2}k_{\rm B}T_{\rm CMB}^2}\frac{e^{h_{\rm p}\nu/k_{\rm B}T}}{(e^{h_{\rm p}\nu/k_{\rm B}T}-1)^{2}}.
\end{equation}
The function $f_{\nu}$ expresses the spectral dependence of the SZ signal and is derived from the Kompaneets equation \citep{1957JETP...4..730}. Relativistic treatments of $f_{\nu}$ have been considered in e.g. \citet{1995ARA&A..33..541R}, \citet{1998ApJ...502....7I}, \citet{1998ApJ...499....1C}, \citet{1998ApJ...508...17N}, and \citet{1998A&A...336...44P}, by incorporating relativistic terms into the Kompaneets equation. Relativistic effects may be important in clusters where the ICM temperatures are high. Indeed \citet{1994ApJ...436L..67A} and \citet{1996ApJ...456..437M} have shown that electrons in the ICM can reach energies above $10$~keV. Challinor \& Lasenby show that these effects lead to a small decrease in the SZ effect. However, Rephaeli argues that the non-relativistic treatment of Compton scattering adopted in \citet{1969Ap&SS...4..301Z} remains valid at frequencies well below the CMB peak value. For the observing frequencies of AMI ($\approx 15$~GHz), it can be assumed that this condition holds. Furthermore Rephaeli claims that for the unmodified Kompaneets equation to be valid, the optical depth of the cluster $\tau$, must be sufficiently large to justify using a diffusion approximation for the scattering process. It is clear that at AMI observing frequencies $h_{\rm p}\nu \ll m_{\rm e}c^{2}$ where $m_{\rm e}$ is the mass of an electron; and so the photons can be assumed to scatter in the Thomson limit. In this limit the scattering rate is $\propto \sigma_{\rm T}n_{\rm e}$ where $\sigma_{\rm T}$ is the Thomson scattering cross-section and $n_{e}$ is the electron number density in the ICM. Thus the optical depth is given by
\begin{equation}\label{e:sz5}
\tau = \int n_{\rm e}(r)\sigma_{\rm T} \mathrm{d}l,
\end{equation}
where $r$ is the radius from the galaxy cluster centre and the integral is along the line of sight. The non-relativistic form for $f_{\nu}$ is given by
\begin{equation}\label{e:sz6}
f_{\nu} = X\coth(X/2) - 4,
\end{equation}
where
\begin{equation}\label{e:sz7}
X = \frac{h_{\rm p}\nu}{k_{\rm B}T_{\rm CMB}}.
\end{equation}
Referring back to equation~\ref{e:sz2}, $y$ is the Comptonisation parameter which is the number of collisions multiplied by the mean fractional change in energy of the photons per collision, integrated along the line of sight. On average the electrons in the ICM transfer an energy $k_{\rm B} T_{\rm e}(r) / m_{\rm e}c^{2}$ to the scattered CMB photons, where $T_{\rm e}(r)$ is the temperature of an electron in the ICM. In the Thomson scattering regime described above this leads to
\begin{equation}\label{e:sz8}
y = \frac{\sigma_{\rm T}k_{\rm B}}{m_{\rm e}c^{2}}\int T_{\rm e}(r) n_{\rm e}(r)\,\mathrm{d}l.
\end{equation}
If the electron gas is assumed to be ideal, then in terms of the gas pressure $P_{e}(r)$, the Comptonisation parameter is given by 
\begin{equation}\label{e:sz9}
y = \frac{\sigma_{\rm T}}{m_{\rm e}c^{2}}\int P_{\rm e}(r)\,\mathrm{d}l.
\end{equation}
Combining equations~\ref{e:sz4},~\ref{e:sz6}, \&~\ref{e:sz9} one obtains the following expression for $\delta I_{\nu , \mathrm{cl}}$ in the non-relativistic limit
\begin{equation}\label{e:sz10}
\delta I_{\mathrm{cl}, \nu} = \frac{2\sigma_{\rm T}(k_{\rm B}T_{\rm CMB})^{3} X^{4}e^{X}}{h_{\rm p}^2c^{4}(e^{X}-1)^{2}}[X\coth(X/2) - 4]\int P_{\rm e}(r)\,\mathrm{d}l.
\end{equation}
Thus for a given cluster $\delta I_{\mathrm{cl}, \nu}$ is independent of $z$. Since the Fourier transform is a linear operator $\delta I_{\nu , \mathrm{cl}}$ can be substituted directly into equation~\ref{e:sz1} to calculate $\widetilde{\delta I}_{\nu , \mathrm{cl}}$.

\citet{1994ApJ...423...12B} noted that the total Comptonisation parameter $Y$, which is the integral of $y$ over the solid angle $d\Omega$ subtended by the galaxy cluster is proportional to the volume integral of the gas pressure. 
$Y$ can be written in terms of spherical coordinates as
\begin{equation}\label{e:sz11}
Y_{\rm sph, phys}(r) = \frac{\sigma_{\rm T}}{m_{\rm e}c^{2}}\int_{0}^{r}P_{\rm e}(r')4\pi r'^{2} \, \mathrm{d}r'.
\end{equation}  
Note that $Y_{\rm sph, phys}(r)$ has dimensions [length$^2$]. $Y_{\rm sph}(r) \equiv Y_{\rm sph, phys}(r) / D_{\rm A}^2$ (where $D_{\rm A}$ is the angular diameter distance to the cluster), which has dimensions [angle$^2$] and is the quantity referred to in this thesis unless stated otherwise. Thus $Y_{\rm sph, phys}$ measured out to large $r$ \textit{is}, with caveats, the total thermal energy of the cluster.


\section{Cluster model selection}\label{s:clus_model_selection}

To determine $\delta I_{\mathrm{cl}, \nu}$, one must select a model which calculates the electron temperature (equation~\ref{e:sz8}) or pressure (equation~\ref{e:sz9}) profile of a cluster. The AMI consortium has implemented a number of cluster models over the years. \citet{2003MNRAS.346..489M} considered the Navarro-Frenk-White (NFW) profile \citep{1995MNRAS.275..720N} as a cluster mass model; their model assumes spherical symmetry and hydrostatic equilibrium, and is used in a joint analysis between SZ and gravitational lensing data (see e.g. \citealt{1994A&A...289L...5S} for how lensing can be used to investigate cluster properties). Marshall also used the Beta model \citep{1976A&A....49..137C, 1978A&A....70..677C} to model the cluster gas profile; the Beta model is another spherically symmetric model, but is purely empirical. 
\citet{2009MNRAS.398.2049F} (from here on FF09) built on this work, but concentrated on modelling multi-frequency SZ data with the Beta model, but using the hydrostatic equilibrium assumption to derive an expression for the cluster mass. Most recently \citet{2012MNRAS.423.1534O} (MO12) presented a new, physical model to describe the baryonic matter as well the dark matter component in order to give a more thorough treatment of the make-up of galaxy clusters; I refer to this as a physical model. 


\section{A physical model for AMI data}\label{s:phys_mod}
\subsection{Model assumptions}
The model presented here is largely based on the one introduced in MO12 but includes the adaptions mentioned in Sections~\ref{s:phys_mod_calcs} and~\ref{s:bayes_ami}. 
For any model it is important to know the underlying assumptions which allow it to be valid. The four main assumptions in the physical model are as follows.
\begin{itemize}
\item The cluster is spherically symmetric. This means that the cluster can be parameterised in terms of the scalar radius $r$ (rather than its vector equivalent $\vec{r}$) from the centre of the cluster.
\item The cluster is in hydrostatic equilibrium up to radius $r_{200}$ (defined below). This means at any radius up to $r_{200}$ the outward pushing pressure force created by the pressure differential at that point must be equal to the gravitational binding force due to the mass enclosed within that radius (see e.g. \citealt{1977ApJ...213L..99B}, and equation~\ref{e:hse} below). 
\item The gas mass fraction $f_{\rm gas}(r)$ is much less than unity up to radius $r_{200}$, so that the total mass is $M(r_{200}) \approx M_{\rm dm}(r_{200})$. Consequently the total mass out to $r_{200}$ is given by the integral of the dark matter density along the radius of the cluster (see equation~\ref{e:nfw_m_tot_2} below). 
\item The cluster gas is assumed to be an ideal gas, so that the electron temperature can be trivially represented in terms of its pressure.
\end{itemize}
\subsection{Dark matter profile}
The model uses an NFW profile \citep{1995MNRAS.275..720N} the dark matter density as a function of cluster radius $r$,
\begin{equation}\label{e:nfw}
\rho_{\rm dm}(r) = \frac{\rho_{\rm s}}{\left(\frac{r}{r_{\rm s}}\right)\left(1+\frac{r}{r_{\rm s}}\right)^{2}},
\end{equation}
where $\rho_{\rm s}$ is an overall density normalisation coefficient and $r_{\rm s}$ is a characteristic radius defined by $r_{\rm s} \equiv r_{200}/c_{200}$ and is the radius at which the logarithmic slope of the profile $ \mathrm{d }\ln \rho(r) / \mathrm{d }\ln r$ is $-2$. $r_{200}$ is the radius at which the average cluster density is $200 \times \rho_{\rm crit}(z)$. $\rho_{\rm crit}(z)$ is the \textit{critical} density of the Universe at the cluster $z$ which is given by $\rho_{\rm crit}(z) = 3H(z)^{2}/8\pi G$ where $H(z)$ is the Hubble parameter (at the cluster redshift) and $G$ is Newton's constant. 
$c_{200}$ is the concentration parameter at this radius. Following \citet{2013MNRAS.430.1344O}, we can calculate $c_{200}$ for an NFW dark matter density profile taken from the expression in \citet{2009MNRAS.393.1235C}
\begin{equation}\label{e:nfw_c200}
c_{200} = \frac{5.26}{1+z} \left( \frac{M(r_{200})}{10^{14}h^{-1}M_{\mathrm{Sun}}} \right)^{-0.1},
\end{equation}
here, $M_{\mathrm{Sun}}$ denotes units of solar mass. The $1/(1+z)$ factor comes from \citet{2001astro.ph.11069W} and is obtained from N-body simulated dark matter halos between $z=0$ and $z=7$. The remainder of the relation was derived in \citet{2007MNRAS.381.1450N} by fitting a power-law for $c_{200}$ to N-body simulated cluster data. Note that the sample used in \citet{2007MNRAS.381.1450N} was assumed to contain clusters that are relaxed.
In equation~\ref{e:nfw_c200} $M(r_{200})$ is the mass enclosed at radius $r_{200}$. Thus for given values of $z$ and $M(r_{200})$, $c_{200}$ can be calculated. 
\subsection{Electron pressure profile}
Following \citet{2007ApJ...668....1N}, the generalised-NFW (GNFW) model is used to parameterise the electron pressure as a function of radius from the cluster centre
\begin{equation}\label{e:gnfw}
P_{\rm e}(r) = \frac{P_{\rm ei}}{\left(\frac{r}{r_{\rm p}}\right)^{c}\left(1+\left(\frac{r}{r_{\rm p}}\right)^{a}\right)^{(b-c)/a}},
\end{equation}
where $P_{\rm ei}$ is an overall pressure normalisation factor and $r_{\rm p}$ is another characteristic radius, defined by $r_{\rm p} \equiv r_{500}/c_{500}$. The parameters $a$, $b$ and $c$ describe the slope of the pressure profile at $ r / r_{\rm p} \approx 1$, $r / r_{\rm p} \gg 1 $ and $r / r_{\rm p} \ll 1$ respectively. For values $r / r_{\rm p} \ll 1$ the logarithmic slope ($ \mathrm{d} \ln P_{\mathrm{e}}(r) / \mathrm{d} \ln r $) converges to $-c$. For values For values $r / r_{\rm p} \gg 1$ the logarithmic slope converges to $-b$. The value of $a$ dictates how quickly (in terms of $r$) the slope switches between these two values, and in the limit that $a$ tends to zero, the logarithmic slope is $-(b+c)/2$ for all $r$. Note that \citet{2007ApJ...668....1N} choose to parameterise the pressure profile with the GNFW model because it closely matches the observed profiles of the Chandra X-ray clusters and results of numerical simulations in their outskirts. In addition to this, the gas pressure distribution is primarily determined by the gravitationally dominant dark matter component (which is fitted with the NFW profile), they argue that it makes sense to parameterise the pressure profile using the generalised NFW model. \\
Consistent with many of the Planck follow-up papers (see e.g. \citealt{2011A&A...536A..11P}) and with MO12 the slope parameters are taken to be $a = 1.0620$, $b=5.4807$ and $c = 0.3292$. These `universal' values are from \citet{2010A&A...517A..92A} and are the GNFW slope parameters derived for the standard self-similar case using scaling relations from a REXCESS sub-sample (of 20 well-studied low-$z$ clusters observed with XMM-Newton), as described in appendix B of the paper \citep{2007A&A...469..363B}. I also use the Arnaud et al. value for the concentration parameter $c_{500} \equiv r_{500} / r_{\rm p}$ of $1.156$. I note however that in \citet{2015A&A...580A..95P} (from here on YP15) using simulations it was shown that the disagreement between Planck and AMI parameter estimates may indicate pressure profiles deviating from the `universal' profile.
\subsection{Model calculations}
\label{s:phys_mod_calcs}
The three cluster model input parameters required to calculate the electron pressure given by equation~\ref{e:gnfw} in the physical model are $M(r_{200})$, $z$, and $f_{\rm gas}(r_{200})$. $f_{\rm gas}(r_{200})$ is the fraction of the total mass attributed to the gas mass up to radius $r_{200}$. Note that in general the total mass out to $r_{\Delta}$ is given by 
\begin{equation}\label{e:nfw_m_tot_1}
M(r_{\Delta}) = \frac{4\pi}{3} \Delta \rho_{\rm crit} (z) r_{\Delta}^{3}. 
\end{equation}
Hence $r_{200}$ can be calculated from $M(r_{200})$, and the mass can be determined at other (known) radii (e.g. $r_{500}$). 
\subsubsection{Total enclosed mass}
Another analytical solution for $M(r)$ can derived using the third assumption stated above. Using equation~\ref{e:nfw}, $M(r)$ is given by
\begin{equation}\label{e:nfw_m_tot_2}
\begin{aligned}
M(r) &= \int_{0}^{r} 4\pi\rho_{\rm dm}(r')r'^{2}\,\mathrm{d}r' \\
&= \int_{0}^{r} 4\pi\frac{\rho_{\rm s}r'^{2}}{\left(\frac{r'}{r_{\rm s}}\right)\left(1+\frac{r'}{r_{\rm s}}\right)^{2}}\,\mathrm{d}r' \\
&= 4\pi\rho_{\rm s}r_{\rm s}^{3}\left[\ln \left(1+\frac{r}{r_{\rm s}}\right)-\left(1+\frac{r_{\rm s}}{r}\right)^{-1}\right].
\end{aligned}
\end{equation}
Hence an expression for $\rho_{\rm s}$ can be obtained by equating~\ref{e:nfw_m_tot_1} and~\ref{e:nfw_m_tot_2}, setting $r = r_{200}$ and solving for $\rho_{\rm s}$
\begin{equation}\label{e:rhos}
\rho_{\rm s} = \frac{200}{3} \left(\frac{r_{200}}{r_{\rm s}}\right)^{3}\frac{\rho_{\rm crit}(z)}{\left[\ln \left(1+\frac{r_{200}}{r_{\rm s}}\right)-\left(1+\frac{r_{\rm s}}{r_{200}}\right)^{-1}\right]}.
\end{equation}
One can then obtain an expression $r_{500}$ as follows. Equating~\ref{e:nfw_m_tot_1} and~\ref{e:nfw_m_tot_2} at $r_{500}$ and substituting in the expression for $\rho_{\rm s}$ gives
\begin{equation}\label{e:r200r5001}
\left( \frac{r_{\rm s}}{r_{500}} \right)^{3} \left[ \ln \left( 1+\frac{r_{500}}{r_{\rm s}} \right) - \left( 1 + \frac{r_{\rm s}}{r_{500}} \right)^{-1} \right] = \frac{5}{2} \left( \frac{r_{\rm s}}{r_{200}} \right)^{3} 
\left[ \ln \left( 1+\frac{r_{200}}{r_{\rm s}} \right) - \left( 1 + \frac{r_{\rm s}}{r_{200}} \right)^{-1} \right].
\end{equation}
Following \citet{2003ApJ...584..702H}, there is an analytic mapping from $r_{200}$ to $r_{500}$. Consider the equation
\begin{equation}\label{e:r200r5002}
g(r_{\rm s}/r_{500}) = \frac{5}{2} g(r_{\rm s}/r_{200}),
\end{equation}
where
\begin{equation}\label{e:r200r5003}
g(x) = x^{3} [ \ln (1 + x^{-1}) - (1 + x)^{-1} ].
\end{equation}
Equation \ref{e:r200r5002} requires that $g(r_{\rm s}/r_{500})$ be inverted so that
\begin{equation}\label{e:r200r5004}
\frac{r_{\rm s}}{r_{500}} = x \left( g_{500}=\frac{5}{2}f(r_{\rm s}/r_{200}) \right),
\end{equation}
where
\begin{equation}\label{e:r200r5005}
x(g_{500}) = \left[ a_{1}g_{500}^{2p} + \frac{9}{16} \right] ^{-1/2} + 2g_{500}.
\end{equation}
Here $p = a_{2} + a_{3} \ln g_{500} + a_{4}(\ln g_{500})^{2}$, and the four fitting parameters correspond to $a_{1} = 0.5116$, $a_{2} = -0.4283$, $a_{3} = -3.13 \times 10^{-3}$ and $a_{4} = -3.52 \times 10^{-5}$. This gives a fit to better than 0.3\% accuracy for $ 0 < c_{200} < 20 $ and is exact in the limit that $ c_{200} \rightarrow 0$. Once $r_{500}$ has been calculated $r_{\rm p}$ can be calculated from  $r_{\rm p} = r_{500}/c_{500}$.
\subsubsection{Hydrostatic equilibrium}
This requires
\begin{equation}\label{e:hse}
\frac{\mathrm{d}P_{\rm g}(r)}{\mathrm{d}r} = -\rho_{\rm g}(r)\frac{GM_{\rm tot}(r)}{r^{2}},
\end{equation}
where $\rho_{\rm g}(r)$ is the gas density and $M(r)$ is the total mass within radius $r$ of the cluster. The gas pressure $P_{\rm g}(r)$ can be related to the electron pressure as
\begin{equation}\label{e:pgas_pelec}
\mu_{\rm g} P_{\rm g}(r) = \mu_{e} P_{\rm e}(r),
\end{equation}
where $\mu_{\rm e}$ is the mean gas mass per electron and $\mu_{\rm g}$ is the mean mass per gas particle. \citet{2000ApJ...540..614M} state that for a plasma with the cosmic helium mass fraction $C_{\rm{He}} = 0.24$ and the solar abundance values in \citet{1989GeCoA..53..197A}, then $\mu_{\rm e} = 1.146$ and $\mu_{\rm g} = 0.592$ in units of proton mass.
\subsubsection{Gas density, mass, and temperature}
Substituting equations~\ref{e:nfw_m_tot_2} and~\ref{e:pgas_pelec} into~\ref{e:hse} and solving for $\rho_{\rm g}(r)$ gives
\begin{equation}
\label{e:nfw_rhog} 
\begin{aligned}
\rho_{\rm g}(r) = \, & \frac{\mu_{\rm e}P_{\rm ei}}{\mu_{\rm g}}\frac{1}{4\pi G \rho_{\rm s}r_{\rm s}^{3}} \\
& \times \frac{r}{\ln \left(1+\frac{r}{r_{\rm s}}\right)-\left(1+\frac{r_{\rm s}}{r}\right)^{-1}}  \\
& \times \left(\frac{r}{r_{\rm p}}\right)^{-c}\left[1+\left(\frac{r}{r_{\rm p}}\right)^{a}\right]^{-(1+(b-c)/a)}\left[b\left(\frac{r}{r_{\rm p}}\right)^{a}+c\right].
\end{aligned}
\end{equation}
From this the gas mass $M_{\rm g}(r)$ can be calculated
\begin{equation}
\label{e:nfw_mgas}
M_{\rm g}(r) = \int_{0}^{r} 4\pi\rho_{\rm g}(r')r'^{2}\,\mathrm{d}r'.
\end{equation}
Note however that this integral must be solved numerically. Nevertheless, we can determine $P_{\rm ei}$ since we know $M(r_{200})$, $f_{\rm gas}(r_{200})$ and $r_{200}$ ($M_{\rm g}(r_{200}) = f_{\rm gas}(r_{200}) M(r_{200})$. Evaluating equations~\ref{e:nfw_rhog} and~\ref{e:nfw_mgas} at $r_{200}$ and solving for $P_{\rm ei}$ gives the following expression
\begin{equation}
\label{e:nfw_pei}
\begin{split}
 P_{\rm ei} = \, & \frac{\mu_{\rm g}}{\mu_{\rm e}}G\rho_{\rm s}r_{\rm s}^{3} M_{\rm g}(r_{200}) \,  \\
& \times \frac{1}{ \displaystyle{\int_{0}^{r_{200}}}\frac{r'^{3}}{\ln \left( 1+\frac{r'}{r_{\rm s}}\right) - \left( 1+\frac{r_{\rm s}}{r'} \right) ^{-1}} \left( \frac{r'}{r_{\rm p}} \right) ^{-c} \left[ 1+ \left( \frac{r'}{r_{\rm p}} \right) ^{a} \right] ^{-(1+(b-c)/a)}\left[ b \left( \frac{r'}{r_{\rm p}} \right) ^{a}+c \right] \,\mathrm{d}r'}.
\end{split}
\end{equation}

The radial profile of the electron number density is given by $n_{\rm e}(r) = \rho_{\rm g}(r) / \mu_{\rm e}$. Assuming an ideal gas equation of state, the electron temperature $T_{\rm e}(r)$ is therefore given by 
\begin{equation}
\label{e:nfw_tgas}
\begin{split}
T_{\rm e}(r) = & \left(\frac{4\pi \mu_{\rm g} G\rho_{\rm s}r_{\rm s}^{3}}{k_{\rm B}}\right) \\
                   & \times \frac{\ln \left( 1 + \frac{r}{r_{\rm s}} \right) - \left( 1 + \frac{r_{\rm s}}{r} \right)^{-1}}{r} \\ 
                   & \times \left [1 + \left(\frac{r}{r_{\rm p}}\right)^{a} \right]\left[b \left(\frac{r}{r_{\rm p}}\right)^{a} + c \right]^{-1},
\end{split}
\end{equation}
which is also equal to the gas temperature $T_{\rm g}(r)$. 

The gas mass can be determined numerically from equation~\ref{e:nfw_mgas} as

\begin{equation}
\label{e:nfw_mgas2} 
\begin{aligned}
M_{\rm g}(r) =& \frac{\mu_{\rm e}P_{\rm ei}}{\mu_{\rm g}}\frac{1}{ G \rho_{\rm s}r_{\rm s}^{3}} \\
& \times \bigintss_{0}^{r} \frac{r'^{3}}{\ln \left(1+\frac{r'}{r_{\rm s}}\right)-\left(1+\frac{r_{\rm s}}{r'}\right)^{-1}}  \\
& \times \left(\frac{r'}{r_{\rm p}}\right)^{-c}\left[1+\left(\frac{r'}{r_{\rm p}}\right)^{a}\right]^{-(1+(b-c)/a)}\left[b\left(\frac{r'}{r_{\rm p}}\right)^{a}+c\right] \mathrm{d}r'.
\end{aligned}
\end{equation}
\subsubsection{Determining $\widetilde{\delta I}_{\mathrm{cl}, \nu }$}
Once $r_{\rm p}$ and $P_{\rm ei}$ have been calculated, the pressure profile can be used in equation~\ref{e:sz9} to calculate the Comptonisation parameter which in turn can be used to calculate $\delta I_{\mathrm{cl}, \nu}$ using equation~\ref{e:sz2}. $\delta I_{\mathrm{cl}, \nu}$ can be Fourier transformed to get the quantity comparable to what an interferometer measures, so that the physical model can be used to analyse data obtained with AMI.


\section{Recognised radio-sources and general noise contributions}\label{s:sources_noise}

In addition to the SZ decrement, visibilities measured by AMI also contain contributions from radio-sources, primordial CMB anisotropies, and instrumental noise. As defined in \citet{2002MNRAS.334..569H}, each visibility measured by an interferometer consists of two components
\begin{equation}\label{e:vis}
V_{\nu}(\vec{u_{i}}) = \tilde{I}_{\nu}(\vec{u_{i}}) + N_{\nu}(\vec{u_{i}}),
\end{equation}
where $\tilde{I}_{\nu}(\vec{u_{i}})$ contains both the contribution from the cluster SZ effect and from the \textit{identified} radio-sources, and $N_{\nu}(\vec{u_{i}})$ contains the contributions from \textit{unidentified} radio-sources, primordial CMB and instrumental noise.


\subsection{Recognised radio-sources}\label{s:identified_sources} 

The LA has been (see e.g. \citealt{2011MNRAS.415.2699A}) and is being used to measure the 15.7-GHz source count.
The LA is used to measure radio-sources (without contamination from the SZ effect since the cluster is resolved out), whilst the SA simultaneously measures the combined SZ and source signals.

The visibility of each recognised radio-source, assuming for illustration that it is unresolved by the LA, is
\begin{equation}\label{e:source_vis}
\begin{aligned}
\tilde{I}_{\mathrm{rs},\nu}(\vec{u}) & = \int A_{\nu}(\vec{x})S_{\nu}(\vec{x})\delta(\vec{x}_{\rm rs})e^{(2\pi i \vec{u}\cdot\vec{x})}\,\mathrm{d}^{2}\vec{x} & = S_{\nu}(\vec{x}_{\rm rs})A_{\nu}(\vec{x}_{\rm rs})e^{i\phi},
\end{aligned}
\end{equation}
where $S_{\nu}(\vec{x})$ is the source flux density at point $\vec{x}$ relative to the phase centre, $\phi = 2 \pi \vec{u} \cdot \vec{x}_{\rm rs}$ 
The variation in source flux density across the AMI observing band is taken account of via the spectral index $\alpha$, where
\begin{equation}\label{e:spec_index}
S_{\nu} = S_{0} \left(\frac{\nu}{\nu_{0}}\right)^{-\alpha},
\end{equation}
where $\nu_{0}$ is some reference frequency and $S_{0}$ is the corresponding source flux density.


\subsection{General Noise Contributions} \label{s:gen_noise}


\subsubsection{Instrumental noise} \label{s:inst_noise}
The main source of instrumental noise is Johnson noise. This refers to the thermal agitation of the charge carriers in any circuit \citep{1928PhRv...32..110N}, and in the context of interferometry, the front-end receivers of the antennas. The antennas are cooled to mitigate this effect, but the remaining contribution is non-negligible. For a given bandwidth $\Delta\nu$ , the root mean square of the Johnson noise voltage from a single antenna is given by (see e.g. \citealt{thompson}) 
\begin{equation}
\label{e:john_noise}
\sigma_{\rm Johnson} = \sqrt{4k_{\rm B}T_{\rm sys}R\Delta\nu},
\end{equation}
where $T_{\rm sys}$ is the system temperature and $R$ is the antenna impedance. Note that when limited to a finite bandwidth, Johnson noise is approximately Gaussian (see e.g. \citealt{Barry2004}). 


\subsubsection{Primordial CMB} \label{s:prim_cmb}

Anisotropies in the temperature of the CMB were predicted as early as \citet{1967Natur.215.1155S} among others, and \citet{1992ApJ...396L...1S} provided the first clear statistical evidence of their existence and \citet{1994Natur.367..333H} provided the first direct evidence of individual spatial structures in the CMB. These anisotropies can be separated into two categories: primordial and late time anisotropies. An example of the latter type is the SZ effect. Primordial anisotropies refer to fluctuations in the CMB that have been present since the surface of last scattering (which occurred at $z \approx 1100$ or $t \approx 4 \times 10^5$~years over a period of $\Delta z \approx 60$). On angular scales visible from the ground the acoustic peaks and troughs are the most significant features in the CMB power spectrum. 
When the Universe was radiation dominated, non-baryonic dark matter began to collapse under gravity to form potential wells\footnote{This only applies to matter that was in causal contact.}, but baryonic matter could not clump due to pressure opposition from Thompson scattering of photons by electrons given that there were $10^9$ photons per baryon. During recombination the acoustic oscillations imprint the CMB, after recombination the atoms fall into the non-baryonic dark matter potential wells.
Acoustic peaks and troughs relate to the waves oscillating in the baryon-photon plasma before recombination occurred. 
Each successive peak refers to the number of times the wave compressed before the radiation-matter decoupling, and is visible at decreasing angular scale. 
In this work, the power spectrum for CMB primordial anisotropies is determined via maximum-likelihood methods as written in \citet{2002MNRAS.334..569H} using the results from \citet{2013ApJS..208...19H}.


\subsubsection{Background unrecognised radio-sources} \label{s:conf_noise}
Although the LA is used to identify radio point sources with flux densities $\geq S_{\rm lim}$ (where $S_{\rm lim}$ is a limiting flux density that is usually taken as $4\times \sigma$ and $\sigma$ is the resultant RMS noise in the summed LA data on the particular sky patch), a large enough number of sources with flux densities $< S_{\rm lim}$ can be a significant contaminant. This type of noise is often referred to as source confusion. 
\citet{1957PCPS...53..764S} showed that if such sources obey a power-law number-flux density relation ($n_{\nu}(S) = \mathrm{d}N_{\nu}(>S)/\mathrm{d}S \propto kS^{\gamma}$ where $k$ \& $\gamma$ are dimensionless constants), then for a random distribution of unresolved radio-sources in the sky, the source confusion noise is given by 
\begin{equation}\label{e:conf_noise}
\sigma_{\mathrm{conf}}^{2} = \int_{0}^{S_{\rm lim}}S^{2}n_{\nu}(S)\,\mathrm{d}S.
\end{equation}
$\gamma$ and $k$ were determined empirically in \citet{2011MNRAS.415.2708A} from the 10C survey to be $\gamma = -1.80$ and $k = 376$ when $n_{\nu}(S)$ is quoted in units of Jy$^{-1}$~sr$^{-1}$, so that when $S_{\rm lim}$ is taken to be $300~\mu$Jy (for a standard length AMI cluster observation) $\sigma_{\mathrm{conf}}^2 = 0.185$~Jy$^2$sr$^{-1}$.  


\section{Bayesian inference}
\label{s:bayes_inf}


\subsection{Parameter estimation}
\label{s:param_est}
Given a model $\mathcal{M}$ and a data vector $\vec{\mathcal{D}}$, one can obtain model parameters (also known as input parameters or sampling parameters) $\vec{\Theta}$ conditioned on $\mathcal{M}$ and $\vec{\mathcal{D}}$ using Bayes' theorem:
\begin{equation}\label{e:bayes}
P\left(\vec{\Theta}|\vec{\mathcal{D}},\mathcal{M}\right) = \frac{P\left(\vec{\mathcal{D}}|\vec{\Theta},\mathcal{M}\right)P\left(\vec{\Theta}|\mathcal{M}\right)}{P\left(\vec{\mathcal{D}}|\mathcal{M}\right)},
\end{equation}
where $P\left(\vec{\Theta}|\vec{\mathcal{D}},\mathcal{M}\right) \equiv \mathcal{P}\left(\vec{\Theta}\right)$ is the posterior distribution of the input parameter set, $P\left(\vec{\mathcal{D}}|\vec{\Theta},\mathcal{M}\right) \equiv \mathcal{L}\left(\vec{\Theta}\right)$ is the likelihood function for the data, $P\left(\vec{\Theta}|\mathcal{M}\right) \equiv \pi\left(\vec{\Theta}\right)$ is the prior probability distribution for the model parameter set, and $P\left(\vec{\mathcal{D}}|\mathcal{M}\right) \equiv \mathcal{Z}\left(\vec{\mathcal{D}}\right)$ is the Bayesian evidence of the data. The evidence can be defined as the factor required to normalise the posterior over the sampling parameter space:
\begin{equation}\label{e:evidence}
\mathcal{Z}\left(\vec{\mathcal{D}}\right) = \int \mathcal{L}\left(\vec{\Theta}\right) \pi\left(\vec{\Theta}\right)\, \mathrm{d}\vec{\Theta},
\end{equation} 
where the integral is carried out over the $N$-dimensional parameter space. For the models using AMI data considered here, the input parameters can be split into two subsets, (which are assumed to be independent of one another): cluster parameters $\vec{\Theta}_{\rm cl}$ and radio-source or `nuisance' parameters $\vec{\Theta}_{\rm rs}$.


\subsection{Model comparison}
\label{s:bayes_model}
While it is the posterior distribution which gives the model parameter estimates from the prior information and data, it is $\mathcal{Z}\left(\vec{\mathcal{D}}\right)$ which is crucial to performing model selection. The nested sampling algorithm, \textsc{MultiNest} \citep{2009MNRAS.398.1601F} is a Monte Carlo algorithm which calculates $\mathcal{Z}\left(\vec{\mathcal{D}}\right)$ by making use of a transformation of the $N$-dimensional evidence integral into a one-dimensional integral that is much easier to evaluate. The algorithm also produces samples from $\mathcal{P}\left(\vec{\Theta}\right)$ as a by-product, meaning that it is suitable for both the parameter estimation and model comparison aspects of this work. Nested sampling will be discussed in more detail in Section~\ref{s:ns}.
Comparing models in a Bayesian way can be done by considering the following. The probability of a model $\mathcal{M}$ conditioned on $\vec{\mathcal{D}}$ can be calculated using Bayes' theorem
\begin{equation}\label{e:bayesmodel}
P\left(\mathcal{M}|\vec{\mathcal{D}}\right) = \frac{P\left(\vec{\mathcal{D}}|\mathcal{M}\right)P\left(\mathcal{M}\right)}{P\left(\vec{\mathcal{D}}\right)}.
\end{equation}
Hence for two models, $\mathcal{M}_{1}$ and $\mathcal{M}_{2}$, the ratio of the models conditioned on the same dataset is given by 
\begin{equation}\label{eqn:bayesmodelcomparison}
\frac{P\left(\mathcal{M}_{1}|\vec{\mathcal{D}}\right)}{P\left(\mathcal{M}_{2}|\vec{\mathcal{D}}\right)} = \frac{P\left(\vec{\mathcal{D}}|\mathcal{M}_{1}\right)P\left(\mathcal{M}_{1}\right)}{P\left(\vec{\mathcal{D}}|\mathcal{M}_{2}\right)P\left(\mathcal{M}_{2}\right)},
\end{equation}
where $P(\mathcal{M}_{2}) / P(\mathcal{M}_{1})$ is the a-priori probability ratio of the models. We set this to one, i.e. we place no bias towards a particular model before performing the analysis. Hence the ratio of the probabilities of the models given the data is equal to the ratio of the evidence values obtained from the respective models (we have defined $\mathcal{Z}_{i}(\mathcal{D}) \equiv P\left(\vec{\mathcal{D}}|\mathcal{M}_{i}\right)$). 
The evidence is simply the average of the likelihood function over the sampling parameter space, weighted by the prior distribution. This means that the evidence is larger for a model with larger areas in its parameter space having higher likelihood values. Moreover, a larger parameter space, either in the form of higher dimensionality or a larger domain, results in a lower evidence value, all other things being equal. Hence the evidence penalises more complex models over basic (lower dimensionality / smaller input parameter space domains) ones which give an equally good fit to the data. Thus the evidence automatically implements Occam's razor: when you have two competing theories that make exactly the same predictions, the simpler one is the better. \citet{jeffreys} provides a scale for interpreting the ratio of evidences as a means of performing model comparison (Table~\ref{t:jeffreys}). A value of $\ln (\mathcal{Z}_{1} / \mathcal{Z}_{2})$ above $5.0$ (less than $-5.0$) presents "strong evidence" in favour of model 1 (model 2). Values $ 2.5 \leq \ln (\mathcal{Z}_{1} / \mathcal{Z}_{2}) < 5.0$ ($ -5.0 < \ln (\mathcal{Z}_{1} / \mathcal{Z}_{2}) \leq -2.5 $) present "moderate evidence" in favour of model 1 (model 2). Values $ 1 \leq \ln (\mathcal{Z}_{1} / \mathcal{Z}_{2}) < 2.5$ ($ -2.5 < \ln (\mathcal{Z}_{1} / \mathcal{Z}_{2}) \leq -1 $) present "weak evidence" in favour of model 1 (model 2). Finally, values $ -1.0 < \ln (\mathcal{Z}_{1} / \mathcal{Z}_{2}) < 1.0 $ require "more information to come to a conclusion" over model preference. \\


\begin{table*}
\centering
\begin{tabular}{{l}{c}{c}}
\hline
$\ln (\mathcal{Z}_{1} / \mathcal{Z}_{2})$  & Interpretation & Probability of favoured model \\ 
\hline
$\leq 1.0$ & better data are needed & $\leq 0.75$ \\
$\leq 2.5 $ & weak evidence in favour of $\mathcal{M}_{1}$ & $0.923$ \\
$\leq 5.0$ & moderate evidence in favour of $\mathcal{M}_{1}$ & $0.993$ \\
$ > 5.0$ & strong evidence in favour of $\mathcal{M}_{1}$ & $ > 0.993 $ \\
\hline
\end{tabular}
\caption{Jeffreys scale for assessing model preferability based on the $\ln \equiv \log_e$ of the evidence ratio of two models.}\label{t:jeffreys}

\end{table*}

\section{Parameter prior distributions}
\label{s:priors}
Prior distributions incorporate the prior knowledge we have on the sampling parameters used in Bayesian inference. The prior parameter space for AMI cluster analysis consists both of parameters associated with the cluster $\pi(\vec{\Theta}_{\rm cl})$ and those associated with each identified radio-source $\pi(\vec{\Theta}_{\rm rs})$. If one assumes that the cluster parameters are separable from those associated with each recognised radio-source, then the total prior distribution is given by
\begin{equation}\label{e:priors}
\pi(\vec{\Theta}_{\rm t}) = \pi(\vec{\Theta}_{cl})\prod\limits_{i} \pi(\vec{\Theta}_{\mathrm{rs},i}),
\end{equation} 
where $i$ labels each recognised radio-source.
The prior distributions assigned to the cluster parameters will be discussed in the Sections where the Bayesian analyses carried out are introduced (i.e. Sections~\ref{s:bayes_ami},~\ref{s:obs_mod},~\ref{s:ein_priors}, and~\ref{s:ap_obs}).


\subsection{Radio-source prior distributions}
\label{s:rs_priors}
\begin{figure}
\centerline{\includegraphics[width=\linewidth]{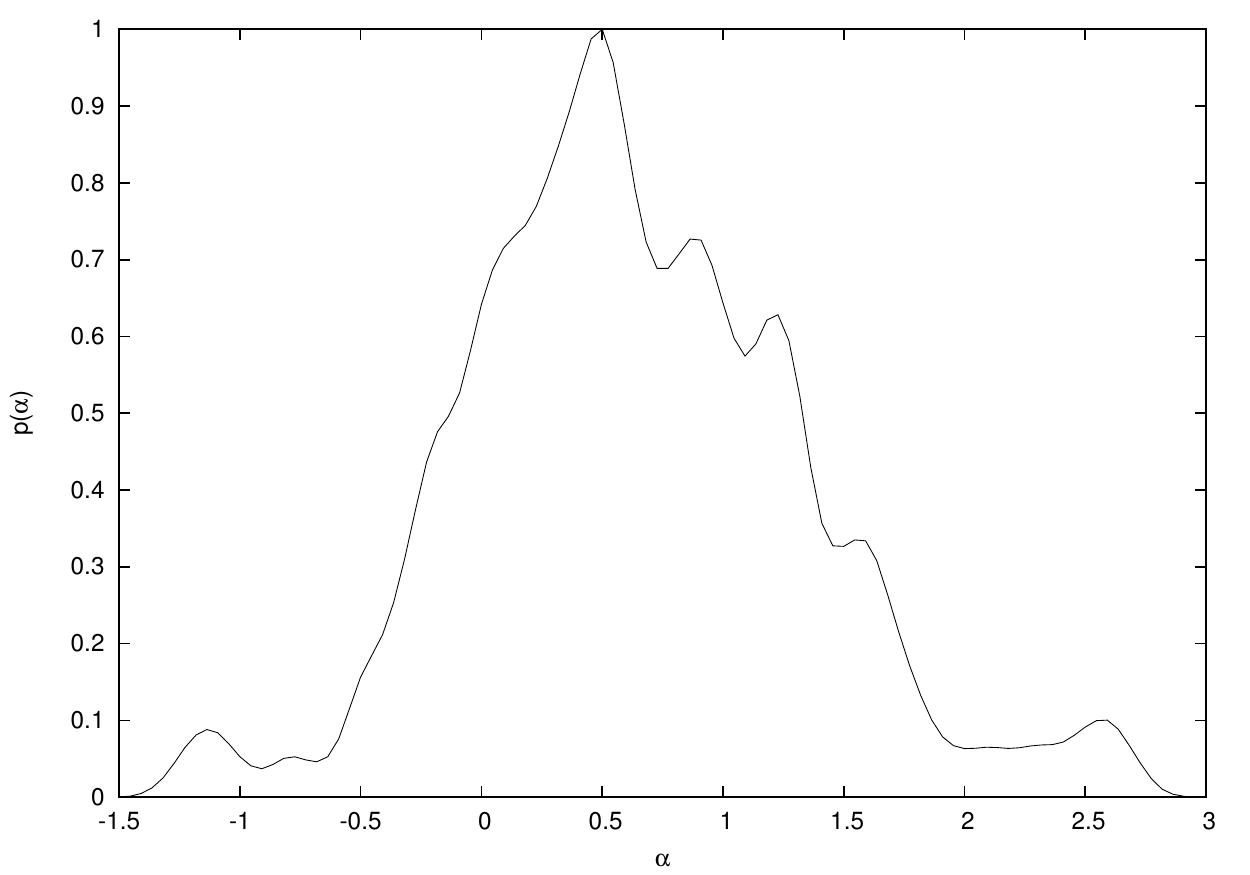}}
\caption{Spectral index distribution adapted from \citet{2007mru..confE.140W} from the 9C survey of radio-sources.}\label{f:waldram}
\end{figure}

Following FF09, each source can be parameterised by four variables: its position on the sky ($x_{\rm rs}$, $y_{\rm rs}$), its measured flux density at some reference frequency $\nu_{0}$, $S_{\rm rs, 0}$, and its spectral index $\alpha_{\rm rs}$. Assuming these are independent, then for source $i$ 
\begin{equation}\label{e:rs_priors}
\pi(\vec{\Theta}_{\mathrm{rs, } i}) = \pi(x_{\mathrm{rs, } i})\pi(y_{\mathrm{rs, } i})\pi(S_{\mathrm{rs, }, 0, i})\pi(\alpha_{\mathrm{rs, } i}).
\end{equation} 
Delta functions are applied to the prior distributions on $x_{\rm rs}$ and $y_{\rm rs}$, due to the LA's ability to measure spatial positions to high accuracy: $\pi(x_{\rm rs}) = \delta(x_{\rm rs, \, LA})$, $\pi(y_{\rm rs}) = \delta(y_{\rm rs, \, LA})$. Delta priors were also set on $S_{\mathrm{rs}, 0}$ \& $\alpha_{\rm rs}$ (centred on the values measured by the LA), if the measured $S_{\mathrm{rs}, 0}$ was less than four times the instrumental noise associated with the observation, and the source was more than 5 arcminutes away from the SA pointing centre: $\pi(S_{\mathrm{rs}, 0}) = \delta(S_{\mathrm{rs}, 0, \, \mathrm{LA}})$, $\pi(\alpha_{\rm rs}) = \delta(\alpha_{\rm rs,  \, LA})$. Otherwise, a Gaussian prior was set on $S_{\mathrm{rs}, 0}$ centred at the LA measured value with a standard deviation equal to $40\%$ of the measured value ($\sigma_{\mathrm{rs}, 0} = 0.4 \times S_{\mathrm{rs}, 0 \,, \mathrm{LA}}$): $\pi(S_{\mathrm{rs}, 0}) = \mathcal{N}(S_{\mathrm{rs}, 0, \, \mathrm{LA}}, \sigma_{\mathrm{rs}, 0})$. The spectral index $\alpha_{\rm rs}$ was modelled using the empirical distribution determined in \citet{2007mru..confE.140W}: $\pi(\alpha_{\rm rs}) = \mathcal{W}(\alpha_{\rm rs})$ and is shown in Figure~\ref{f:waldram}.

\section{The likelihood function}\label{s:likelihood}

The likelihood function gives the probability of observing data given a set of parameter values. In the case of AMI observations, the data are visibilities observed by AMI and the parameters are those described in the previous Section. Following \citet{2002MNRAS.334..569H} and FF09, it is convenient first to place the $N_{\mathrm{vis},\nu}$ observed complex visibilities $V_{\nu}(\vec{u}_{i})$ into a data vector $d_{\nu}$ for each frequency channel (six channels in the case of the analogue correlator AMI data), ordered such that
\begin{equation}\label{e:data_arrays} 
\vec{d}_{\nu,i} = \left\{ 
\begin{array}{ll}
\mathrm{Re}[V_{\nu}(\vec{u}_i)] & \mbox{$(i \leq N_{\mathrm{vis},\nu})$} \\
 & \\
\mathrm{Im}[V_{\nu}(\vec{u}_{i-N_{\mathrm{vis},\nu}})] & \mbox{$(N_{\mathrm{vis},\nu}+1 \leq i \leq 2N_{\mathrm{vis},\nu})$}.
\end{array}
\right.
\end{equation} 
Similarly, one can define the noise vectors $\vec{n}_{\nu}$ containing only the contributions to the noise components $N_{\nu}(\vec{u}_{i})$. Section~\ref{s:gen_noise} explains the three contributors to $N_{\nu}(\vec{u}_{i})$. We take the likelihood to be Gaussian
\begin{equation}\label{e:lhood} 
\mathcal{L}(\vec{\Theta}) = \frac{1}{Z_{N}}e^{-\frac{1}{2}\chi^{2}}.
\end{equation}
Here $\chi^{2}$ is a measure of the goodness-of-fit of the model to the data (which is simply the concatenation of data vectors $d_{\nu}$ for all  $\nu$)  $\vec{d}$ and the predicted data $\vec{d}^{\rm p}(\vec{\Theta})$:
\begin{equation}\label{e:chisq} 
\chi^{2} = \sum\limits_{\nu,\nu'} (\vec{d}_{\nu} - \vec{d}_{\nu}^{\rm p}(\vec{\Theta}))^{\rm T} \boldsymbol{\boldsymbol{\mathsf{C}}}_{\nu,\nu'}^{-1} (\vec{d}_{\nu'} - \vec{d}_{\nu'}^{\rm p}(\vec{\Theta})).
\end{equation} 
$\vec{d}_{\nu}^{\rm p}(\vec{\Theta})$ is assumed to consist of the signal measured from the cluster and \textit{recognised} radio-sources. $\boldsymbol{\mathsf{C}}_{\nu,\nu'} \equiv \langle \vec{n}_{\nu}\vec{n}^{\rm T}_{\nu'} \rangle$ is the covariance matrix of the visibilities. Assuming instrumental (Section~\ref{s:inst_noise}), CMB (Section~\ref{s:prim_cmb}), and confusion (Section~\ref{s:conf_noise}) noise are independent of each other, $\boldsymbol{\mathsf{C}}_{\nu,\nu'}$ can be written as
\begin{equation}\label{e:covariance}
\boldsymbol{\mathsf{C}}_{\nu,\nu'} = \boldsymbol{\mathsf{C}}^{\rm ins}_{\nu,\nu'} + \boldsymbol{\mathsf{C}}^{\rm CMB}_{\nu,\nu'} + \boldsymbol{\mathsf{C}}^{\rm conf}_{\nu,\nu'}.
\end{equation}
Note that the instrumental noise associated with AMI observations is measured, and so does not need to be predicted. For further information on all three sources of noise, see FF09 Section~5.3 and \citet{2002MNRAS.334..569H}. 
$Z_{N}$ is a normalisation factor given by
\begin{equation}\label{e:lhood_norm}
Z_{N} = (2\pi)^{N_{\rm vis}}|\boldsymbol{\mathsf{C}}|^{\frac{1}{2}},
\end{equation}
where $N_{\rm vis}$ is the total number of visibilities observed over all six frequency channels.


%% file: CHAP-3/chapter3.tex
\chapter{Physical modelling of clusters detected by Planck}\label{c:third} 

YP15 present the results of the AMI follow-up of clusters detected by Planck-- this follow-up is analysed using the `observational model', which parameterises a cluster in terms of its integrated Comptonisation parameter $Y$ and angular scale $\theta$. YP15 find that these AMI estimates for $Y$ are consistently lower than the values obtained from Planck data, and conclude that this may indicate that the cluster pressure profiles are deviating from the `universal' one. 
I use the physical model described in Section~\ref{s:phys_mod} with data obtained from AMI of clusters detected by Planck (including ones which were detected after the analysis in YP15 was carried out). 
I also consider the cluster mass estimates given in the PSZ2 Planck cluster catalogue \citep{2016A&A...594A..27P} and compare them with the values obtained using AMI data. 
Furthermore I use the PSZ2 mass estimates as inputs to simulations which are then analysed in the same way as real AMI observations. The work discussed in this Chapter has been published in MNRAS \citep{2019MNRAS.483.3529J}, and has been modified post-referee comments.


\section{Selection and observation of the cluster sample}
\label{s:Planck_samp}
PSZ2 contains 1653 cluster candidates detected in the all-sky 29 month mission. The initial cluster selection criteria for AMI closely resembles that described in YP15, with a few modifications as follows. 
\begin{itemize}
\item The lower $z$ limit $ z \leq 0.100 $ was relaxed here, to see how well AMI data can constrain physical model parameters at low redshift. However it is important to realise that the sample at $z \leq 0.100$ were not observed specifically for the purpose of this work, but were part of other observation projects. 
\item The Planck signal-to-noise ratio (S/N) lower bound was reduced to $4.5$.
\item The automatic radio-source environment rejection remained the same. However the manual rejection was done on a map-by-map basis-- see Section~\ref{s:planck_phys_results_i}.
\item Note that the observation declination limits $20^{\circ} < \delta < 87^{\circ}$ were kept. 
\end{itemize}
This led to an initial sample size of 199 clusters, 
The maximum and minimum values of some key parameters for this sample from the Planck catalogue are given in Table~\ref{t:initial_sample}. Note that $M_{\rm{SZ}}$ is taken in PSZ2 as the hydrostatic equilibrium mass $M(r_{500})$, assuming the best-fit $Y-M$ relation.
\begin{table}
\centering
\begin{tabular}{{l}{c}{c}}
\hline
Parameter & Minimum value & Maximum value \\
\hline 
Declination & $20.31^{\circ}$ & $86.24^{\circ}$ \\
$z$ & $0.045$ & $0.83$ \\
S/N & 4.50 & 28.40 \\
$M_{\rm{SZ}}$~($\times 10^{14}~M_{\mathrm{Sun}}$) & $1.83$ & $10.80$ \\
\hline
\end{tabular}
\caption{Minimum and maximum values for a selection of parameters taken from PSZ2 for the AMI sample of 199 clusters.}\label{t:initial_sample}
\end{table}

The pointing strategy for each cluster was as follows. Clusters were observed using a single pointing centre on the SA, which has a primary beam of size $\approx 20~$arcmin FWHM, to noise levels of $\lessapprox 120~\mu \rm{Jy}~\rm{beam}^{-1}$. To cover the same area with the LA, which has a primary beam of size $\approx $ 6~arcmin FWHM, the cluster field was observed as a 61-point hexagonal raster. The noise level of the raster was $\lessapprox 100~\mu \rm{Jy}~\rm{beam}^{-1}$ in the central 19 pointings, and slightly higher in the outer regions. The observations for a given cluster field were carried out simultaneously on both arrays, and the average observation time per cluster was $\approx 30~$ hours.
The observations were carried out between 2013 and 2015, and so they began before the PSZ2 catalogue was published. This means that the AMI pointing centre coordinates in general were not the same as those published in the final Planck catalogue which was released in 2015. This is discussed in the context of the cluster centre offset parameters in Section~\ref{s:bayes_ami}. 
Data from both arrays were flagged for interference and calibrated using the AMI in-house software package \textsc{REDUCE}. Flux calibration was applied using contemporaneous observations of the primary calibration sources 3C 286, 3C 48, and 3C 147. The assumed flux densities for 3C 286 were converted from Very Large Array total-intensity measurements \citep{2013ApJS..204...19P} and are consistent with the \citet{1987Icar...71..159R} model of Mars transferred onto an absolute scale, using results from the Wilkinson Microwave Anisotropy Probe. The assumed flux densities for 3C 48 and 3C 147 were based on long-term monitoring with the SA using 3C 286 for flux calibration. Phase calibration was applied using interleaved observations of a nearby bright source selected from the VLBA Calibrator survey \citep{2008AJ....136..580P}; in the case of the LA, a secondary amplitude calibration was also applied using contemporaneous observations of the phase calibration source on the SA.


\section{AMI data analysis} \label{s:bayes_ami}

The likelihood function given by equation~\ref{e:lhood}, along with all the preceeding calculational steps covered in Chapter~\ref{c:second} are calculated using our AMI Bayesian data analysis pipeline, \textsc{McAdam}. 
Referring back to the prior distributions defined in Section~\ref{s:priors}, the cluster sampling parameters for the physical model are 
\begin{equation}\label{e:clus_priors}
\pi(\vec{\Theta}_{\rm cl}) = \pi(M(r_{200}))\pi(f_{\rm gas}(r_{200}))\pi(z)\pi(x_{\rm c})\pi(y_{\rm c}).
\end{equation} 
$x_{\rm c}$ and $y_{\rm c}$ are the cluster centre offsets from the SA pointing centre, measured in arcseconds. The prior distributions assigned to the cluster parameters are the same as the ones used in \citet{2013MNRAS.430.1344O}, but with an alteration to the mass limits. Upon running \textsc{McAdam} on data from a few of the Planck clusters, it was found that the posterior distributions of $M(r_{200})$ were hitting the lower bound $1 \times 10^{14}~M_{\mathrm{Sun}}$ used in \citet{2013MNRAS.430.1344O}. Hence for this analysis the lower limit on $M(r_{200})$ was decreased. Table~\ref{t:phys_priors} lists the type of prior used for each cluster parameter and the probability distribution parameters.

\begin{table}
\centering
\begin{tabular}{{l}{c}}
\hline
Parameter & Prior distribution \\ 
\hline
$x_{\rm c}$ & $\mathcal{N}(0'', 60'')$ \\
$y_{\rm c}$ & $\mathcal{N}(0'', 60'')$ \\
$z$ & $\delta(z_{\rm Planck})$ \\
$M(r_{200})$ & $\mathcal{U} [ \log (0.5\times 10^{14} M_{\rm{Sun}}),\log (50\times 10^{14} M_{\rm{Sun}})]$ \\
$f_{\rm gas}(r_{200})$ & $\mathcal{N}(0.13, 0.02)$ \\
\hline
\end{tabular}
\caption{Cluster parameter prior distributions. $\delta$ denotes a Dirac delta function, $\mathcal{U}$ is a uniform distribution and $\mathcal{N}$ is a normal distribution (parameterised by its mean and standard deviation).}\label{t:phys_priors}
\end{table}

I note here that $M(r_{500})$ (the AMI mass estimate I compare with those obtained in PSZ2) is not a sampling parameter of the physical model, but it can be calculated by evaluating equation~\ref{e:nfw_m_tot_1} at $r = r_{500}$. $r_{500}$ is calculated as part of the steps to determine the pressure profile given by equation~\ref{e:gnfw}, and so this does not cause any calculation overheads.


\section{PSZ2 redshift values}
\label{s:pl_z}
The values of $z_{\rm Planck}$ used for each cluster's $z$ prior distribution were taken to be the values stated in PSZ2. Catalogue $z$ values are measured in the optical / infrared or X-ray, with major input from the Sloan Digital Sky Survey \citep{2000AJ....120.1579Y}. A number of cluster catalogues have been extracted from these data (see e.g. \citealt{2010ApJS..191..254H}, \citealt{2012ApJS..199...34W}, and \citealt{2014ApJ...785..104R}), providing estimates of both spectroscopic and photometric $z$ values, the reliability of the latter values falls as $z$ increases. In the X-ray part of the spectrum, the Meta-Catalogue of X-ray detected Clusters of galaxies, or MCXC \citep{2011A&A...534A.109P} has a substantial number of matches with the Planck-catalogue clusters. The MCXC is from the available catalogues based on the ROSAT All-Sky Survey \citep{1999A&A...349..389V} as well as serendipitous X-ray catalogues (see e.g. \citealt{1990ApJS...72..567G}). MCXC contains only clusters with measured $z$, but does not state the redshift type or source.
Further sources of Planck catalogue clusters candidate $z$s are the Russian-Turkish Telescope \citep{2015A&A...582A..29P} and the ENO telescopes in the Canary Islands \citep{2016A&A...586A.139P}; for each $z$ these state whether it was obtained photometrically or spectroscopically. 


\section{PSZ2 methodology for deriving cluster mass estimates}
\label{s:psz2_m} 

For comparison with the mass values obtained with AMI data, I look at the PSZ2 mass estimates obtained from Planck data and the requisite scaling relations. The mass values published in PSZ2 are derived from data from one of three detection algorithms: MMF1, MMF3 (both of which are extensions of the matched multi-filter algorithm suitable for SZ studies (MMF, see \citealt{1996MNRAS.279..545H}, \citealt{2002MNRAS.336.1057H} and \citealt{2006A&A...459..341M}), over the whole sky) \& PowellSnakes (PwS, \citealt{2012MNRAS.427.1384C}). The former two rely on multi-frequency matched-filter detection methods, whilst PwS is a fully Bayesian method. Since the PwS methodology most closely matches the Bayesian analysis pipeline used for AMI data, I focus on the cluster parameter values from PwS. 
PwS will described in more detail in Section~\ref{s:ap_PwSlhood} where I carry out Bayesian analysis on AMI and Planck datasets simultaneously, which requires extensive use of the algorithm.

The observable quantity measured by Planck is the integrated Comptonisation parameter $Y$. As described in Section~5 of the PSZ2 paper \citep{2016A&A...594A..27P}, for each cluster candidate there is a two-dimensional posterior of the integrated Comptonisation parameter within the radius $5r_{500}$, $Y(5r_{500})$ and the angular scale radius of the GNFW pressure, $\theta_{\rm p}$ ($= r_{\rm p}/D_{A}$). The values for $Y(5r_{500})$ published in PSZ2 are obtained by marginalising over $\theta_{\rm p}$ and then taking the expected value of $Y(5r_{500})$. I refer to this value as $Y_{\rm marg}(5r_{500})$.
As described in Sections~5.2 and~5.3 of \citet{2016A&A...594A..27P}, this `blind' measurement of the integrated Comptonisation parameter may not be reliable when the underlying cluster pressure distribution deviates from that given by the GNFW model.
To overcome this, a function relating $Y(5r_{500})$ and $\theta_{\rm p}$ is derived in an attempt to provide prior information on the angular scale of the cluster based on X-ray measurements and earlier Planck mission samples. I refer to this function as the slicing function. 


\subsection{Derivation of the slicing function}

The scaling relations considered here are given in \citet{2014A&A...571A..20P}. Of particular importance to deriving the slicing function, are the $Y(r_{500}) - M(r_{500})$ and $\theta_{500} - M(r_{500})$ relations. The first of these is given by
\begin{equation}\label{e:y500m500}
E(z)^{-2/3}\left[\frac{D_{A}^{2}Y(r_{500})}{10^{-4}\rm{Mpc}^{2}}\right] = 10^{-0.19 \pm 0.02} \left[\frac{(1-b)M(r_{500})}{6 \times 10 ^{14}~M_{\rm{Sun}}}\right] ^{1.79 \pm 0.08},
\end{equation}
where $E(z) = \sqrt{\Omega_{\rm M}(1 + z)^{3} + \Omega_{\Lambda}}$ and is equal to the ratio of the Hubble parameter evaluated at redshift $z$ to its value now for a flat $\Lambda$CDM Universe. The factor in the exponent $-2/3$ arises from the scaling relations between mass, temperature and Comptonisation parameter given by equations~1--5 in \citet{2006ApJ...650..128K}. $(1-b)$ represents a bias factor, which is assumed in \citet{2014A&A...571A..20P} to contain four possible observational biases of departure from hydrostatic equilibrium, absolute instrument calibration, temperature inhomogeneities and residual selection bias. Its value is calculated to be $(1 - b) = 0.80 ^{+0.02}_{- 0.01}$ from numerical simulations as described in Appendix~A.4 of \citet{2014A&A...571A..20P}. Equation~\ref{e:y500m500} uses the fitting parameters from the relation between $Y_{\rm X}$ (the X-ray `analogue' of the integrated Comptonisation parameter see e.g. \citealt{2006ApJ...650..128K}, $Y_{\rm X}(r_{500}) \equiv M_{\rm g}(r_{500}) T_X$ where $M_{\rm g}$ is the cluster gas mass within $r_{500}$ and $T_X$ is the spectroscopic temperature in the range $[0.15,0.75]r_{500}$) and the X-ray hydrostatic mass, $M_{\rm HE}(r_{500})$ (which is equal to $(1-b) M(r_{500})$), established for 20 local \emph{relaxed} clusters by \citet{2010A&A...517A..92A} to give the relation between the X-ray mass proxy $M_{Y_{X}}(r_{500})$ and $M(r_{500})$. Finally, the fitting parameters for the $Y(r_{500}) - M_{Y_{X}}(r_{500})$ relation are obtained empirically from a 71-cluster sample consisting of SZ data from the Planck Early SZ clusters \citep{2011A&A...536A..11P}, of Planck-detected LoCuSS clusters \citep{2013A&A...550A.129P} and from the XMM-Newton validation programme \citep{2011A&A...536A...9P}, all with X-ray data taken from XMM-Newton observations (\citealt{2013MNRAS.430..134W} and \citealt{2012MNRAS.423.1024M}).

The $\theta_{500} - M(r_{500})$ relation is based on the equation $M(r_{500}) = 500 \times \frac{4\pi}{3} \rho_{\rm crit}(z) r_{500}^{3}$ and is given by
\begin{equation}\label{e:theta500m500}
\theta_{500} = 6.997 \left[\frac{h}{0.7}\right]^{-2/3} \left[ \frac{(1-b)M_{500}}{3 \times 10^{14}~M_{\rm{Sun}}} \right]^{1/3} E(z)^{-2/3}\left[ \frac{D_{A}}{500~\rm{Mpc}} \right].
\end{equation}
Equations~(\ref{e:y500m500}) and~(\ref{e:theta500m500}) can be solved for $(1-b)M(r_{500})$ and equated to give $Y(r_{500})$ as a function of $\theta_{500}$
\begin{equation}\label{e:y500theta500}
Y(r_{500}) = \left[ \frac{\theta_{500}}{6.997} \right] ^{5.4 \pm 0.2} \left[ \frac{h}{0.7} \right] ^{3.60 \pm 0.13} \left[\frac{E(z)^{4.26 \pm 0.13} D_{A}^{3.4 \pm 0.2}}{10^{19.29 \pm 0.54} ~ \rm{Mpc}^{3.4 \pm 0.2}} \right],
\end{equation}
where $Y(r_{500})$ is in $\rm{sr}$. Assuming a GNFW pressure profile, $Y(r_{500})$ can be converted to the corresponding value of $Y(5r_{500})$, through the relation
\begin{equation}\label{e:yr500y5r500}
\frac{Y(r_{500})}{Y(5r_{500})} = \frac{B \left( \frac{(c_{500})^{a}}{1+(c_{500})^{a}}; \frac{3 - c}{a}, \frac{b - 3}{a} \right)}{B \left( \frac{(5c_{500})^{a}}{1+(5c_{500})^{a}}; \frac{3 - c}{a}, \frac{b - 3}{a} \right)},
\end{equation} 
where $B(x,y,z) = \int_{0}^{x} t^{y-1}(1-t)^{z-1}\rm{d}t$ is the incomplete beta function. For the GNFW parameter values used in equation~\ref{e:gnfw}, equation~\ref{e:yr500y5r500} gives a value of $0.55$. Similarly, $\theta_{500}$ can be related to $\theta_{\rm p}$ through the relation $\theta_{\rm p} = \theta_{500} / c_{500} $. 


\subsection{Mass estimates}
\label{s:planckmass}
\begin{figure}
  \begin{center}
  \includegraphics[ clip=, width=\linewidth]{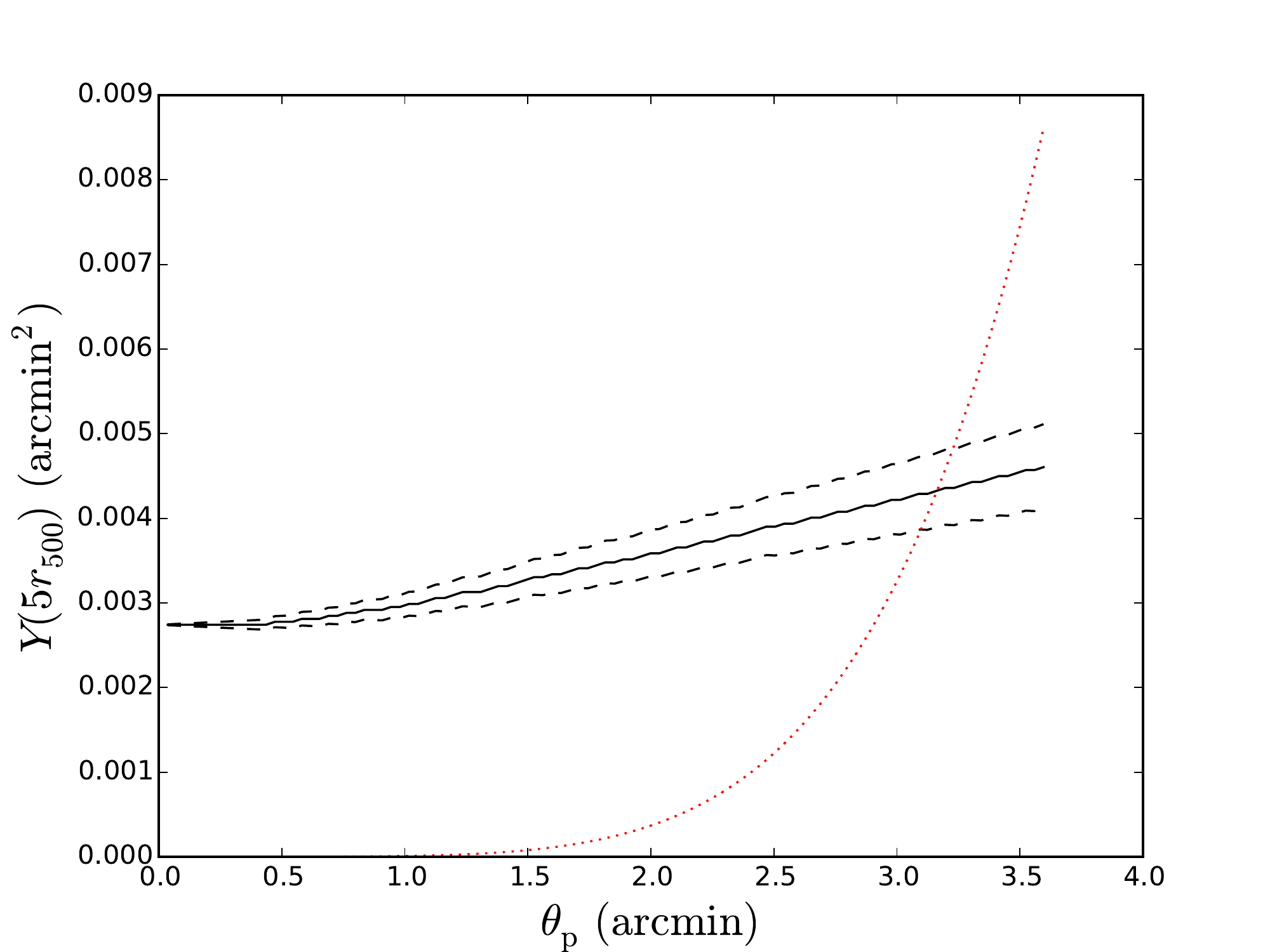}
  \caption{Example of the posterior slicing methodology for cluster PSZ2G228.16+75.20. The black solid line represents the `ridge' (i.e. the most probable value of $Y(5r_{500})$ for each $\theta_{\rm p}$) of the posterior. The upper dashed curve represents the upper boundaries of the 68\% maximum likelihood confidence interval on $Y(5r_{500})$ for each value of $\theta_{\rm p}$, and the lower dashed curve corresponds to the lower boundaries. The red dotted curve is the slicing function.}
  \label{f:slicing}
  \end{center}
\end{figure}

For a given cluster, the resulting $Y(5r_{500})$ function is used to `slice' the posterior, and the value where the function intersects the posterior `ridge' is taken to be the most reliable estimate of $Y(5r_{500})$, given the external information. The posterior ridge (see Figure~\ref{f:slicing}) is defined to be the value of $Y(5r_{500})$ which gives the highest probability density for a given $\theta_{\rm p}$. The error estimates are obtained by considering where the slicing function intersects with the ridges defined by the 68\% maximum likelihood  confidence intervals for $Y(5r_{500})$ at each $\theta_{\rm p}$. $Y(5r_{500})$ is then converted to $Y(r_{500})$ using the the reciprocal of the value given by equation~\ref{e:yr500y5r500}, and this is used to derive a value for $M(r_{500})$ using equation~\ref{e:y500m500}, but with the $(1-b)$ term excluded. The bias term is not included in the $M(r_{500})$ calculation because it has already been accounted for in the slicing function. Note that this value of $M(r_{500})$ is what is referred to as $M_{\rm SZ}$ in PSZ2.

\section{Obtaining AMI mass estimates}
\label{s:planck_phys_results_i}
First I describe how I arrived at a final sample of clusters for which the AMI mass estimates are compared with those derived from Planck data.


\subsection{Final cluster sample}


\subsubsection{Well constrained posterior sample}
\textsc{McAdam} was used on data from the initial sample of 199 clusters. \textsc{MultiNest} failed to produce posterior distributions for two clusters. These clusters were surrounded by high flux, extended radio-sources. Of the 197 clusters for which posterior distributions were produced, 73 clusters show good constraints (adjudged by physical inspection) on the sampling parameters $M(r_{200})$, $f_{\rm gas}(r_{200})$, $x_{\rm c}$ and $y_{\rm c}$; with $z$s ranging from $0.089$ to $0.83$. 

\begin{figure*}
  \begin{center}
  \includegraphics[ width=0.45\linewidth]{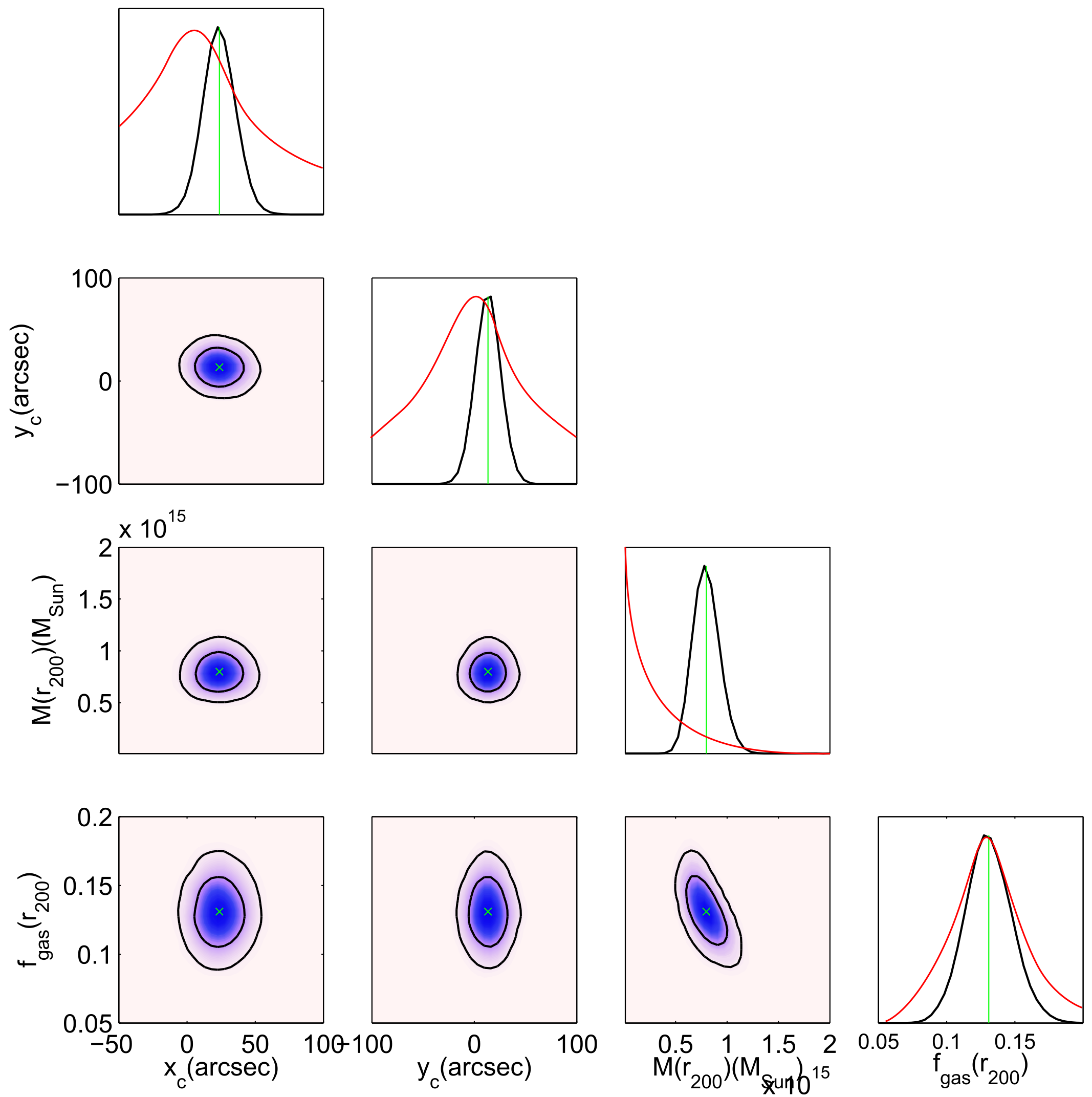}
  \includegraphics[ width=0.45\linewidth]{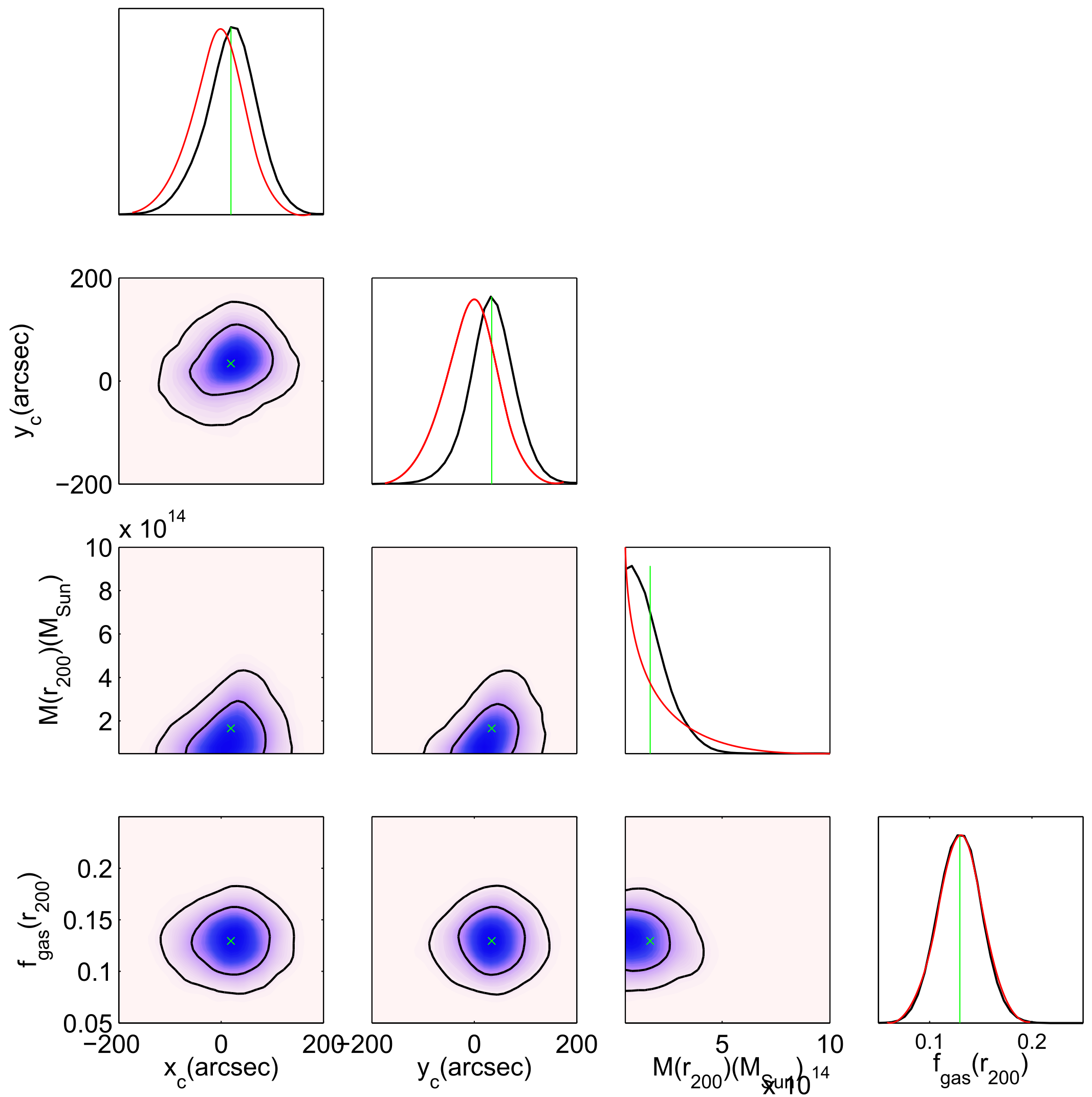}
  \medskip
  \centerline{(a) \hskip 0.45\linewidth (b)}
  \caption{Posterior distributions derived from AMI data for the sampling parameters: $M(r_{200})$; $f_{\rm gas}(r_{200})$; $x_{\rm c}$ \& $y_{\rm c}$. The contoured maps show the two-dimensional posteriors for the different pairs of parameters. The contours represent the 95\% and 68\% mean confidence intervals, with the green crosses denoting the expected value of the joint distributions. The four one-dimensional plots are the marginalised posteriors corresponding to the variable given at the bottom of the respective column. The red curves are the prior distributions on the relevant parameters. Each green line is the expected value of the distribution. Posterior distributions in (a) show narrow distributions on the cluster mass, with the domain spanning feasible mass values for a galaxy cluster (cluster PSZ2G184.68+28.91). In such cases the posteriors are said to be well constrained. The mass posteriors in (b) show that the data imply unphysical values for its mass, as the posterior distribution is hitting the lower bound of the prior ($5 \times 10^{13} M_{\rm{Sun}}$) at almost its peak value (cluster PSZ2G121.77+51.75). The distribution also resembles the uniform in log-space prior assigned to $M(r_{200})$. In such cases the posteriors are said to be poorly constrained with respect to the mass estimates. }
\label{f:pl_phys_post}
  \end{center}
\end{figure*}

I illustrate a `well constrained' posterior distribution (for cluster PSZ2G184.68+28.91) in the first half of Figure~\ref{f:pl_phys_post}, plotted using \textsc{GetDist}\footnote{\url{http://getdist.readthedocs.io/en/latest/}.} (a kernel density estimation algorithm, which is described in Section~\ref{s:kde}). In contrast the second half of Figure~\ref{f:pl_phys_post} is an example of a cluster (PSZ2G121.77+51.75) which shows poor constraints on mass as the posterior distribution is peaked at the lower boundary of the mass sampling range ($5\times 10^{13} M_{\rm{Sun}}$) which could not be classed as a detection within our mass prior range. I also note that in the latter case the mass posterior largely resembles the prior distribution.

\subsubsection{Moderate radio-source environment sample}
\begin{figure*}
  \begin{center}
  \includegraphics[ width=0.45\linewidth]{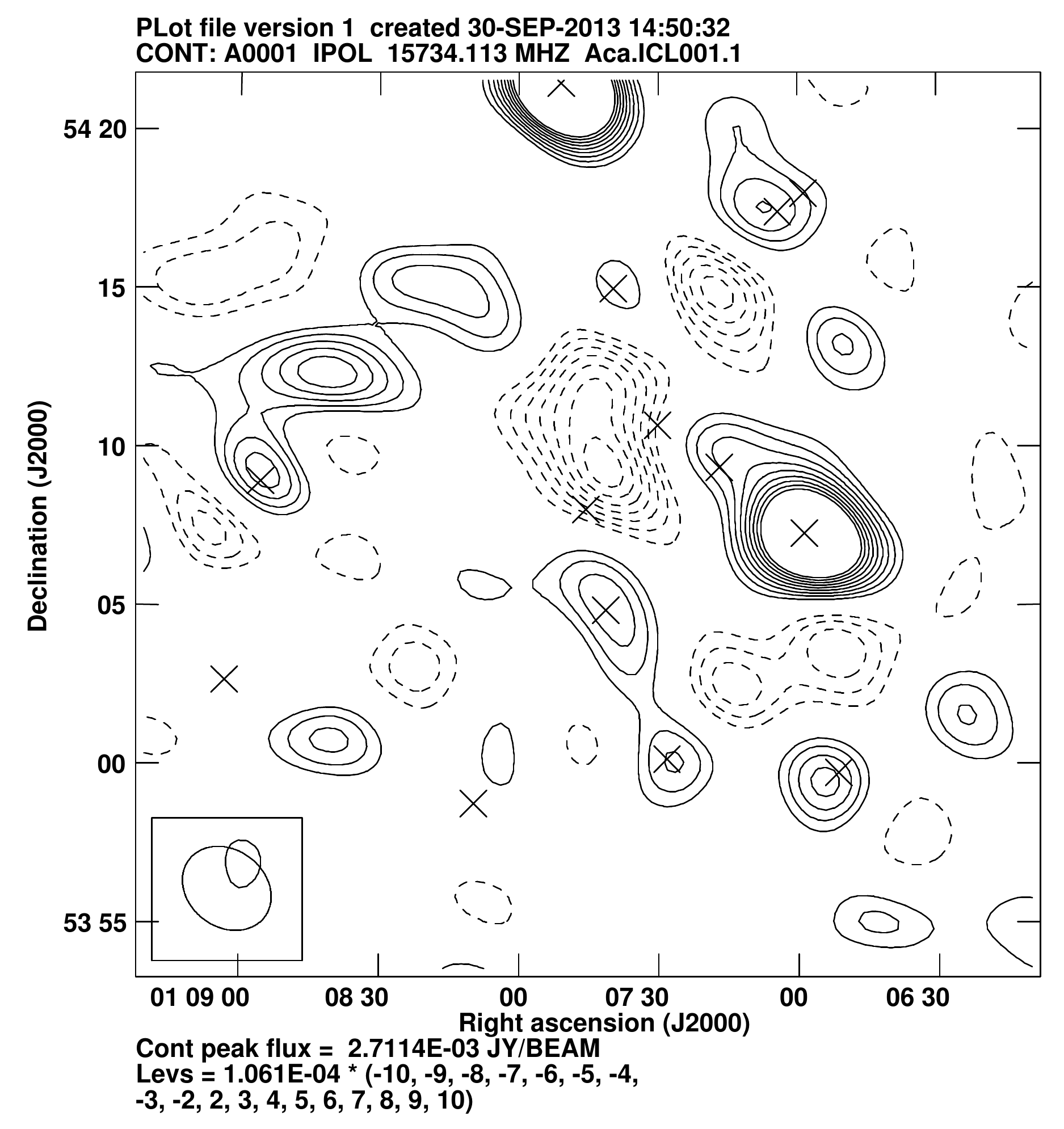}
  \includegraphics[ width=0.45\linewidth]{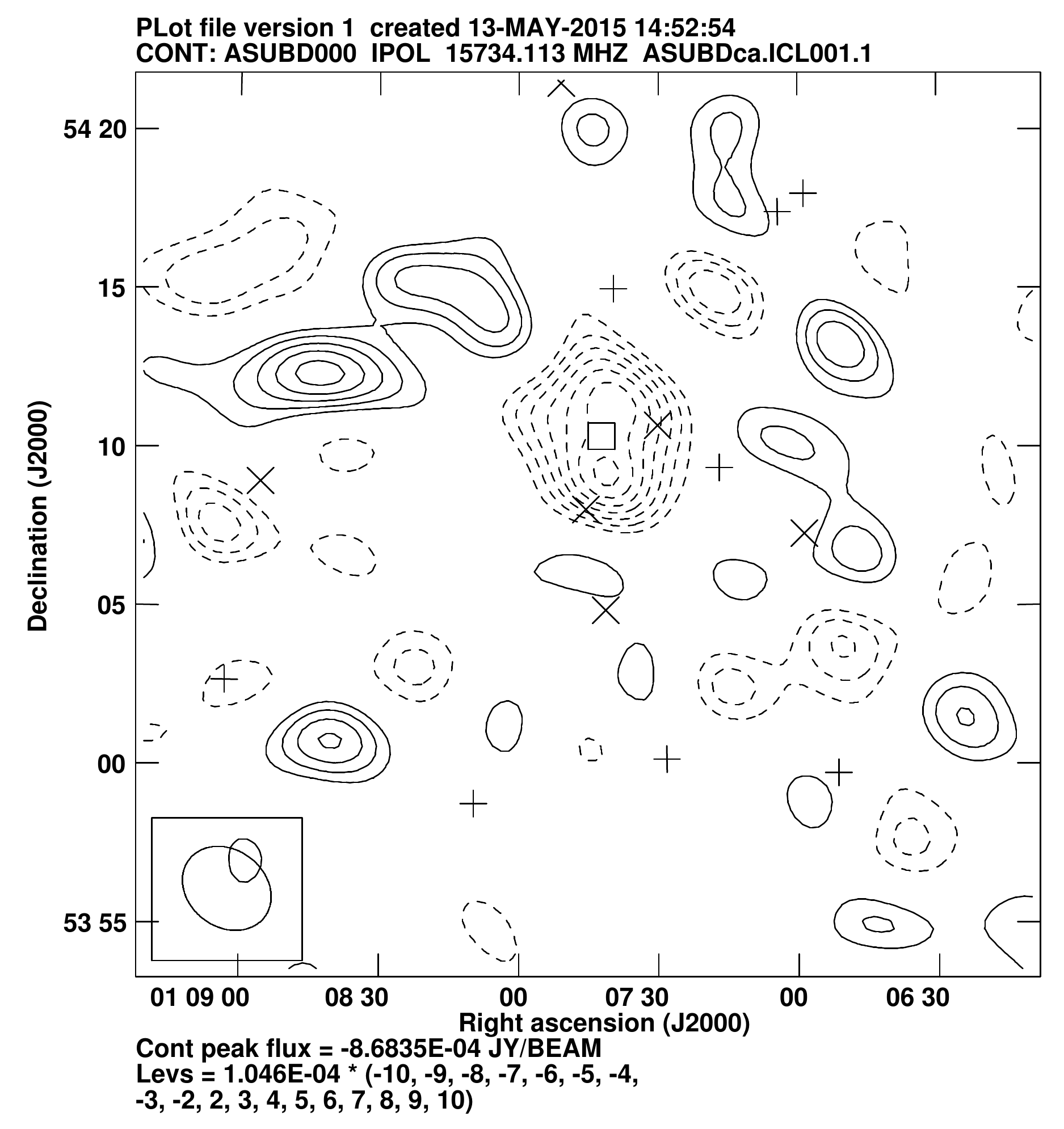}
  \medskip
  \centerline{(a) \hskip 0.45\linewidth (b)}
  \caption{(a) Unsubtracted map produced from AMI observation. Contours are plotted at $\pm (2, 3, 4, ..., 10)\times$ the r.m.s. noise level, and dashed contours are negative. (b) Source subtracted map produced from AMI observation. The $\Box$ denotes the \textsc{McAdam}-determined centre of the cluster (posterior mean values for $x_{\rm c}$ and $y_{\rm c}$). Here `$+$' signs denote radio-source positions as measured by the LA which were assigned delta priors on their parameters, whilst `$\times$' denote sources which were assigned priors as described in Section~\ref{s:rs_priors}.} 
  \label{f:bad_rs}
  \end{center}
\end{figure*}

For the 197 cluster sample, AMI data maps were produced using the software package \textsc{AIPS}\footnote{\url{http://aips.nrao.edu/}.} using the automated \textsc{CLEAN} procedure with a limit determined using \textsc{IMEAN}. Source-finding was carried out at four $\sigma$ on the LA continuum map, as described in \citet{2011MNRAS.415.1883D} and \citet{2011MNRAS.415.2699A}. For each cluster both a non-source-subtracted and a source-subtracted map was produced. The values used to subtract the sources from the maps were the mean values of the one-dimensional marginalised posterior distributions of the sources' position, flux and spectral index produced by \textsc{McAdam}.
Maps of the 73 cluster sample were inspected in detail. It was found that for seven of these clusters, even though the posterior distributions were well constrained, that the radio-source and primordial CMB contamination could bias the cluster parameter estimates in an unpredictable way. In these cases it was found that the subtracted maps contained residual flux close to the cluster centre, from either radio-sources (some of which were extended), radio-frequency interference, or CMB. PSZ2G125.37-08.67 is an example of one of these clusters and its non-source-subtracted and source-subtracted maps are shown in Figure~\ref{f:bad_rs}. I thus arrived at a 66 cluster sample.

\subsubsection{Well defined cluster-centre sample}
\begin{figure}
  \begin{center}
  \includegraphics[ width=0.9\linewidth]{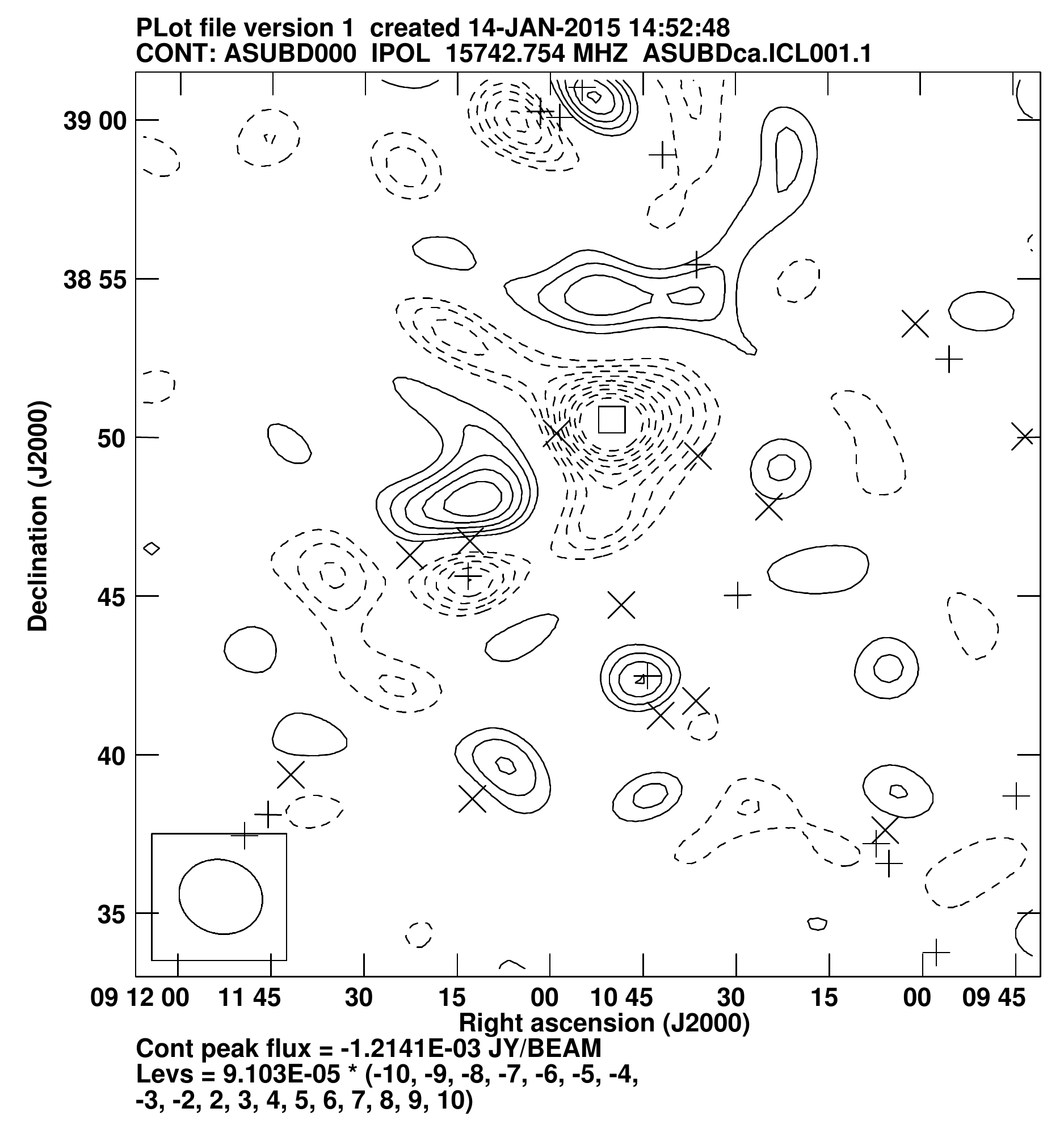}
  \caption{Subtracted map of cluster with ill-defined centre. The cluster is clearly offset from the observation pointing centre (middle of the map), and the lobes to the bottom and the top left of the cluster cause the centre position to be ambiguous.} 
  \label{f:offset_map}
  \end{center}
\end{figure}

The posteriors of $x_{\rm c}$ and $y_{\rm c}$ give the position of the modelled cluster centre relative to the actual SA pointing centre used for the observation. For seven of the 66 cluster sample, it was found that the mean posterior values of $x_{\rm c}$ and $y_{\rm c}$ changed dramatically between different runs of \textsc{McAdam} (on the same cluster data), by up to $70$ arcseconds in either direction, leading to differences in mass estimates of up to $70\%$. The estimates for these clusters are not reliable, since the model was creating a completely different cluster between runs, and so these clusters were excluded leaving a 59 cluster sample. For the remaining clusters, the change in $M(r_{200})$ between runs was much smaller than the standard deviation of the corresponding posterior distributions. Figure~\ref{f:offset_map} shows the subtracted map for PSZ2G183.90+42.99, which we consider to be an example of a cluster with an ill-defined centre. The map shows three flux decrement peaks close to the cluster centre. Movement of the centre between these peaks with the current source environment modelling would lead to a change in the size of the predicted cluster, and consequently different mass estimates each time.

\subsubsection{PwS detected cluster sample}
For five of the 59 cluster sample, the data available on the Planck website\footnote{\url{https://pla.esac.esa.int/pla/catalogues}.} did not contain a detection using the PwS algorithm, and so no mass estimates based on PwS data could be calculated. Hence the final sample size for which I present the mass estimates from both AMI and Planck data is 54. \\

It is important to realise that selection biases are introduced in reducing the sample size from $199$ to $54$. In particular, selecting only the clusters which showed good AMI posterior constraints means that clusters corresponding to a signal too faint for AMI to detect, clusters with large enough angular size for AMI's shortest baselines not to be able to measure the signal from the outskirts of the cluster ("resolved clusters"), and clusters where the radio-source and CMB contamination dwarfs the signal of the cluster, are all likely to have been excluded from the sample to some extent. In addition, removing the seven clusters with an ill defined centre likely removes some unrelaxed clusters from the sample. 


\section{AMI and PSZ2 mass estimates}
 
\begin{figure}
  \begin{center}
  \includegraphics[ width=\linewidth]{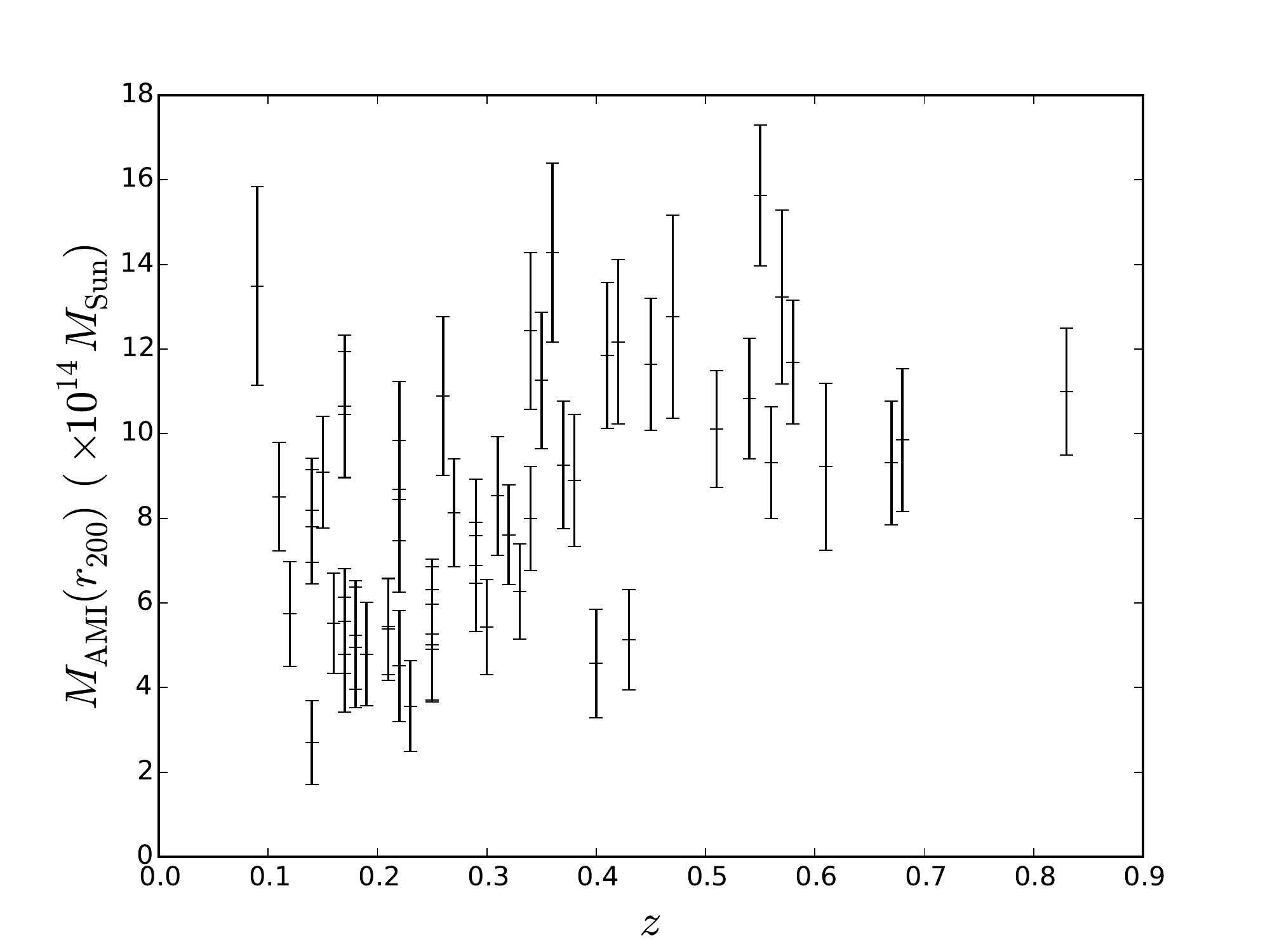}
  \caption{Plot of $M(r_{200})$ derived from AMI data using physical modelling vs redshift for the sample of 54 clusters.}
\label{f:m200z}
  \end{center}
\end{figure}

The AMI and PSZ2 parameter estimates for the 54 clusters are given in Table~\ref{t:pl_phys_results} in Appendix~\ref{c:appendixa}.
The clusters are listed in ascending order of $z$. Note that whether a redshift is photometric or spectroscopic is stated in the fifth column.
All AMI values are the mean values of the corresponding parameter posterior distributions, with the error taken as the standard deviation. The estimates of the sampling parameters are included for comparison with each other, and with the sampling prior ranges and associated parameters given in Table~\ref{t:phys_priors}. The AMI values for $M(r_{500})$ are given for comparison with the corresponding PSZ2 estimates. 
Two values for the PSZ2 mass estimates are given, $M_{\rm Pl,\, marg}(r_{500})$ and $M_{\rm Pl,\, slice}(r_{500})$. $M_{\rm Pl,\, marg}(r_{500})$ corresponds to the mass given by the $Y(r_{500}) - M(r_{500})$ relation when the marginalised integrated Comptonisation parameter is used as described in Section~\ref{s:psz2_m}. The uncertainties associated with these $Y$ values are taken as the standard deviations of the marginalised posteriors. $M_{\rm Pl,\, slice}(r_{500})$ is detailed in Section~\ref{s:planckmass}; its associated errors are calculated from the $Y(5r_{500})$ values where the slicing function intersects with the two ridges formed by the 68\% maximum likelihood confidence interval values of the $Y(5r_{500})$ probability densities over the posterior domain of $\theta_{\rm p}$. \\
Figure~\ref{f:m200z} shows $M(r_{200})$ as a function of $z$. Excluding the clusters at $z = 0.089,\, 0.4$ and $0.426$, there is a steepening in mass between $ 0.1 \lessapprox z \lessapprox 0.5$ before it flattens off at higher $z$. This result is consistent with the PSZ2 mass estimates presented in \citet{2016A&A...594A..27P}.

\begin{figure*}
  \begin{center}
  \includegraphics[ width=\linewidth]{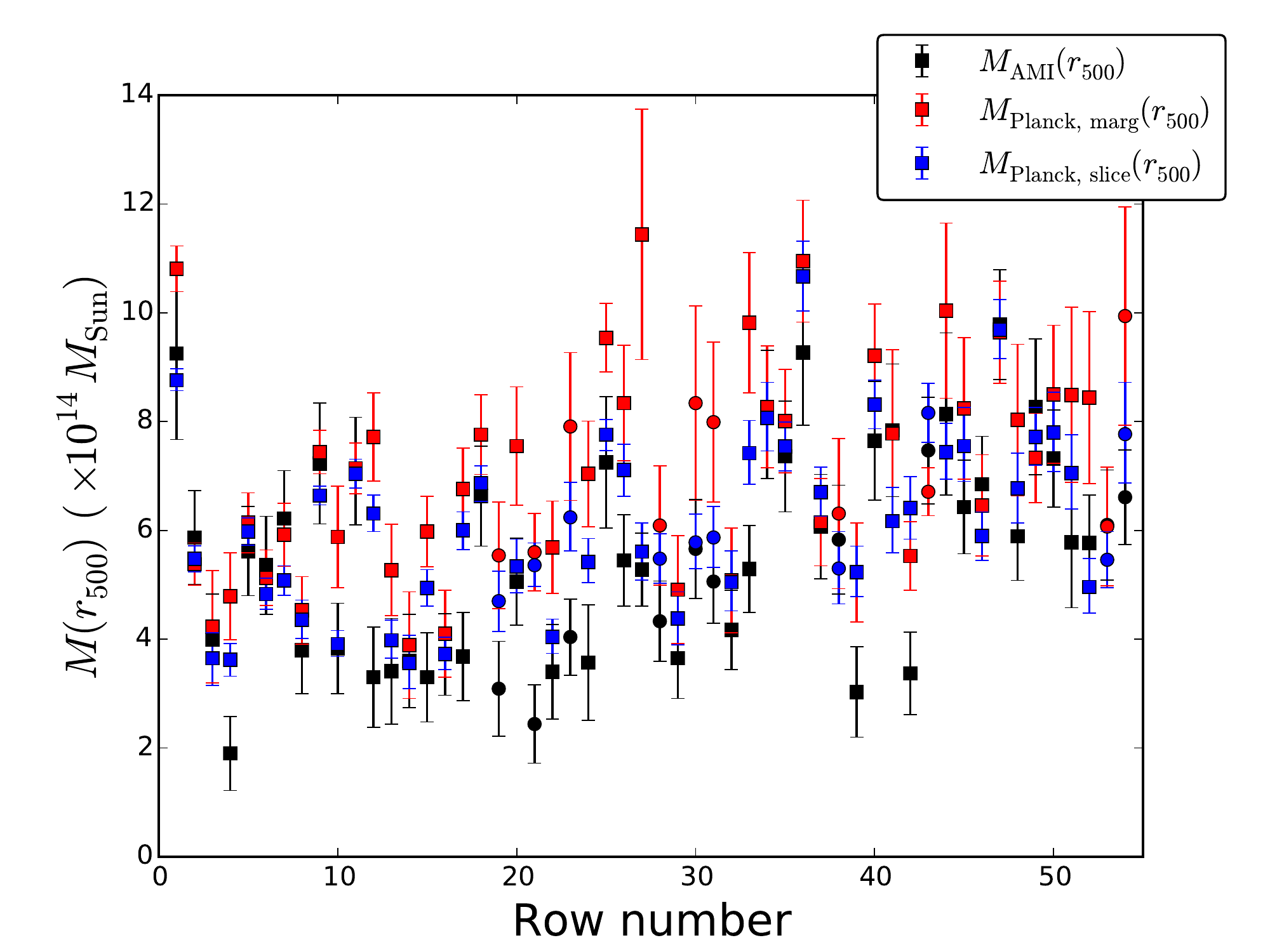}
  \caption{Plot of $M(r_{500})$ vs row number of Table~\ref{t:pl_phys_results}
for three different cases: the value derived from AMI data using the physical model, $M_{\rm AMI}(r_{500})$; the value derived from Planck data using the marginalised value for $Y(5r_{500})$, $M_{\rm Pl,\, marg}(r_{500})$ and the value derived from Planck data using the slicing function value for $Y(5r_{500})$, $M_{\rm Pl,\, slice}(r_{500})$. The row number is monotonically related to $z$, as Table~\ref{t:pl_phys_results} is sorted by ascending $z$. The points with circular markers correspond to clusters whose redshifts were measured photometrically (as listed in Table~\ref{t:pl_phys_results}). }
\label{f:m500planck}
 \end{center}
\end{figure*}

\begin{figure*}
  \begin{center}
  \includegraphics[ width=\linewidth]{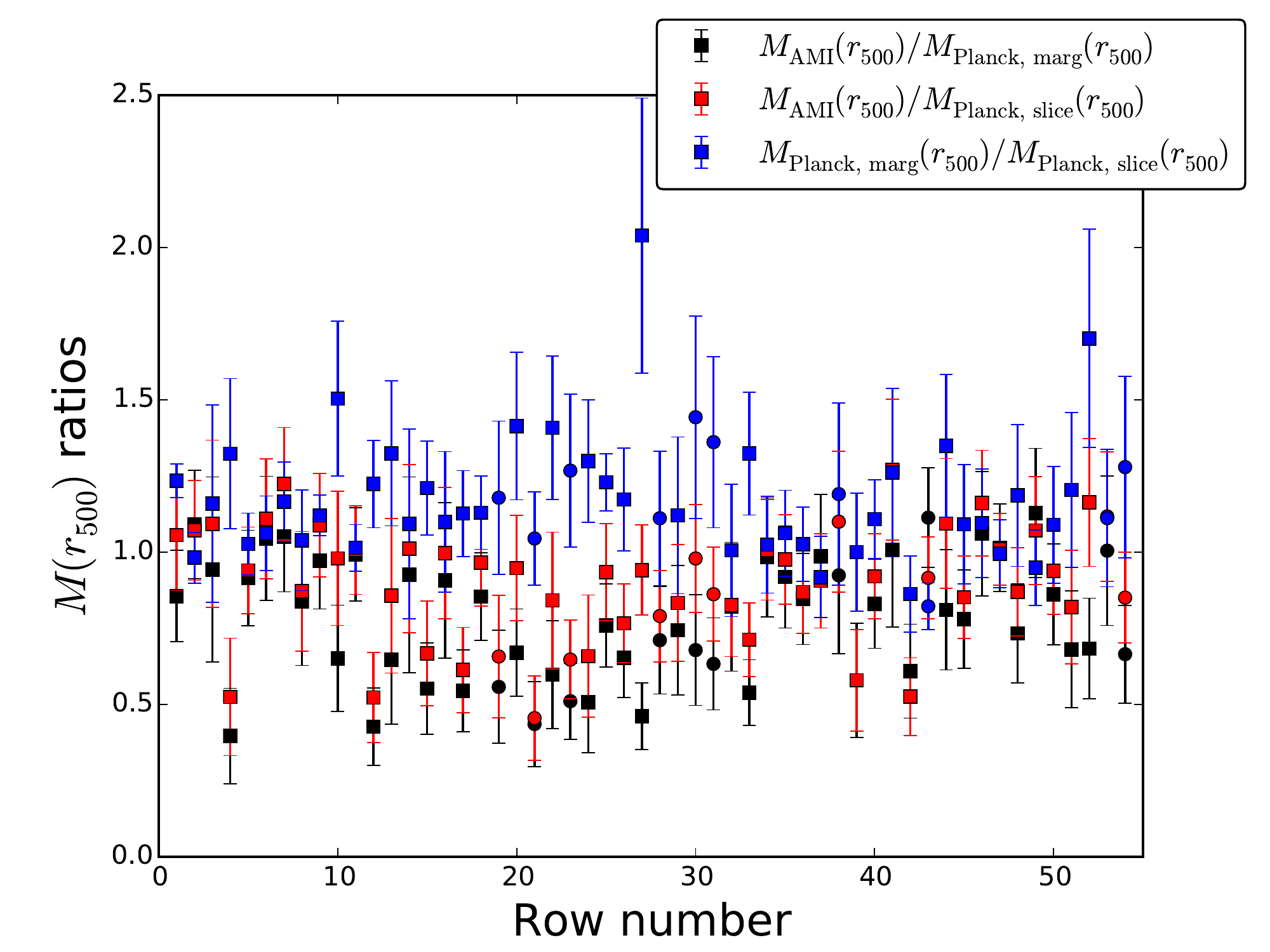}
  \caption{Plot of $M(r_{500})$ ratios vs row number of Table~\ref{t:pl_phys_results} 
   for three different cases: $M_{\rm AMI}(r_{500}) / M_{\rm Pl,\, marg}(r_{500})$; $M_{\rm AMI}(r_{500}) / M_{\rm Pl,\, slice}(r_{500})$ and $M_{\rm Pl,\, marg}(r_{500}) / M_{\rm Pl,\, slice}(r_{500})$. The points with square markers correspond to clusters whose redshifts were measured spectroscopically, and the circular markers correspond to photometric redshifts (as listed in Table~\ref{t:pl_phys_results}).
}
\label{f:m500planckfrac}
  \end{center}
\end{figure*}

I now focus on the comparison between AMI and Planck mass estimates. Note that \citet{2016A&A...594A..27P} do not provide any means for estimating $M(r_{200})$ from their data, as $r_{200}$ is the distance related to the scale radius ($r_{200} = c_{200} \times r_{\rm s}$) for the NFW dark matter profile given by equation~\ref{e:nfw}, which they do not incorporate into their modelling process.
Figure~\ref{f:m500planck} gives the AMI and two Planck estimates for $M(r_{500})$ vs the row number, in Table~\ref{t:pl_phys_results}. I have not used $z$ as the independent variable in this plot for clarity. The row number is monotonically related to $z$, as Table~\ref{t:pl_phys_results} is sorted by ascending $z$. 
From Figure~\ref{f:m500planck} it is clear that AMI underestimates the mass relative to both PSZ2 values. In fact $M(r_{500})$ is lower than $M_{\rm Pl,\, slice}(r_{500})$ in 37 out of 54 cases. $M(r_{500})$ is lower than $M_{\rm Pl,\, marg}(r_{500})$ in 45 out of 54 cases.
31 of the AMI masses are within one combined standard deviation of $M_{\rm Pl,\, slice}(r_{500})$, while 46 are within two. Four clusters have discrepancies larger than three combined standard deviations. Three of these clusters are at relatively low redshift ($\leq 0.25$), whilst one is at $z=0.43$. \\ 
It is also noteworthy that $M_{\rm Pl,\, marg}(r_{500})$ is larger than $M_{\rm Pl,\, slice}(r_{500})$ in 47 out of 54 cases. This implies that the additional information obtained from X-ray data incorporated in the slicing function consistently predicts a lower mass cluster than from the Planck SZ data alone. \\
Figure~\ref{f:m500planckfrac} shows the ratios of the mass estimates between the three different methods. The most obvious thing to note is that the ratio of PSZ2 masses is consistently greater than one, which again emphasises the fact that the marginalisation method attributes a much higher mass to the clusters than the slicing method. Furthermore, the ratio of AMI mass to the marginalised mass is small at medium redshift, which suggests that the marginalised mass is systematically high in this range. This graph also emphasises that the AMI mass and the slicing methodology mass are the most consistent with one another.


\section{AMI simulations with PSZ2 mass inputs}
\label{s:pl_phys_results_ii}
To investigate further the discrepancies between the mass estimates, it was decided to create simulated data based on the PSZ2 mass estimates obtained from the slicing methodology, which were then `observed' by AMI. The data from these simulated observations were analysed the same way as the real data.
The simulations were carried out using the in-house AMI simulation package \textsc{Profile}, which has been used in various forms in e.g. \citet{2002MNRAS.333..318G}, \citet{2011MNRAS.415.2708A}, \citet{2012MNRAS.421.1136A} and \citet{2013MNRAS.430.1344O}.
The input parameters for the simulation-- which uses the physical model to create the cluster-- are the sampling parameters of the model. Since \citet{2016A&A...594A..27P} does not give a method for calculating $M(r_{200})$ it was calculated as follows. First $r_{500}$ was calculated by solving equation~\ref{e:nfw_m_tot_1} with $M(r_{\Delta}) = M_{\rm SZ}$ and $r_{\Delta} = r_{500}$. $r_{200}$ can be determined from $r_{500}$, but we note that the function mapping from $r_{200}$ to $r_{500}$ is non-invertible, thus $r_{200}$ had to be calculated by solving equation~\ref{e:r200r5001} iteratively. $M(r_{200})$ can then be calculated by evaluating equation~\ref{e:nfw_m_tot_1} at $r_{200}$. \\ 
As well as the values of $M(r_{200})$ derived from PSZ2 mass estimates, values for the other inputs were also required. I used $f_{\rm gas}(r_{200}) = 0.13$, $z = z_{\rm Planck}$, and $x_{\rm c} = y_{\rm c} = 0$~arcsec.\\
The objective of these simulations was to see whether we could recover the mass input into the simulation to create a cluster using the physical model, `observed' by AMI and then analysed using the same model. I tried this for the four sets of simulations described below.  \\
For each simulation different noise / canonical radio-source environment realisations (where relevant) were used each time. Due to the large sample size this should not affect any systematic trends seen in the results, and it avoids having to pick a particular realisation to be used in all the simulations. 


\subsection{Simulations of clusters plus instrumental noise}
\label{s:NSNBsims}
\begin{figure}
  \begin{center}
  \includegraphics[ width=0.9\linewidth]{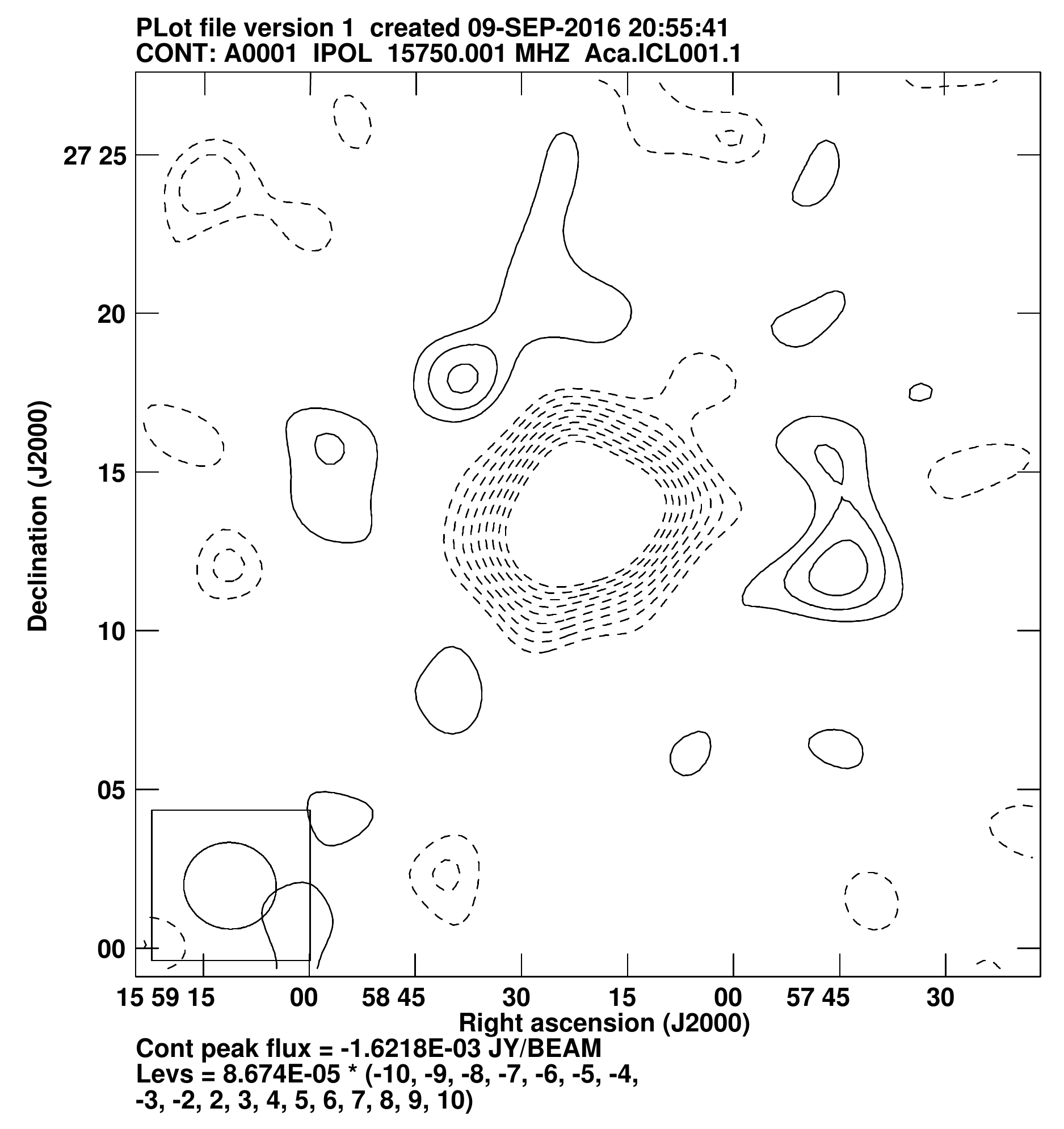}
  \caption{Unsubtracted map produced from simulated AMI data of cluster PSZ2G044.20+48.66, including instrumental noise.}
  \label{f:sim_NSNB_map}
  \end{center}
\end{figure}

\begin{figure}
  \begin{center}
  \includegraphics[ width=\linewidth]{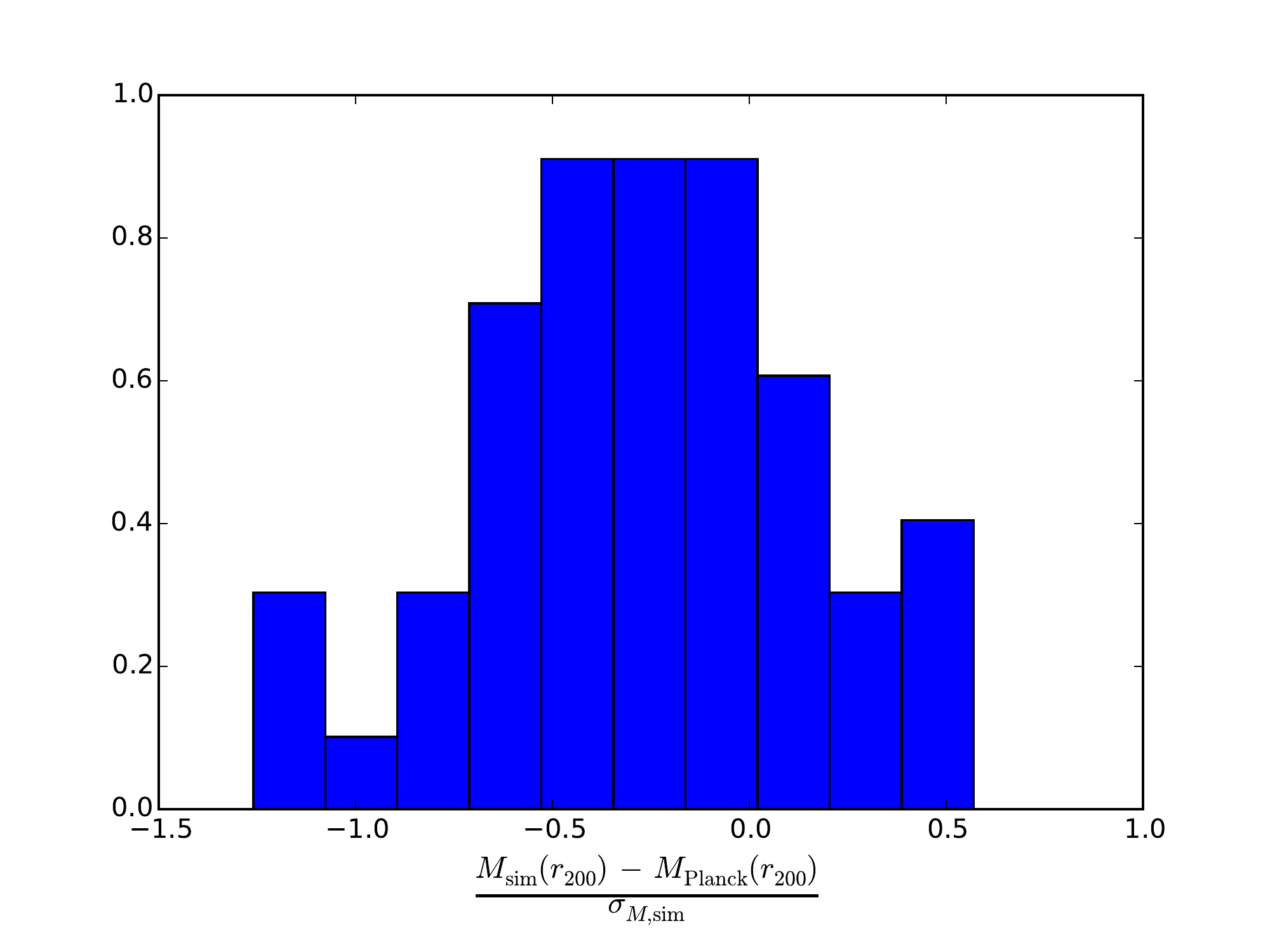}
  \caption{Normalised histogram of the differences between the input and output masses of the AMI simulations including the cluster and instrumental noise only, in units of standard deviations of the output mass.}
\label{f:simulatednsnb}
  \end{center}
\end{figure}

For each cluster, $M(r_{200})$ was calculated and Gaussian instrumental noise (Section~\ref{s:inst_noise}) was added to the sky. The RMS of the noise added was $0.7~ \rm{Jy}$ per channel per baseline per second, a value typical of an AMI cluster observation. 
Figure~\ref{f:sim_NSNB_map} shows the map produced from the simulated data of cluster PSZ2G044.20+48.66 plus this instrumental noise. The mass estimate derived from the Bayesian analysis of this cluster is 0.56 standard deviations above the input value.\\
Figure~\ref{f:simulatednsnb} shows the difference between the input masses and the ones recovered from running the simulated observations through \textsc{McAdam}, visualised using a histogram. All but three of the clusters lie within one standard deviation of the input mass, and even these clusters (PSZ2G154.13+40.19, PSZ2G207.88+81.31 and PSZ2G213.39+80.59) give an output mass 1.01, 1.26 and 1.08 standard deviations below the input mass. 

\subsection{Simulations further adding confusion noise and primordial CMB}
\label{s:noisesims}
\begin{figure}
  \begin{center}
  \includegraphics[ width=0.9\linewidth]{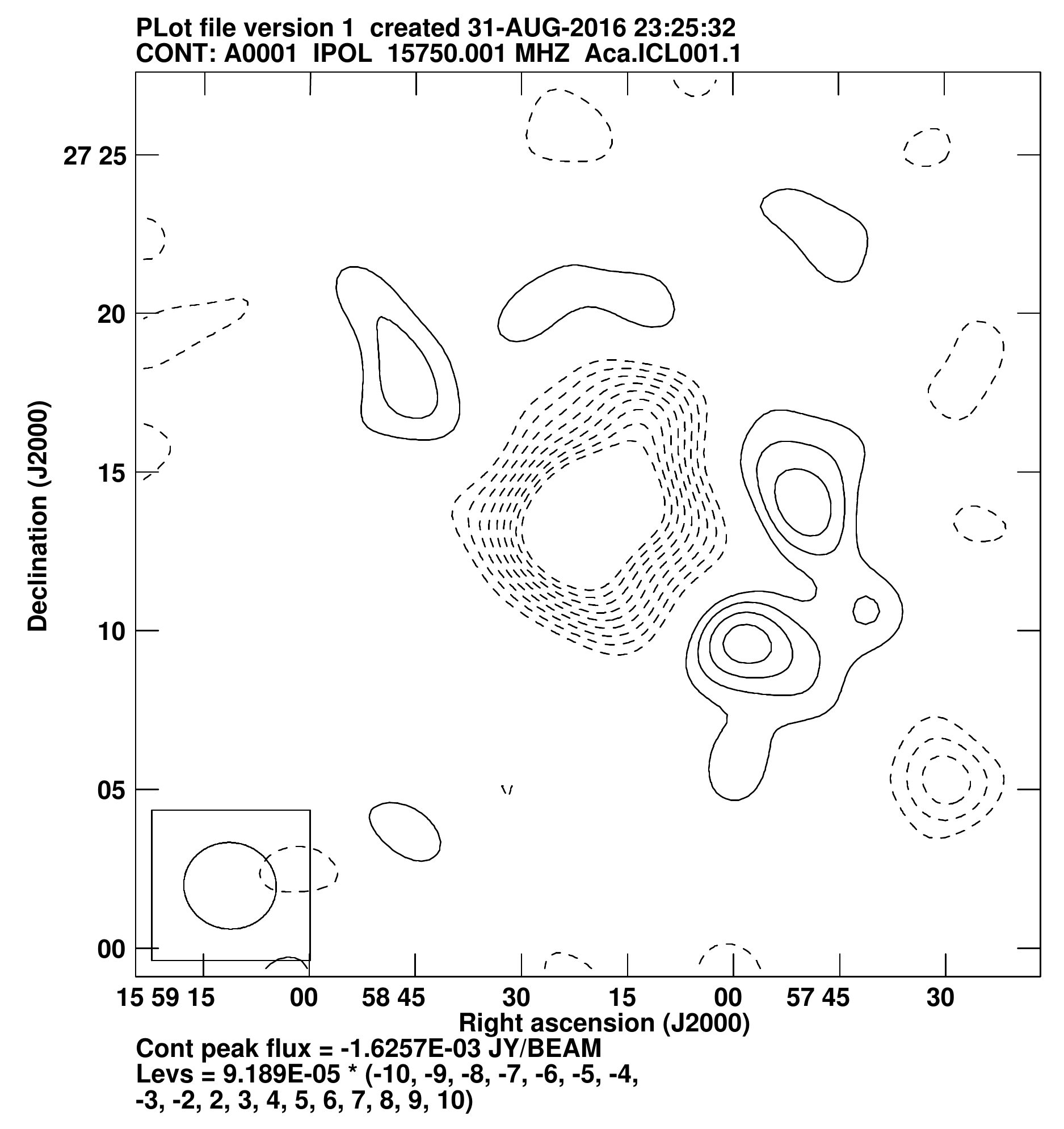}
  \caption{Unsubtracted map produced from simulated AMI data of cluster PSZ2G044.20+48.66, including instrumental, confusion and CMB noise.}
  \label{f:sim_NS_map}
  \end{center}
\end{figure}

\begin{figure}
  \begin{center}
  \includegraphics[ width=\linewidth]{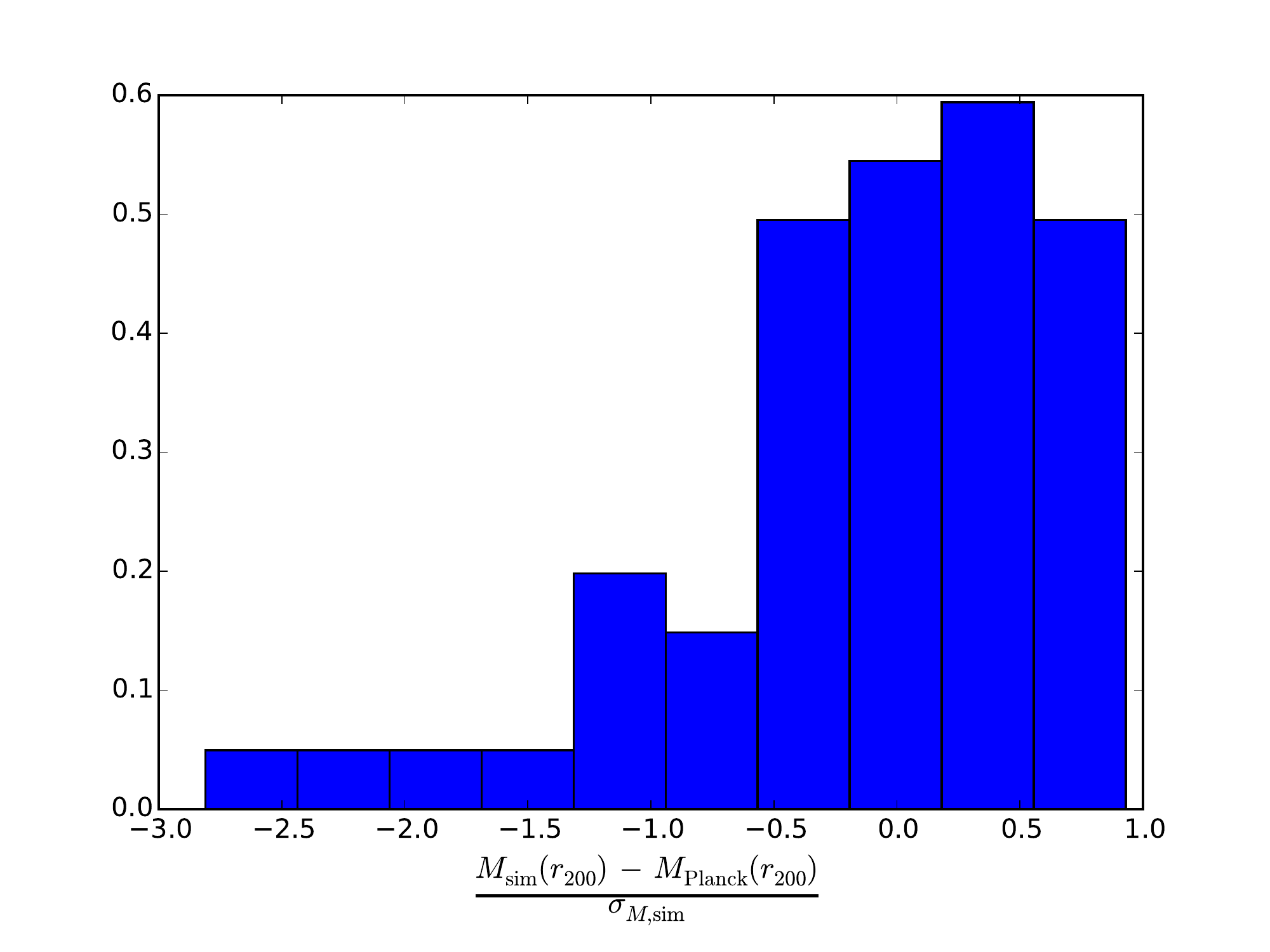}
  \caption{Normalised histogram of the differences between the input and output masses of the AMI simulations, in units of standard deviations of the output mass. This is the case for instrumental, confusion and CMB noise contributions.}
\label{f:simulatedns}
  \end{center}
\end{figure}

Confusion noise is defined to be the flux from radio-sources below a certain limit (see Section~\ref{s:conf_noise}, here $S_{\rm{conf}} = 0.3~\rm{mJy}$). In this Section all radio-source realisations only contribute to the confusion noise. However in Sections~\ref{s:CSsims} and~\ref{s:sourcesims} sources above $S_{\rm{conf}}$ are included. The confusion noise contributions (see e.g. Section~5.3 of FF09) were sampled from the probability density function corresponding to the 10C source counts given in \citet{2011MNRAS.415.2708A}, and placed at positions chosen at random. Similarly, the primordial CMB (Section~\ref{s:prim_cmb}) realisations were sampled from an empirical distribution \citep{2013ApJS..208...19H}, and randomly added to the maps. \\
Figure~\ref{f:sim_NS_map} shows the map produced from the simulated data of cluster PSZ2G044.20+48.66, including the three noise contributions. The mass estimate derived from the Bayesian analysis of this cluster is 0.22 standard deviations above the input value.
The differences between output and input masses are shown in Figure~\ref{f:simulatedns}. This time eight out of the 54 clusters cannot recover the input mass to within one standard deviation. In all eight of these cases, the mass is underestimated with respect to the input value. Five of the outlier values correspond to clusters at low redshift ($z < 0.2$). This suggests that the confusion and CMB noise may be causing AMI to systematically underestimate the cluster masses, and may explain why AMI mass estimates were consistently lower than those obtained by Planck for the real data. 

\subsection{Simulations further adding a canonical radio-source environment}
\label{s:CSsims}
\begin{figure}
  \begin{center}
  \includegraphics[ width=0.9\linewidth]{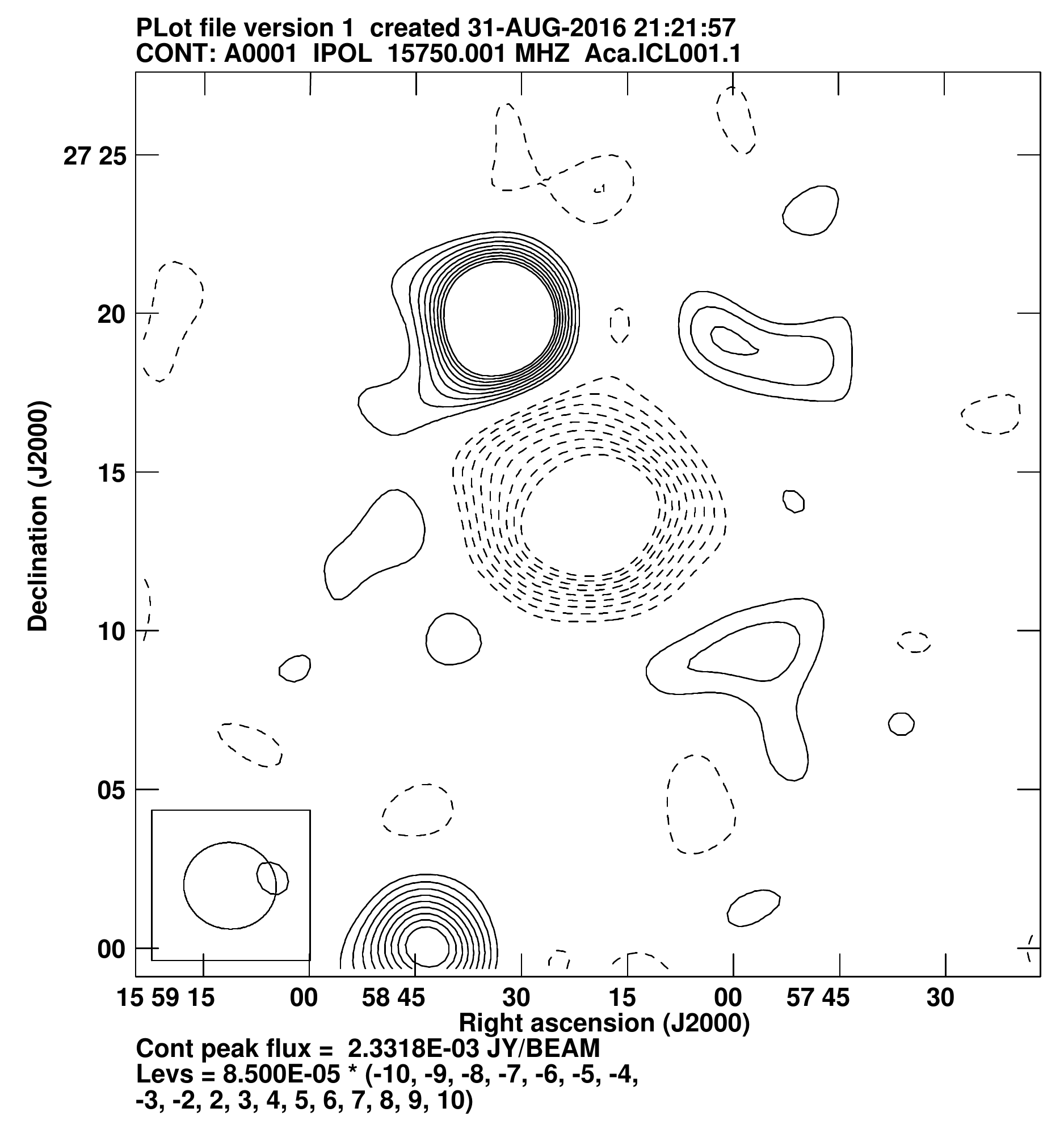}
  \caption{Unsubtracted map produced from simulated AMI data of cluster PSZ2G044.20+48.66, including a canonical radio-source environment as well as instrumental, confusion and CMB noise.}
  \label{f:sim_CS_map}
  \end{center}
\end{figure}

\begin{figure}
  \begin{center}
  \includegraphics[ width=\linewidth]{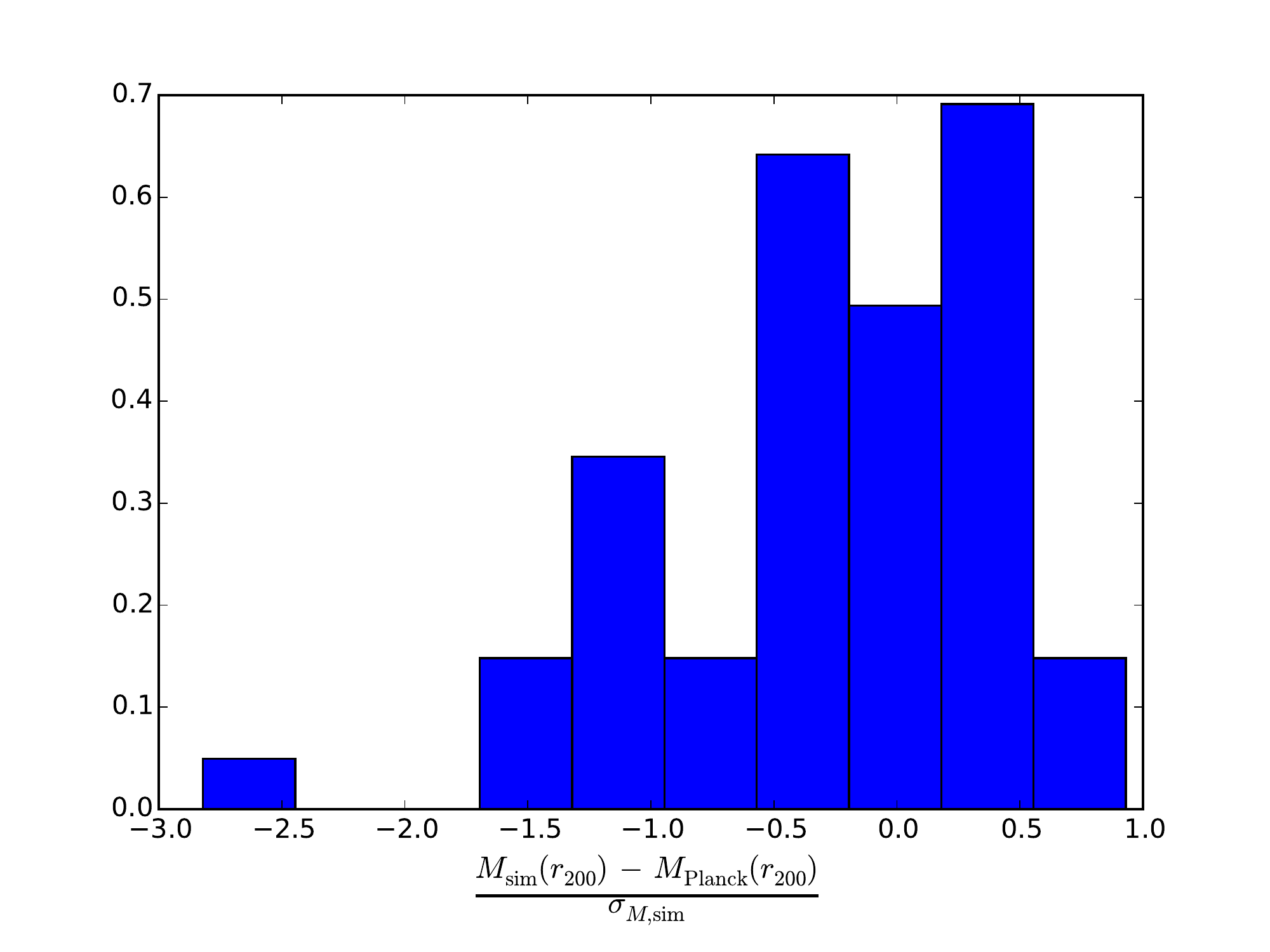}
  \caption{Normalised histogram of the differences between the input and output masses of the AMI simulations, in units of standard deviations of the output mass. This is the case for a canonical radio-source environment as well instrumental, confusion and CMB noise contributions.}
\label{f:simulatedcs}
  \end{center}
\end{figure}

The third set of simulations included detectable radio-sources (Section~\ref{s:identified_sources}, which formed a canonical radio-source environment. They were created in the same way as with the confusion noise described above, but with higher flux limits so that in reality, the LA would have been able to detect them. The upper flux limit was set to $25~\rm{mJy}$.\\
Figure~\ref{f:sim_CS_map} shows the map produced from the simulated data of cluster PSZ2G044.20+48.66, including a canonical source environment and background noise. The mass estimate derived from the Bayesian analysis of this cluster is 0.51 standard deviations below the input value.
Figure~\ref{f:simulatedcs} shows that the canonical radio-source environment have little effect on the mass estimation relative to Section~\ref{s:noisesims}, as there are still 8 clusters which give mass estimates greater than one standard deviation away from the input value. 
Note that in this case, the outliers occurred across the entire range of redshifts, which suggests that in Section~\ref{s:noisesims} the low redshift trend was just a coincidence. 

\subsection{Simulations with LA observed radio-source environment plus instrumental, confusion and CMB noise}
\label{s:sourcesims}
\begin{figure*}
  \begin{center}
  \includegraphics[ width=0.45\linewidth]{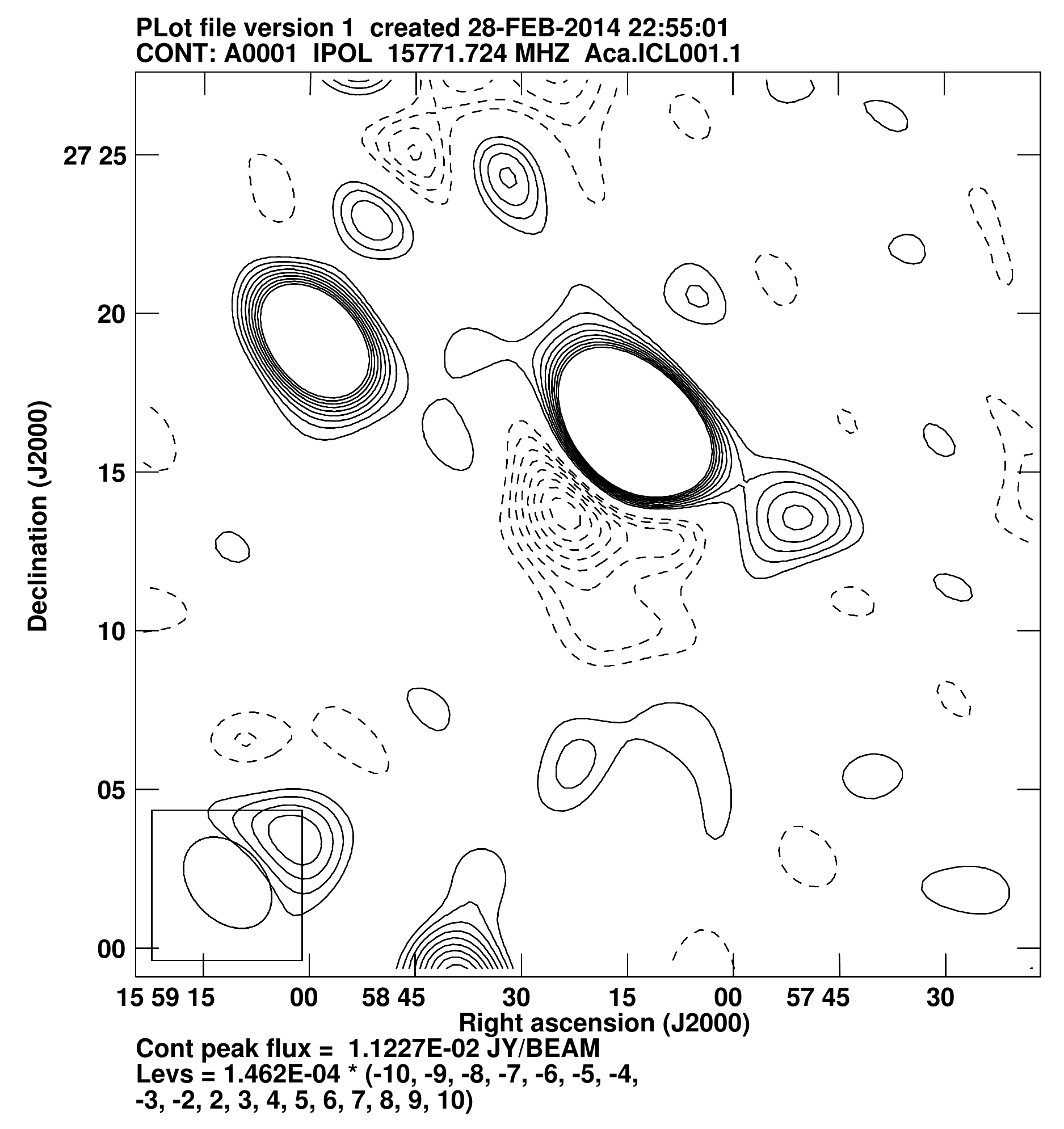}
  \includegraphics[ width=0.45\linewidth]{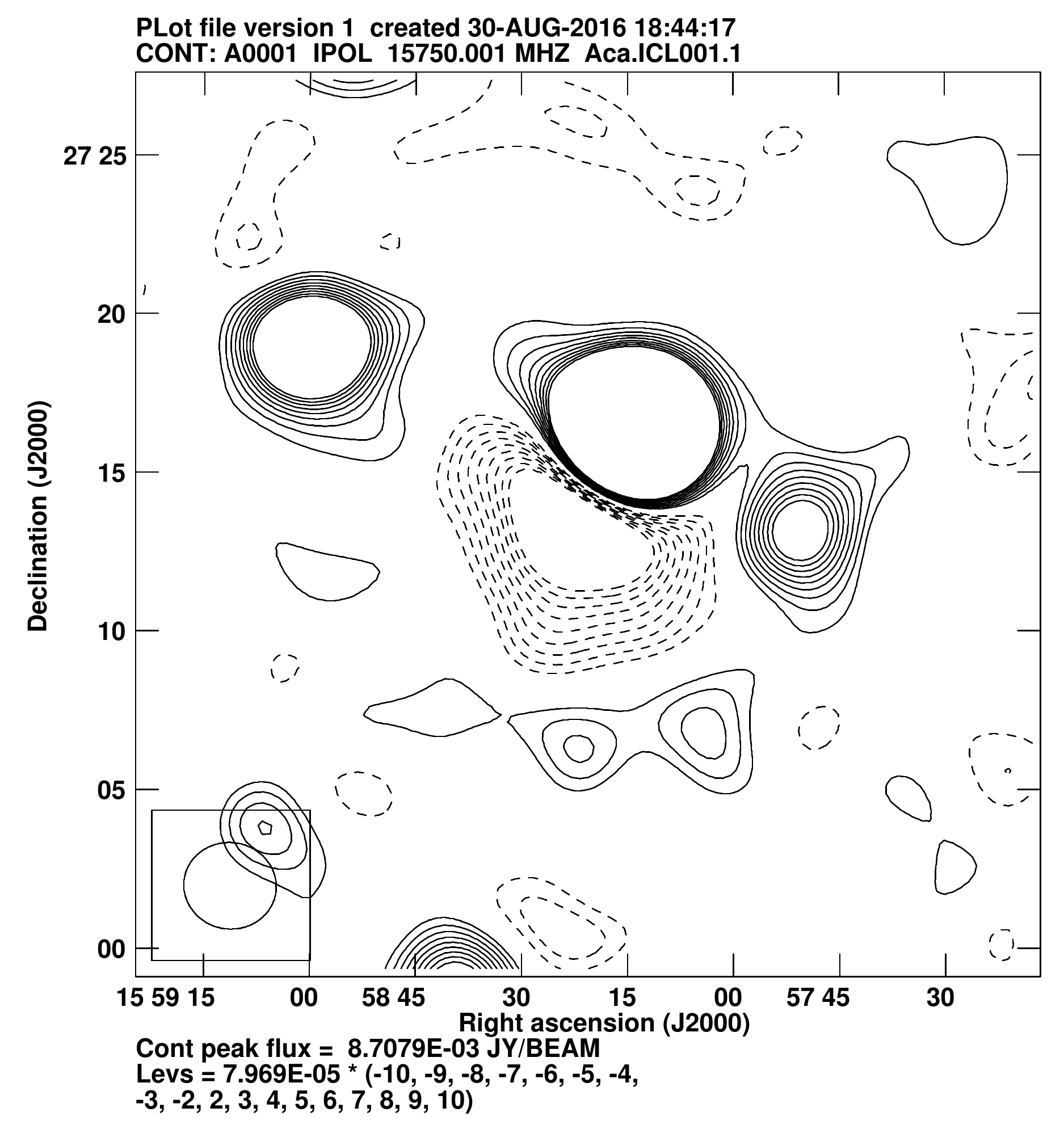}
  \medskip
  \centerline{(a) \hskip 0.45\linewidth (b)}
  \caption{(a) Unsubtracted map produced from real AMI data of cluster PSZ2G044.20+48.66. (b) Unsubtracted map produced from simulated AMI data of PSZ2G044.20+48.66, including the real source environment (as measured by the LA) as well as instrumental, confusion and CMB noise. The peak flux in the simulation has been underestimated relative to the real observation by $\approx 25\%$. This could be due to the source sitting on a negative decrement caused by background noise, or it could be from the cluster decrement.}
  \label{f:sim_rs_map}
  \end{center}
\end{figure*}

\begin{figure}
  \begin{center}
  \includegraphics[ width=\linewidth]{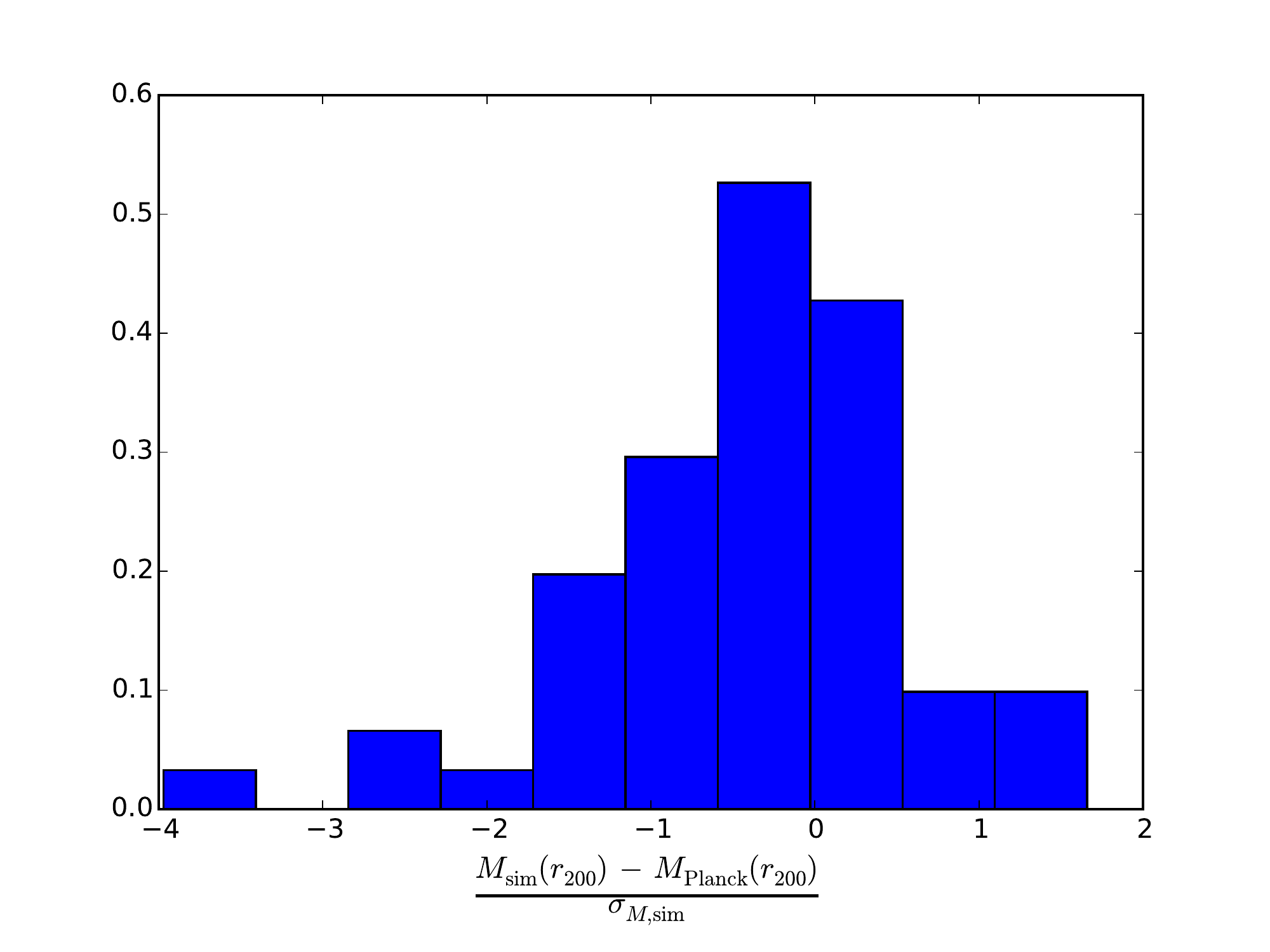}
  \caption{Normalised histogram of the differences between the input and output masses of the AMI simulations, in units of standard deviations of the output mass. This is the case for the real radio-source environment as measured by the LA, with instrumental, confusion and CMB noise contributions.}
\label{f:simulatedrs}
  \end{center}
\end{figure}

The final set of simulations included the radio-source environment measured by the LA during the real observation for each cluster. These are only estimates of the actual source environments, and are only as reliable as the LA's ability to measure them. 
Figure~\ref{f:sim_rs_map} shows the maps produced from the real \& simulated data of cluster PSZ2G044.20+48.66. The mass estimate derived from the Bayesian analysis of the simulated dataset is just 0.08 standard deviations above the input value. \\
Figure~\ref{f:simulatedrs} shows that including the LA observed radio-source environment has a large effect on the results, as this time there are 16 clusters which are more than one standard deviation away from the input mass. Furthermore, three of these overestimated the mass relative to the input, the first time we have seen this occur in any of the simulations. 
A possible source of bias could be due to for example, the empirical prior on the spectral index incorrectly modelling some radio-sources. 
Another source of bias could be the position of a source relative to the cluster, and the magnitude of the source flux. For example, if a high flux radio-source is close to the centre of the galaxy cluster, then even a slight discrepancy between the real and the modelled values for the source could have a large effect on the cluster parameter estimates. 

I now compare these results to the simulations in YP15 (which concluded that the underestimation of the simulation input values could be due to deviation from the `universal' profile, see Figure~23a in the paper). The results of the large cluster simulations (total integrated Comptonisation parameter $= 7 \times 10^3$~arcmin$^2$ and $\theta_{\rm p} = 7.4$~arcmin) in YP15 seem biased low at a more significant level than those in Figure~\ref{f:simulatedrs}, as in the former case less than half of the clusters recover the true value within two standard deviations. For the smaller clusters however, YP15 found a slight upward bias in the simulation results, but this is probably smaller in magnitude than the bias found in this Section. \\

\subsection{Statistics of results of real and simulated data}
\label{s:simsstats}

Looking at the histograms produced in Sections~\ref{s:NSNBsims},~\ref{s:noisesims},~\ref{s:CSsims}, and~\ref{s:sourcesims}, in the last three cases it is apparent that there is a negative skew in the data, i.e. the output masses are negatively biased relative to the input masses. The skews calculated from the samples associated with the four histograms are $-0.17$, $-1.30$, $-0.91$, and $-0.96$ respectively in units of standard deviations of the output mass. This suggests that the inclusion of confusion and CMB noise bias the cluster mass. 
I also calculate the median values of these histograms, and compare them with the medians corresponding to the real AMI and PSZ2 masses given in Figure~\ref{f:m500planck}. The median values for the four histograms are $-0.24$, $0.09$, $-0.27$ and $-0.34$ respectively in units of standard deviations of the output mass. For the real data the median values for $(M_{\rm AMI}(r_{500}) - M_{\rm Pl,\, marg}(r_{500})) / \sigma_{\rm AMI}$ and $(M_{\rm AMI}(r_{500}) - M_{\rm Pl,\, slice}(r_{500})) / \sigma_{\rm AMI}$ are $-1.57$ and $-0.56$. It makes sense to compare the second of these real data values with those obtained from the simulations, as it was $M_{\rm Pl,\, slice}(r_{500})$ which was used to derive the input masses. The fact that the median from the real data is greater in magnitude than the values from the simulations implies in general, our simulations can recover their input values with better agreement than that obtained between real AMI estimates and those obtained from Planck data using the slicing function methodology. This seems plausible as you would expect that inferring results from data which was created using the same model used in the inference would be more accurate than results from data taken from two different telescopes, which use different models in their inference. 
Furthermore the simulation medians tell us that AMI is capable of inferring the masses derived with the slicing methodology, if the cluster is created using the model used in the inference and assuming there are no large discrepancies between the real and simulated AMI observations. 


\section{Conclusions}
\label{s:pl_phys_summary}

We have made observations of galaxy clusters detected by the Planck space telescope, with the Arcminute Microkelvin Imager (AMI) radio interferometer system in order to compare mass estimates obtained from their data. I analysed this data using the physical model described in Section~\ref{s:phys_mod}, following largely the data analysis method outlined in \citet{2009MNRAS.398.2049F}. This allowed us to derive physical parameter estimates for each cluster, in particular the total mass out to a given radius. I have also calculated two mass estimates for each cluster from Planck's PowellSnakes detection algorithm \citep{2012MNRAS.427.1384C} data following \citet{2016A&A...594A..27P} (PSZ2), and found the following.

\begin{itemize}
\item For the AMI mass estimates of Planck selected clusters there is generally a steeping in the mass of galaxy clusters as a function of redshift, which flattens out at around $z \approx 0.5$.
\item AMI $M(r_{500})$ estimates are within one combined standard deviation of the PSZ2 slicing function mass estimates for 31 out of the final sample of 54 clusters. However, the AMI masses are lower than both PSZ2 estimates for 37 out of the 54 cluster sample.
\item The PSZ2 mass estimates derived from the marginalised $Y-\theta$ posteriors are larger than those which use the slicing function in 47 out of 54 cases. This suggests that the X-ray data which form the basis of the slicing procedure predict lower cluster masses relative to what the SZ Planck data alone find.
\end{itemize}

To investigate further the possible biasing of AMI mass estimates, I created simulations of AMI data with input mass values from the PSZ2 slicing methodology. I considered four different cases for the simulations: 1) galaxy cluster plus instrumental noise; 2) galaxy cluster plus instrumental plus confusion \& CMB noise; 3) galaxy cluster plus instrumental, confusion \& CMB noise, plus a randomly positioned radio-source environment; 4) galaxy cluster plus instrumental, confusion \& CMB noise, plus the radio-source environment recognised by the LA in the real observations. These simulated datasets were analysed in the same way as the real datasets, and I found the following.
\begin{itemize}
\item For case 1), the physical model recovered the input mass to within one standard deviation for 51 of the 54 clusters. The three which did not give an underestimate relative to the masses input to the simulation.
\item For case 2), eight of the simulations gave results which were more than one standard deviation lower than the input values. This highlights the effect of incorporating the noise sources into the error covariance matrix rather than trying to model the associated signals explicitly. 
\item Case 3) shows similar results to case 2), which implies that `ideal' radio-sources placed randomly in the sky have little effect on cluster mass estimates.
\item However in case 4) with real source environments, 16 simulations did not recover the input mass to within one standard deviation. This suggests that real radio-source environments, which can include sources with high flux values, and often sources which are located very close to the cluster centre, introduce biases in the cluster mass estimates. In real observations there are also additional issues (the sources are not `ideal'), such as sources being extended and emission not being circularly symmetric on the sky.
\item Cases 2), 3) and 4) give distributions of output $-$ input mass which are negatively skewed. Thus AMI mass estimates are expected to be systematically lower than the PSZ2 slicing methodology values.
\item The median values of the distributions of output $-$ input mass of the simulations in each of the four cases are smaller in magnitude than those obtained from comparing AMI and PSZ2 estimates from real data. This is expected as I used the same model to simulate and analyse the clusters in all four cases. 
\item Compared to the results of simulations of large clusters carried out in \citet{2015A&A...580A..95P}, which test the robustness of the `universal' pressure profile, the case 4) bias appears relatively small in magnitude, and in the same direction (downward). When comparing the case 4) results with the small cluster simulations of \citet{2015A&A...580A..95P}, the latter shows a relatively small bias in the opposite direction.
\item The simulated and real data medians also indicate that while the simulations have shown that AMI mass estimates are systematically low, this does not fully accommodate for the discrepancies in the results obtained from the real data. This suggests that there is a systematic difference between the AMI \& Planck data and / or the cluster models used to determine the mass estimates (which generally leads to PSZ2 estimates being higher than those obtained from AMI data).
\end{itemize}

%% file: CHAP-4/chapter4.tex
\chapter{Comparison of physical and observational galaxy cluster modelling using AMI data}\label{c:fourth}

This Chapter 
provides a follow-up to Chapter~\ref{c:third} in which I performed Bayesian inference on data obtained with the Arcminute Microkelvin Imager (AMI) array to derive estimates of physical properties of clusters that have been detected by Planck. 
I now focus on the observational properties of clusters obtained from telescopes such as AMI and Planck which measure the SZ effect: the angular radius $\theta$, and the integrated Comptonisation parameter $Y$. 
For the sample considered in the previous Chapter, we compare observational parameters derived from the physical model with those obtained from two observational models similar to the one described in YP15 and \citet{2012MNRAS.421.1136A}, using data from AMI. 
I also compare the different models using Bayesian analysis as described in Section~\ref{s:bayes_model}, as well as with another technique presented here (see Section~\ref{s:emd}). The work discussed in this Chapter has been submitted to MNRAS and is under review \citep{2018arXiv180501968J}.


\section{Physical model estimates of observational parameters}
\label{s:phys_obs}

$Y$ can be calculated using the physical model (PM from here on in this chapter) presented in Section~\ref{s:phys_mod}, by first calculating $P_{\rm e}(r)$ and then calculating $Y(r)$ using equation~\ref{e:sz11}. $\theta$ and $r$ are related through $\theta = r / D_{\rm A}$. The prior distributions used are the same as the ones used in the previous Chapter.


\section{Observational models}
\label{s:obs_mod}

Here I consider two observational models, observational model I (OM I) and observational model II (OM II). They are based on the model used in YP15. They use the same GNFW profile (given by equation~\ref{e:gnfw}) to model the gas content, but with the slope parameters stated in Section~\ref{s:phys_mod}; they take into account only the cluster gas -- they do not explicitly model the dark matter component. They work in \textit{angular} rather than physical sizes.
Like the PM, they also use equation~\ref{e:sz11} to calculate $Y$. However, the calculation steps are different.
We start be evaluating equation~\ref{e:sz11} in the limit that $r \rightarrow \infty$. It can be shown that for the GNFW pressure profile this gives (see Appendix~\ref{s:pl_obs_ytot} for a derivation of this result) 
\begin{equation}\label{e:obs_int}
\lim_{r\to\infty} Y_{\rm sph}(r) \equiv Y_{\rm tot} = \lim_{r\to\infty} \frac{\sigma_{\rm T}}{m_{\rm e}c^{2}}\int_{0}^{r}P_{\rm e}(r')4\pi r'^{2} \, \mathrm{d}r' = \frac{4 \pi P_{\rm{ei}} D_{\rm{A}} \theta_{\rm p}^3 \sigma_{\rm T}}{m_{\rm e}c^{2}} \frac{\Gamma \left( \frac{3 - c}{a} \right) \Gamma \left( \frac{b - 3}{a} \right)}{a \Gamma \left( \frac{b - c}{a} \right)}
\end{equation}  
where $\Gamma(x)$ is the Gamma function and $\theta_{\rm p} = r_{\rm p} / D_{\rm A}$. Note that for finite $r$ (and thus $\theta$)
\begin{equation}\label{e:obs_Y}
Y_{\rm sph}(\theta) = \frac{4 \pi P_{\rm{ei}} D_{\rm{A}} \theta_{\rm p}^2 \sigma_{\rm T}}{m_{\rm e}c^{2}}\int_{0}^{\theta} \left( \frac{\theta'}{\theta_{\rm p}} \right)^{2-c} \left( 1 + \left( \frac{\theta'}{\theta_{\rm p}} \right)^a \right)^{(c-b)/a}\, \mathrm{d}\theta'.
\end{equation}  
Both equations have a common (unknown) factor $D_{\rm A} P_{\rm{ei}}$. Hence for given (i.e. input) values of $Y_{\rm tot}$ and $\theta_{\rm p}$, equation~\ref{e:obs_int} can be solved for $D_{\rm A} P_{\rm{ei}}$ and then equation~\ref{e:obs_Y} can be solved for finite $\theta$ numerically. Furthermore the OMs assume that the cluster is spherically symmetric and that the cluster gas can be described by the equation of state of an ideal gas.  
The OMs have four cluster input parameters: $Y_{\rm tot}$, $\theta_{\rm p}$ , $x_{\rm c}$ and $y_{\rm c}$. They differ only in the prior distributions they use. 


\subsection{Observational model I prior}
\label{s:obs_i_priors}
The priors used on $Y_{\rm tot}$ and $\theta_{\rm p}$ are the same as the `new' priors used in YP15. These were derived from the Planck completeness simulations \citep{2014A&A...571A..29P} as follows. The simulations were produced by drawing a cluster population from the Tinker mass function \citep{2008ApJ...688..709T} and using the scaling relations in \citet{2011A&A...536A..11P} to obtain observable quantities. This cluster population was injected into the real Planck data and a simulated union catalogue was created by running the Planck detection pipelines on this simulated dataset. An elliptical Gaussian function was then fitted to the posterior of $Y_{\rm tot}$ and $\theta_{\rm p}$ in log space. Hence the prior has the Planck selection function implicitly included in it. \\
For consistency, the same cluster centre priors were used in both observational models as in the PM. The priors for OM I are summarised in Table~\ref{t:obs_i_priors}.
\begin{table}
\centering
\begin{tabular}{{l}{c}}
\hline
Parameter & Prior distribution \\ 
\hline
$x_{\rm c}$ & $\mathcal{N}(0'', 60'')$ \\
$y_{\rm c}$ & $\mathcal{N}(0'', 60'')$ \\
$\log  (Y_{\rm tot}), \, \log (\theta_{\rm p})$ & $\mathcal{N}((-2.7, 0.62), (0.29, 0.12), 40.2^{\circ})$ \\
\hline
\end{tabular}
\caption{Observational model I input parameter prior distributions. Note that the Gaussian elliptical function on $\log  (Y_{\rm tot}) - \log (\theta_{\rm p})$ is parameterised in terms of the mean in both dimensions, the respective standard deviations and the offset of the principle axes from the vertical and horizontal axes measured clockwise.}\label{t:obs_i_priors}
\end{table}


\subsection{Observational model II}
\label{s:obs_ii_priors}

The priors on $Y_{\rm tot}$ and $\theta_{\rm p}$ in OM II incorporate the spectroscopic or photometric redshift of each cluster. 
From the $z$ and $M(r_{200})$ priors of the PM and for $f_{\rm{gas}}(r_{200}) = 0.13$, upper and lower bounds on $Y_{\rm tot}$ and $\theta_{\rm p}$ are calculated using the PM.
Note that $Y_{\rm tot}$ and $\theta_{\rm p}$ are assumed to be a-priori \textit{uncorrelated}, unlike in OM I. For the lowest redshift cluster ($z = 0.0894$), these limits are $ \theta_{\rm p,\, min} = 4.24~\rm{arcmin}$, $ \theta_{\rm p,\, max} = 19.04~\rm{arcmin}$, $ Y_{\rm tot,\, min} =  1.06 \times 10^{-4}~\rm{arcmin}^{2}$ and $ Y_{\rm tot,\, max} = 0.19~\rm{arcmin}^{2}$; for the highest redshift ($z = 0.83$) cluster these limits are $ \theta_{\rm p,\, min} = 0.67~\rm{arcmin}$, $ \theta_{\rm p,\, max} = 3.01~\rm{arcmin}$, $ Y_{\rm tot,\, min} =  5.7 \times 10^{-6}~\rm{arcmin}^{2}$ and $ Y_{\rm tot,\, max} = 0.01~\rm{arcmin}^{2}$. It clear that $z$ has a large effect on the PM calculations, as it is used to calculate the angular scale from $r$ through $\theta = r / D_{\rm A}(z)$ where $D_{\rm A}(z)$ is the angular diameter distance of the cluster at redshift $z$, and to convert the units of $Y$. It is also used to calculate $c_{200}$ which affects the scale of the self-similar dark matter density profile, and the normalisation constant $\rho_{\rm{s}}$ in equation~\ref{e:nfw} is proportional to $\rho_{\rm{crit}}(z)$.
The priors for OM II are summarised in Table~\ref{t:obs_ii_priors}.
\begin{table}
\centering
\begin{tabular}{{l}{c}}
\hline
Parameter & Prior distribution \\ 
\hline
$x_{\rm c}$ & $\mathcal{N}(0'', 60'')$ \\
$y_{\rm c}$ & $\mathcal{N}(0'', 60'')$ \\
$\theta_{\rm p}$ & $\mathcal{U} [ \log ( \theta _{\rm p,\, min}(z) ),\log ( \theta _{\rm p,\, max}(z) )]$ \\
$Y_{\rm tot}$ & $\mathcal{U} [ \log ( Y_{\rm tot, \, min}(z) ),\log ( Y_{\rm tot,\, max}(z) )]$ \\
\hline
\end{tabular}
\caption{Observational model II input parameter prior distributions.}\label{t:obs_ii_priors}
\end{table}
Note that in using the PM calculations to calculate the prior limits, we have made the assumptions underlying the PM that OM I is not subject to (i.e. hydrostatic equilibrium up to radius $r_{200}$ and $f_{\rm{gas}}$ is much less than unity up to the same radius).


\section{AMI model comparisons}
\label{s:obs_results_i}

I now use AMI data to compare the PM, OM I and OM II, and begin by comparing their observational parameter estimates.
Secondly I introduce a metric which measures the `distance' between probability distributions. In this context the distance is measured between the $\left(Y(r_{500}), \,\theta_{500}\right)$ posterior distributions of the three models. Finally the models are compared using the evidence ratios introduced in Section~\ref{s:bayes_model}.
The results obtained from these analyses are given in Appendix~\ref{c:appendixb}, which lists the values obtained for the 54 cluster sample in ascending order of $z$. \\
I emphasise the notation used for $Y$. For consistency I parameterise $Y$ by $r$ for all three models ($Y \equiv Y(r)$). For the PM, $Y(r)$ has units [length$^{2}$]; to convert this to the more conventional [angle$^{2}$] we divide by $D_{A}^{2}$: $ Y(r) \rightarrow Y(r)/D_{A}^{2} $ as mentioned in Section~\ref{s:sz_theory}. The $Y$ value given by an OM is naturally in units of [angle$^{2}$]; when I refer to $Y(r)$ in the context of the OMs I equivalently mean $Y(\theta)$.


\subsection{Physical and observational models Y values comparison}
\label{s:amiycomparison}


\begin{figure*} 
  \begin{center}
  \includegraphics[ width=0.90\linewidth]{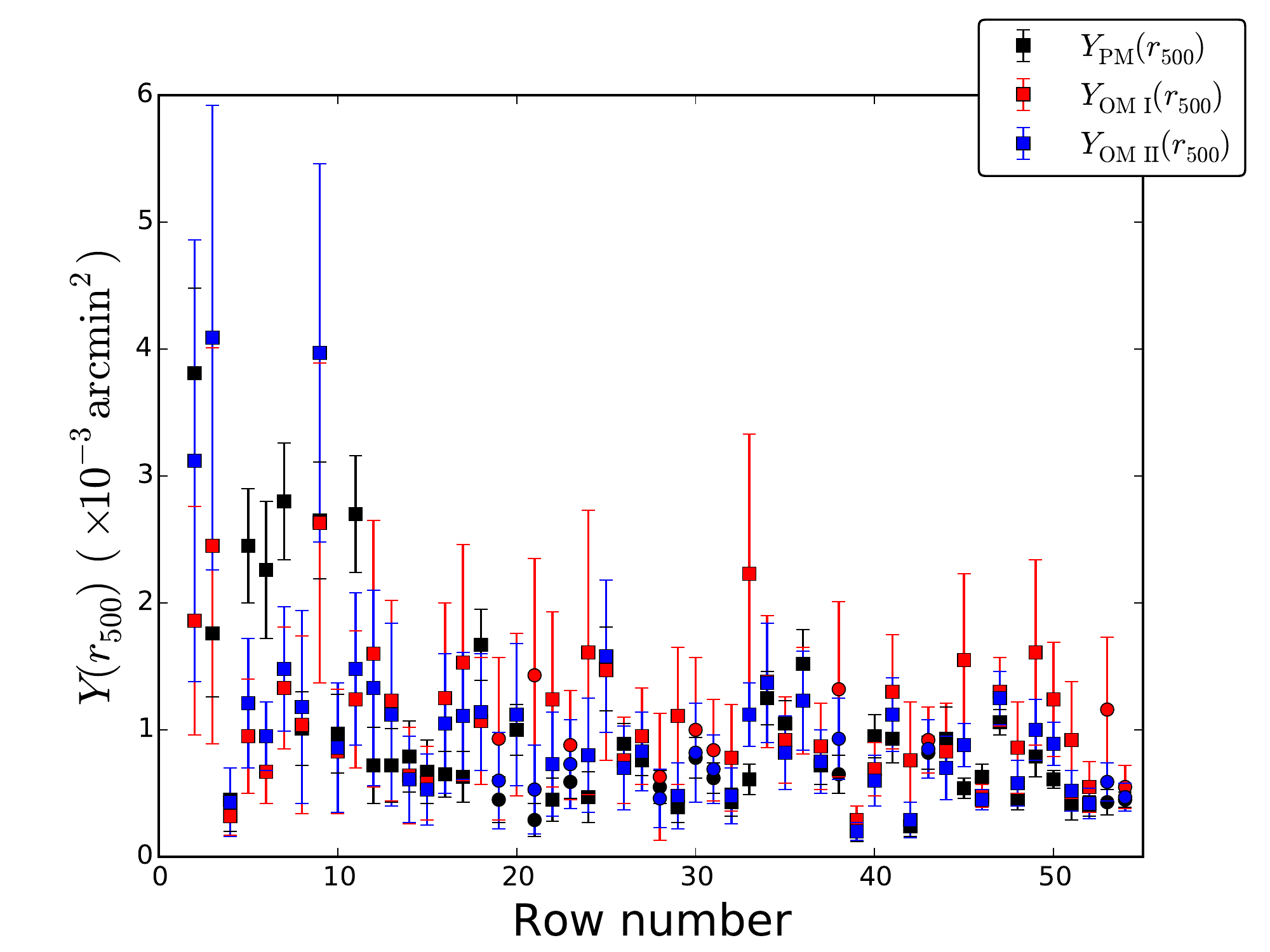}
  \caption{Plot of $Y(r_{500})$ obtained from AMI data using the physical and observational models vs row number of Table~\ref{t:pl_obs_results1}. The points with circular markers correspond to clusters whose redshifts were measured photometrically as opposed to spectroscopically. For clarity purposes the first row is not plotted due to its relatively large value ($Y(r_{500}) \approx 10~\rm{arcmin}^{2}$).}
\label{f:y500ami}
  \end{center}
\end{figure*}


\begin{figure*} 
  \begin{center}
  \includegraphics[ width=0.90\linewidth]{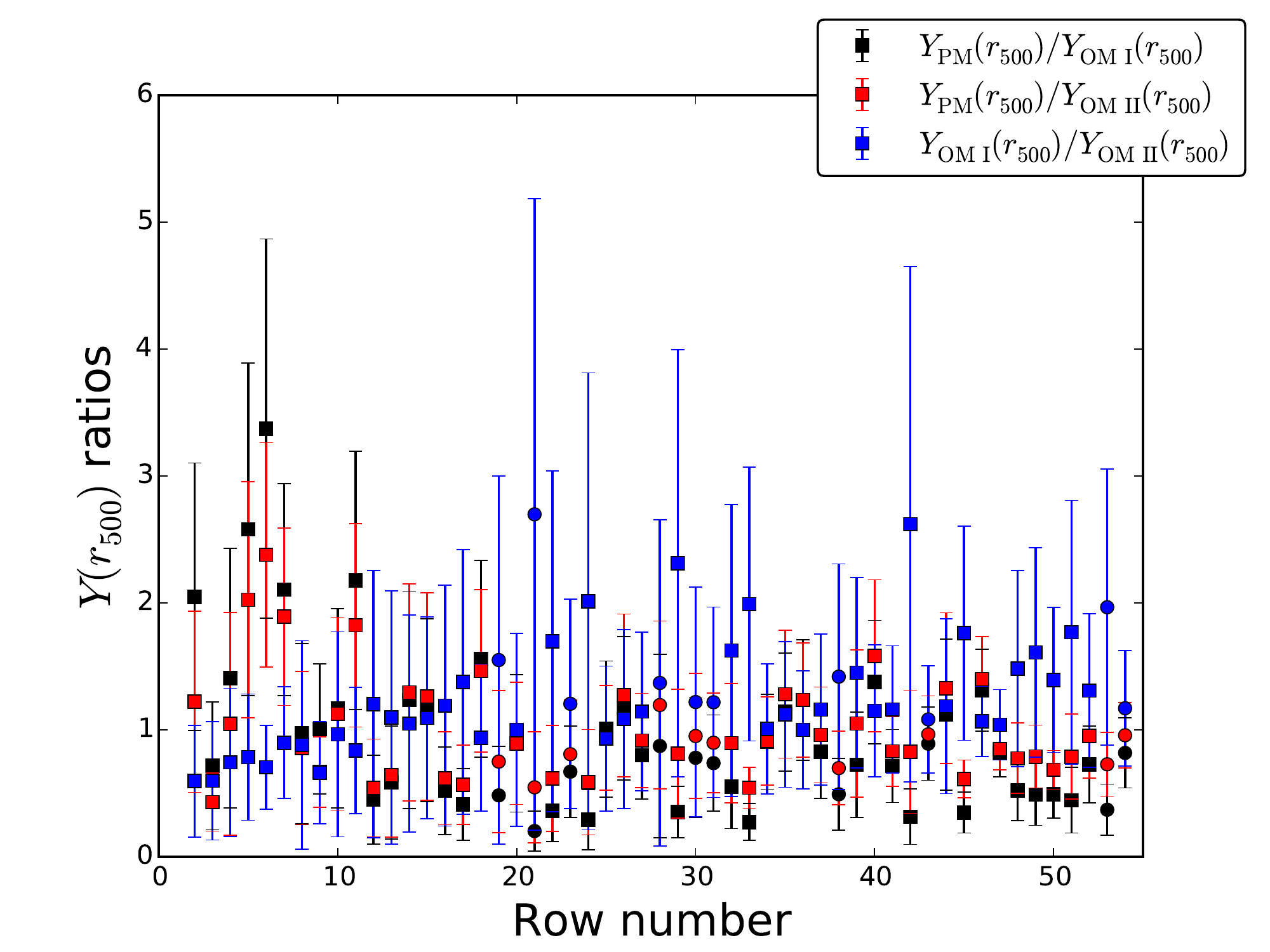}
  \caption{Plot of $Y(r_{500})$ ratio vs row number of Table~\ref{t:pl_obs_results1} for three different cases: $Y_{\rm PM}(r_{500}) / Y_{\rm OM~I}(r_{500})$; $Y_{\rm PM}(r_{500}) / Y_{\rm OM~II}(r_{500})$ and  $Y_{\rm OM~I}(r_{500}) / Y_{\rm OM~II}(r_{500})$. The points with square markers correspond to clusters whose redshifts were measured spectroscopically, and the circular markers photometrically (as listed in Table~\ref{t:pl_obs_results1}).}
\label{f:y500amifrac}
  \end{center}
\end{figure*}

Figure~\ref{f:y500ami} shows the posterior mean values for $Y(r_{500})$ for the three models used on the same AMI datasets. I first note that the errors associated with the OM estimates are generally larger than those with the PM. Secondly it appears that the OM I $Y$ are less strongly correlated with $z$ than those from the PM and OM II. This may be because OM I contains no explicit $z$-information, and in fact its only reliance on $z$ is from the simulated and empirical datasets used to fit its prior distribution, but the same prior is used for all clusters, and so the dependence on redshift is very weak. \\
I now compare the results from the three models pairwise. Note that when we refer to the dispersion between values in units of standard deviations, 
we are referring to the combined standard deviation of the two $Y$ values.
When comparing PM and OM I values of $Y$, just 15 clusters are within one standard deviation, 27 within two and 18 are more than three standard deviations away from each other. The same comparison between PM and OM II gives corresponding values of 23, 40 and 5. This implies that the dispersion between OM II and PM is much smaller (especially in the extreme cases), and shows the importance in the choice of priors. 
Table~\ref{t:physdispersion} gives a summary of the dispersion of the PM with respect to the OMs.
Figure~\ref{f:y500amifrac} shows the fractional difference between the $Y$ values for the three models, and shows that the PM estimates are generally much higher than both OM values at low $z$. However, in general the PM yields lower $Y$ estimates compared to the OMs (PM underestimates $Y$ relative to OM I and OM II 35 and 36 times respectively).

\begin{table*}
\small
\centering
\begin{tabular}{{l}{c}{c}{c}}
\hline
 Model comparison ($Y_{\mathcal{M}_{i}} \equiv$) & $|Y_{\rm{PM}} - Y_{\mathcal{M}_{i}}| / \sigma_{\rm{PM\&} \mathcal{M}_{i}} < 1$ & $|Y_{\rm{PM}} - Y_{\mathcal{M}_{i}}| / \sigma_{\rm{PM\&} \mathcal{M}_{i}} < 2$ & $|Y_{\rm{PM}} - Y_{\mathcal{M}_{i}}| / \sigma_{\rm{PM\&} \mathcal{M}_{i}} > 3$ \\ 
\hline
$Y_{\rm OM \, I}$ & $15$ & $27$ & $18$ \\
$Y_{\rm OM \, II}$ & $23$ & $40$ & $5$ \\
\hline
\end{tabular}
\caption{Difference between physical model mean values for $Y(r_{500})$ \& observational model mean values, measured in units of the physical model $Y(r_{500})$ standard deviation. The numbers in the columns correspond to the number of clusters out of the sample of 54 which satisfy the criterion specified in the respective header.}\label{t:physdispersion}
\end{table*}
Looking at the dispersion between OM I and OM II, 36 clusters are within one standard deviation, four within two and just four are more than three standard deviations away from each other. This implies that OM II seems to be in reasonable agreement with the two other models (usually in between the values from the other models). 



\subsection{Earth Mover's distance}
\label{s:emd}
The Earth Mover's distance (EMD), first introduced in \citet{rubner} is a "distance" function defined between two distributions. In the case where these distributions integrate over all space to the same value (e.g. they are probability distributions), the EMD is given in terms of the first Wasserstein distance \citep{levina}. 
A common analogy used to describe the EMD is the following: if the probability distributions are interpreted as two different ways of piling up a certain amount of earth, and the amount of earth at position $\vec{x}_{i}$ and $\vec{x}_{j}$ belonging to each probability distribution at those points are $P_{1}(\vec{x}_{i})$ and $P_{2}(\vec{x}_{j})$, then the EMD is the minimum cost of moving one pile into the other, where the cost of moving each "spadeful" is taken to be the mass of each spadeful ($f_{ij}$) $\times$ the distance by which it is moved ($|\vec{x_{i}} - \vec{x_{j}}|$).
For discrete two-dimensional probability distributions $P_{1}$ \& $P_{2}$, with two-dimensional domains $\vec{x}_{i}$ \& $\vec{y}_{j}$, then the EMD between these probability distributions $d_{\rm EMD}(P_{1}, P_{2})$ is defined to be the minimum value of
\begin{equation}\label{e:emd}
W(P_{1}, P_{2}) = \sum_{i = 1}^{m} \sum_{j = 1}^{n} f_{ij} |\vec{x_{i}} - \vec{y_{j}}|
\end{equation}
with respect to distance and $f_{ij}$.
Here $m$ and $n$ are the number of values in the domains of $P_{1}$ and $P_{2}$ respectively and $f_{ij}$ are the `flow' of probability density from $P_{1}(\vec{x}_{i})$ to $P_{2}(\vec{y}_{j})$. Different implementations of the algorithm use different distance measures, but we use the Euclidean distance in equation~\ref{e:emd}. The $f_{ij}$ are subject to the following constraints
\begin{equation}\label{e:emdconstraint1}
f_{ij} \geq 0, \, 1 \leq i \leq m, \, 1 \leq j \leq n;
\end{equation}
\begin{equation}\label{e:emdconstraint2}
\sum _{j=1}^{n}f_{ij} = P_{1}(\vec{x}_{i}), \, 1 \leq i \leq m;
\end{equation}
\begin{equation}\label{e:emdconstraint3}
\sum _{i=1}^{m}f_{ij} = P_{2}(\vec{y}_{j}), \, 1 \leq j \leq n;
\end{equation}
\begin{equation}\label{e:emdconstraint4}
\sum _{i = 1}^{m} \sum _{j = 1}^{n} f_{ij} = \sum _{i = 1}^{m} P_{1}(\vec{x}_{i}) = \sum _{j = 1}^{n}P_{2}(\vec{y}_{j}) = 1.
\end{equation}
For a more detailed account of the EMD see \citet{levina}.


\subsection{Application of EMD}
\label{s:app_emd}

\begin{figure*}
  \begin{center}
  \includegraphics[ width=0.45\linewidth]{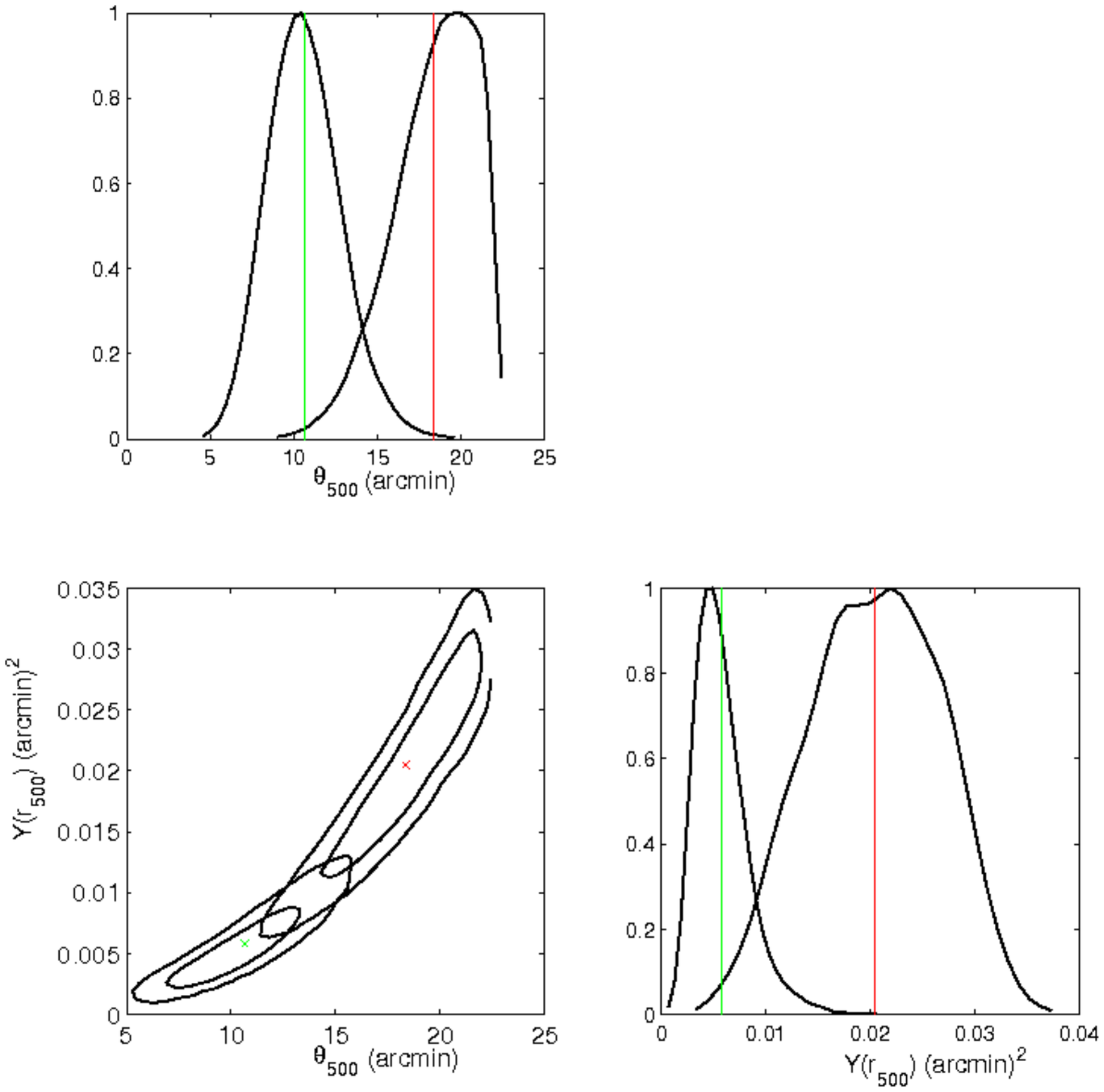}
  \includegraphics[ width=0.45\linewidth]{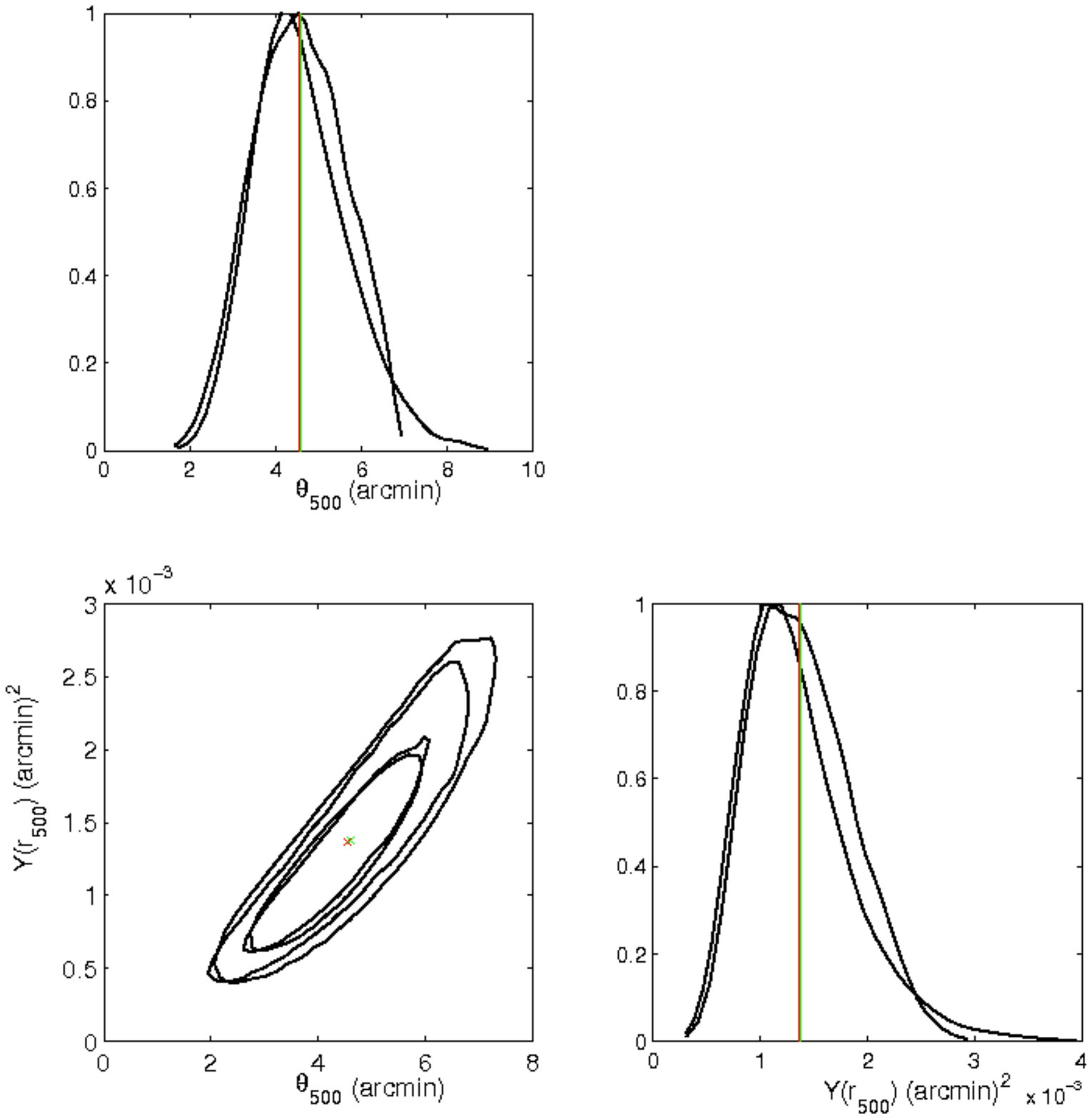}
  \medskip
  \centerline{(a) \hskip 0.45\linewidth (b)}
  \caption{(a) Highest $d_{\rm EMD}$ value $Y(r_{500})-\theta_{500}$ posteriors for cluster PSZ2G044.20+48.66 at $z = 0.0894$. (b) Lowest $d_{\rm EMD}$ value $Y(r_{500})- \theta_{500}$ posteriors for cluster PSZ2G132.47-17.27 at $z = 0.341$. For both triangle plots, the top graph shows the marginalised $\theta_{500}$ posteriors for OM II and OM I. The bottom right graph shows the marginalised $Y(r_{500})$ posteriors. The bottom left graph shows the two-dimensional $Y(r_{500})-\theta_{500}$ posteriors from which the EMD is calculated. The contours represent the 95\% and 68\% confidence intervals. Note that the parameters in the plots are not normalised, but the ones in the distance calculations are normalised by transforming the parameters as discussed in the text. For all of the plots, the green crosses / lines are the mean values of the OM I posteriors (the smaller values in (a)) and the red crosses / lines are the mean values of the OM II posteriors (the larger values in (a)). For Figure (b), the mean values for $Y(r_{500})$ are so close together that the lines cannot be distinguished.}
  \label{f:highlow_emd}
  \end{center}
\end{figure*}

\begin{figure}
  \begin{center}
  \includegraphics[ width=0.90\linewidth]{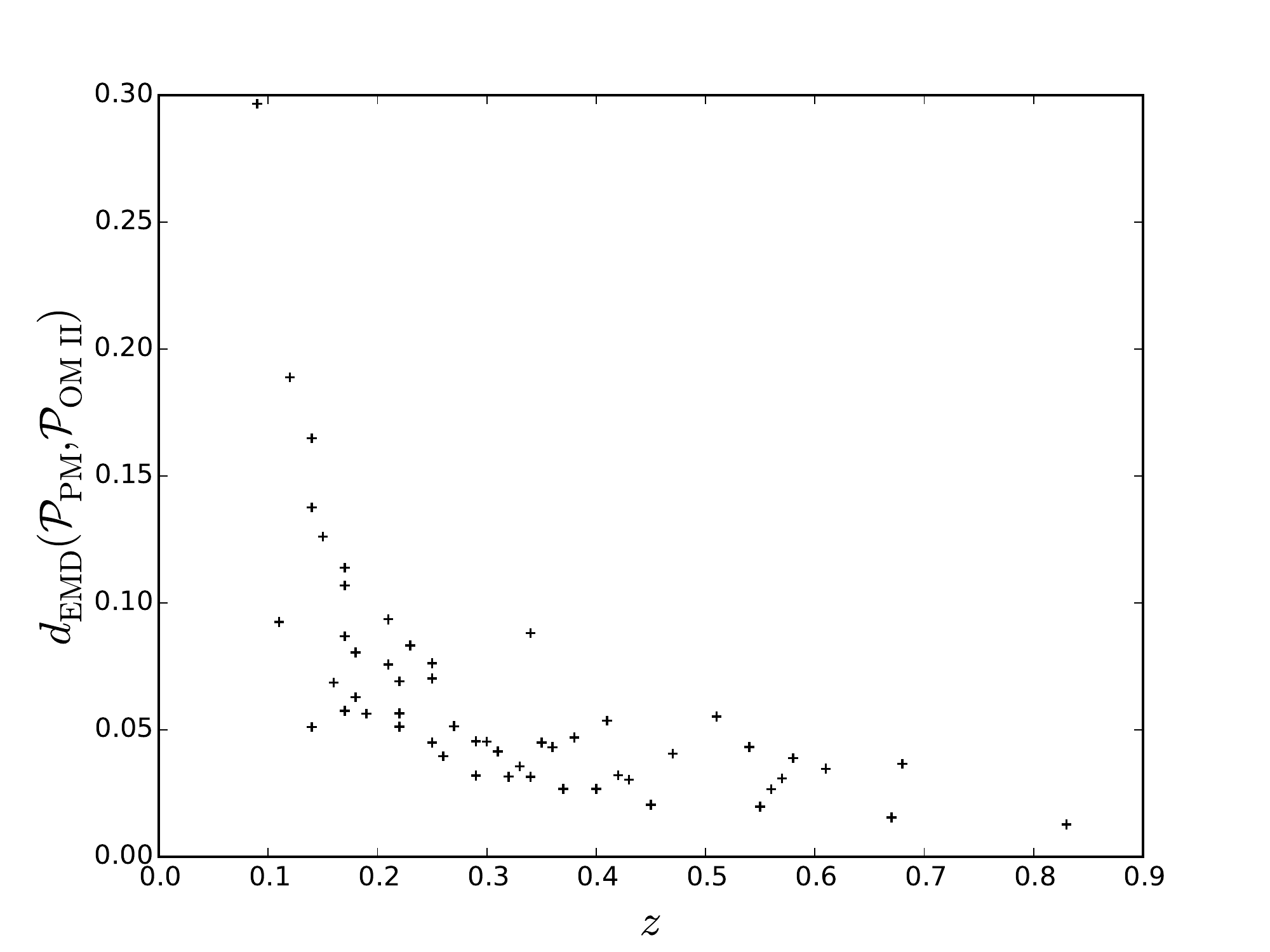}
  \caption{Earth Mover's distance calculated between $Y(r_{500})-\theta_{500}$ posteriors for PM and OM II, versus $z$ for the 54 clusters. The crosses indicate the point-- they are not error bars.}
\label{f:physobsii_emd}
  \end{center}
\end{figure}

The EMD metric is applied to the different pairs of models using Gary Doran's wrapper\footnote{\url{https://github.com/garydoranjr/pyemd}.} for Yossi Rubner's algorithm \citep{rubner}. Before running the algorithm the $\left(Y(r_{500}), \, \theta_{500} \right)$ posteriors are normalised so that the metric is not skewed towards $\theta_{500}$ (the use of Euclidean distances in the EMD algorithm, are obviously misrepresentative if the dimensions are not normalised). Each dimension is normalised to the range $[0,1]$ by performing the following transformations
\begin{equation}\label{e:emdtransformations}
\theta_{500} \rightarrow \frac{\theta_{500} - \theta_{500, \, \rm min}}{\theta_{500, \, \rm max} - \theta_{500, \, \rm min}} ; Y(r_{500}) \rightarrow \frac{Y(r_{500}) - Y_{\rm min}(r_{500})}{Y_{\rm max}(r_{500}) - Y_{\rm min}(r_{500})}.
\end{equation}
The values for $\theta_{500, \, \rm min}$, $\theta_{500, \, \rm max}$, $Y_{\rm min}(r_{500})$ and $Y_{\rm max}(r_{500})$ are deduced by considering all of the values of $Y(r_{500})$ and $\theta_{\rm p}$ from the posteriors obtained from the three models at once, to ensure that all posterior values are normalised by the same factor. The larger the value of the EMD, the `further away' the distributions are from each other. The EMD was calculated for each cluster with each pair of models (giving $3 \times 54 = 162$ distances in total). The full set of EMD values calculated can be found in Table~\ref{t:pl_obs_results2} in Appendix~\ref{c:appendixb}. Table~\ref{t:summary_emd} provides a summary of $d_{\rm EMD}(\mathcal{P}_{\rm PM}, \mathcal{P}_{\rm OM\, I})$, $d_{\rm EMD}(\mathcal{P}_{\rm OM\, I}, \mathcal{P}_{\rm OM\, II})$, $d_{\rm EMD}(\mathcal{P}_{\rm PM}, \mathcal{P}_{\rm OM\, II})$, and the union of the three. 
\begin{table*}
\centering
\begin{tabular}{{l}{c}{c}{c}{c}}
\hline
Statistic & $d_{\rm EMD}(\mathcal{P}_{\rm PM}, \mathcal{P}_{\rm OM\, I})$ & $d_{\rm EMD}(\mathcal{P}_{\rm PM}, \mathcal{P}_{\rm OM\, II})$ & $d_{\rm EMD}(\mathcal{P}_{\rm OM\, I}, \mathcal{P}_{\rm OM\, II})$ & union \\
\hline
mean & $ 0.093$ & $ 0.067$ & $ 0.057$ & $ 0.072$ \\
standard deviation & $ 0.057$ & $ 0.050$ & $ 0.077$ & $ 0.064$ \\
median & $ 0.076$ & $ 0.051$ & $ 0.027$ & $ 0.051$ \\
min & $ 0.020$ & $ 0.013$ & $ 0.006$ & $ 0.006$ \\
max & $ 0.225$ & $ 0.297$ & $ 0.514$ & $ 0.514$ \\
\hline
\end{tabular}
\caption{Summary of EMD values calculated between the $Y(r_{500})-\theta_{500}$ posterior distributions from all three model pairs, and their union.}\label{t:summary_emd}
\end{table*}
Concerning both mean and median, the posteriors are most discrepant between the PM and OM I, followed by PM and OM II. However it is interesting to note that the two largest EMD values come from $d_{\rm EMD}(\mathcal{P}_{\rm OM\, II}, \mathcal{P}_{\rm OM\, I})$ and $d_{\rm EMD}(\mathcal{P}_{\rm PM}, \mathcal{P}_{\rm OM\, II})$ cases, with values $0.514$ and $0.297$ respectively. Furthermore these are from the same cluster, which is at the lowest $z$ ($= 0.0894$). This suggests that incorporating $z$ information into an observational model for very low redshift clusters has a significant effect. 
Ignoring the lowest redshift cluster (or by looking at the median value, which is skewed less by outliers), it is clear that of the three models, OM I and OM II posteriors are most in agreement with each other. Figure~\ref{f:highlow_emd} shows the $Y(r_{500}), \, \theta_{500}$ posterior distributions created using \textsc{GetDist} (with the 95\% and 68\% confidence intervals plotted), for the highest and lowest EMD values obtained from the 162 values calculated. Both of these come from OM II $-$ OM I comparisons. \\
Figure~\ref{f:physobsii_emd} shows $d_{\rm EMD}(\mathcal{P}_{\rm PM}, \mathcal{P}_{\rm OM\, II})$ vs $z$ from which it is apparent that there is a negative correlation between $d_{\rm EMD}$ and $z$. 

\subsection{Physical and observational models comparison}
\label{s:phys_obs_comp}

As described in Section~\ref{s:bayes_model}, one can perform a model comparison, by comparing the Bayesian evidence values calculated when the models were applied to the same (AMI) datasets. We can also define the detection ratio of a model as the ratio of the evidences of the `data' and `null-data' runs. The first of these corresponds to modelling the cluster, background and detectable radio-sources. The null-data run models everything but the cluster. The ratio of these evidences therefore gives a measure of the significance that the cluster has in modelling the data. Note that the null-data run is the same for all three models considered here, as they only differ in the way they model the galaxy cluster itself. Table~\ref{t:pl_obs_results2} in Appendix~\ref{c:appendixb} gives the log of a detection ratio, $\ln (\mathcal{Z}_{i} / \mathcal{Z}_{\rm null})$ for each of the three models, and the ratios between the different pairs of models, $\ln (\mathcal{Z}_{i} / \mathcal{Z}_{j})$ where $\mathcal{Z}_{i}$ and $\mathcal{Z}_{j}$ are one of $\mathcal{Z}_{\rm{PM}}$, $\mathcal{Z}_{\rm{OM I}}$ or $\mathcal{Z}_{\rm{OM II}}$, for each cluster.


\subsubsection{Physical model and observational model I} 

\begin{figure*}
  \begin{center}
  \includegraphics[ width=0.45\linewidth]{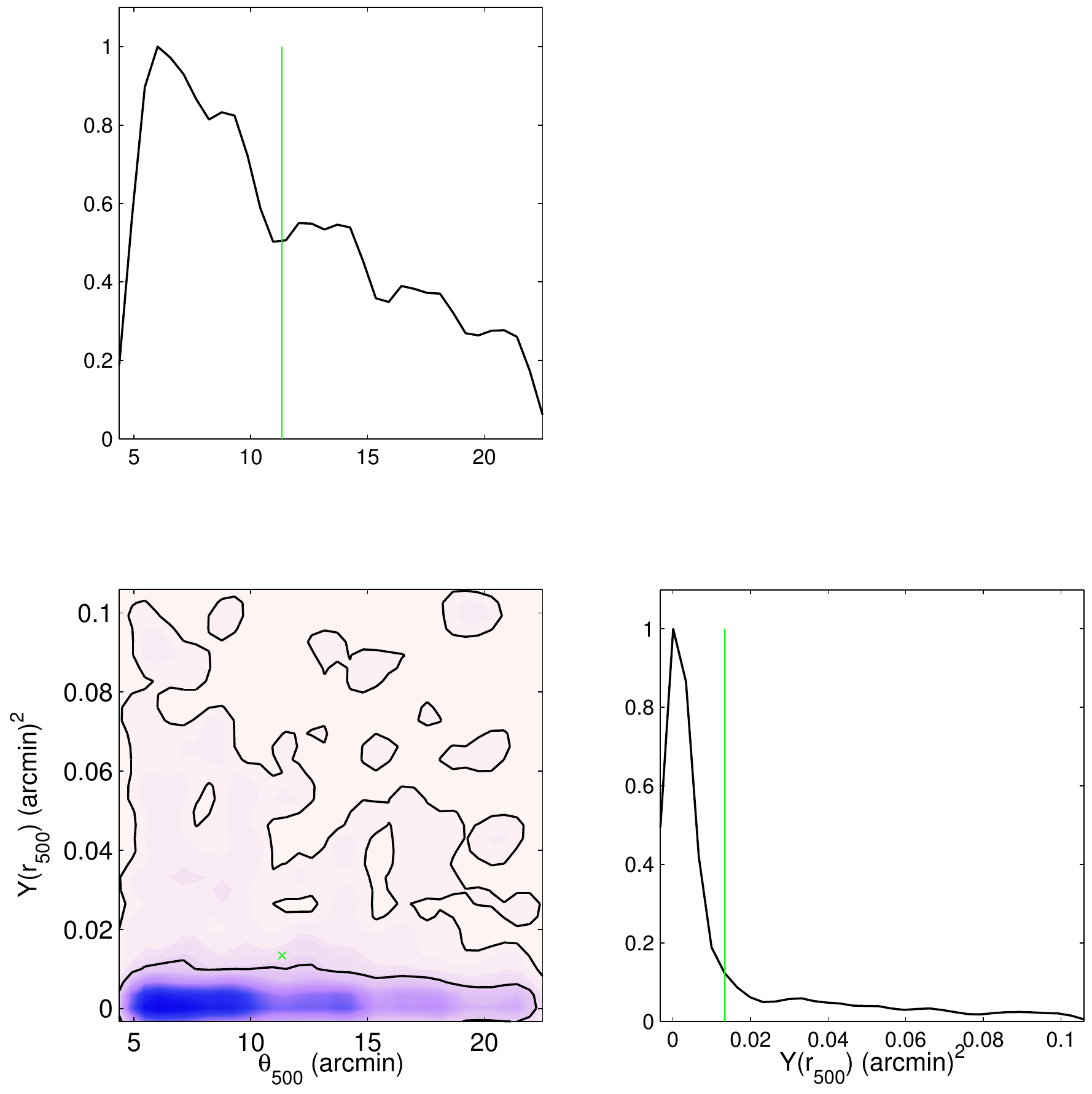}
  \includegraphics[ width=0.45\linewidth]{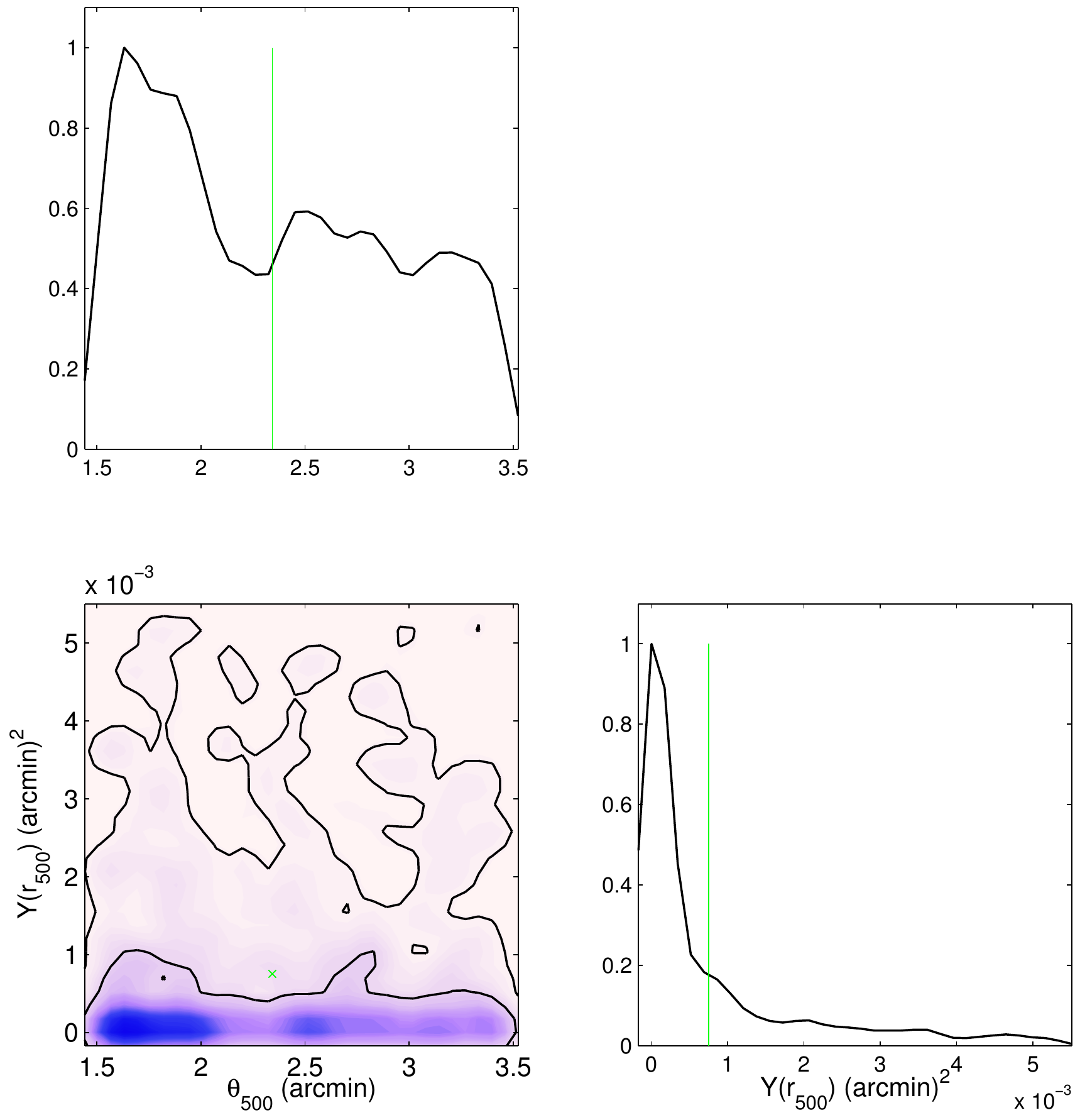}
  \medskip
  \centerline{(a) \hskip 0.45\linewidth (b)}
  \caption{(a) Lowest $z$ ($= 0.0894$) prior parameter space for $Y(r_{500})-\theta_{500}$ using the PM. (b) Highest $z$ ($= 0.83$) prior parameter space for $Y(r_{500})-\theta_{500}$ for the PM and OM II. Note the scales on the axes are different for each plot, and the green vertical lines represent the mean values.}
  \label{f:phys_priors}
  \end{center}
\end{figure*}

The data favour OM I over the PM for 50 of the 54 clusters. Though in 36 of the 50 cases $\log(\mathcal{Z}_{\rm{PM}} / \mathcal{Z}_{\rm{OM I}})$ is between minus one and zero, which according to the Jeffreys scale means "more data are needed to come to a meaningful conclusion". (see Table~\ref{t:jeffreys}). A further 12 of these had $\log(\mathcal{Z}_{\rm{PM}} / \mathcal{Z}_{\rm{OM I}})$ values between $-2.5$ and $-1$ which can be interpreted as "weak preference" in favour of OM I, whilst no clusters had a value of $\log(\mathcal{Z}_{\rm{PM}} / \mathcal{Z}_{\rm{OM I}})$ less than minus five ("strong preference" in favour of OM I). The largest absolute value for the ratio was actually in favour of the PM with $\ln (\mathcal{Z}_{\rm PM} / \mathcal{Z}_{\rm OM\, I}) = 4.73 \pm 0.23$ (for the lowest $z$ cluster) which suggests "moderate preference" towards the PM. There is no correlation between $\log(\mathcal{Z}_{\rm{PM}} / \mathcal{Z}_{\rm{OM I}})$ and $z$. \\ Figure~\ref{f:phys_priors} shows the prior space for the observational parameters corresponding to the PM with the lowest and highest $z$ values in the sample.

\subsubsection{Observational models I \& II}

\begin{figure}
  \begin{center}
  \includegraphics[ width=0.90\linewidth]{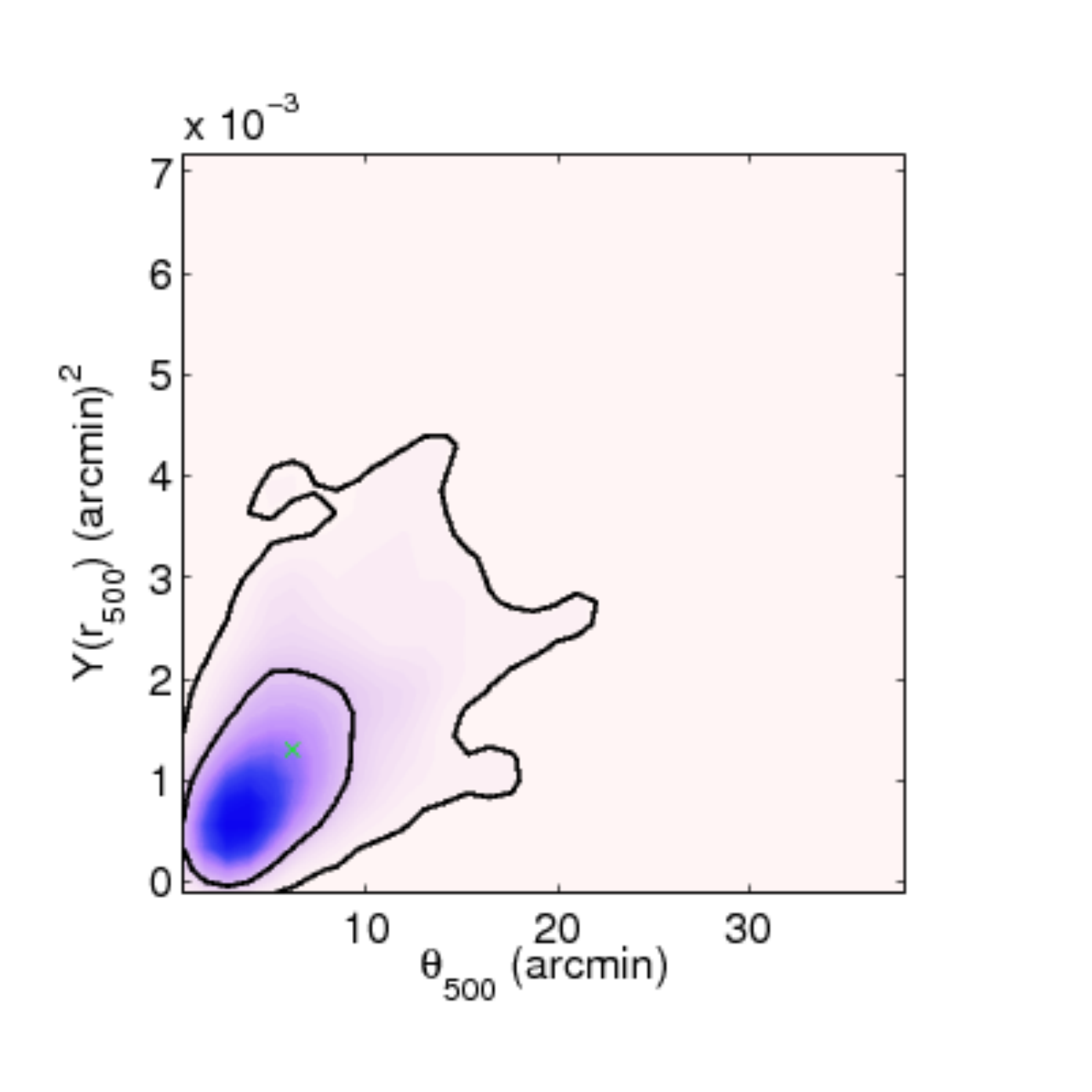}
  \caption{Two-dimensional prior probability distribution of $Y(r_{500})$ and $\theta_{500}$ for OM I, which is based on Planck data as detailed in Section~\ref{s:obs_i_priors}.}
\label{f:obsiobsii_priors}
  \end{center}
\end{figure}

Similarly, OM I is favoured over OM II for 53 clusters, but with 14 cases having  $0 \leq \log(\mathcal{Z}_{\rm{OM I}} / \mathcal{Z}_{\rm{OM II}}) \leq 1$. Again the highest absolute value came from the lowest redshift cluster, highlighting the importance of $z$ information at such a low $z$ value.
Since these models have the same input parameters, it is easier to compare their sampling parameter spaces.
Figure~\ref{f:obsiobsii_priors} shows the prior range of $\left(Y(r_{500}), \, \theta_{500}\right)$ for OM I. Around 68\% of the prior mass (i.e. the inner contour in the Figure) is bounded roughly by $Y(r_{500}) = 2\times 10^{-3}~\rm{arcmin}^{2}$ and $\theta_{500} = 10~\rm{arcmin}$. The 95\% contour gives upper bounds of $Y(r_{500}) \approx 4\times10^{-3}~\rm{arcmin}^{2}$ and $\theta_{500} \approx 15~\rm{arcmin}$. In comparison the OM II prior ranges for the lowest redshift cluster are $\theta_{500} = [4.9,~19.0]~\rm{arcmin}$ and $Y(r_{500}) = [0.006,~1.0]\times 10^{-1}~\rm{arcmin}^{2}$, and for the highest redshift cluster are $ \theta_{500} = [0.8,~3.5]~\rm{arcmin}$, $ Y(r_{500}) = [0.003,~5.0]\times 10^{-3}~\rm{arcmin}^{2}$. The ratio of the upper and lower limits for $\theta$ and $Y$ are approximately $4.5$ and $1.8 \times 10^{3}$ across all clusters. This suggests that the ratio of the bounds of the parameter space for each cluster does not change for the OM II, but that the sampling space is shifted depending on $z$.
Note that even though the sampling parameters for the observational models are $Y_{\rm tot}$ and $\theta_{\rm p}$, these are related to $Y(r_{500})$ and $\theta_{500}$ by constant factors, and so comparisons made on both are equivalent.

\subsubsection{Physical model and observational model II}

Comparison of PM and OM II, the models which incorporate redshift information into their priors leads to interesting results. For 43 clusters, the PM is preferred over OM II. However for all of these clusters $\log(\mathcal{Z}_{\rm{PM}} / \mathcal{Z}_{\rm{OM II}})$ is less than one, meaning that none of them give "conclusive" model preference. There are only three clusters which give "weak evidence" in favour of a model (OM II). These are the clusters at redshift $z = 0.144, \, 0.341, \, 0.5131$ with ratio values $ -1.88, \, -1.06, \, -1.16$ respectively. The fact that data from 51 clusters do not provide any "conclusive" preference between PM and OM II, suggests that these models are equally well suited for the current data, even though their parameter estimates are often not in such agreement.


\section{Conclusions}
\label{s:pl_obs_summary}
For the cluster sample analysed in the previous Chapter, I compare the parameter estimates obtained from different physical and observational models applied to AMI data using Bayesian analysis. 
The physical model (PM) used is as described in Section~\ref{s:phys_mod}, and the observational models (OM I and OM II) are based on the one described in \citet{2015A&A...580A..95P}. 
I have focused on comparisons of $Y(r_{500})$ and found the following.
\begin{itemize}
\item The PM generally yields lower estimates of $Y$ relative to the observational models, apart from at low $z$ where the reverse is true. 
\item For two thirds of the sample, the OM I and OM II estimates are within one combined standard deviation of each other.
\end{itemize}
To investigate further the discrepancies between the three models, we computed the Earth Mover's distance between the two-dimensional posterior distributions in $Y(r_{500}), \, \theta_{\rm 500}$ space, for each model pair. This gives a measure of the `distance' between the respective probability distributions. I then compared the evidence values obtained from the Bayesian analysis of the AMI data using the different models, referring to the Jeffreys scale to form conclusions on model preference, and found the following.
\begin{itemize}
\item Based on the Earth Mover's distances calculated for each cluster, the posteriors are most discrepant between the PM and OM I models when the sample was considered as a whole, followed by PM and OM II.
\item The two largest discrepancies come from the lowest-$z$ cluster, one between PM \& OM I and one between OM II \& OM I, suggesting that $z$ information at very low $z$ can have a large effect on the different models.
\item The distance between posteriors from PM and OM II clearly decreases with increasing $z$. This suggests that the difference between physical and observational model parameter estimates, provided the latter also includes $z$ information, is reduced at higher $z$.
\item When comparing Bayesian evidence values, OM I is preferred over PM for 50 of the clusters, although only 14 of these showed either "weak" or "moderate" preference to OM I (the remaining 36 being "inconclusive"); however the highest $\log(\mathrm{evidence} \, \mathrm{ratio})$ actually favours the PM ("moderate" preference) and occurs for the lowest-$z$ cluster.
\item Similarly, OM I is preferred to OM II in 53 of the cases. 14 suggested more data are needed to come to a "meaningful" conclusion, while the remaining 39 clusters showed "weak" or "moderate" preference for OM I. This suggests that OM I is the preferred model in more cases relative to OM II than when OM I is compared with PM.
\item For 43 of the clusters, PM is preferred over OM II; however in all of these cases, the Jeffreys scale suggests "no conclusion can be made without more data", and only three clusters give any "conclusive" preference (a "weak" preference in favour for OM II).
\end{itemize}

%% file: CHAP-5/chapter5.tex
\chapter{Physical modelling of galaxy clusters using Einasto dark matter profiles}\label{c:fifth}

This Chapter 
provides an alternative to the physical model presented in Section~\ref{s:phys_mod}. The physical model described previously uses an NFW profile \citep{1995MNRAS.275..720N} for the dark matter component of the galaxy cluster, which is derived from N-body simulations of galaxy clusters. \citet{1965TrAlm...5...87E} gives an empirical profile for dark matter halos. Previous investigations comparing the two dark matter profiles using simulated data (see e.g. \citealt{2014MNRAS.441.3359D}, \citealt{2014ApJ...797...34M}, \citealt{2016MNRAS.457.4340K} and \citealt{2016JCAP...01..042S}) have shown that the Einasto model provides a better fit. In particular, \citet{2016JCAP...01..042S} showed for weak lensing analysis of clusters that the NFW profile can overestimate virial masses of very massive halos ($\geq 10^{15}M_{\mathrm{Sun}} / h$ where $M_{\mathrm{Sun}}$ is units of solar mass and $h$ is the reduced Hubble constant) by up to 10\%.

It is these previous analyses which have motivated us to derive a physical galaxy cluster model for interferometric SZ data which uses the Einasto profile to model the dark matter component of the cluster. I also compare the parameter estimates and fits of the NFW \& Einasto models for the cluster A611 with data obtained with AMI, and with simulations created with both Einasto and NFW profiles. The work discussed in this Chapter has been published in MNRAS \citep{2019MNRAS.489.3135J}. Note the paper includes post-referee changes.


\section{Einasto physical model} 
\label{s:ein_mod}

The physical model presented here (PM II) follows the same calculational steps as the model presented in Section~\ref{s:phys_mod} (PM I) to calculate $\delta I_{\mathrm{cl}, \nu}$, but with an Einasto profile replacing the NFW one used for the dark matter component. Below we derive the relevant equations for the Einasto case.
Furthermore PM II is subject to the same assumptions as PM I listed in Section~\ref{s:phys_mod}.

The three input parameters required to calculate $\delta I_{\nu , \mathrm{cl}}$ for either PM are $M(r_{200})$, $f_{\rm gas}(r_{200})$, and $z$. A fourth input parameter is required for the PM II which we call the Einasto parameter $\alpha_{\rm Ein}$, which is also described below. 


\subsubsection{Dark matter profile}
\label{s:dm_models}

\begin{figure}
  \begin{center}
  \includegraphics[ width=0.90\linewidth]{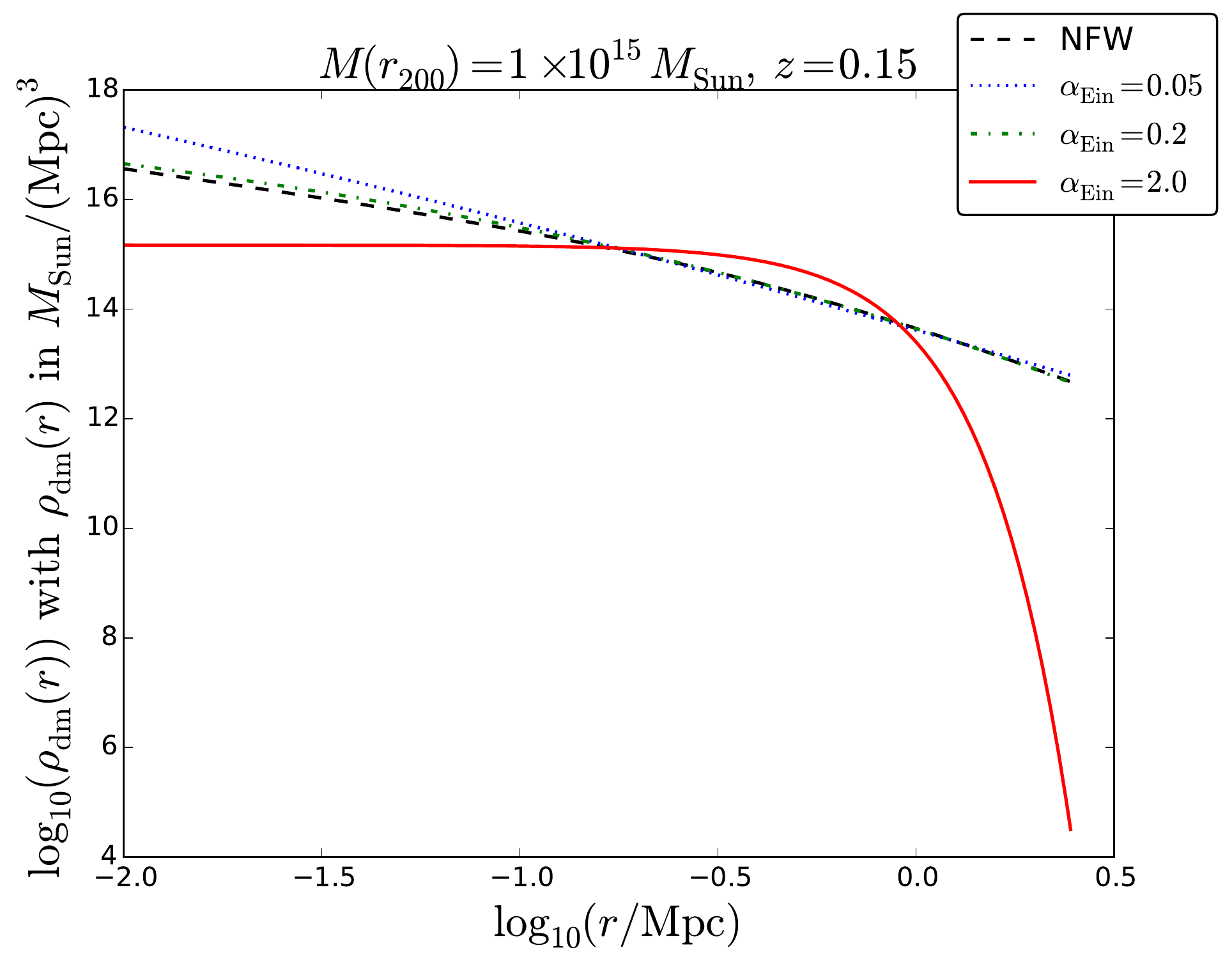}
  \caption{Logarithmic dark matter density profiles as a function of log cluster radius using NFW and Einasto models. Three values of the Einasto profile are used: $0.05, \, 0.2,$ and $2.0$. The additional input parameters used to generate these profiles are: $z =  0.15$, $M(r_{200}) = 1\times 10^{15} M_{\mathrm{Sun}})$ and $f_{\rm gas}(r_{200}) = 0.12$.}
\label{f:einnfwdmdens}
  \end{center}
\end{figure}

Assuming an Einasto profile \citep{1965TrAlm...5...87E}, the dark matter density profile for a cluster $\rho_{\rm dm, PM \, II}$ is given by 
\begin{equation}
\label{e:einasto}
\rho_{\rm dm, PM \ II} = \rho_{-2} \exp \left[ -\frac{2}{\alpha_{\rm Ein}} \left( \left(\frac{r}{r_{-2}}\right)^{\alpha_{\rm Ein}} - 1 \right) \right],
\end{equation}
where $\alpha_{\rm Ein}$ is a shape parameter, $r_{-2}$ is the scale radius where the logarithmic derivative of the density is $-2$ (analogue to $r_{\rm s}$ in the NFW model, but note that in general $ r_{-2} \neq r_{\rm s}$), and $\rho_{-2}$ is the density at this radius. The parameter $\alpha_{\rm Ein}$ controls the degree of curvature of the profile. The larger its value, the more rapidly the slope varies with respect to $r$. In the limit that $\alpha_{\rm Ein} \rightarrow  0$, the logarithmic derivative is $-2$ for all $r$.
It is tempting to assume that the Einasto profile is capable of providing a better fit due to the fact that the Einasto profile has an extra degree of freedom (three for the Einasto profile, two for the NFW), the shape parameter. However \citet{2016MNRAS.457.4340K} claims that this is not strictly true, as the Einasto profile was seen to give a better fit to simulated dark matter haloes even with $\alpha_{\rm Ein}$ fixed. The asymptotic values of the logarithmic slope for the two profiles are as follows: as $r \rightarrow 0$ then $\mathrm{d}\ln \rho_{\mathrm{dm, PM \, I}}(r) / \mathrm{d}\ln r \rightarrow -1$ and $\mathrm{d}\ln \rho_{\mathrm{dm, PM \, II}}(r) / \mathrm{d}\ln r \rightarrow 0$. As $r \rightarrow \infty$ then $\mathrm{d}\ln \rho_{\mathrm{dm, PM \, I}}(r) / \mathrm{d}\ln r \rightarrow -3$ and $\mathrm{d}\ln \rho_{\mathrm{dm, PM \, II}}(r) / \mathrm{d}\ln r \rightarrow - \infty$. The magnitude of $\alpha_{\mathrm{Ein}}$ determines how quickly the slope changes between the two asymptotic values. Throughout this work when I refer to the NFW or Einasto model, I really mean the physical model which uses the NFW or Einasto model when considering the dark matter density profile. 
\\
Referring back to equation~\ref{e:einasto}, the ratio $r_{200}/r_{-2}$ is defined as the concentration parameter $c_{200}$. \citet{2014MNRAS.441.3359D} determines an analytical form for $c_{200}$ as a function of total mass and redshift for Einasto profiles based on simulations similar to those described in \citet{2007MNRAS.378...55M} and \citet{2008MNRAS.391.1940M}
\begin{equation}
\label{e:einastoc200}
\log_{10}\left(c_{200}\right) = j(z) + k(z) \log_{10} \left[ \frac{M\left(r_{200}\right)}{10^{12}h^{-1}M_{\mathrm{Sun}}} \right],
\end{equation}
where $j(z) = 0.459 + 0.518\exp(-0.49z^{1.303})$ and $k(z) = -0.13 + 0.029z$. The fitting is said to be accurate in the redshift range $[0,5]$. 
To calculate $\rho_{-2}$ we must make the assumption stated for PM I, that the total mass enclosed at $r_{200}$ is approximately equal to the enclosed dark matter mass. That is
\begin{equation}
\label{e:massapprox}
M(r_{200}) = M_{\rm dm}(r_{200}) + M_{\rm g}(r_{200}) \approx M_{\rm dm}(r_{200}),
\end{equation}
where $M_{\rm dm}(r_{200})$ and $M_{\rm g}(r_{200})$ are the dark matter and gas masses. With this assumption we can say that for any $r \leq r_{200}$
\begin{equation}
\label{e:ein_mass}
\begin{split}
M(r) &\approx \int_{0}^{r} 4\pi r'^{2} \rho_{\rm dm, PM \, II}(r') \, \rm{d}r' \\
     &= \frac{4 \pi \rho_{-2} r_{-2}^{3}}{\alpha_{\rm Ein}} \exp \left( 2 / \alpha_{\rm Ein} \right) \left( \frac{\alpha_{\rm Ein}}{2} \right) ^{3 / \alpha_{\rm Ein}} \\
     & \quad \times \gamma \left[ \frac{3}{\alpha_{\rm Ein}}, \frac{2}{\alpha_{\rm Ein}} \left( \frac{r}{r_{-2}} \right)^{\alpha_{\rm Ein}} \right],
\end{split}
\end{equation}
where $\gamma \left[a, x \right] = \int_{0}^{x} t^{a-1} e^{-t} \rm{d} t $ is the incomplete lower gamma function. The steps taken to get this result are given in Appendix~\ref{s:einastointegral}. Equation~\ref{e:nfw_m_tot_1} can be evaluated at $r_{200}$ and equated with equation~\ref{e:ein_mass} evaluated at the same radius to obtain the following solution for $\rho_{-2}$
\begin{equation}
\label{e:rhom2}
\begin{split}
\rho_{-2} = & \frac{200}{3} \left(\frac{r_{200}}{r_{-2}} \right)^{3} \rho_{\rm crit}(z) \times \frac{1}{\left[1 / \alpha_{\rm Ein} \exp \left( 2 / \alpha_{\rm Ein} \right) \left( \frac{\alpha_{\rm Ein}}{2} \right) ^{3 / \alpha_{\rm Ein}}\right]} \\
          & \times \frac{1}{\gamma \left[ \frac{3}{\alpha_{\rm Ein}}, \frac{2}{\alpha_{\rm Ein}} \left( \frac{r_{200}}{r_{-2}} \right)^{\alpha_{\rm Ein}} \right]}.
\end{split}
\end{equation}
Equivalently, equation~\ref{e:ein_mass} can be evaluated at $r_{200}$ and set equal to the known value of $M(r_{200})$ to determine $\rho_{-2}$.
Figure~\ref{f:einnfwdmdens} shows the logarithmic dark matter density profiles as a function of $r$ for a cluster at $z =  0.15$ with $M(r_{200}) = 1\times 10^{15} M_{\mathrm{Sun}}$ and $f_{\rm gas}(r_{200}) = 0.12$ for PM I and PM II for the $\alpha_{\rm Ein}$ values: $0.05, \, 0.2, \, 2.0$. It is clear that the Einasto profiles diverge the most from each other at low $r$ and for the high $\alpha_{\rm Ein}$ value at high $r$ as well.


\subsubsection{Gas density and pressure profiles}
\label{s:p_models}

Calculating the pressure normalisation constant (defined below) again requires the assumption that the cluster is in hydrostatic equilibrium up to radius $r_{200}$. This means at any radius equal to or below $r_{200}$ the outward pushing pressure force created by the pressure differential at that point must be equal to the gravitational binding force due to the mass enclosed within that radius, i.e. that equation~\ref{e:hse} holds.
Furthermore I follow \citet{2007ApJ...668....1N} and assume the GNFW model given by equation~\ref{e:gnfw} for the pressure profile, as in PM I.
However, for all analysis presented in this Chapter (both PMs), the GNFW slope parameters are taken to be $a = 1.0510$, $b=5.4905$ and $c = 0.3081$. These `universal' values were taken from \citet{2010A&A...517A..92A} and are the best fit GNFW slope parameters derived from the REXCESS sub-sample (observed with XMM-Newton, \citealt{2007A&A...469..363B}), as described in Section~5 of Arnaud et al.. I also take the Arnaud et al. value of $c_{500}$ which is $1.177$. Note that in the previous Chapters (as well as in MO12) 
 slightly different values derived for the standard self-similar case (Appendix~B of Arnaud et al.) were used ($a = 1.0620$, $b=5.4807$, $c = 0.3292$ and $c_{500}=1.156$). It was shown in \citet{2013MNRAS.430.1344O} that PM I is not affected by which of these two sets of parameters is used. \\
The analytical function used to convert from $r_{200}$ to $r_{500}$ in PM I is specific to the NFW dark matter profile case and so is not applicable to PM II. I have not found an analytic fitting function for the conversion in the case of an Einasto dark matter profile and so I obtain $r_{500}$ iteratively as described in Appendix~\ref{s:r500newton}. 
As in PM I, the pressure profile can be substituted into the hydrostatic equilibrium equation to derive an expression for the gas density. Using equation~\ref{e:ein_mass} for $M(r)$ gives
\begin{equation}
\label{e:ein_rhog} 
\begin{split}
\rho_{\rm g}(r) =& \, \frac{\mu_{e}}{\mu_{\rm g}}\frac{P_{\rm ei}}{4\pi G \rho_{-2}r_{-2}^{3}} \frac{1}{\left[\left(1 / \alpha_{\rm Ein} \right) \exp( 2 / \alpha_{\rm Ein}) \left( \alpha_{\rm Ein} / 2 \right) ^{3 / \alpha_{\rm Ein}}\right]} \\
                 & \times \frac{r}{\gamma \left[ \frac{3}{\alpha_{\rm Ein}}, \frac{2}{\alpha_{\rm Ein}} \left( \frac{r_{200}}{r_{-2}} \right)^{\alpha_{\rm Ein}} \right]} \\
                 & \times \left(\frac{r}{r_{\rm p}}\right)^{-c}\left[1+\left(\frac{r}{r_{\rm p}}\right)^{a}\right]^{-\left(\frac{a+b-c}{a}\right)}\left[b\left(\frac{r}{r_{\rm p}}\right)^{a}+c\right].
\end{split}
\end{equation}
Note that like PM I, the gas mass $M_{\rm g}(r)$ given by 
\begin{equation}
\label{e:ein_gasmass}
M_{\rm g}(r) = \int_{0}^{r} 4\pi\rho_{\rm g}(r')r'^{2}\,\mathrm{d}r'
\end{equation}
must be integrated numerically. Hence $f_{\rm gas}(r) = M_{\rm g}(r)/M(r)$ does not have a closed form solution. Nevertheless, we can use equations~\ref{e:ein_rhog} and~\ref{e:ein_gasmass} to determine $P_{\rm ei}$ since we know $M(r_{200})$, $f_{\rm gas}(r_{200})$ and $r_{200}$. Evaluating equations~\ref{e:ein_rhog} and~\ref{e:ein_gasmass} at $r_{200}$ and solving for $P_{\rm ei}$ gives the following expression
\begin{equation}
\label{e:ein_pei}
\begin{split}
 P_{\rm {ei}} = & \left(\frac{\mu_{\rm g}}{\mu_{e}}\right)
(G\rho_{-2}r^3_{-2})\left[\frac{\exp \left(2/\alpha_{\rm Ein} \right)}{\alpha_{\rm Ein}} \left(\alpha_{\rm Ein}/2\right)^{3/\alpha_{\rm Ein}} \right]M_{\rm g}(r_{200}) \\
              & \times  \frac{1}{ \bigint_{0}^{r_{200}} r'^{3}  \frac{\left[b \left(\frac{r'}{r_{\rm p}}\right)^{a} + c \right]}{\gamma\left[\frac{3}{\alpha_{\rm Ein}},\frac{2}{\alpha_{\rm Ein}} \left(\frac{r'}{r_{-2}}\right)^{\alpha_{\rm Ein}}\right]  \left(\frac{r'}{r_{\rm p}}\right)^c \left[1 + \left(\frac{r'}{r_{\rm p}}\right)^a\right]^{\left(\frac{a + b - c}{a}\right)} } {\rm d}r'},
\end{split}
\end{equation}
which must be evaluated numerically. Once $P_{\rm ei}$ and $r_{\rm p}$ have been calculated, the Comptonisation parameter and therefore $\delta I_{\nu , \mathrm{cl}}$ can be calculated the same way as in PM I. 


\subsubsection{Additional cluster parameters}
\label{s:tm_models}

As stated in Section~\ref{s:phys_mod}, the radial profile of the electron number density is given by $n_{e}(r) = \rho_{\rm g}(r) / \mu_{e}$. Using the ideal gas assumption, the electron temperature is therefore given by 
\begin{equation}
\label{e:ein_tgas}
\begin{split}
T_{e}(r) = & \left(\frac{4\pi \mu_{\rm g} G\rho_{-2}r_{-2}^{3}}{k_{\rm B}}\right)
\left[\left(1/\alpha_{\rm Ein}\right) \exp \left(2/\alpha_{\rm Ein}\right) \left(\alpha_{\rm Ein}/2 \right)^{3/\alpha_{\rm Ein}}\right] \\
                   & \times \frac{\gamma\left[\frac{3}{\alpha_{\rm Ein}},\frac{2}{\alpha_{\rm Ein}}\left (\frac{r}{r_{-2}}\right)^ {\alpha_{\rm Ein}}\right]}{r} \\ 
                   & \times \left [1 + \left(\frac{r}{r_{\rm p}}\right)^{a} \right]\left[b \left(\frac{r}{r_{\rm p}}\right)^{a} + c \right]^{-1}
\end{split}
\end{equation}
which also equals $T_{\rm g}(r)$. \\
The gas mass can be determined numerically from equation~\ref{e:ein_gasmass},
\begin{equation}
\label{e:ein_gasmass2}
\begin{split}
M_{\rm g}(r) =& \left(\frac{\mu_{e}}{\mu_{\rm g}}\right) \frac{1}{G} \frac{P_{\rm ei}}{\rho_{-2}} \frac{1}{\left[\left(1/\alpha_{\rm Ein}\right)\exp \left(2/\alpha_{\rm Ein} \right) \left(\alpha_{\rm Ein}/2\right)^{3/\alpha_{\rm Ein}}\right]r_{-2}^{3}} \\
              & \times \bigintss_{0}^{r} r'^{3} \frac{\left[b \left(\frac{r'}{r_{\rm p}}\right)^{a} + c \right]}{\gamma\left[\frac{3}{\alpha_{\rm Ein}},\frac{2}{\alpha_{\rm Ein}} \left(\frac{r'}{r_{-2}}\right)^{\alpha_{\rm Ein}}\right]} \\
              & \times \left(\frac{r'}{r_{\rm p}}\right)^c \left[1 + \left(\frac{r'}{r_{\rm p}}\right)^a\right]^{\left(\frac{a + b - c}{a}\right)} {\rm d}r'.
\end{split}
\end{equation}


\subsubsection{Prior probability distributions}
\label{s:ein_priors}
For both PM I and PM II I adopt the following approach (excluding any mention of $\alpha_{\rm Ein}$ in the former case). \\
As in Section~\ref{s:bayes_ami}, the cluster parameters are assumed to be independent of one another, so that
\begin{equation}\label{e:cluspriors}
\pi(\vec{\Theta}_{\rm cl}) = \pi(\alpha_{\rm Ein})\pi(M(r_{200}))\pi(f_{\rm gas}(r_{200}))\pi(z)\pi(x_{\rm c})\pi(y_{\rm c}).
\end{equation} 
Table~\ref{t:ein_clust_priors} lists the type of prior used for each cluster parameter and the probability distribution parameters. The values used for $z$ and $\alpha_{\rm Ein}$ will be specified on a case by case basis in Section~\ref{s:ein_results2}.

\begin{table}
\begin{center}
\begin{tabular}{{l}{c}}
\hline
Parameter & Prior distribution \\ 
\hline
$x_{\rm c}$ & $\mathcal{N}(0'', 60'')$ \\
$y_{\rm c}$ & $\mathcal{N}(0'', 60'')$ \\
$z$ & $\delta(z)$ \\
$M(r_{200})$ & $\mathcal{U} [ \log (0.5\times 10^{14} M_{\mathrm{Sun}}), \log (50\times 10^{14} M_{\mathrm{Sun}})]$ \\
$f_{\rm gas}(r_{200})$ & $\mathcal{N}(0.12, 0.02)$ \\
$\alpha_{\rm Ein}$ & $\delta(\alpha_{\rm Ein})$ \\
\hline
\end{tabular}
\caption{Cluster parameter prior distributions, where the normal distributions are parameterised by their mean and standard deviations.}
\label{t:ein_clust_priors}
\end{center}
\end{table}


\section{Results}
\label{s:ein_results}


\subsection{Cluster parameter profiles}
\label{s:ein_results1}

I first present the results of using the Einasto model in the profiling of cluster dark matter for a range of different cluster input parameters, along with the equivalent results from PM I. \\
I consider two input masses, $M(r_{200}) = 1\times 10^{14} M_{\mathrm{Sun}}$ and $M(r_{200}) = 1\times 10^{15} M_{\mathrm{Sun}}$, which roughly span the range of galaxy cluster masses. I use $z$-values of $0.15$ and $0.9$, take $f_{\rm gas}(r_{200}) = 0.12$ following \citet{2011ApJS..192...18K}, and consider $\alpha_{\rm Ein}$ values of $0.05, \, 0.2,$ and $ 2.0$ -- see Figure~\ref{f:einnfwdmdens}. I note that the same $r$ range ($-2 \leq \log_{10}(r) \leq 0.5$ where $r$ is in units of Mpc) is considered for each cluster, and thus even though each parameter profile is self-similar in $r$ with respect to mass and redshift, they are different for each cluster over the range of $r$ considered here.


\subsubsection{Dark matter mass profiles}
\label{s:dmmresults}

\begin{figure*}
  \begin{center}
    \begin{tabular}{@{}cc@{}}
     \includegraphics[ width=0.50\linewidth]{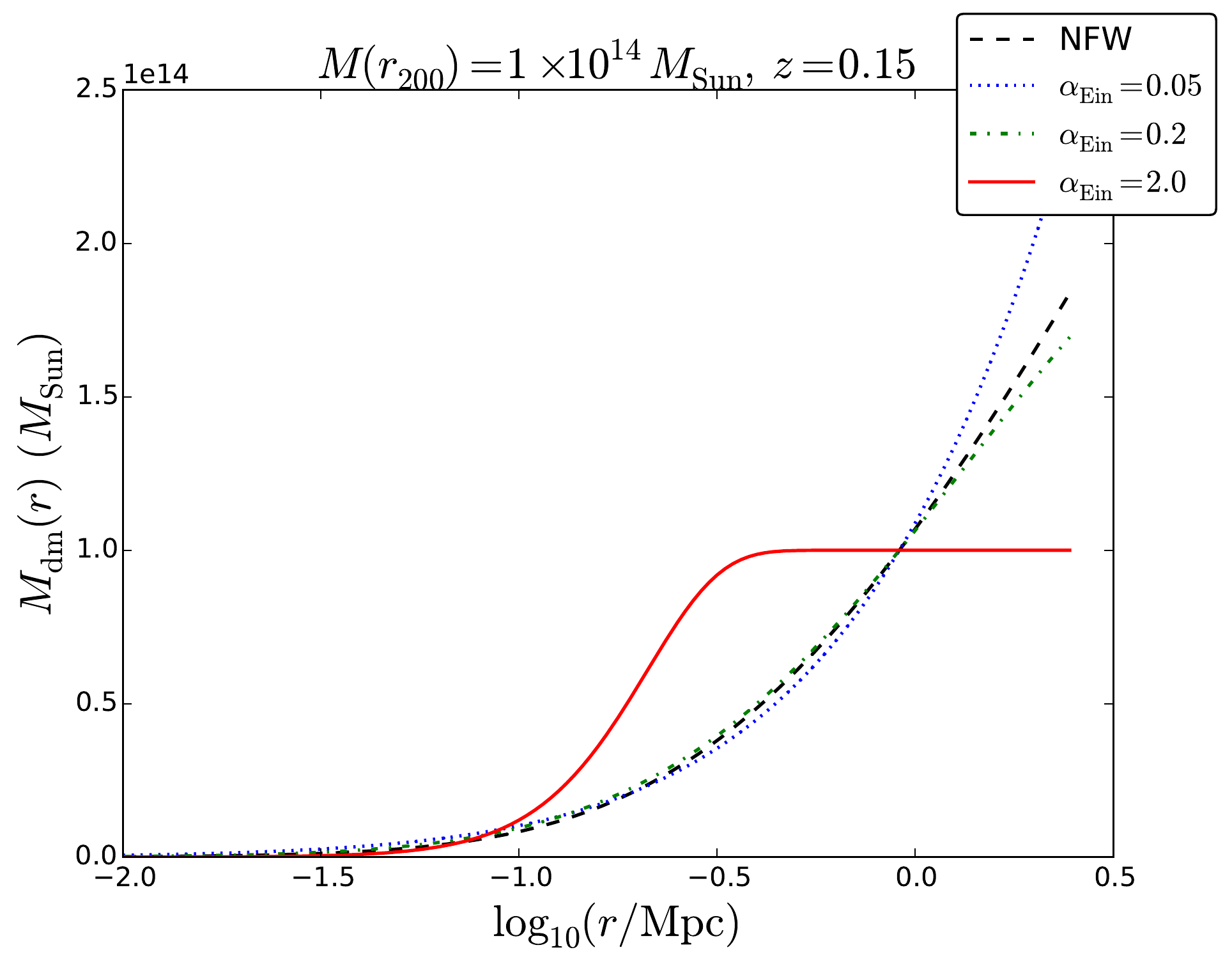} &
     \includegraphics[ width=0.50\linewidth]{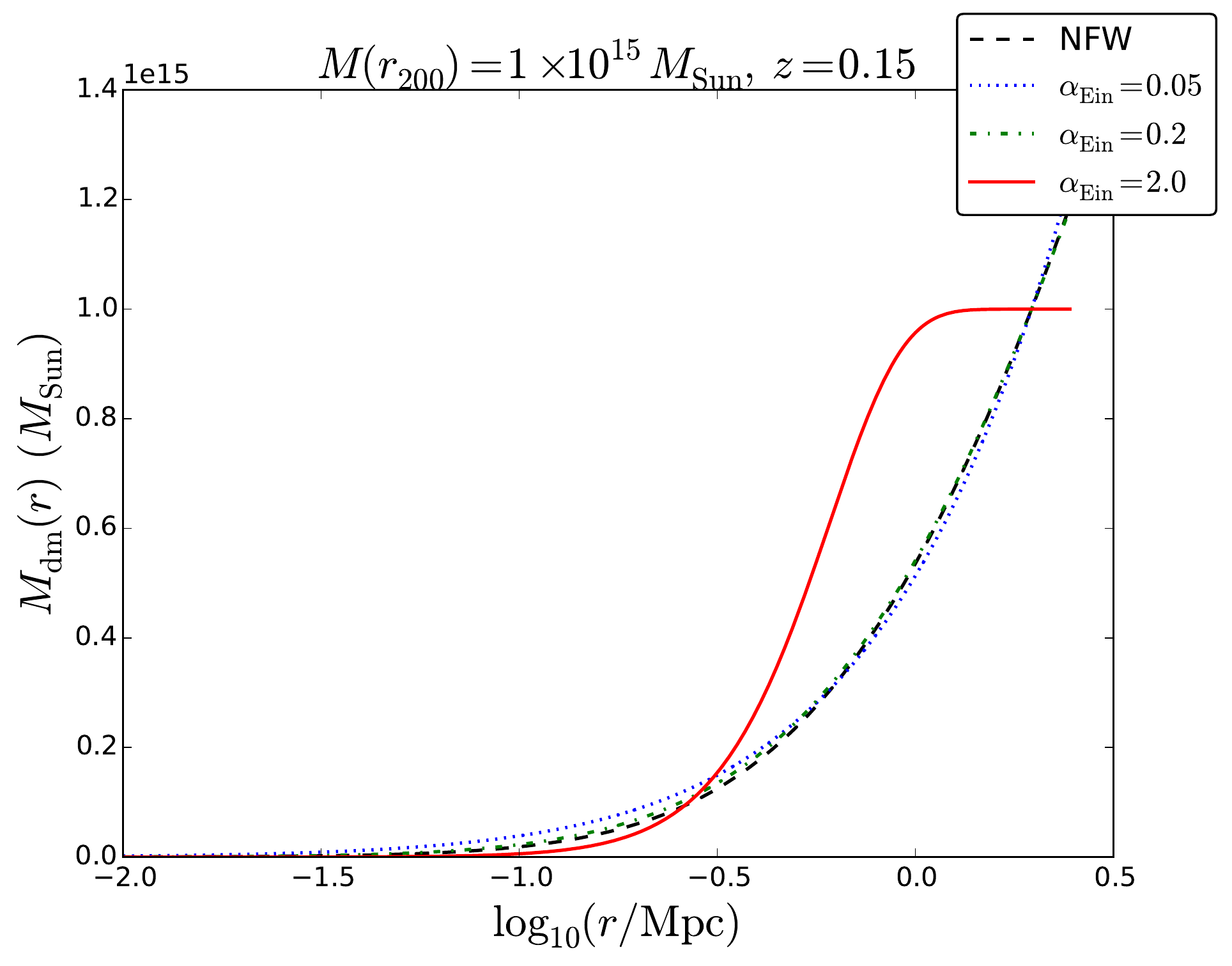} \\
     \includegraphics[ width=0.50\linewidth]{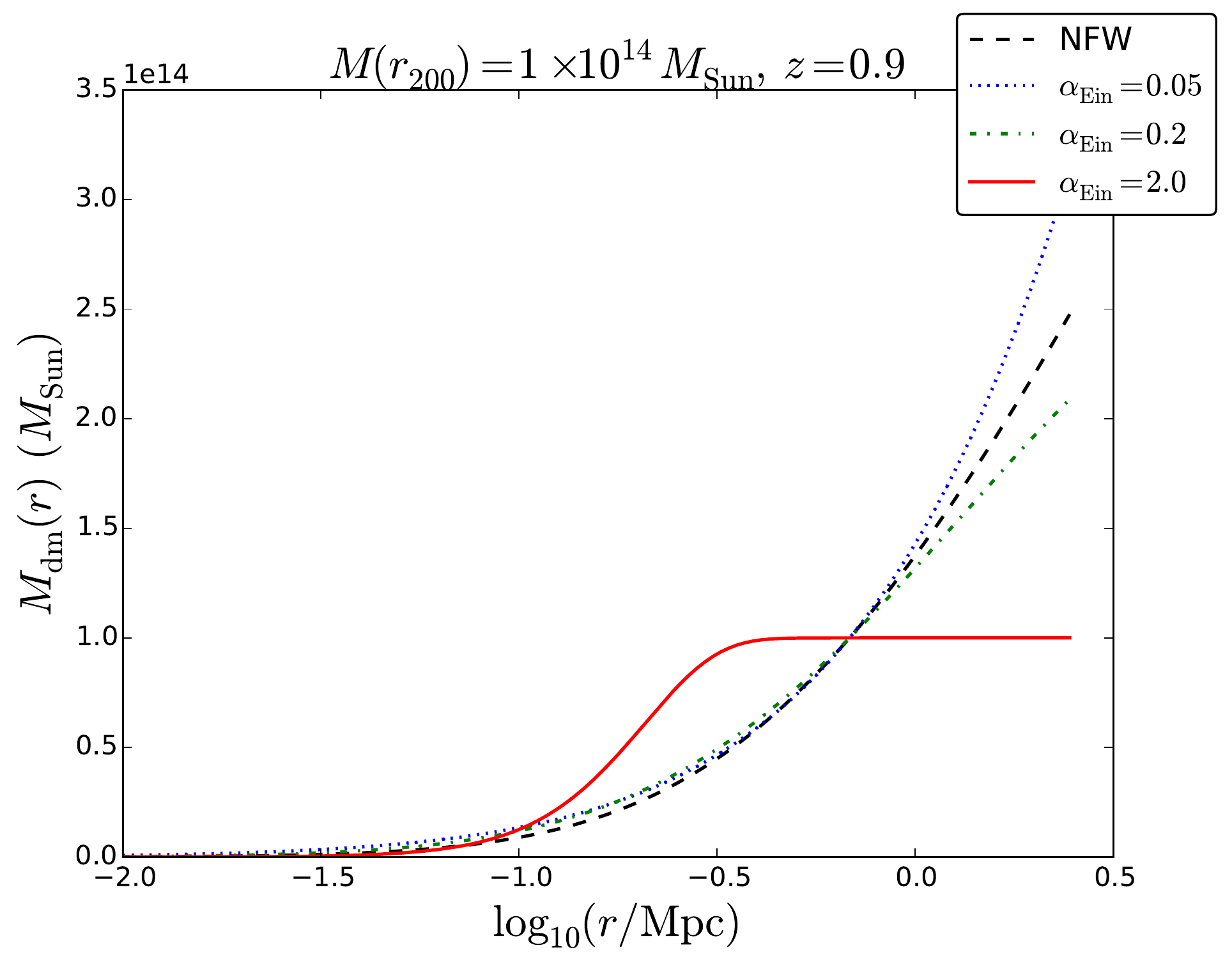} &
     \includegraphics[ width=0.50\linewidth]{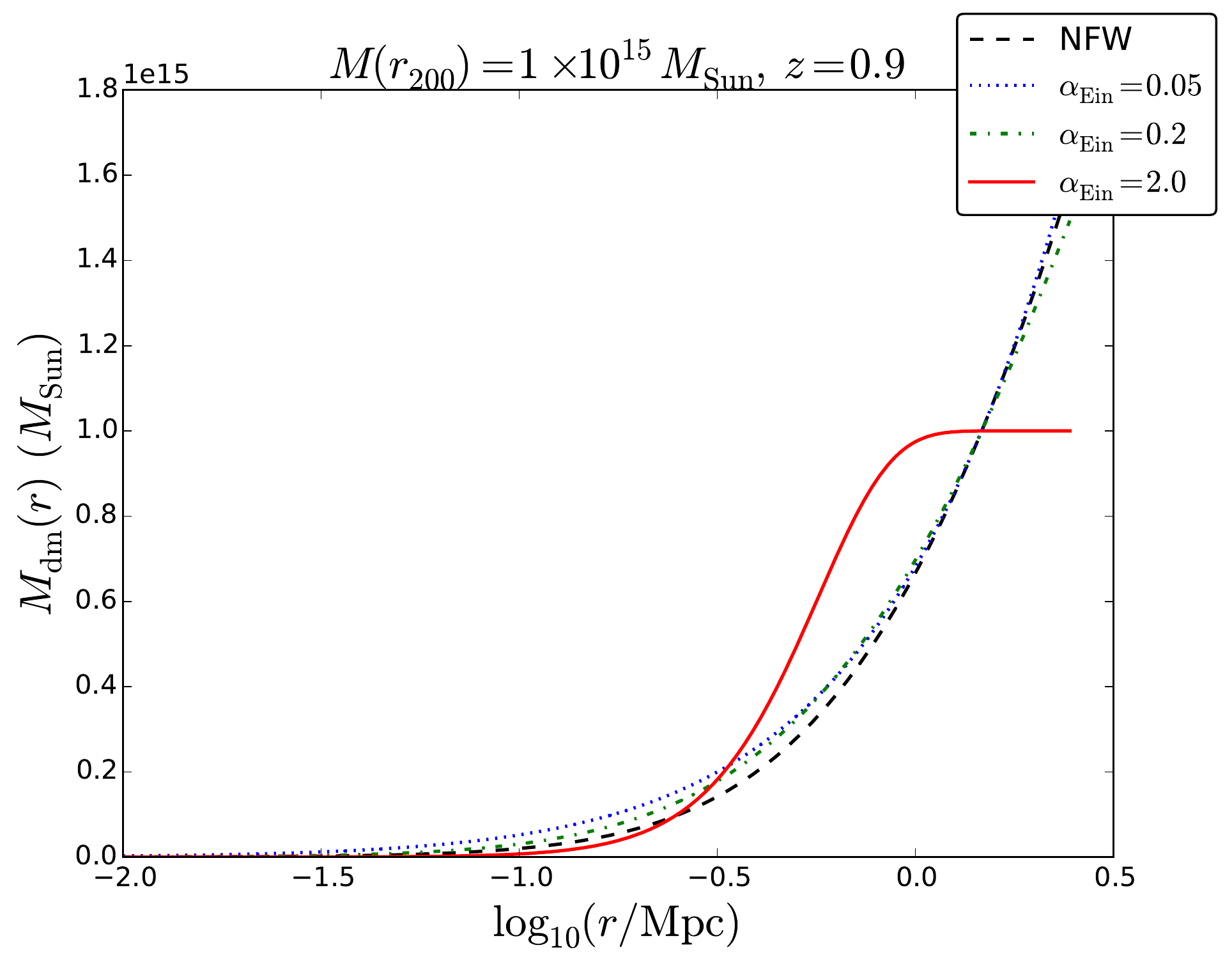} \\
    \end{tabular}

  \caption{Dark matter mass profiles as a function of log cluster radius using NFW and Einasto models. Values of $\alpha_{\mathrm{Ein}} = 0.05$, $0.2$, and $2.0$ are used as inputs. Top row has $z = 0.15$, bottom row has $z = 0.9$. Left column has $M(r_{200}) = 1\times 10^{14} M_{\mathrm{Sun}}$, right column has $M(r_{200}) = 1\times 10^{15} M_{\mathrm{Sun}}$.}
\label{f:einnfwdmmass}
  \end{center}
\end{figure*}

Figure~\ref{f:einnfwdmmass} shows the dark matter mass profiles. The Einasto profiles are calculated using equation~\ref{e:ein_mass} and the NFW profile from the equivalent relation given by equation~\ref{e:nfw}. Note that even though the notation in these equations corresponds to the total mass, this is in fact just the dark matter mass as we have used the approximation $M(r) \approx M_{\rm dm}(r)$ in deriving them.
The $\alpha_{\rm Ein} = 2$ case always converges quickly as the density rapidly falls to zero, while the other three profiles including the NFW show divergent behaviour at the largest radii considered here. The high mass inputs result in similar profiles for the $\alpha_{\rm Ein} = 0.05$, $\alpha_{\rm Ein} = 0.2$ and NFW cases, whereas the low mass inputs result in the $\alpha_{\rm Ein} = 0.05$ case diverging somewhat more rapidly than the others.

\subsubsection{Gas density profiles}
\label{s:grhoresults}

\begin{figure*}
  \begin{center}
    \begin{tabular}{@{}cc@{}}
     \includegraphics[ width=0.50\linewidth]{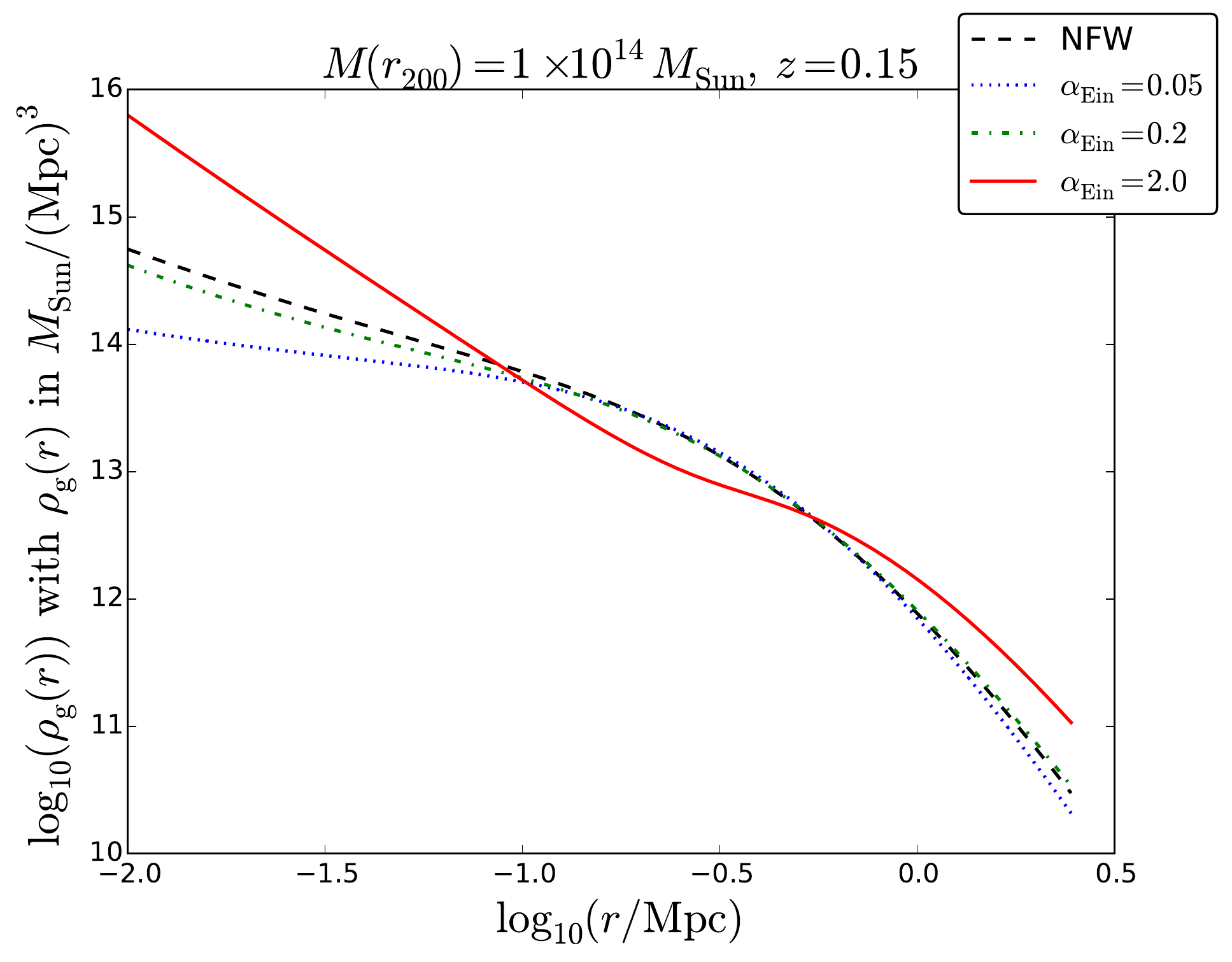} &
     \includegraphics[ width=0.50\linewidth]{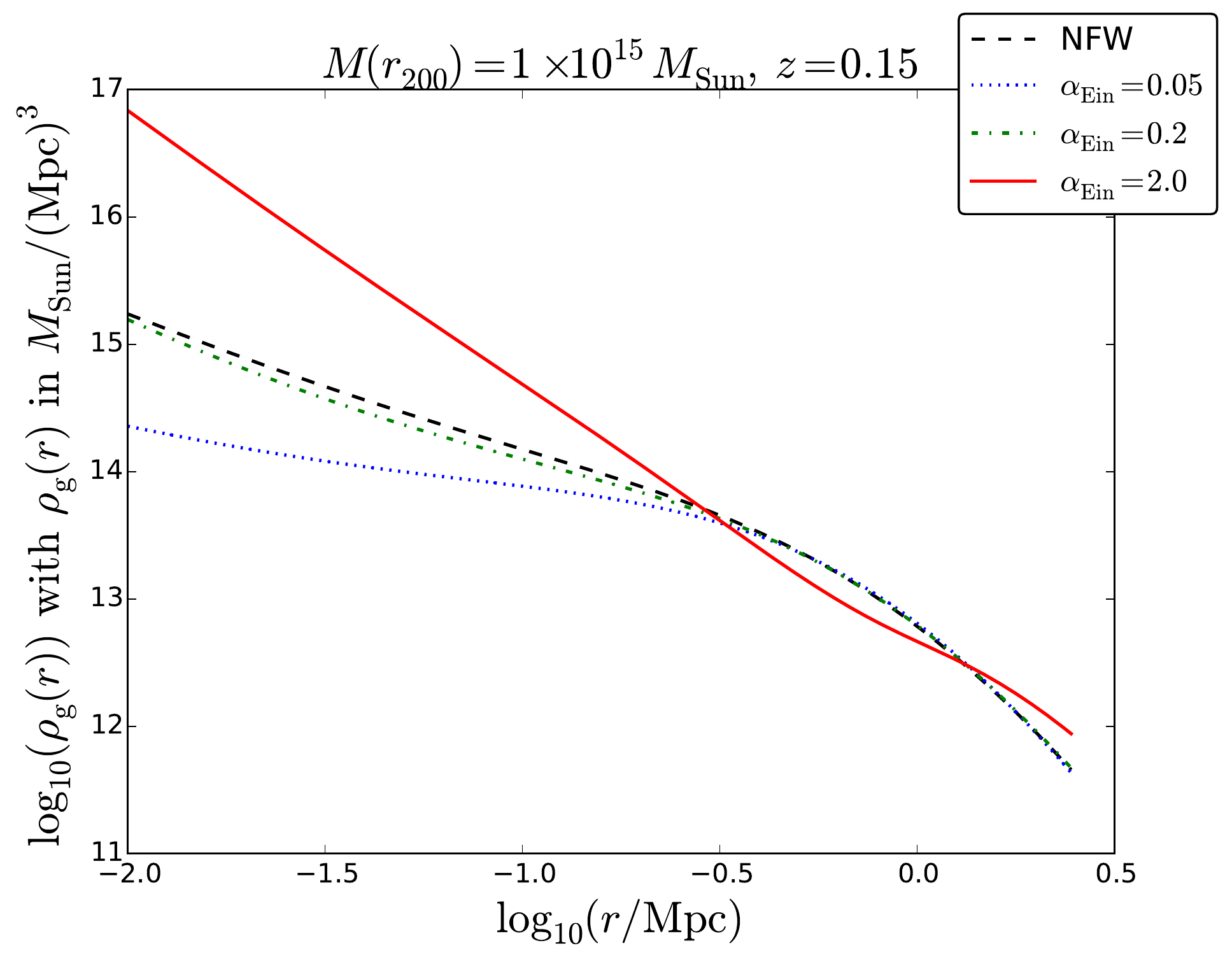} \\
     \includegraphics[ width=0.50\linewidth]{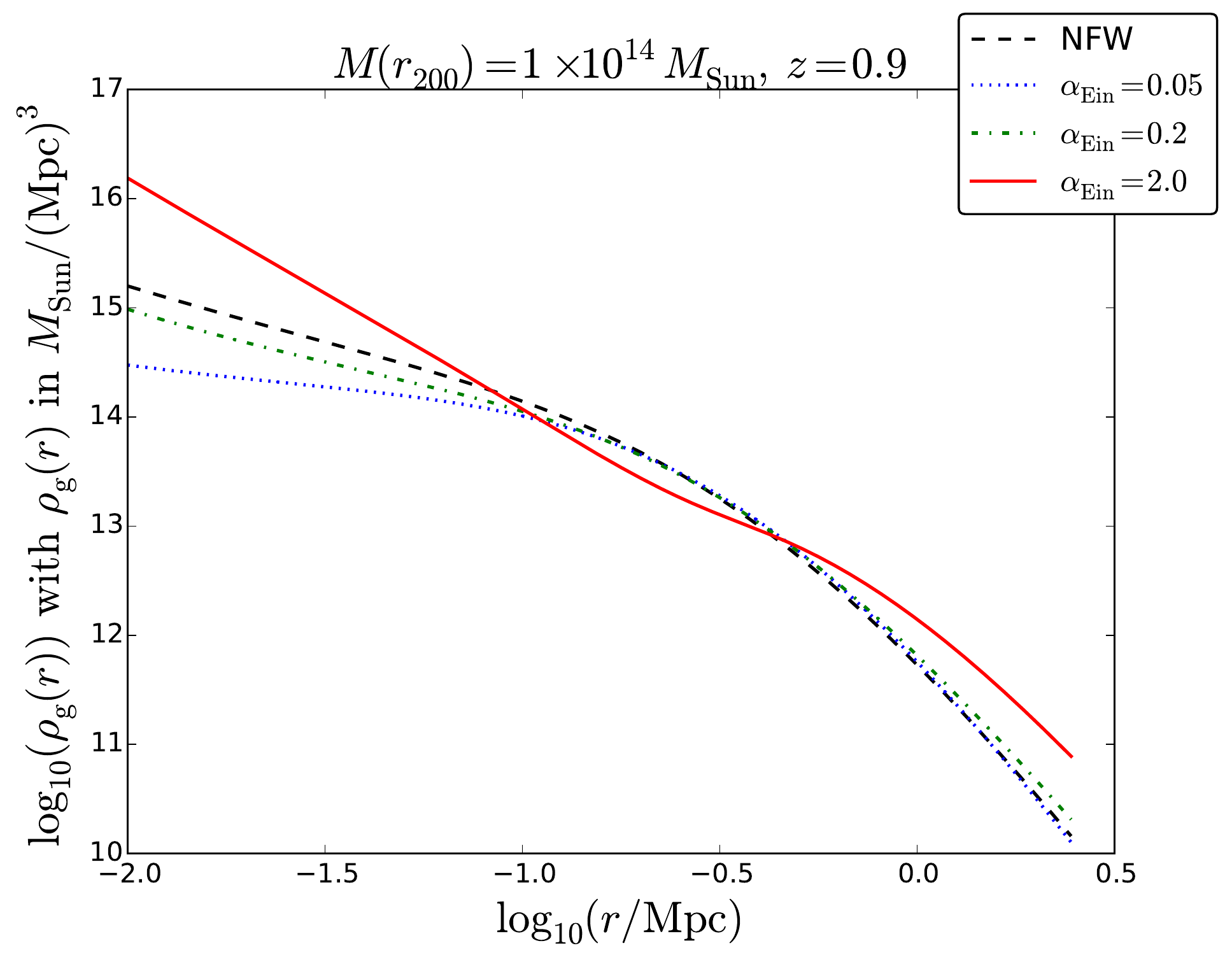} &
     \includegraphics[ width=0.50\linewidth]{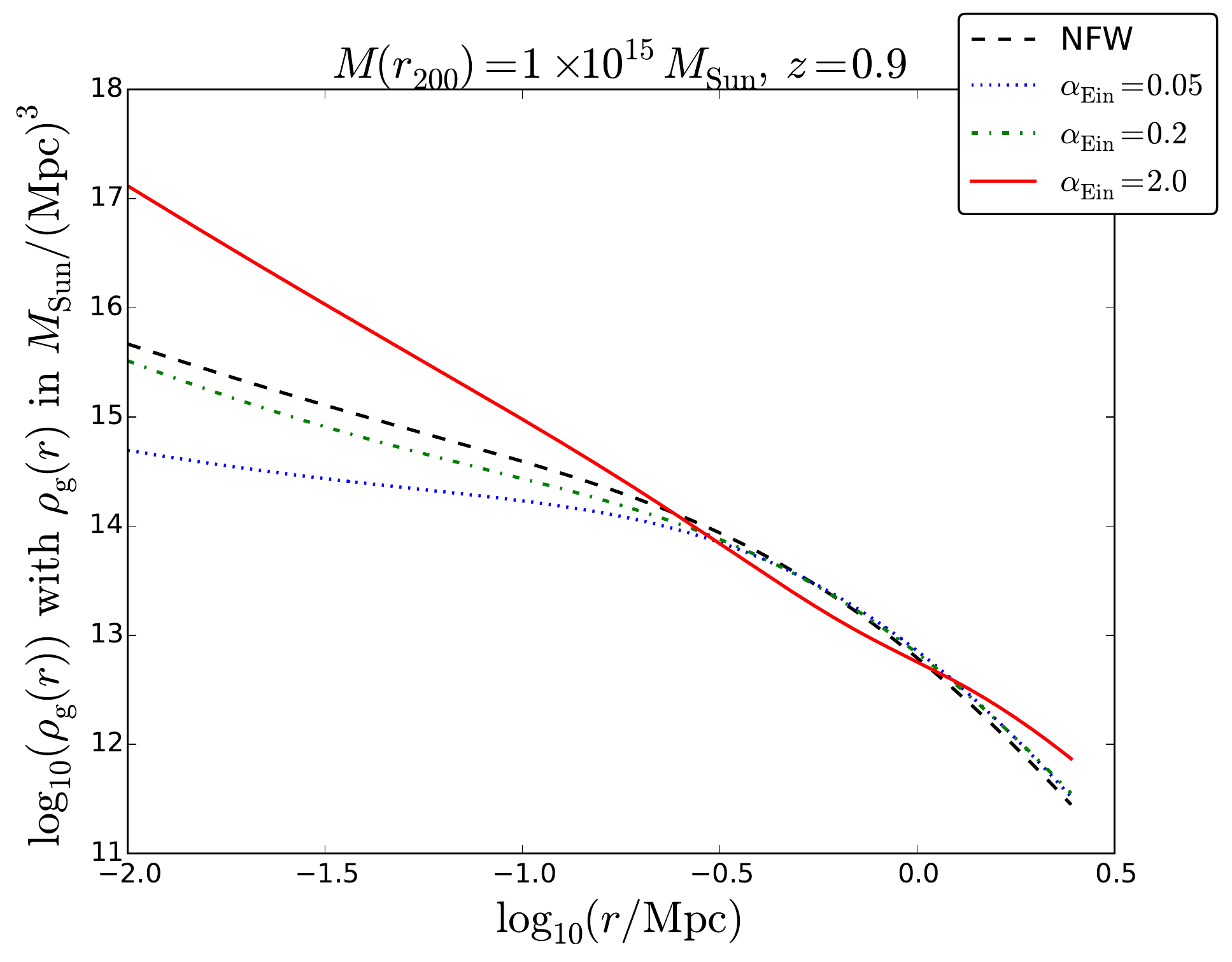} \\
    \end{tabular}
  \caption{Logarithmic gas density profiles as a function of log cluster radius using NFW and Einasto models. Values of $\alpha_{\mathrm{Ein}} = 0.05$, $0.2$, and $2.0$ are used as inputs. Top row has $z = 0.15$, bottom row has $z = 0.9$. Left column has $M(r_{200}) = 1\times 10^{14} M_{\mathrm{Sun}}$, right column has $M(r_{200}) = 1\times 10^{15} M_{\mathrm{Sun}}$.}
\label{f:einnfwgrho}
  \end{center}
\end{figure*}

Figure~\ref{f:einnfwgrho} shows the gas density profiles. The Einasto profiles are calculated using equation~\ref{e:ein_rhog} and the NFW profile from the equivalent relation given in MO12 (equation~6). Note that when calculating $\rho_{\rm g}(r)$ for arbitrary $r$, we are assuming hydrostatic equilibrium at that radius so that equation~\ref{e:hse} holds, and we have to assume that $f_{\rm gas}(r') \approx 0$ for all $r' \leq r$ so that $M(r) \approx M_{\rm dm}(r)$ at this radius.
The plots show that the profiles are similar for all inputs of mass and redshift, with the $\alpha_{\rm Ein} = 0.2$ Einasto profile again most resembling the NFW profile. However, the $\alpha_{\rm Ein} = 2.0$ profile has the highest gas density at high $r$ for both masses and both $z$ values.

\subsubsection{Gas mass profiles}
\label{s:gmassresults}

\begin{figure*}
  \begin{center}
  \begin{tabular}{@{}cc@{}}
     \includegraphics[ width=0.50\linewidth]{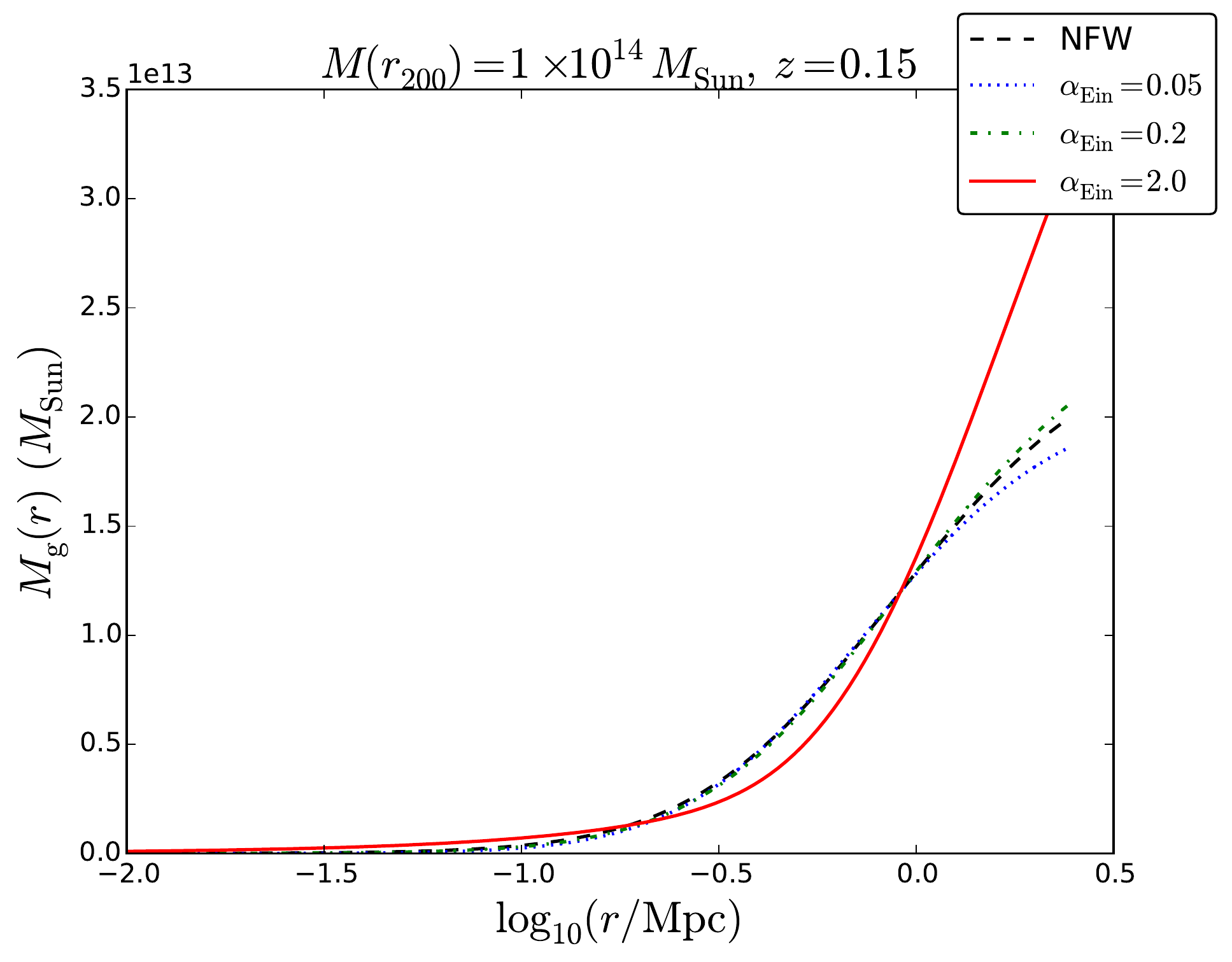} &
     \includegraphics[ width=0.50\linewidth]{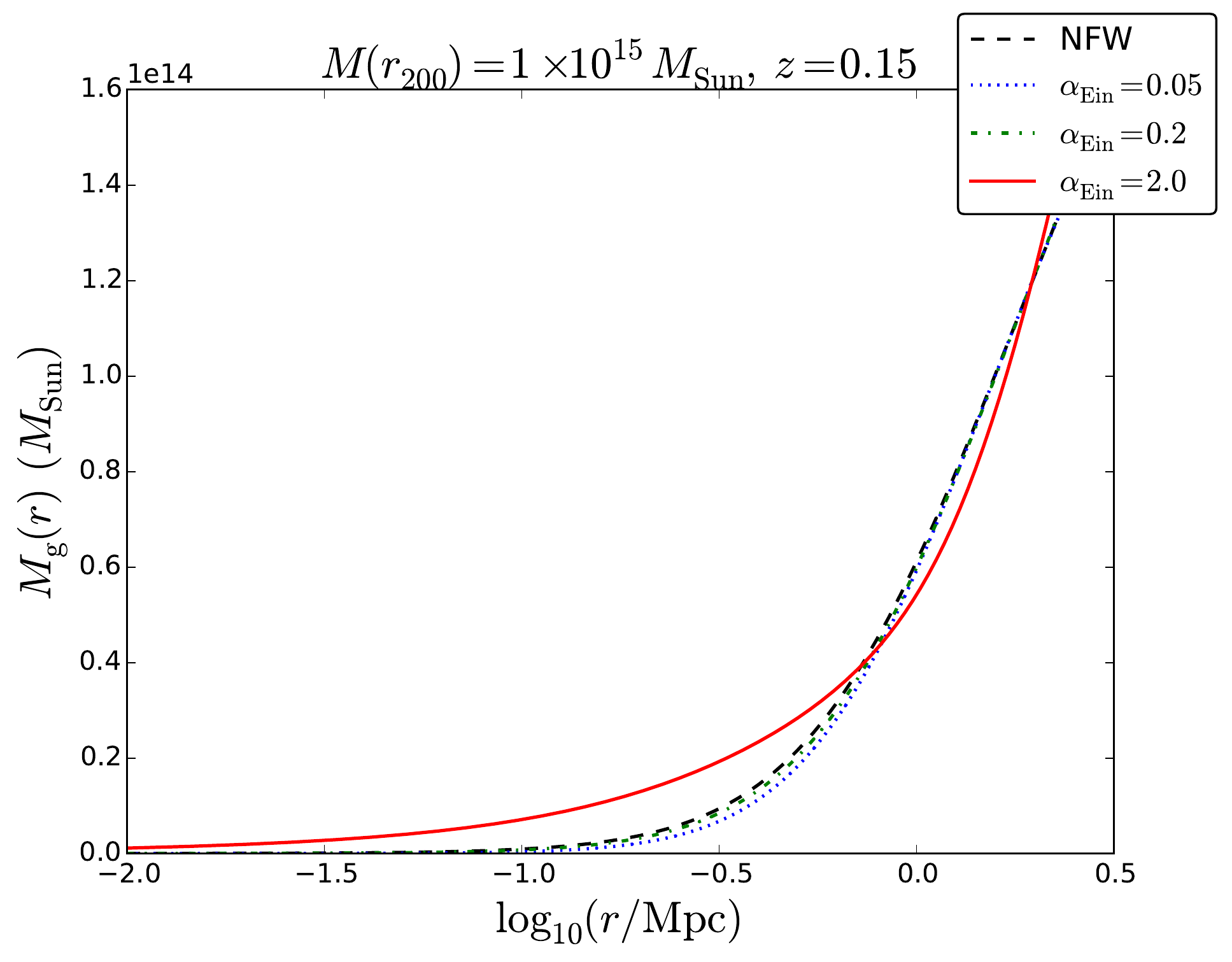} \\
     \includegraphics[ width=0.50\linewidth]{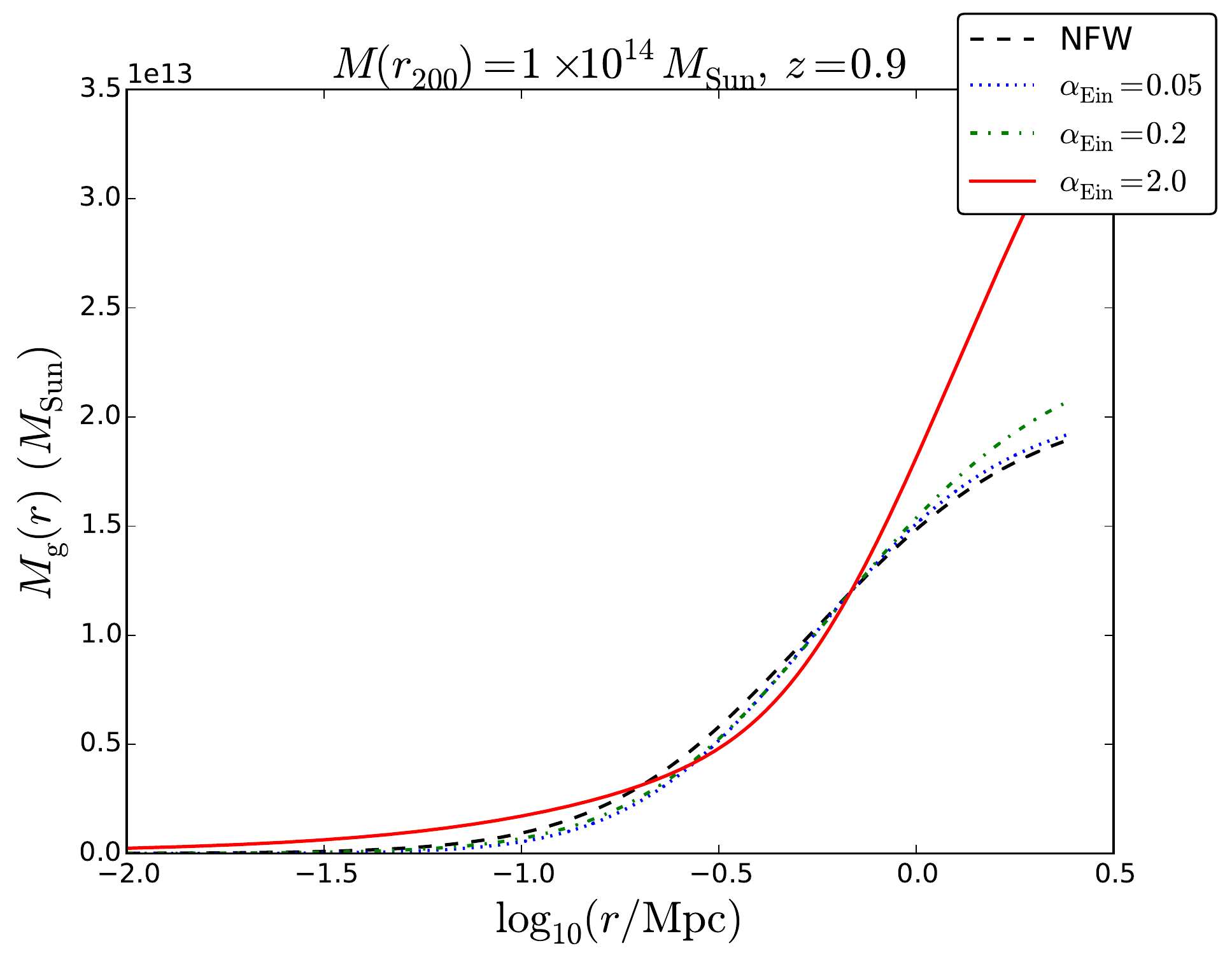} &
     \includegraphics[ width=0.50\linewidth]{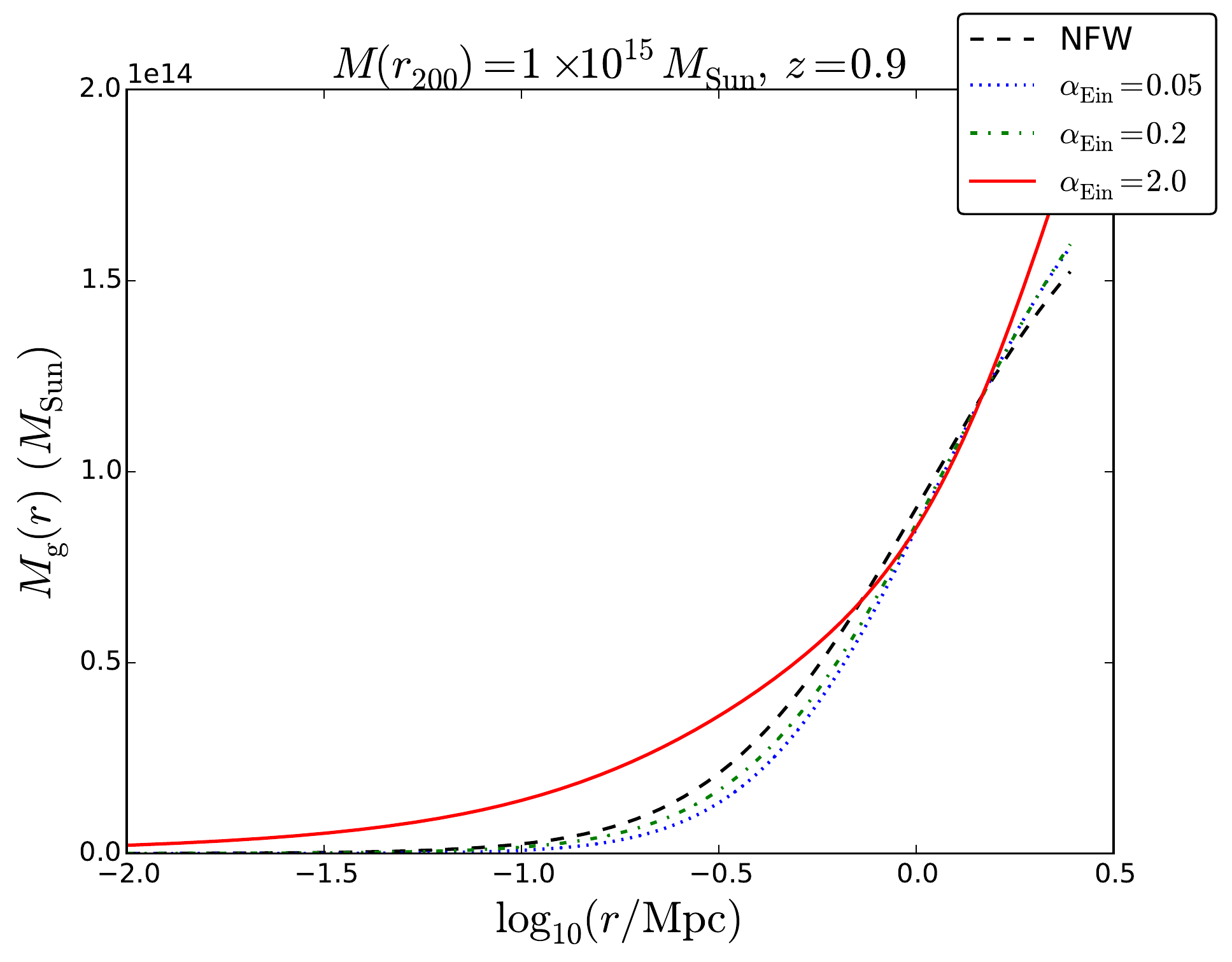} \\
    \end{tabular}
  \caption{Gas mass profiles as a function of log cluster radius using NFW and Einasto models. Values of $\alpha_{\mathrm{Ein}} = 0.05$, $0.2$, and $2.0$ are used as inputs. Top row has $z = 0.15$, bottom row has $z = 0.9$. Left column has $M(r_{200}) = 1\times 10^{14} M_{\mathrm{Sun}}$, right column has $M(r_{200}) = 1\times 10^{15} M_{\mathrm{Sun}}$.}
\label{f:einnfwgm}
  \end{center}
\end{figure*}

Figure~\ref{f:einnfwgm} shows $M_{\rm g}(r)$ as a function of cluster radius. As in Figure~\ref{f:einnfwdmmass} with the dark matter mass profiles, the high mass inputs correspond to divergent behaviour at large $r$. But for $\alpha_{\rm Ein} = 2.0$ the profile of $M_{\rm g}(r)$ also shows a more noticeable such divergence. 
Furthermore, in all four input parameter cases, $\alpha_{\rm Ein} = 2.0$ shows more divergent behaviour than other values of $\alpha_{\rm Ein}$ and the NFW profile in gas mass, which is in contrast to the dark matter mass profiles. 

\subsubsection{Gas temperature profiles}
\label{s:gtempresults}

\begin{figure*}
  \begin{center}
  \begin{tabular}{@{}cc@{}}
     \includegraphics[ width=0.50\linewidth]{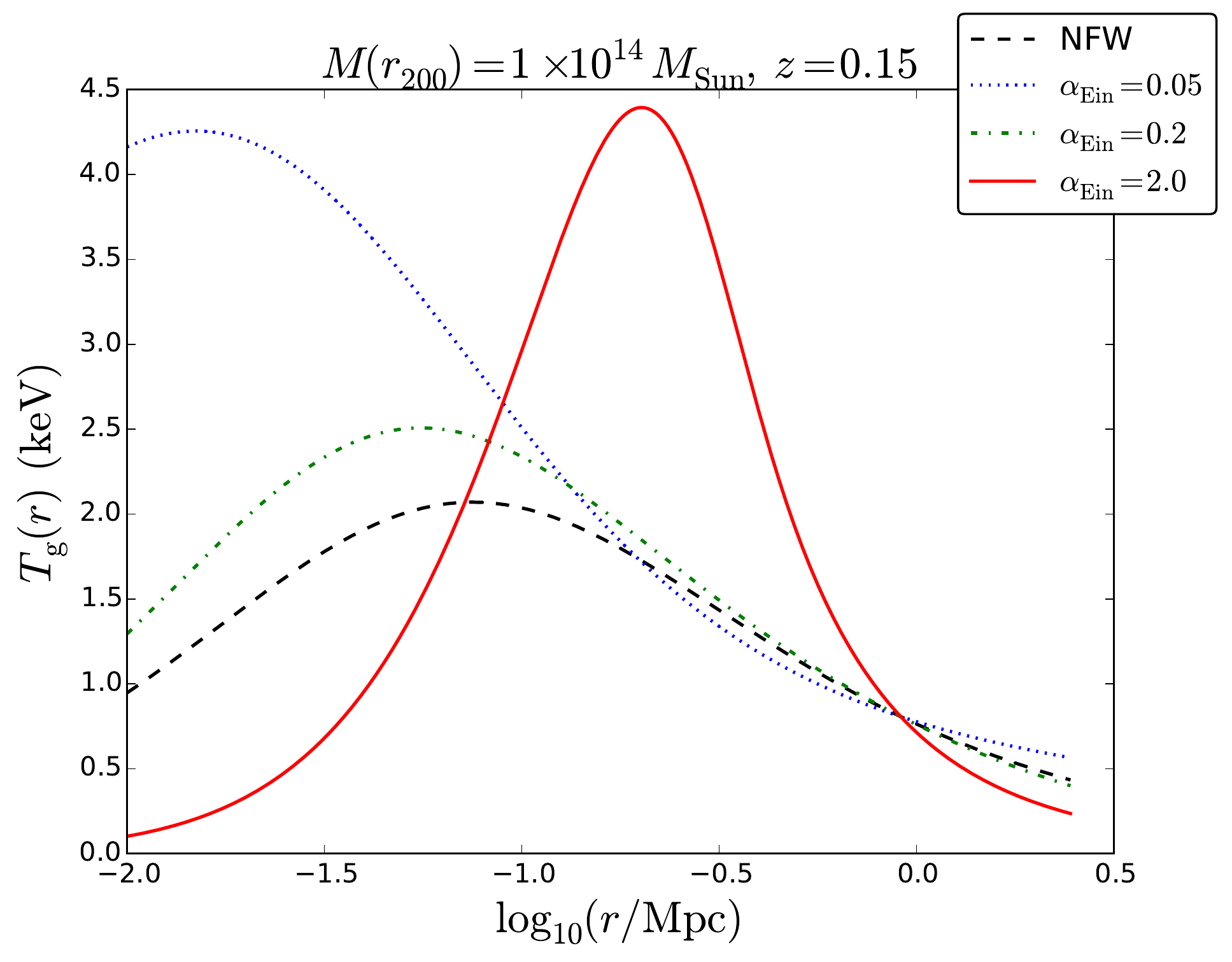} &
     \includegraphics[ width=0.50\linewidth]{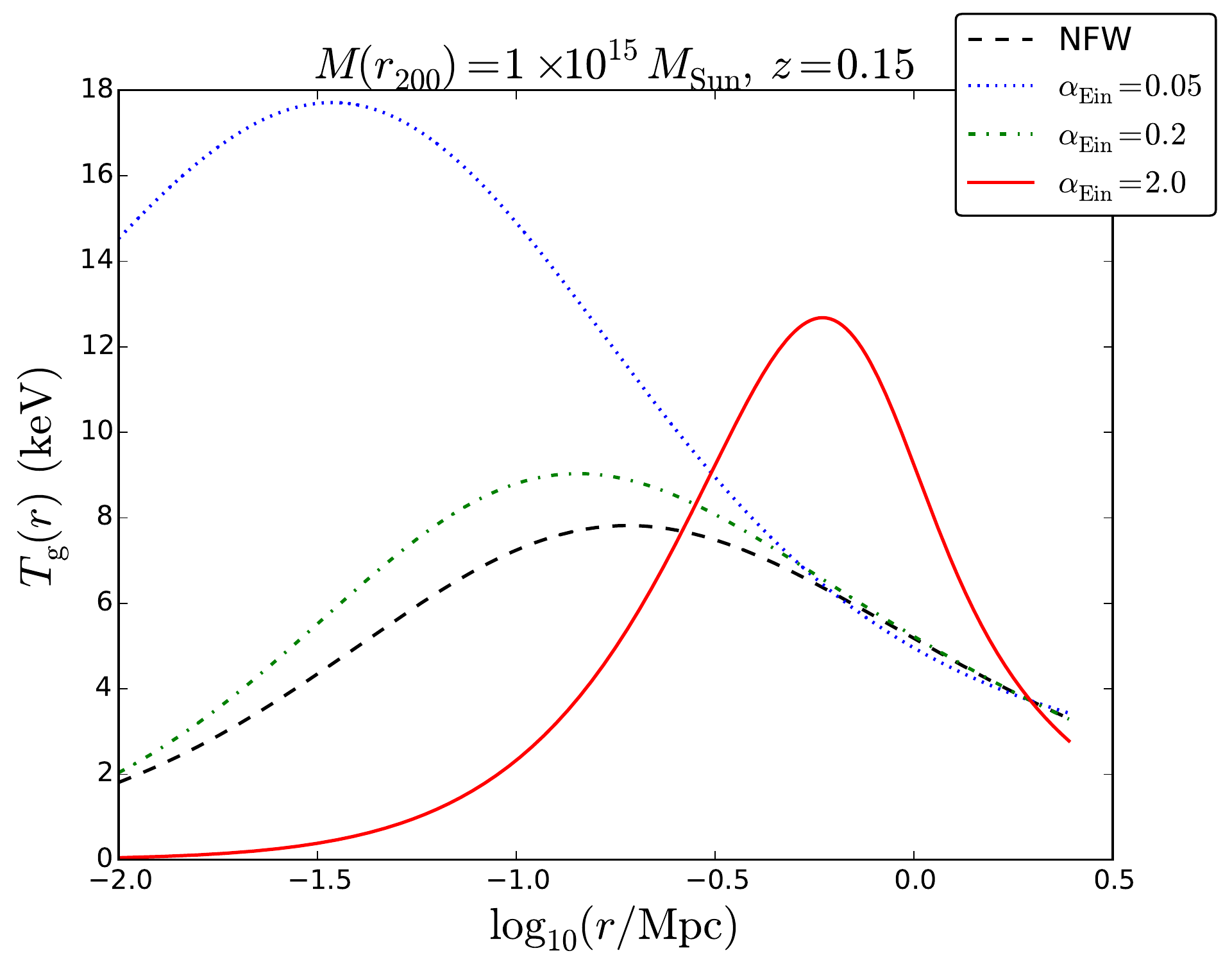} \\
     \includegraphics[ width=0.50\linewidth]{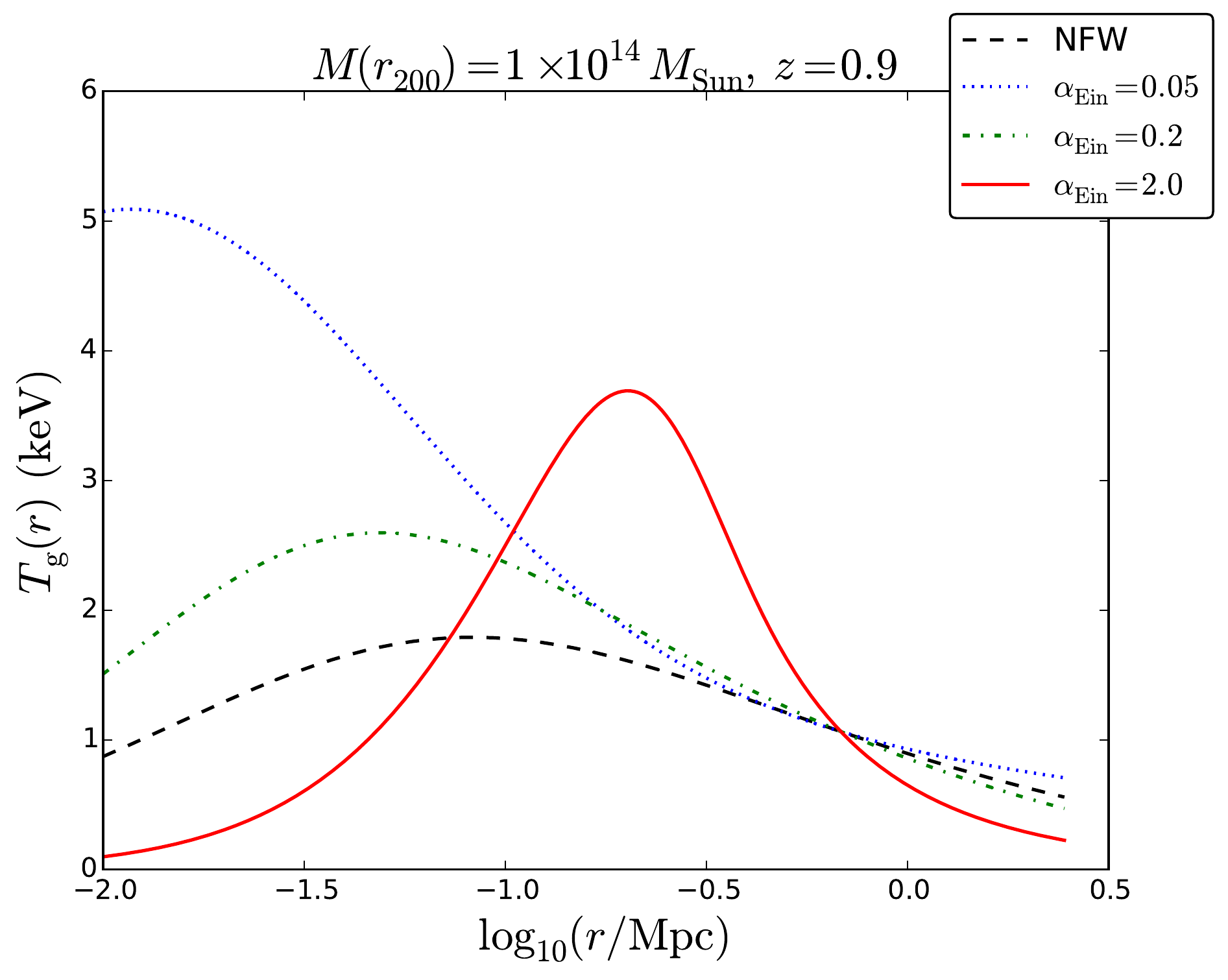} &
     \includegraphics[ width=0.50\linewidth]{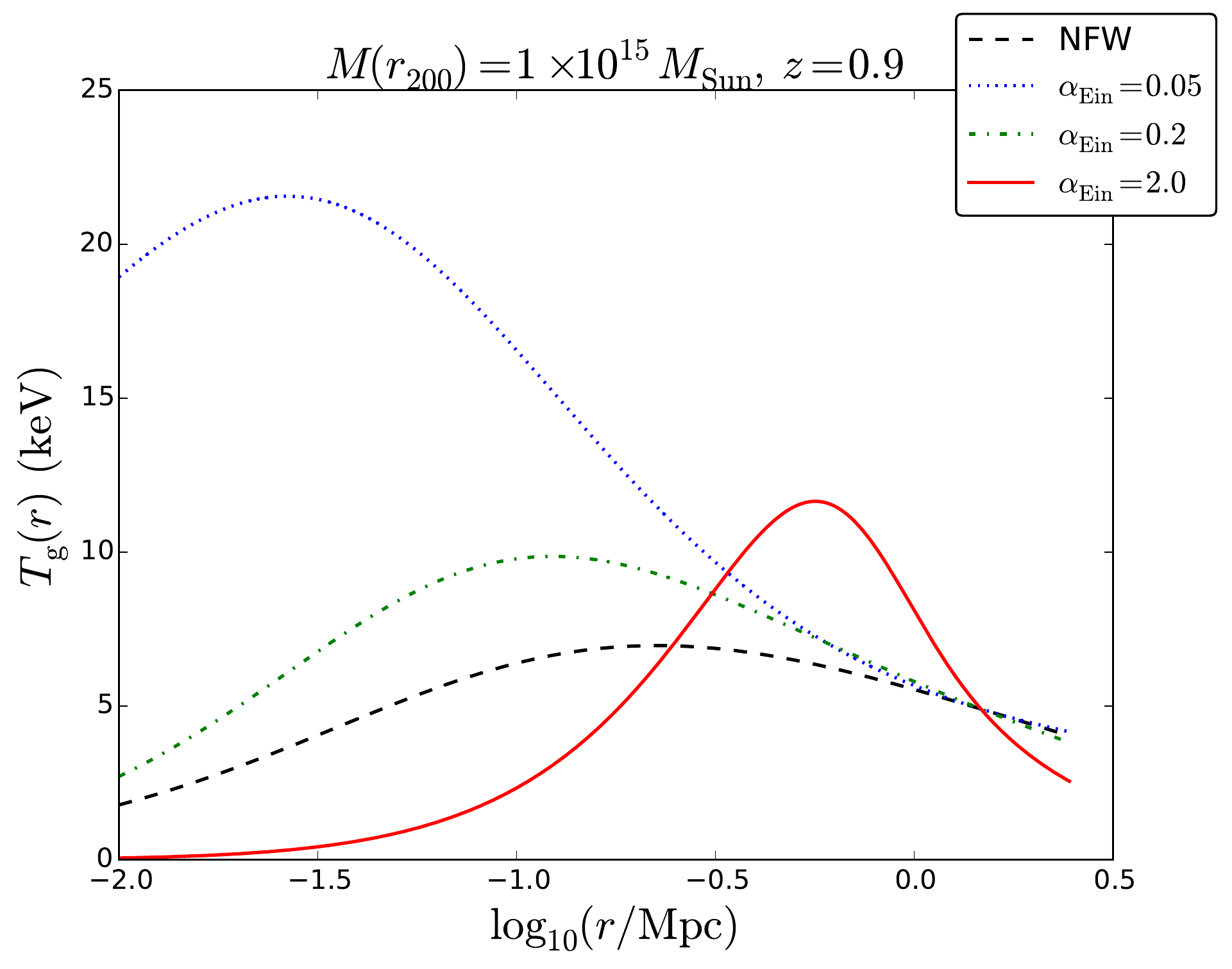} \\
    \end{tabular}
  \caption{Gas temperature profiles as a function of log cluster radius using NFW and Einasto models. Values of $\alpha_{\mathrm{Ein}} = 0.05$, $0.2$, and $2.0$ are used as inputs. Top row has $z = 0.15$, bottom row has $z = 0.9$. Left column has $M(r_{200}) = 1\times 10^{14} M_{\mathrm{Sun}}$, right column has $M(r_{200}) = 1\times 10^{15} M_{\mathrm{Sun}}$.}
\label{f:einnfwgt}
  \end{center}
\end{figure*}

Gas temperature profiles are shown in Figure~\ref{f:einnfwgt}. The $\alpha_{\rm Ein} = 2.0$ is very distinctive, always peaking at much higher $r$ than the other three and also always much more sharply.

\subsection{Bayesian analysis of AMI data}
\label{s:ein_results2}

I now focus on applying the PM II to real and simulated AMI data, to compare the parameter estimates and Bayesian evidences with those obtained from the PM I.


\subsubsection{Analysis of real AMI observations of A611}
\label{s:ein_a611}

\begin{figure}
  \begin{center}
  \includegraphics[ width=0.90\linewidth]{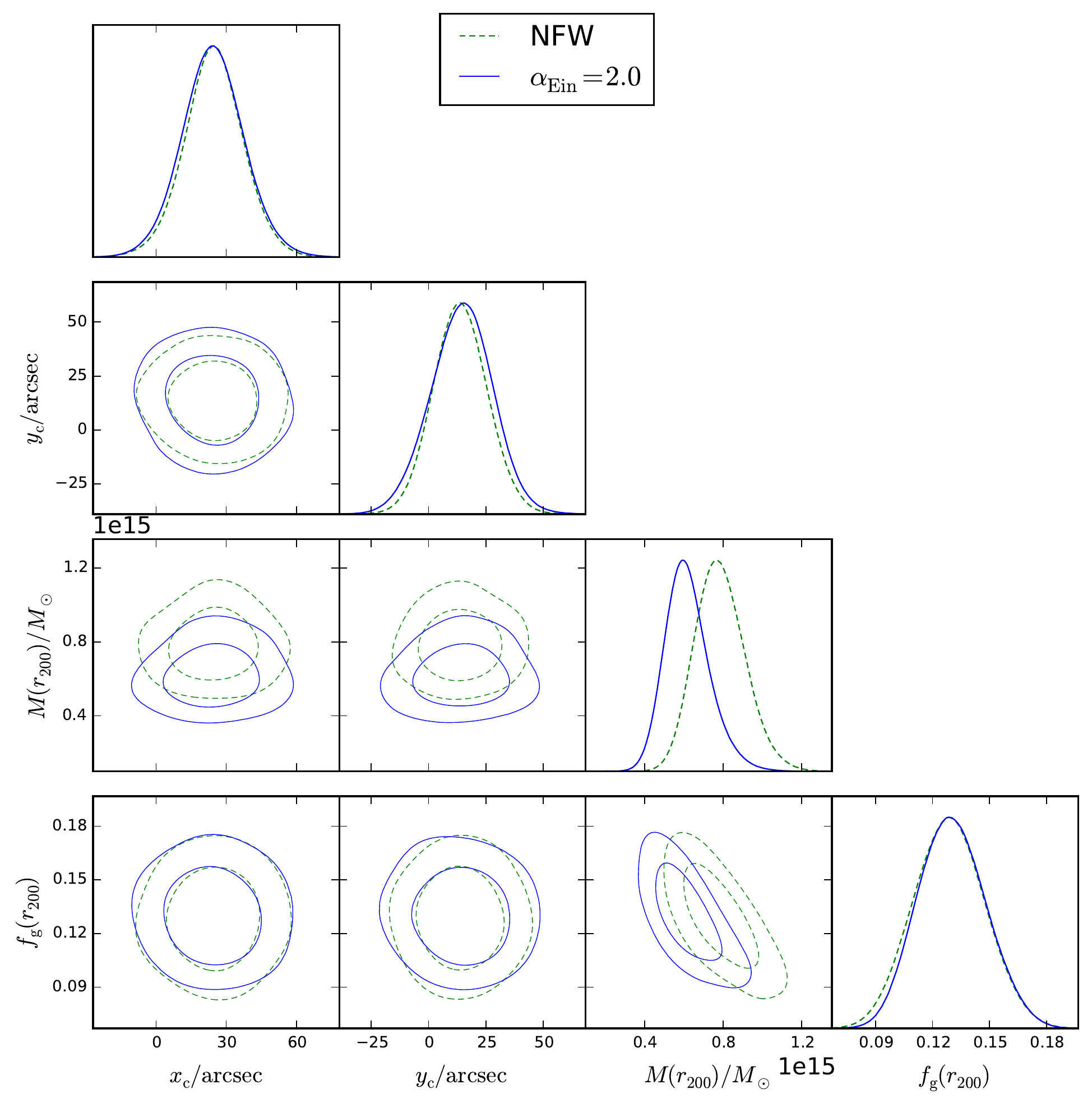}
  \caption{Marginalised posterior distributions of physical model input parameters for the NFW and $\alpha_{\rm Ein} = 2.0$ models applied to real A611 data. The contour plots are the two dimensional marginalised plots of the parameters named in the corresponding row / column. The line plots are the fully marginalised posterior distributions.}
\label{f:a611posterior}
  \end{center}
\end{figure}

I conduct Bayesian analysis on data from observations with AMI of the cluster A611 at $z = 0.288$, which has been studied through its X-ray emission, strong lensing, weak lensing and SZ effect (see \citealt{2007MNRAS.379..209S}, \citealt{2011A&A...528A..73D}, \citealt{2010A&A...514A..88R} and \citealt{2016MNRAS.460..569R} respectively). These studies suggest that there is no significant contamination from radio-sources and that the cluster is close to the $T_{\rm{X}}$--$T_{\rm{SZ}}$ relation for clusters in hydrostatic equilibrium. \\
I first compare the posterior distributions for the input parameters (except those with $\delta$-function priors). The means and standard deviations of the four analyses are given in Table~\ref{t:a611results}. As in Section~\ref{s:ein_results1}, $\alpha_{\rm Ein} = 0.05$ and $\alpha_{\rm Ein} = 0.2$ show similar results to PM I. $\alpha_{\rm Ein} = 2$ gives a different estimate for $M(r_{200})$, and its posterior distribution is shown in Figure~\ref{f:a611posterior} along with that obtained with the NFW profile. These posterior distributions are plotted using \textsc{GetDist} and the contours on the two-dimensional plots represent the 95\% and 68\% confidence intervals. The mean mass estimates are within one combined standard deviation away from each other. However, as seen in Table~\ref{t:a611results} the value of $\ln(\mathcal{Z}_{\rm{Ein}} / \mathcal{Z}_{\rm NFW})$ imply that `no model is favoured by the data' according to the Jeffreys scale. 

\begin{table*}
\begin{center}
\begin{tabular}{{l}{c}{c}{c}{c}{c}}
\hline
Model & $x_{\rm c}$~(arcsec) & $y_{\rm c}$~(arcsec) & $M(r_{200})$~($\times 10^{14}M_{\mathrm{Sun}}$) & $f_{\rm gas}(r_{200})$ & $\ln \left(\mathcal{Z}\right)$ \\ 
\hline
NFW & $24.7 \pm 12.4$ & $13.9 \pm 11.5$ & $7.84 \pm 1.24$ & $0.129 \pm 0.020$ & $3.862944\times10^{4} \pm 0.25$ \\
$\alpha_{\rm Ein} = 0.05$  & $22.7 \pm 12.5$ & $13.1 \pm 12.6$ & $7.45 \pm 1.24$ & $0.130 \pm 0.019$ & $3.862921\times10^{4}  \pm 0.25$ \\
$\alpha_{\rm Ein} = 0.2$  & $25.5 \pm 12.8$ & $14.9 \pm 13.0$ & $7.67 \pm 1.27$ & $0.127 \pm 0.017$ & $3.862967\times10^{4}  \pm 0.24$ \\
$\alpha_{\rm Ein} = 2.0$  & $24.3 \pm 12.4$ & $14.3 \pm 13.2$ & $6.17 \pm 1.12$ & $0.130 \pm 0.017$ & $3.862924\times10^{4} \pm 0.24$ \\
\hline
\end{tabular}
\caption{Marginalised posterior distribution mean values and standard deviations of physical model input parameters and Bayesian evidences associated with each model, applied to real A611 data.}
\label{t:a611results}
\end{center}
\end{table*}

\subsubsection{Simulated AMI data}
\label{s:ein_simulated}
\citet{2016JCAP...01..042S} study the errors associated with fitting NFW profiles to Einasto dark matter halos and vice versa for weak lensing studies. I conduct similar work in the context of simulated SZ observations. The simulations were carried out using the in-house AMI simulation package \textsc{Profile}, which has been used in various forms in e.g. \citet{2002MNRAS.333..318G} and \citet{2013MNRAS.430.1344O}.\\
As before I consider Einasto profiles with the $\alpha_{\rm Ein}$ values $0.05$, $0.2$, and $2.0$ plus an NFW profile. each with $M(r_{200}) = 1\times 10^{14} M_{\mathrm{Sun}}$ or $M(r_{200}) = 1\times 10^{15} M_{\mathrm{Sun}}$, $z = 0.15$ or $z = 0.90$ and $f_{\rm gas}(r_{200}) = 0.12$. These 16 simulations, were analysed as in Section~\ref{s:ein_a611}. Note for all of these simulations no radio-sources, primordial CMB or confusion noise were included, and instrumental noise was set to a negligible level.
Table~\ref{t:ein_simrestab} in Appendix~\ref{s:ein_simrestab} summarises the input and output values of the 16 simulations. The first column gives the model used to \textit{simulate} the cluster, with the following two columns giving the mass and $z$ input values. For each simulation, I \textit{analysed} the data using two models, one using the NFW profile and one using an Einasto profile. For data simulated using an NFW profile, when analysing the data with an Einasto profile I used $\alpha_{\rm Ein} = 0.2$. For data simulated using an Einasto profile, when analysing the data with an Einasto profile I set $\alpha_{\rm Ein}$ equal to the value used as the input for the simulation. 

\begin{figure*}
  \begin{center}
  \includegraphics[ width=0.45\linewidth]{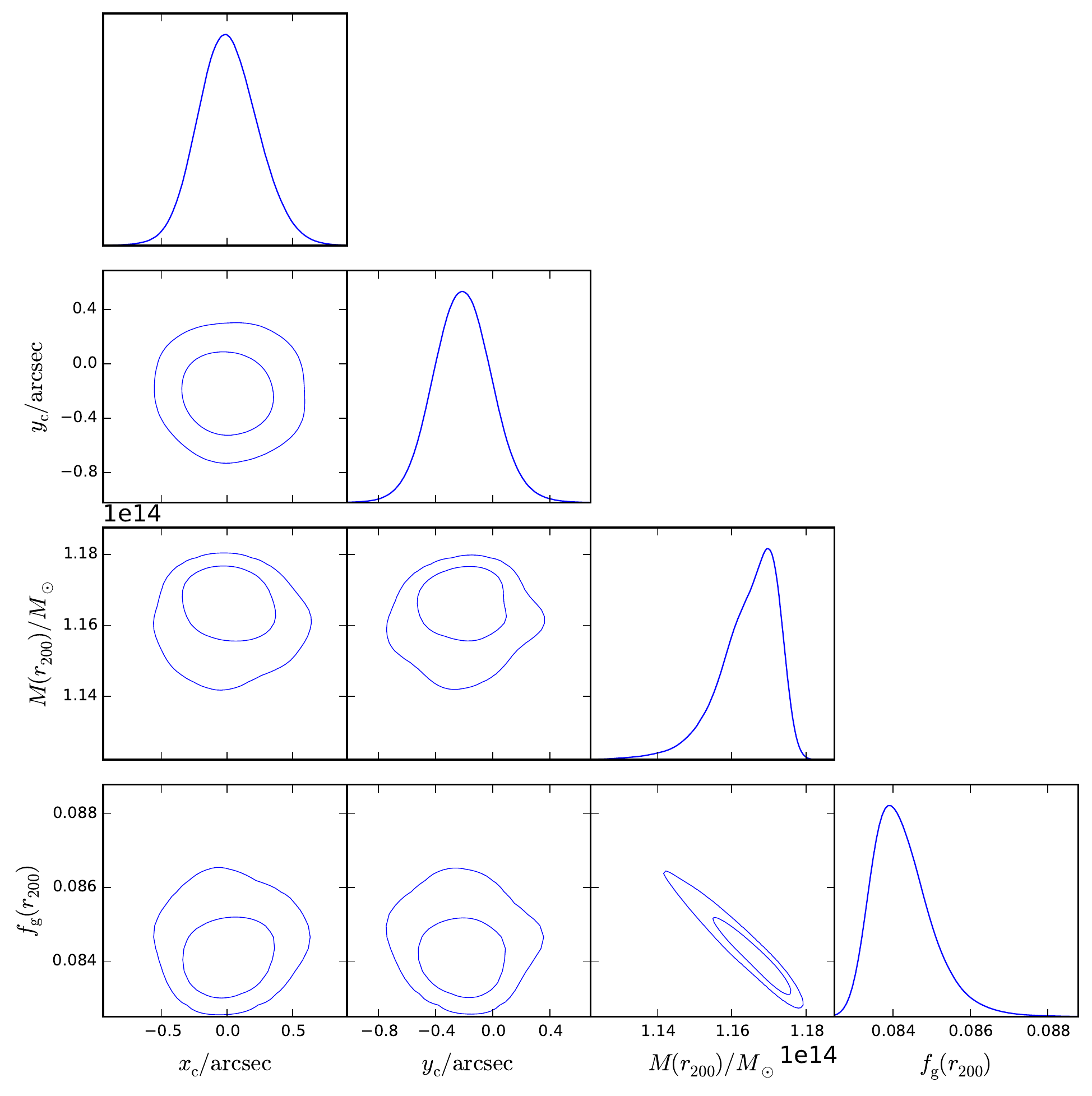}
  \includegraphics[ width=0.45\linewidth]{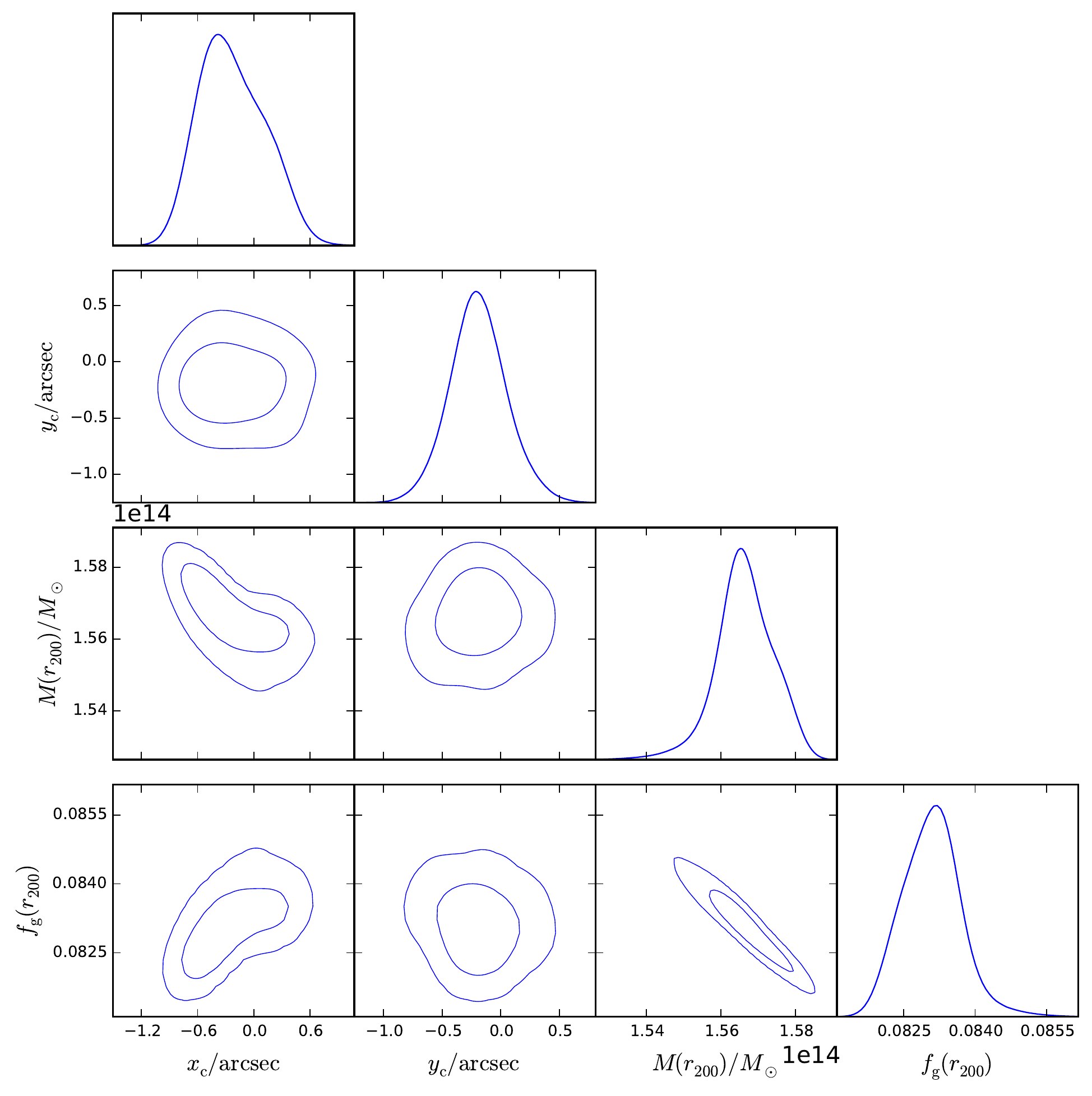}
  \medskip
  \centerline{(a) \hskip 0.45\linewidth (b)}
  \caption{Posterior distributions for cluster simulated with $\alpha_{\rm Ein} = 2.0$, $M(r_{200}) = 1\times 10^{14} M_{\mathrm{Sun}}$ and $z=0.9$, modelled with: (a) Einasto dark matter profile, and (b) NFW dark matter profile.}
\label{f:einnfwbadbadposts}
  \end{center}
\end{figure*}

In all but one of the simulations (NFW simulated with $M(r_{200}) = 1\times 10^{14} M_{\mathrm{Sun}}$ and $z=0.9$), the Einasto posterior mean mass value was closer to the input value than the corresponding NFW value. It's worth nothing that a more thorough statistical treatment would involve repeating the Bayesian analyses many times to see if these results held consistently, but this was not considered here. In 11 out of 16 cases the Einasto profile recovers the input mass to within 10\% (interestingly, it does so for all the NFW simulated clusters). However, in only two of 16 cases does the Einasto model recover the input value within three standard deviations. This could be due to errors associated with the simulated `observing' of the cluster on a pixelated grid, binning the data in u-v space/ frequency and then modelling the data by creating another pixelated grid. These effects are not accounted for in the Bayesian analysis, thus leading to an underestimate in the associated errors. Furthermore, the fact that the Einasto model recovers the NFW simulated clusters better than when those simulations are analysed with the NFW profile for three of the four NFW simulated clusters, could be down to the fact that the Einasto model is more robust to the imperfections associated with the generation of the simulations. Another source of error underestimation could be the sampling errors being underestimated in the nested sampling algorithm as studied in \citet{2017arXiv170309701H}. 
Looking at the individual evidence values for both Einasto and NFW models, the value is considerably lower for the high mass simulations, $ \ln(\mathcal{Z}_{\rm low \, mass} / \mathcal{Z}_{\rm high \, mass}) \approx 3000$ suggests the models fit the low mass datasets much better when averaged over the (same) parameter sampling spaces. 
It is crucial to note that when comparing evidences calculated from different datasets (specifically their ratio), we are not looking at $Pr(\mathcal{M}|\vec{\mathcal{D}}_{1}) / Pr(\mathcal{M}|\vec{\mathcal{D}}_{2})$, since the $Pr(\vec{\mathcal{D}})$-like terms on the right hand side of equation~\ref{e:bayesmodel} do not cancel in this case. Nevertheless for the same model, the evidence ratio between two different datasets does give a measure of the relative goodness of fit of the datasets to the model.
\begin{figure*}
  \begin{center}
  \includegraphics[ width=0.45\linewidth]{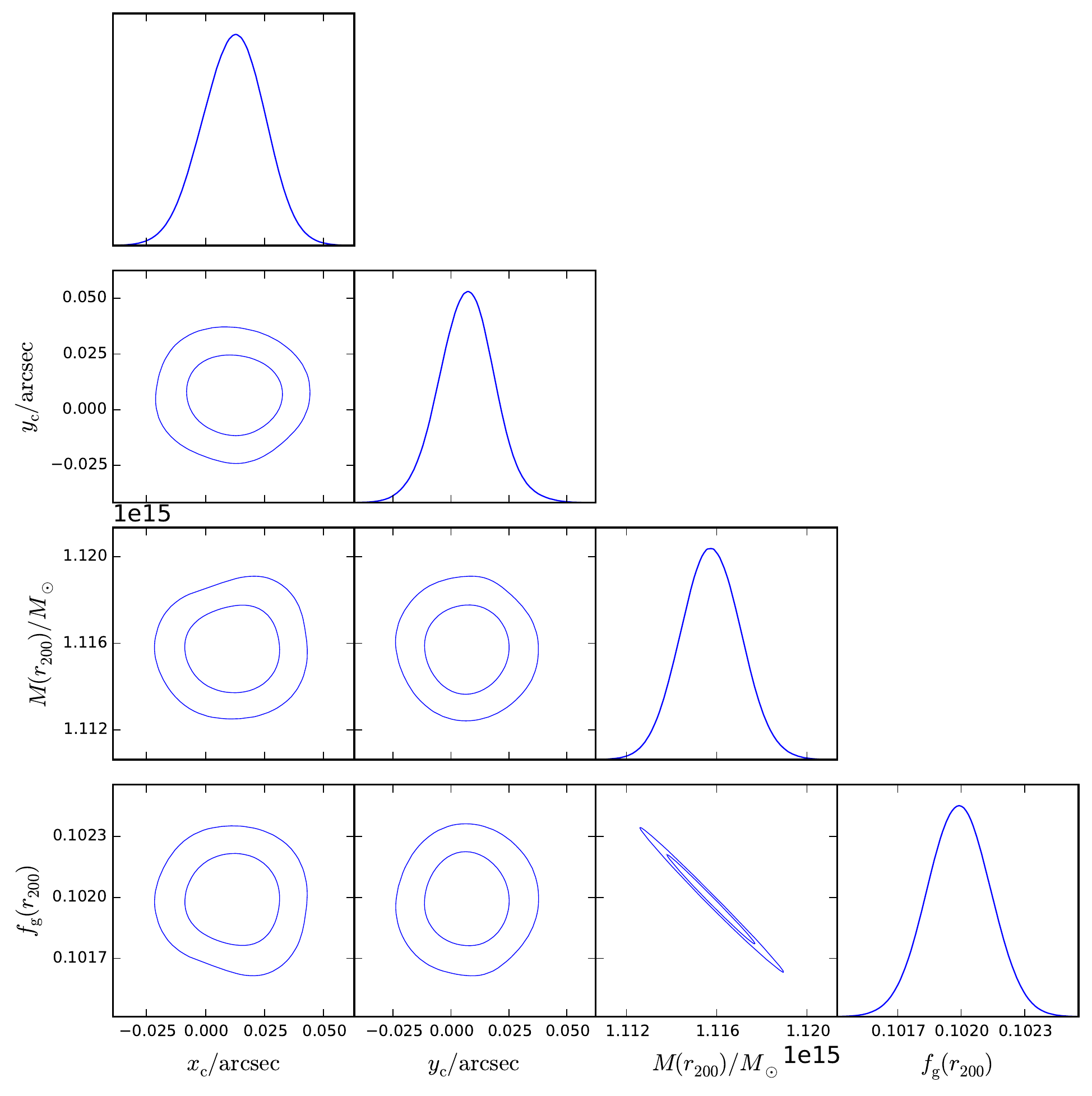}
  \includegraphics[ width=0.45\linewidth]{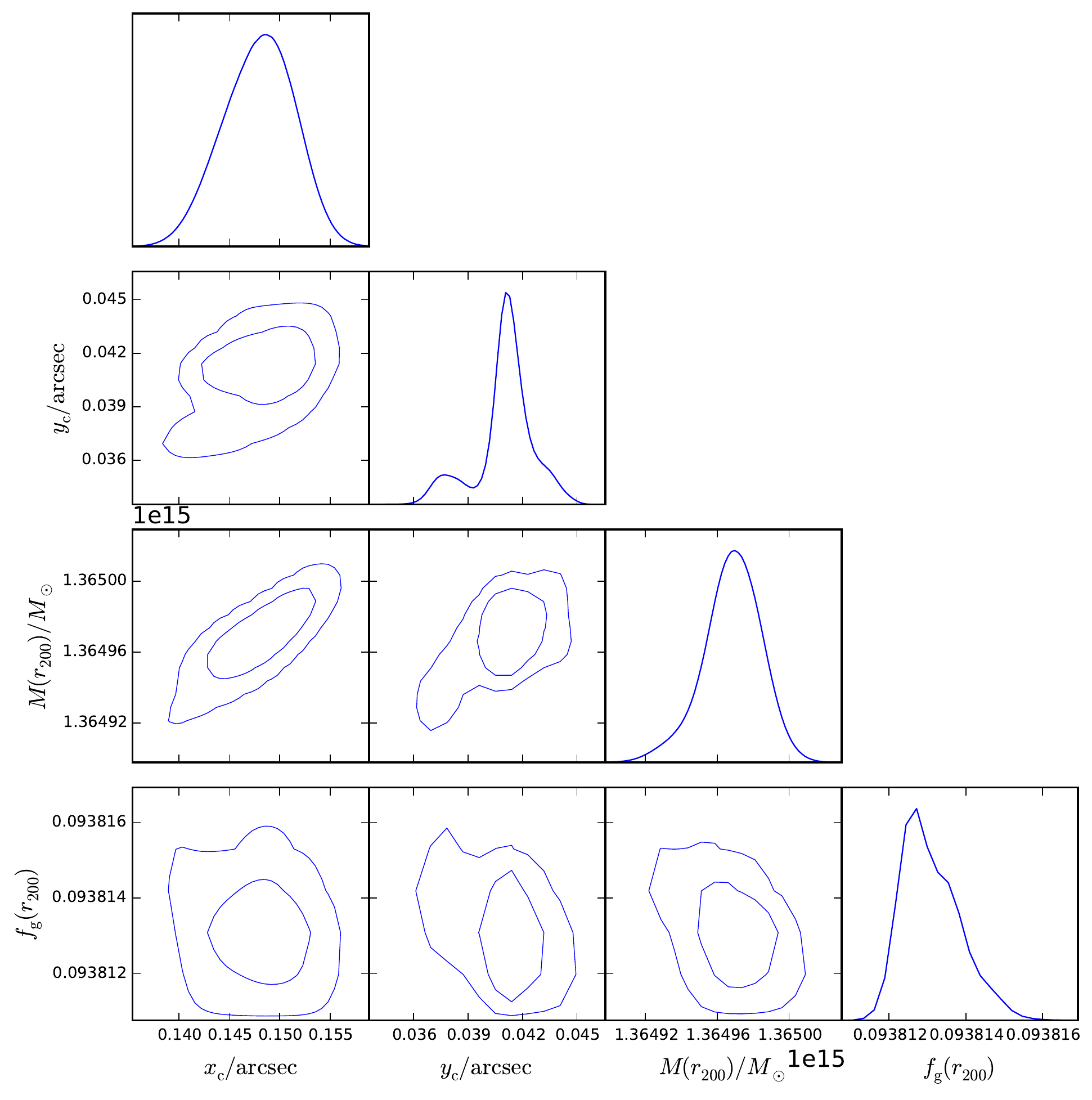}
  \medskip
  \centerline{(a) \hskip 0.45\linewidth (b)}
  \caption{Posterior distributions for cluster simulated with $\alpha_{\rm Ein} = 2.0$, $M(r_{200}) = 1\times 10^{15} M_{\mathrm{Sun}}$ and $z=0.9$, modelled with: (a) Einasto dark matter profile, and (b) NFW dark matter profile.}
\label{f:einnfwgoodbadposts}
  \end{center}
\end{figure*}
Looking at the evidence ratios between the Einasto and NFW models for a given simulation, more data is needed to come to a conclusive decision over model preference in $10$ of the simulations. Three simulations lead to `substantial preference' in favour of the Einasto model $\left(\ln\left(\mathcal{Z}_{\rm Ein} / \mathcal{Z}_{\rm NFW}\right) \geq 5\right)$. In two of these cases ($\alpha_{\rm Ein} = 0.2$ with $M(r_{200}) = 1\times 10^{15} M_{\mathrm{Sun}}$ and $z=0.9$, and $\alpha_{\rm Ein} = 2.0$ with $M(r_{200}) = 1\times 10^{15} M_{\mathrm{Sun}}$ and $z=0.9$) the posteriors show reasonable constraints in both the Einasto and NFW analyses (Figure~\ref{f:einnfwgoodbadposts} shows posterior distributions for $\alpha_{\rm Ein} = 2.0$ with $M(r_{200}) = 1\times 10^{15} M_{\mathrm{Sun}}$ and $z=0.9$), with the former giving better estimates of mass and $f_{\rm{gas}}(r_{200})$. The third case however ($\alpha_{\rm Ein} = 2$ simulated with $M(r_{200}) = 1\times 10^{14} M_{\mathrm{Sun}}$ and $z=0.9$) leads to low estimates of $f_{\rm{gas}}(r_{200})$ in both cases (Figure~\ref{f:einnfwbadbadposts}), and a very high mass estimate in the case of the NFW model. 
The two cases where the NFW model is preferred over the Einasto also produce posteriors similar to those in Figure~\ref{f:einnfwbadbadposts}.

\begin{figure*}
  \begin{center}
  \includegraphics[ width=0.45\linewidth]{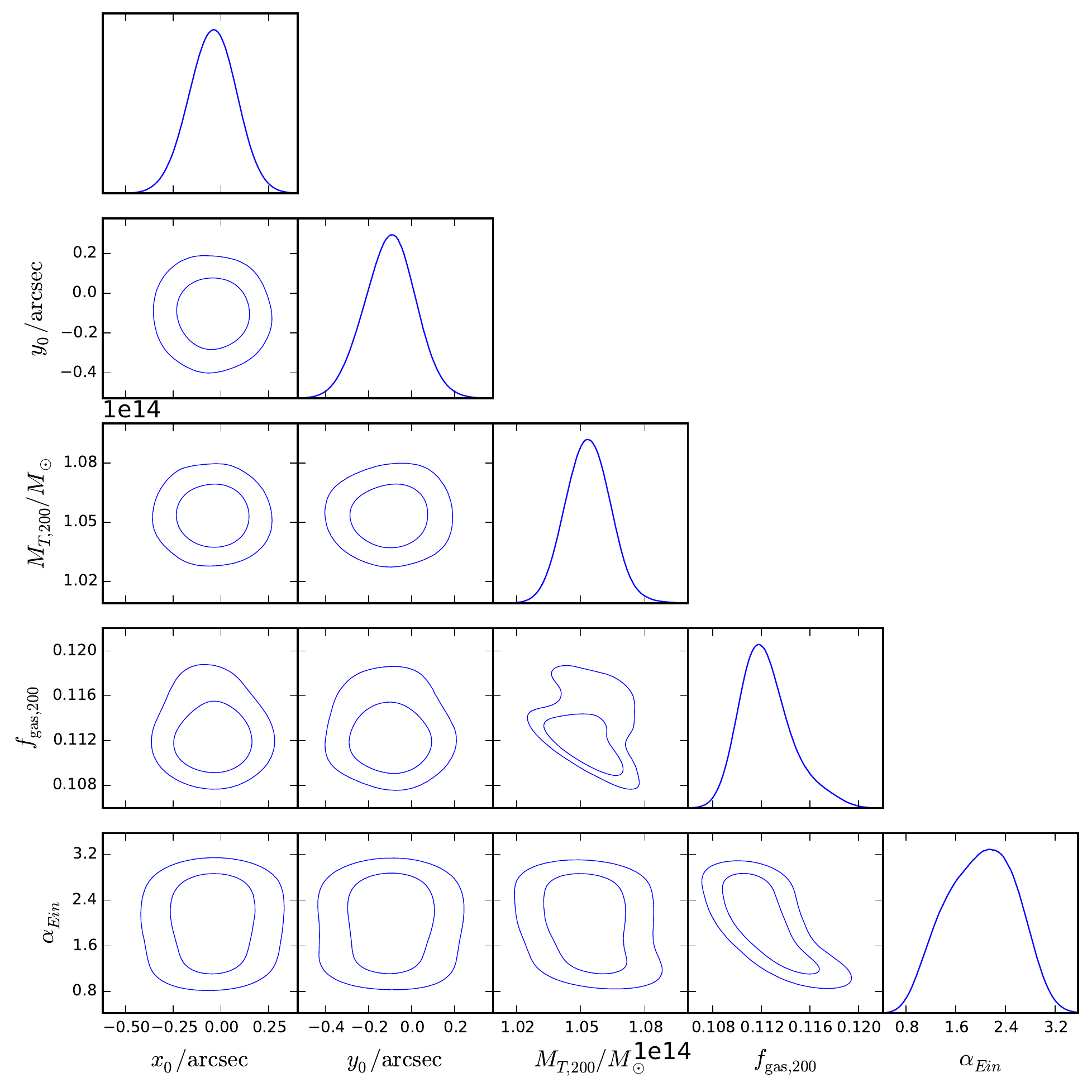}
  \includegraphics[ width=0.45\linewidth]{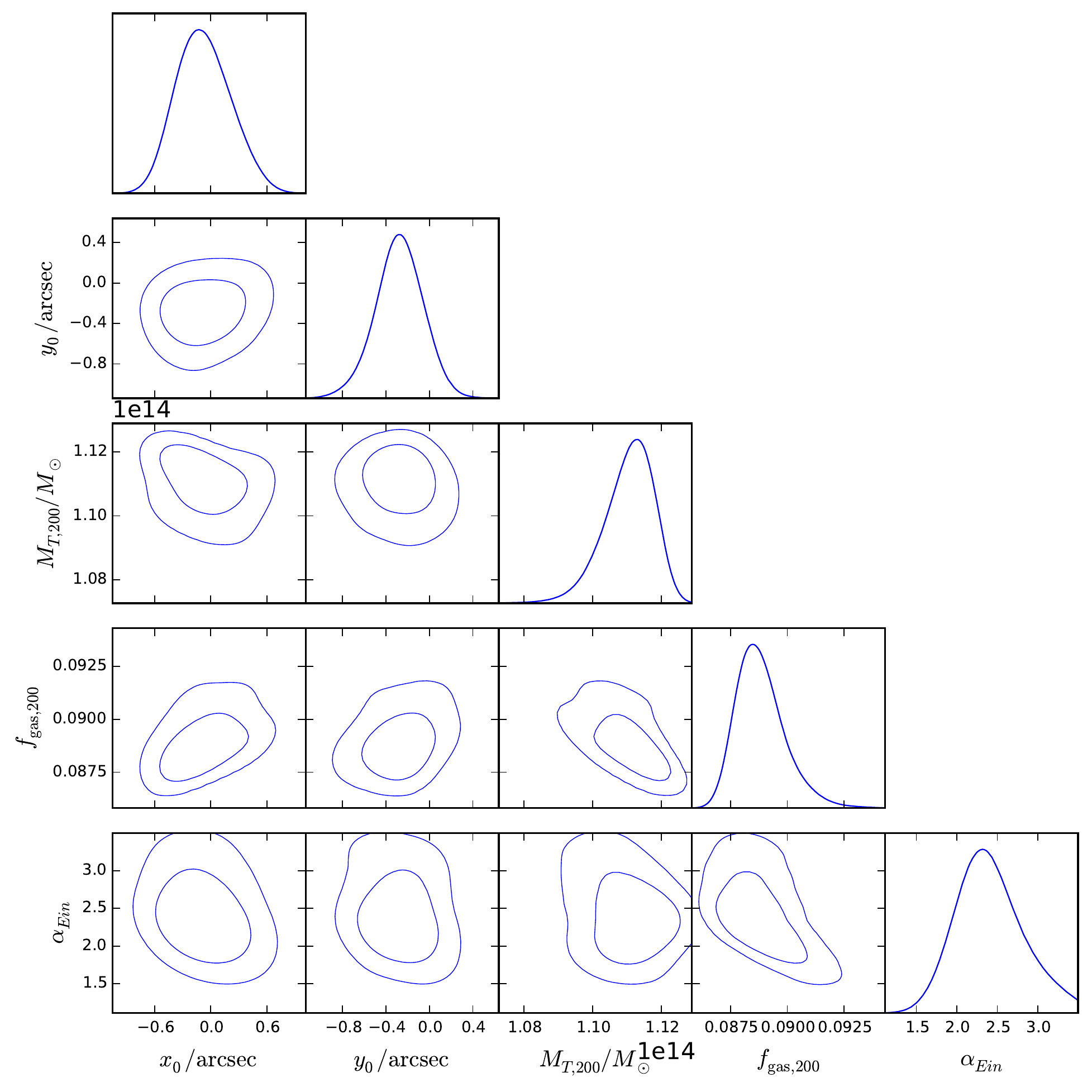}
  \medskip
  \centerline{(a) \hskip 0.45\linewidth (b)}
  \caption{Posterior distributions of Einasto model input parameters for: (a) $\alpha_{\rm Ein} = 2.0$, $M(r_{200}) = 1\times 10^{14} M_{\mathrm{Sun}}$ and $z=0.15$ simulated cluster, and (b) $\alpha_{\rm Ein} = 2.0$, $M(r_{200}) = 1\times 10^{14} M_{\mathrm{Sun}}$ and $z=0.9$ simulated cluster.}
\label{f:eingoodalpha}
  \end{center}
\end{figure*}

Finally, I tried running the Bayesian analysis on eight of the Einasto simulated clusters with uniform analysis priors on $\alpha_{\rm Ein}$. These clusters corresponded to the simulations with input values of either $\alpha_{\rm Ein} = 0.2$ or $\alpha_{\rm Ein} = 2.0$. For the former value of $\alpha_{\rm Ein}$ I assigned the uniform prior $\mathcal{U}[0.05, 0.35]$ and $\mathcal{U}[0.5, 3.5]$ for the latter. For two of these simulations the posterior distributions did not show much degeneracy between any of the input parameters, including $\alpha_{\rm Ein}$. Both of these clusters had $\alpha_{\rm Ein}=2$, $M(r_{200}) = 1 \times 10^{14} M_{\mathrm{Sun}}$ and $z=0.15$ or $z=0.9$ as inputs. Their posterior distributions are shown in Figure~\ref{f:eingoodalpha}. Both posteriors give a mean value for the shape parameter within one standard deviation of the input value ($2.01 \pm 0.54$ and $2.39 \pm 0.40$), but looking at the distributions they are not sharply peaked, meaning the errors on the estimates are quite large. Nevertheless these simulations do show the Einasto profile is capable of recovering some information about $\alpha_{\rm Ein}$, in contrast to the efforts in MO12 to recover $c_{200}$ which led to large $c_{200}-M(r_{200})$ degeneracies (although $c_{200}$ relates to the scale of the dark matter profile, not its shape). 

\section{Conclusions}
\label{s:ein_summary}

Based on the physical model introduced in Section~\ref{s:phys_mod} (PM I) which uses an NFW profile \citep{1995MNRAS.275..720N} to model the dark matter content of galaxy clusters, I derive a new physical model (PM II) which models the dark matter with an Einasto profile \citep{1965TrAlm...5...87E}.
The Einasto profile has an additional degree of freedom compared to the NFW profile, which dictates the shape of the dark matter density as a function of radius. For different values of $\alpha_{\rm Ein}$ we have investigated the profiles of several physical properties of a cluster, namely the dark matter density, dark matter mass, gas density, gas mass and gas temperature. I have also provided the equivalent profiles in the NFW case. From this I found the following.
\begin{itemize}
\item Of the three values of $\alpha_{\rm Ein}$ considered, $\alpha_{\rm Ein} = 0.2$ gave the most similar profile to that given by the NFW model (as discussed in \citealt{2014MNRAS.441.3359D}), with the main discrepancy between the two arising in the peak amplitude of the gas temperature.
\item $\alpha_{\rm Ein} = 2.0$ showed the most convergent behaviour in $M_{\rm dm}(r)$ at high $r$, but the most divergent in $M_{\rm g}(r)$ in the same limit.
\item The gas temperature profiles were somewhat different for the $\alpha_{\rm Ein}$ values considered here. This suggests that if one can carefully measure the temperature profile of a cluster, then one could infer $\alpha_{\rm Ein}$ and use this in the model presented here (though one has to be aware of cooling flow and merger activity).
\end{itemize}
Next we applied Bayesian analysis to real and simulated AMI datasets using PM I and PM II, to compare the models' parameter estimates and fits to the data. Using real data from cluster A611 I found the following.
\begin{itemize}
\item The $\alpha_{\rm Ein} = 0.05$ and $\alpha_{\rm Ein} = 0.2$ models gave very similar results to the NFW model; the $\alpha_{\rm Ein} = 2$ model however underestimates $M(r_{200})$ relative to the other three models.
\item The Bayesian evidence values calculated from these four analyses were roughly equal, suggesting no model provided a statistically significant fit relative to the others.
\end{itemize}
Simulating clusters with either NFW or Einasto dark matter profiles, which were then `observed' by AMI, I found the following.
\begin{itemize}
\item For 15 out of 16 clusters, the Einasto model recovered the input mass better than the NFW model. The only cluster where this was not the case (NFW simulated with $M(r_{200}) = 1\times 10^{14} M_{\mathrm{Sun}}$ and $z=0.9$), the posterior distributions do not show good constraints on the sampling parameters, and so the parameter estimates should not be used.
\item The evidence values of both Einasto and NFW models are considerably lower for the high mass simulations. 
\item Considering the evidence ratios between the Einasto and NFW models for a given simulation, more data is needed to come to a conclusive decision over model preference in $10$ of the cases. However according to the Jeffreys scale \citep{jeffreys}, three of the simulations gave `substantial' preference towards the Einasto model; and in two of these cases the NFW analysis did not constrain the sampling parameters as well as the Einasto analysis. In the third case neither analysis constrained the parameters well. 
\item The two clusters where the evidence ratio was in favour of the NFW model also showed poor posterior distribution constraints. 
\item When allowing $\alpha_{\rm Ein}$ to vary in the analysis, in two out of eight of the Einasto simulations used the posterior distributions showed some constraints on the value of $\alpha_{\rm Ein}$ which gave estimates close to the input values. 
\end{itemize}

%% file: CHAP-6/chapter6.tex
\chapter{Enhanced physical modelling I: relaxing the $f_{\rm gas}$ assumption}\label{c:sixth}

As stated in Section~\ref{s:phys_mod}, one of the key assumptions of the physical model (for both PM I and PM II) is that the gas mass fraction $f_{\rm gas}(r)$ is much smaller than unity up to $r_{200}$, so that we can say the total mass at $r_{200}$ is equal to dark matter mass enclosed up to this radius. In this Chapter we relax this assumption for both models, so that the total mass is the sum of the dark matter and gas contributions. We refer to these new models as PMT I and PMT II which respectively use NFW and Einasto profiles to model the dark matter content. 


\section{Total mass equations} \label{s:tot_mass_eqn}

Dropping the assumption that $f_{\rm gas}(r) \ll 1$ 
we can no longer assume that $M(r) \approx \int_{0}^{r} 4\pi\rho_{\rm dm}(r')r'^{2}\,\mathrm{d}r'$, but instead
\begin{equation}\label{e:tot_mass1}
M(r) = \int_{0}^{r} 4\pi\rho_{\rm dm}(r')r'^{2}\,\mathrm{d}r' + \int_{0}^{r} 4\pi\rho_{\rm g}(r')r'^{2}\,\mathrm{d}r'.
\end{equation}
Using the hydrostatic equilibrium assumption given by equation~\ref{e:hse} to substitute for $\rho_{\rm g}(r')$, we get the following integral equation
\begin{equation}\label{e:tot_mass2}
M(r) = \int_{0}^{r} 4\pi\rho_{\rm dm}(r')r'^{2}\,\mathrm{d}r' - \frac{4\pi}{G} \int_{0}^{r} \frac{\mathrm{d}P_{\mathrm{g}}(r')}{\mathrm{d}r'} \frac{r'^{4}}{M(r')}\,\mathrm{d}r'.
\end{equation} 
Differentiating equation~\ref{e:tot_mass2} with respect to $r$ gives the differential equation
\begin{equation}\label{e:mass_diff}
\frac{\mathrm{d} M(r)}{\mathrm{d} r} = 4\pi\rho_{\rm dm}(r)r^{2} - \frac{4\pi}{G} \frac{\mathrm{d}P_{\mathrm{g}}(r)}{\mathrm{d}r} \frac{r^{4}}{M(r)}.
\end{equation}
Assuming a GNFW profile (equation~\ref{e:gnfw}) for $P_{\mathrm{e}}$, and relating it to $P_{\mathrm{g}}$ using equation~\ref{e:pgas_pelec} the second term on the RHS of equation~\ref{e:mass_diff} becomes
\begin{equation}\label{e:gas_mass_diff}
 \frac{\mu_{\rm e}}{\mu_{\rm g}} \frac{4\pi P_{\rm ei}}{G} \frac{r^3}{M(r)} \left(\frac{r}{r_{\rm p}}\right)^{-c}\left[1+\left(\frac{r}{r_{\rm p}}\right)^{a}\right]^{-(1+(b-c)/a)}\left[b\left(\frac{r}{r_{\rm p}}\right)^{a}+c\right] \equiv \frac{K(r)}{M(r)}.
\end{equation}
Hence for PMT I (NFW dark matter profile) 
\begin{equation}\label{e:nfw_mass_diff}
\frac{\mathrm{d} M(r)}{\mathrm{d} r} = 4 \pi \rho_{\rm s} r_{\rm s}^3 \frac{r}{\left( r + r_{\rm s} \right)^2} + \frac{K(r)}{M(r)},
\end{equation}
and for PMT II (Einasto dark matter profile)
\begin{equation}\label{e:ein_mass_diff}
\frac{\mathrm{d} M(r)}{\mathrm{d} r} = 4 \pi \rho_{-2} r^2 \exp \left[ -\frac{2}{\alpha_{\rm Ein}} \left( \left( \frac{r}{r_{-2}}\right)^{\alpha_{\rm Ein}} - 1 \right) \right] + \frac{K(r)}{M(r)}.
\end{equation}


\section{Determining cluster profile parameters} \label{s:tot_mass_pars}

Equations~\ref{e:nfw_mass_diff} and~\ref{e:ein_mass_diff} are first order non-linear differential equations with dependent variable $M$ and independent variable $r$. They are subject to the boundary condition that $M(r_{200}) = $ the value input to the model. Each equation has four unknown parameters: $r_{\rm s}$ for PMT I ($r_{-2}$ for PMT II), $\rho_{\rm s}$ for PMT I ($\rho_{-2}$ for PMT II), $r_{\rm p}$ and $P_{\rm ei}$. $r_{\rm s}$ ($r_{-2}$) can be calculated the same way as previously. $\rho_{\rm s}$ ($\rho_{-2}$) can be calculated in a similar way to previously (i.e. as in Section~\ref{s:phys_mod} for PM I and Section~\ref{s:dm_models} for PM II), but we now solve 
\begin{equation}\label{e:rho_dm_mass}
M_{\rm dm}(r) = (1 - f_{\rm gas}(r))\int_{0}^{r} 4\pi\rho_{\rm dm}(r')r'^{2}\,\mathrm{d}r', 
\end{equation}
at $r = r_{200}$ for known $M(r_{200})$ and $f_{\rm gas}(r_{200})$. However $r_{\rm p}$ can no longer be determined, since the mapping from $r_{200}$ to $r_{500}$ explicitly requires the assumption $M(r) = M_{\rm dm}(r)$ for both dark matter models. Thus $P_{\rm ei}$ cannot be uniquely determined from the ODEs, as there is a family of solutions of ($r_{\rm p}$, $P_{\rm ei}$) which satisfy the ODEs, and therefore the pressure profile is no longer uniquely defined for a given set of cluster input parameters. I have thought of three ways to overcome this issue, only one of which I pursue. Nevertheless I now give a brief note on all three ideas.


\subsection{Determining $r_{\rm p}$ and $P_{\rm ei}$ directly from constraints on $M$ and its derivative}
\label{s:tot_mass_pars_det1}
If we knew the value of $M(r)$ and $\mathrm{d} M(r) / \mathrm{d} r$ at two different radii then we would be able to determine unique values of $r_{\rm p}$ and $P_{\rm ei}$ directly from the ODEs. However I have not been able to think of any sensible conditions to impose on $\mathrm{d} M(r) / \mathrm{d} r$ other than $\mathrm{d} M(r \rightarrow \infty) / \mathrm{d} r = 0$.
Furthermore, evaluating equation~\ref{e:mass_diff} asymptotically (i.e. $r \rightarrow 0$ and $r \rightarrow \infty$) does not yield any useful results. I therefore have not been able to use this method successfully in determining $r_{\rm p}$ and $P_{\rm ei}$.


\subsection{Determining $r_{\rm p}$ and $P_{\rm ei}$ using Lagrange multipliers}
\label{s:tot_mass_pars_det2}
Consider the function
\begin{equation}\label{e:lagr_constr}
g(P_{\rm ei}, r_{\rm p}, r) = M(r) - M_{\rm dm}(r) - M_{\rm g}(P_{\rm ei}, r_{\rm p}, r),
\end{equation}
which tells us that $(P_{\rm ei}, r_{\rm p})$ must satisfy $g(P_{\rm ei}, r_{\rm p}, r) = 0$ for all $r$. Since the ODEs in Section~\ref{s:tot_mass_eqn} are derived from $g(P_{\rm ei}, r_{\rm p}, r) = 0$, they share the same family of solutions of $(P_{\rm ei}, r_{\rm p})$. Thus finding values of $(P_{\rm ei}, r_{\rm p})$ which satisfy the ODEs (subject to their boundary condition on $M(r_{200})$) also satisfies $g = 0$ (subject to the same boundary condition).
We can formulate a constrained optimsation problem using Lagrange multipliers
\begin{equation}\label{e:lagr_eqn}
f(P_{\rm ei}, r_{\rm p}) - \lambda g(P_{\rm ei}, r_{\rm p}, r)
\end{equation}
to find stationary points in $f(P_{\rm ei}, r_{\rm p})$ subject to the constraint $g = 0$ for arbitrary $\lambda$. The form of $f(P_{\rm ei}, r_{\rm p})$ dictates the nature of $(P_{\rm ei}, r_{\rm p})$ at which the stationary point(s) of equation~\ref{e:lagr_constr} are observed. For example $f(P_{\rm ei}, r_{\rm p}) = \left( P_{\rm ei} r_{\rm p} \right)^2$ would find the minimum value of the product $P_{\rm ei} r_{\rm p}$ which satisfies $g = 0$.\\
I do not pursue this idea any further however, since I cannot justify using a particular form for $f(P_{\rm ei}, r_{\rm p})$, and because I suspect that finding the stationary points of equation~\ref{e:lagr_eqn} is difficult numerically.


\subsection{Determining $r_{\rm p}$ and $P_{\rm ei}$ using approximate methods}
\label{s:tot_mass_pars_det3}
Since $\rho_{\rm s}$ ($\rho_{-2}$) can be (correctly) calculated from equation~\ref{e:rho_dm_mass} for the PMTs, we can use it in the calculational steps given by the PMs to get approximate values for $r_{\rm p}$ and $P_{\rm ei}$. The issue with this method is that it is difficult to quantify the assumptions made, as we start off considering dark matter and gas contributions to the total mass to calculate $\rho_{\rm s}$ ($\rho_{-2}$), but then have to resort to the $M(r_{200}) \approx M_{\rm dm}(r_{200})$ to calculate $r_{\rm p}$ and $P_{\rm ei}$. Despite this issue, I have adopted this method (due to its simplicity) to plot the mass profiles of clusters with a range of input parameters for illustrative purposes below. Note however that I have not implemented the PMTs into the Bayesian analysis software \textsc{McAdam}, since not being able to quantify the assumptions of the models invalidates their use in Bayesian inference.


\section{Mass profile plots} 
\label{s:tot_mass_plots}
\begin{figure*}
  \begin{center}
    \begin{tabular}{@{}cc@{}}
     \includegraphics[width=0.5\linewidth]{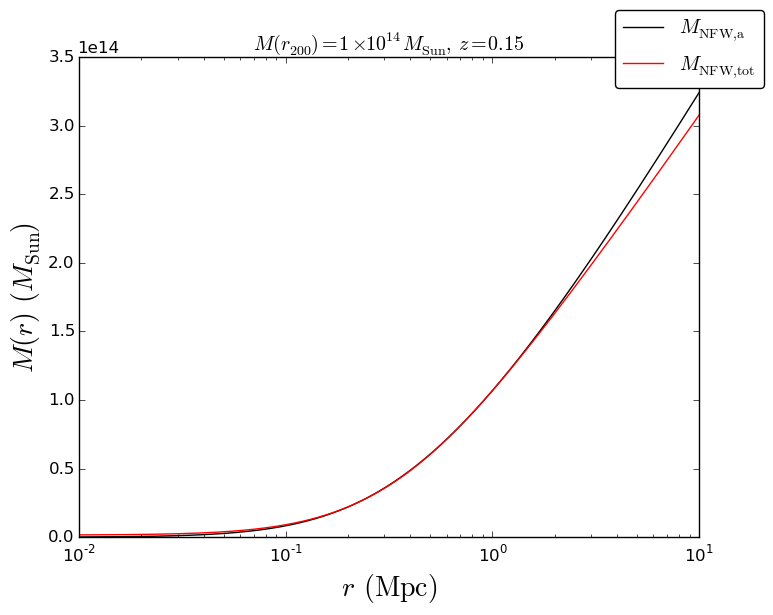} &
     \includegraphics[width=0.5\linewidth]{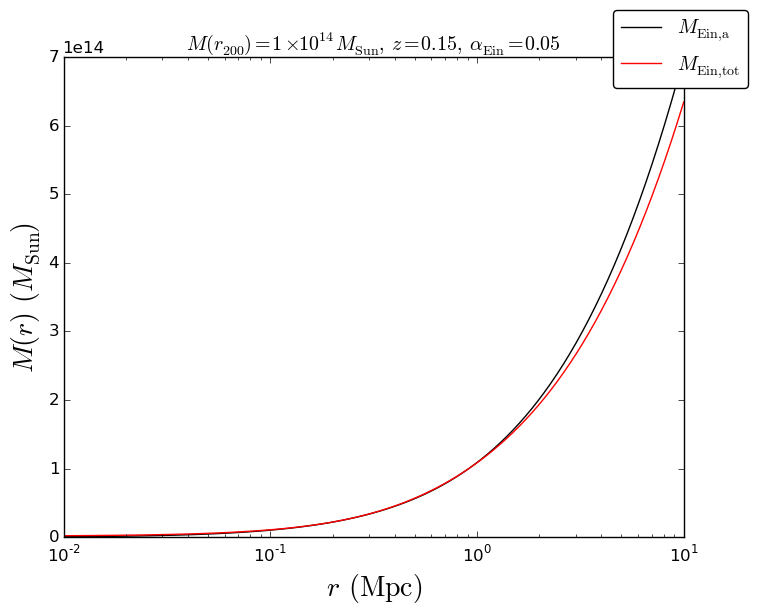} \\
     \includegraphics[ width=0.5\linewidth]{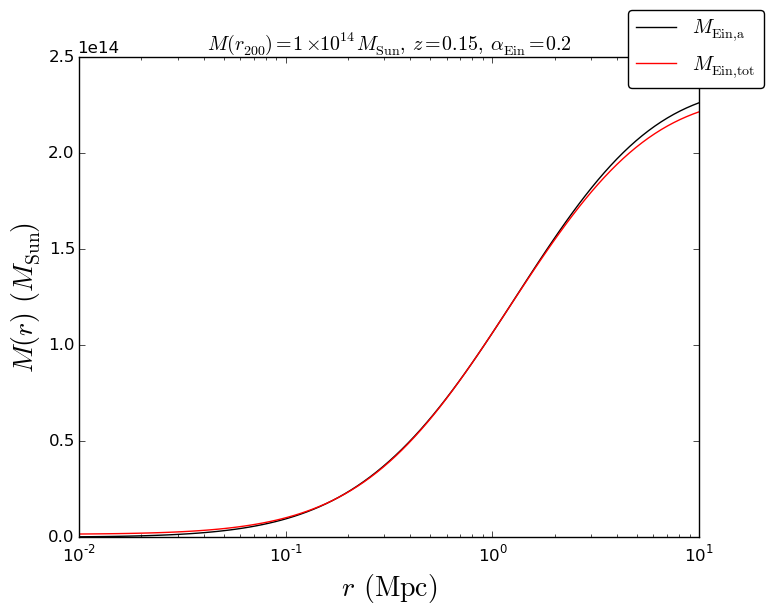} &
     \includegraphics[ width=0.5\linewidth]{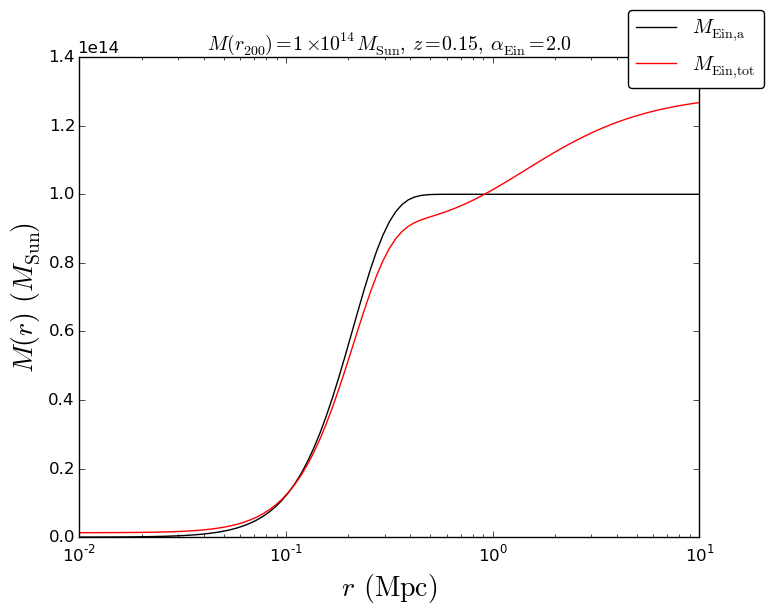} \\
    \end{tabular}

  \caption{Mass profiles of cluster with input parameters given in titles. PM I (NFW dark matter profile, $M(r) \approx M_{\rm dm}(r)$ approximation) and PMT I (NFW dark matter profile, $M(r) = M_{\rm dm}(r) + M_{\rm g}(r)$) are shown in the top left graph by black and red curves respectively. The other three graphs plot PM II (Einasto dark matter profile, $M(r) \approx M_{\rm dm}(r)$ approximation), and PMT II (Einasto dark matter profile, $M(r) = M_{\rm dm}(r) + M_{\rm g}(r)$) in black and red respectively, for $\alpha_{\rm Ein}$ values of $0.05$ (top right),  $0.2$ (bottom left), and $ 2.0$ (bottom right).}
\label{f:tm_lowz_lowm}
  \end{center}
\end{figure*}
\begin{figure*}
  \begin{center}
    \begin{tabular}{@{}cc@{}}
     \includegraphics[width=0.5\linewidth]{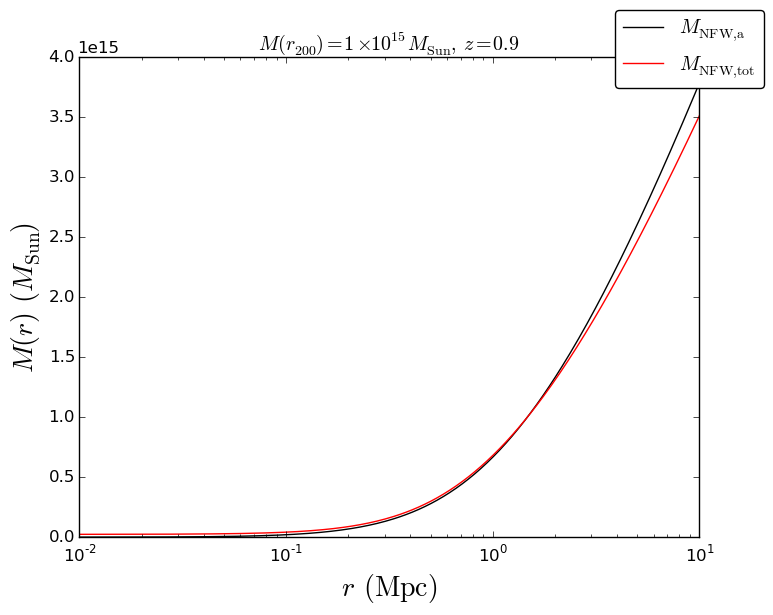} &
     \includegraphics[width=0.5\linewidth]{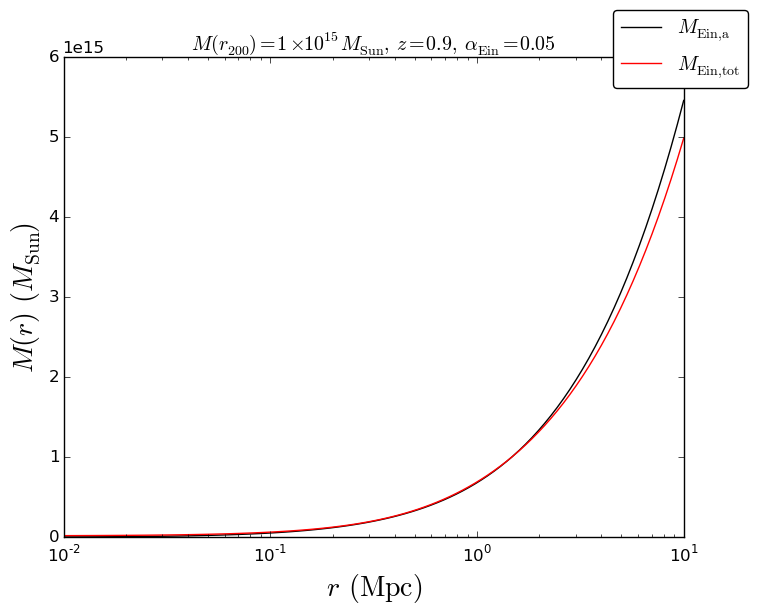} \\
     \includegraphics[ width=0.5\linewidth]{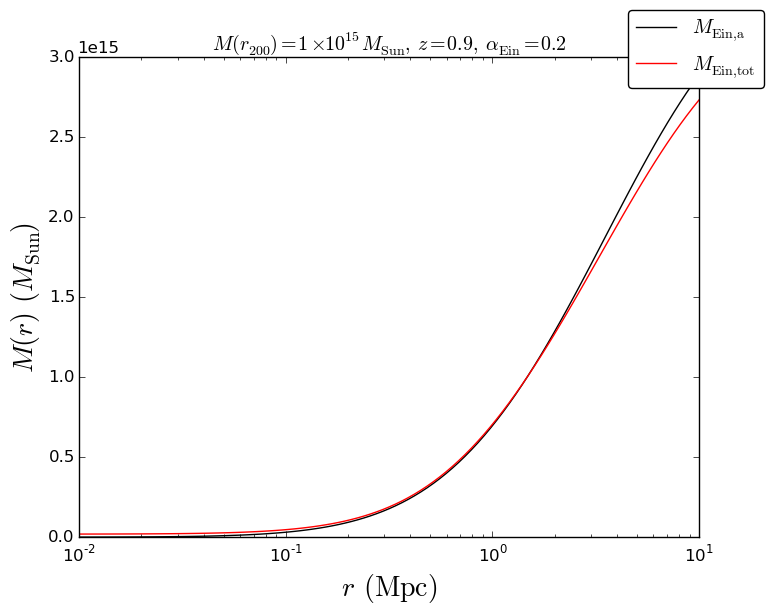} &
     \includegraphics[ width=0.5\linewidth]{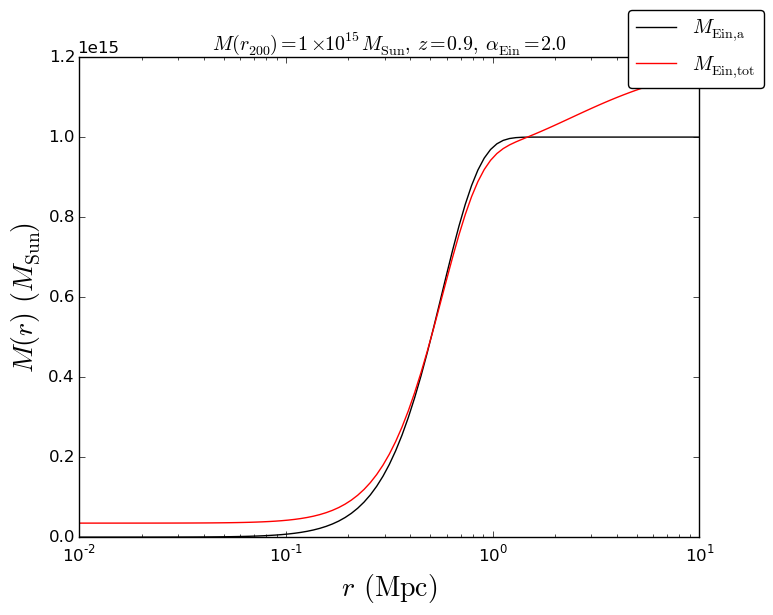} \\
    \end{tabular}

  \caption{Mass profiles of cluster with input parameters given in title, for models given in Figure~\ref{f:tm_lowz_lowm}.}
\label{f:tm_highz_highm}
  \end{center}
\end{figure*}

We now compare the mass profiles of PM I and PM II (calculated using equations~\ref{e:nfw_m_tot_2} and~\ref{e:ein_mass} respectively), with those obtained from PMT I and PMT II (calculated using equations~\ref{e:nfw_mass_diff} and~\ref{e:ein_mass_diff} respectively), using values for $r_{\rm p}$ and $P_{\rm ei}$ obtained using the method outlined in Section~\ref{s:tot_mass_pars_det3} for the PMTs.
As in Section~\ref{s:ein_results1} we consider two input masses, $M(r_{200}) = 1\times 10^{14} M_{\mathrm{Sun}}$ and $M(r_{200}) = 1\times 10^{15} M_{\mathrm{Sun}}$, which roughly span the range of galaxy cluster masses. We use $z$-values of $0.15$ and $0.9$, and take $f_{\rm gas}(r_{200}) = 0.12$ following \citet{2011ApJS..192...18K}. For PM II and PMT II we consider $\alpha_{\rm Ein}$ values of $0.05, \, 0.2,$ and $ 2.0$. \\
\begin{figure*}
  \begin{center}
    \begin{tabular}{@{}cc@{}}
     \includegraphics[width=0.5\linewidth]{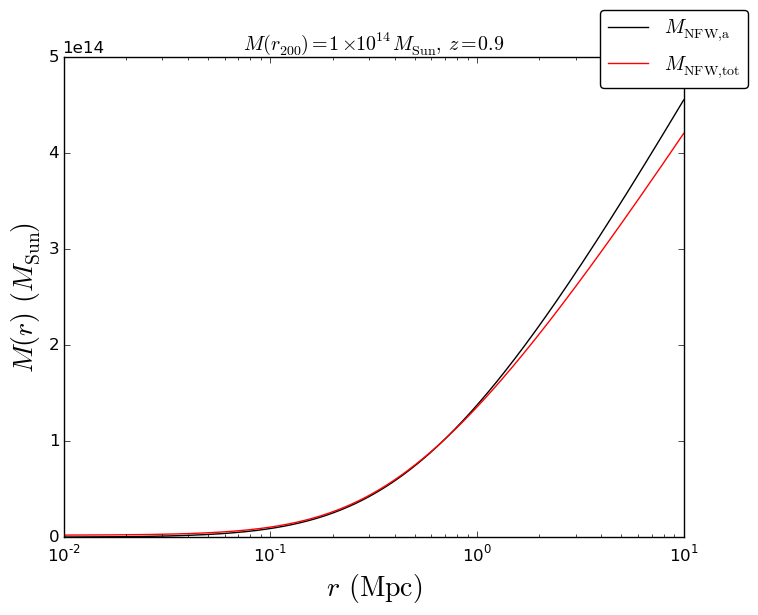} &
     \includegraphics[width=0.5\linewidth]{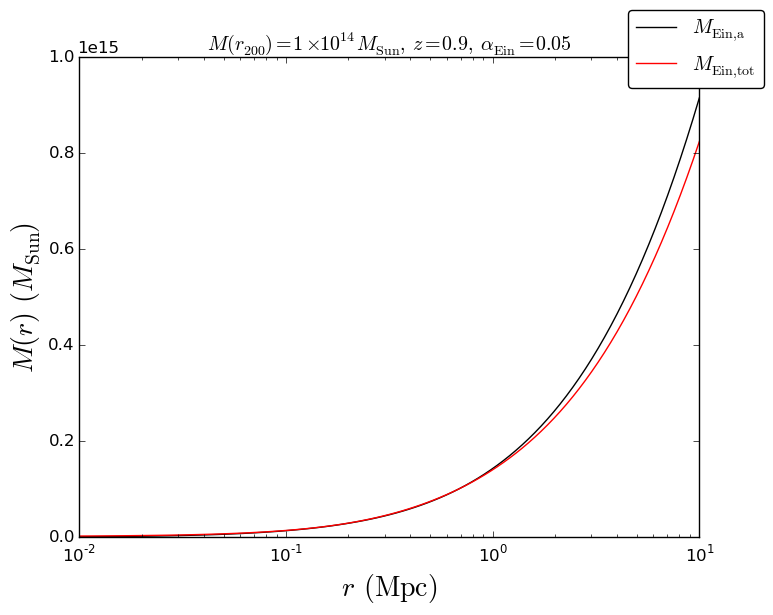} \\
     \includegraphics[ width=0.5\linewidth]{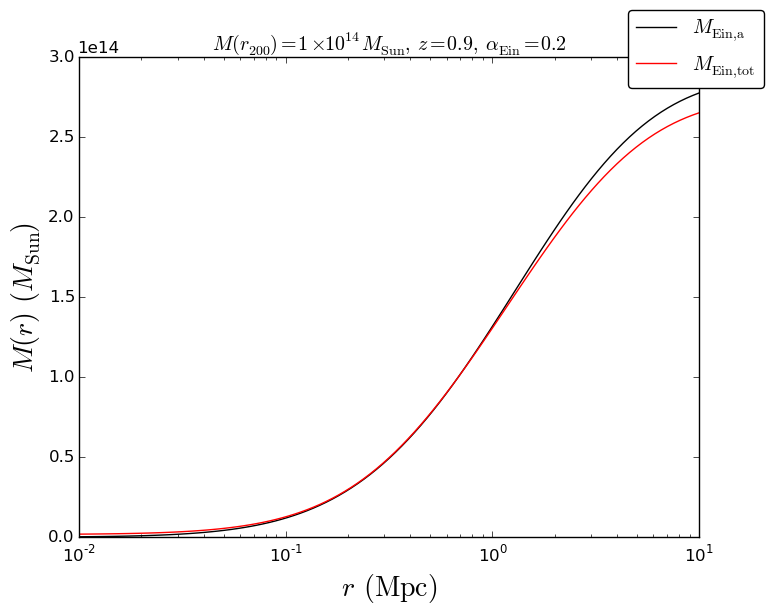} &
     \includegraphics[ width=0.5\linewidth]{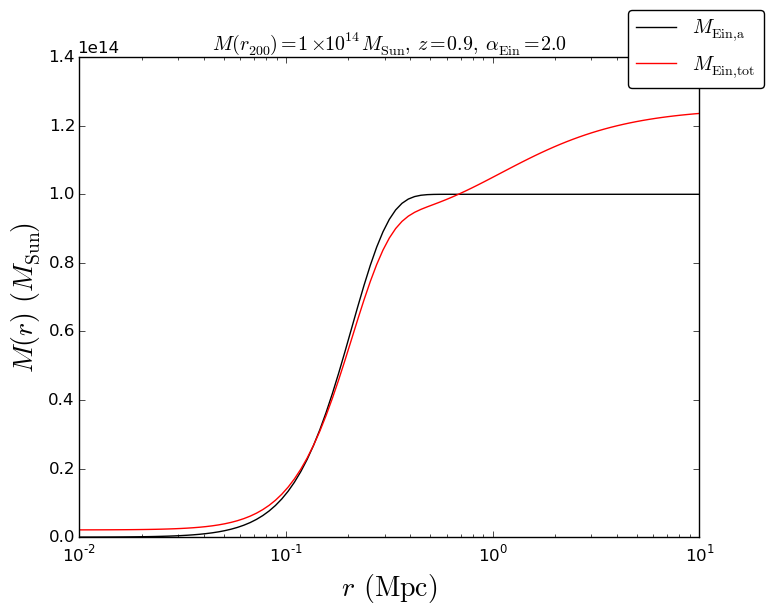} \\
    \end{tabular}

  \caption{Mass profiles of cluster with input parameters given in title, for models given in Figure~\ref{f:tm_lowz_lowm}.}
\label{f:tm_highz_lowm}
  \end{center}
\end{figure*}
Figure~\ref{f:tm_lowz_lowm} shows the profiles for low $M(r_{200})$ and $z$. All four profiles are similar up to $r \approx 1 $~Mpc (which is also $\approx r_{200}$), after which the NFW and Einasto profiles diverge. The two NFW profiles (PM I and PMT I) have roughly the same shape, but start to diverge slightly at high $r$ ($~ 10$ Mpc) with PM I taking higher values than PMT I. In the case of $\alpha_{\rm Ein} = 0.05$, both PM II and PMT II diverge to large mass values at high $r$, with PMT II taking smaller values than PM II. $\alpha_{\rm Ein} = 0.2$ shows a similar relationship between PM II and PMT II, but with the two taking lower values than PM I and PMT I at high $r$. $\alpha_{\rm Ein} = 2$ presents an interesting result as PMT II does not appear to converge at high $r$ like PM II does. Note that for the Einasto dark matter profile,
\begin{equation}
\lim_{r\to\infty} M_{\rm dm}(r) = 4 \pi \rho_{-2} \exp(2 / \alpha_{\rm Ein}) r_{-2}^3 \left( \frac{\alpha_{\rm Ein}}{2} \right)^{\frac{3}{\alpha_{\rm Ein}}} \frac{1}{\alpha_{\rm Ein}} \Gamma \left(\frac{3}{\alpha_{\rm Ein}} \right),
\end{equation}
and so the first term on the right hand side of equation~\ref{e:tot_mass2} is roughly constant at high $r$, meaning the increase in mass must be from the gas component. It seems unphysical that the gas content would contribute so much to the total mass at high $r$ and thus questions the validity of the model (at least for the values of $r_{\rm p}$ and $P_{\rm ei}$ used here).
\begin{figure*}
  \begin{center}
    \begin{tabular}{@{}cc@{}}
     \includegraphics[width=0.5\linewidth]{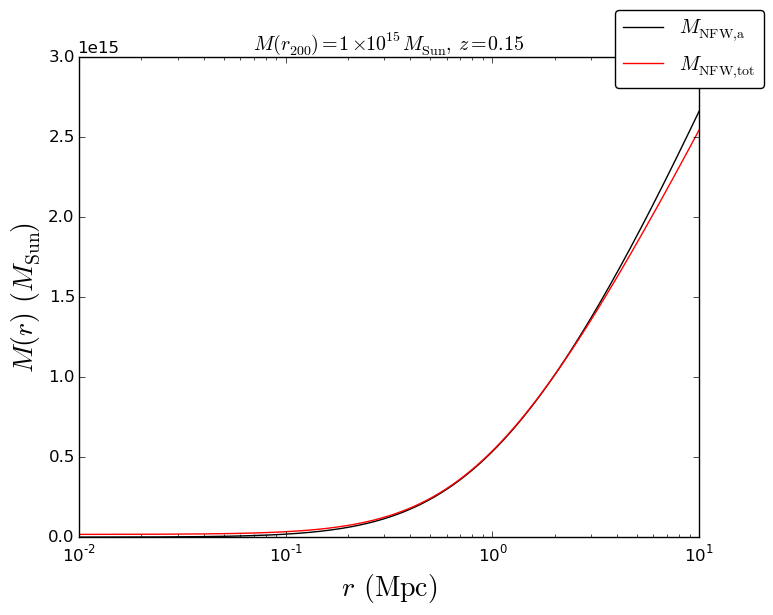} &
     \includegraphics[width=0.5\linewidth]{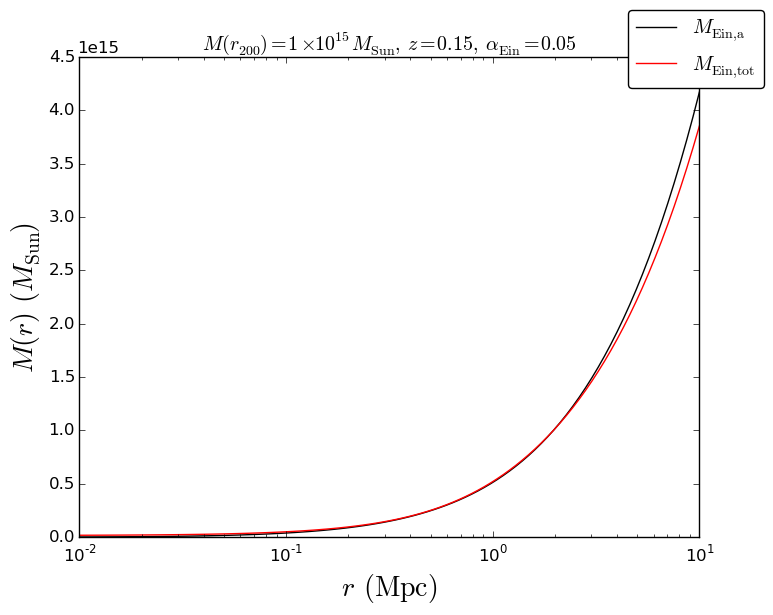} \\
     \includegraphics[ width=0.5\linewidth]{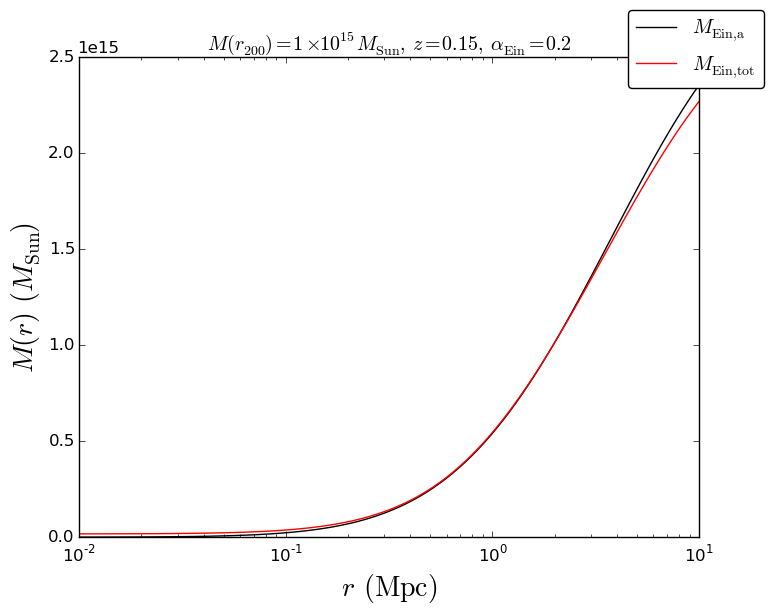} &
     \includegraphics[ width=0.5\linewidth]{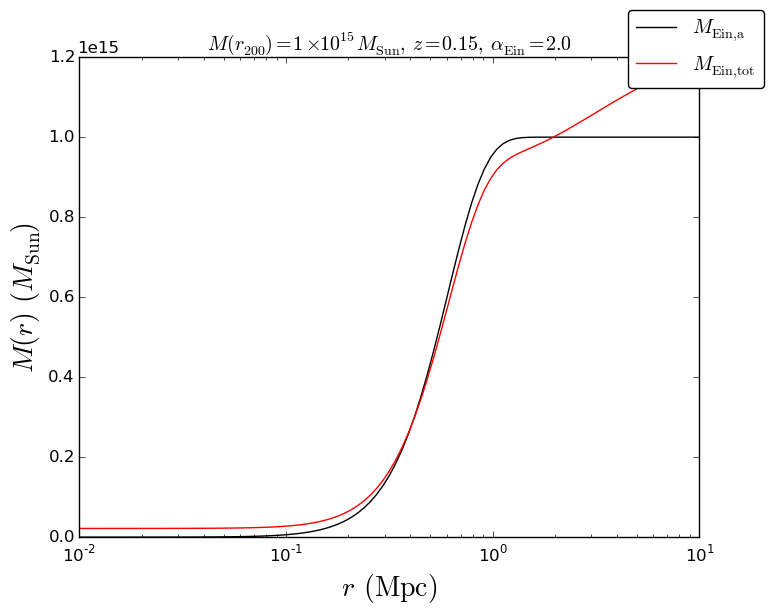} \\ 
    \end{tabular}

  \caption{Mass profiles of cluster with input parameters given in title, for models given in Figure~\ref{f:tm_lowz_lowm}.}
\label{f:tm_lowz_highm}
  \end{center}
\end{figure*}
Figures~\ref{f:tm_highz_lowm},~\ref{f:tm_lowz_highm}, and~\ref{f:tm_highz_highm} show the profiles for the other three combinations of $M(r_{200})$ and $z$ inputs: low mass \& high $z$; high mass \& low $z$ and high mass \& high $z$, respectively. All three cases show similar results between the approximate and full mass results to the previous case, which implies that the desparity between the two sets of results is not dependent on the input parameters (boundary conditions imposed on the ODEs).


\section{Conclusions}
\label{s:tot_mass_conc}

This Chapter relaxes the $M(r_{200}) \approx M_{\rm dm}(r_{200})$ assumption present in the physical model presented in Section~\ref{s:phys_mod} (PM I) and the equivalent Einasto physical model (Section~\ref{s:ein_mod}, PM II), to see if this would produce more physically plausible models for clusters. I derive two new models PMT I and PMT II based on the equation $M(r) = M_{\rm dm}(r) + M_{\rm g}(r)$. Both PMTs require non-linear ordinary differential equations in $M(r)$ to be solved. But to do this, values for $r_{\rm p}$ and $P_{\rm ei}$ need to be determined and this turns out to be a non-trivial process. I investigated three possible ways of calculating $r_{\rm p}$ \& $P_{\rm ei}$ and found the following.
\begin{itemize}
\item Determining $r_{\rm p}$ and $P_{\rm ei}$ directly from constraints on $M$ and its derivative seems inplausible as we could not justify assigning a particular value to the derivative at any finite value of $r$.
\item Treating the problem of finding a value of $r_{\rm p}$ and $P_{\rm ei}$ from the family of solutions as a constrained optimsation problem (with a function $f(r_{\rm p}, P_{\rm ei})$ dictating the nature of the values of $r_{\rm p}$ and $P_{\rm ei}$ obtained, and the differential equations in $M(r)$ providing the constraints) seems promising in theory. However in practice, justifying a particular form for $f(r_{\rm p}, P_{\rm ei})$ isn't straightforward, and I anticipate that the optimisation is difficult numerically.
\item The third method relied on using the calculational steps of PM I and PM II to determine `approximate' values for $r_{\rm p}$ and $P_{\rm ei}$ from `true' values of $r_{\rm s}$ for PMT I ($r_{-2}$ for PMT II) and $\rho_{\rm s}$ for PMT I ($\rho_{-2}$ for PMT II) calculated without the assumption that $\rho_{\rm g} \ll \rho_{\rm dm}$.
\end{itemize}
The third method was by far the simplest and the one I used to plot the mass profiles for the PMTs to compare with the equivalent PM profiles. From plots of the profiles we found the following.
\begin{itemize}
\item The values of the input parameters $M(r_{200})$ and $z$ had very little effect on the shape or scale of the PMT I or PMT II profiles when compared with the corresponding PM profiles.
\item For the $\alpha_{\rm Ein} = 2$ case PMT II does not show the convergence in mass at high $r$ that PM II does. Since $M_{\rm dm}(r)$ asymptotically converges as $r \rightarrow \infty$ this implies that it is the gas which is contributing to the mass increase, which seems unphysical for large $r$.
\end{itemize}


%% file: CHAP-7/chapter7.tex
\chapter{Enhanced physical modelling II: Inclusion of non-thermal pressure}\label{c:seventh}

All physical models presented so far in this thesis assume that the cluster gas pressure comes solely from the \textit{thermal} gas pressure. Cosmological simulations have long predicted that magnetic fields, gas bulk motion and turbulence contribute to pressure support (see e.g. \citealt{2009ApJ...705.1129L}, \citealt{2010ApJ...725...91B}, \citealt{2011AAS...21710903B}, \citealt{2011ApJ...731L..10N}, \citealt{2011MmSAI..82..594N}, \citealt{2012ApJ...758...74B}, and \citealt{2012MNRAS.419L..29P}).\\ 
Observational studies of clusters using the Chandra, Suzaku and XMM-Newton satellites have long invoked (see e.g. \citealt{2009PASJ...61.1117B} for cluster A1795, \citealt{2009A&A...501..899R} A2204, \citealt{2009MNRAS.395..657G} PKS0745-191, \citealt{2010PASJ...62..371H} A1413, \citealt{2010ApJ...714..423K} A1689, \citealt{2011MNRAS.414.2101U} Virgo and \citealt{2011Sci...331.1576S} Perseus) these additional pressure sources to explain their observations. So including a non-thermal contribution to the hydrostatic equilibrium (HSE) relation given by equation~\ref{e:hse}, and altering the succeeding calculational steps of the PMs accordingly should be interesting.

In this Chapter I first give an overview of the contributors to non-thermal pressure. I then derive physical models for both NFW and Einasto (dark matter) models and incorporate non-thermal pressure into the HSE equation. We refer to these two models as PMN I and PMN II. We then plot the cluster parameter profiles of PMN I \& PMN II and compare with those already obtained for PM I and PM II. Note that I do not include any modifications discussed in Chapter~\ref{c:sixth} here.


\section{Non-thermal gas pressure}\label{s:nt_p}

Galaxies orbiting or infalling onto clusters not only stir the gas, but also make the ICM clumpier.
In the dense inner regions of clusters, these clumps only exist on short timescales as the ram pressure (pressure exerted on a body as it moves through a fluid medium) acting on the gas is high. At higher radii where the average cluster density is lower, orbital times are longer and accretion of new cluster material is ongoing, clumpiness can have significant effects on the total pressure profile. 
The clumpiness of the ICM depends on a number of physical processes, such as efficient feedback, which removes gas from merging structures, and thermal conduction, which homogenises the ICM temperature (see e.g.\citealt{2004ApJ...606L..97D}).
Cosmic rays can originate from accretion shocks and supernova explosions, active galactic nuclei (AGN), and radio galaxies (see \citealt{2014IJMPD..2330007B} for a review). 

\section{Modelling non-thermal gas pressure} \label{s:nt_mod}


\subsection{Analytic expression for non-thermal gas pressure} \label{s:nt_eqn}

\citet{2016arXiv160804388M} (from here on DM16) derive an analytic expression for the non-thermal pressure $P_{\rm nt}$ component in galaxy clusters. They derive the function $P_{\rm nt}(r)$ by considering a subset of ten cosmological hydrodynamical zoom-in simulations of galaxy clusters from the sample of \citet{2014MNRAS.440.2290M}. The ten simulations were performed using the \textsc{ramses} code \citep{2002A&A...385..337T} and have total masses $> 10^{14} M_{\mathrm{Sun}}$. Half of the subsample are relaxed according to the criteria outlined in Section~2.1 of DM16 (based on the ratio of the velocity dispersion of dark matter particles to the velocity dispersion of an equivalent virialised system). These simulations do not include non-thermal contributions from cosmic rays and magnetic fields.

DM16 derive an expression for $P_{\rm nt}$ by evolving a cluster from high $z$ and measuring its $\rho_{\rm g}(r)$, $M(r)$, thermal pressure $P_{\rm th}$ and thermal mass $M_{\rm th}(r)$. From these four quantities the form of $P_{\rm nt}$ can be determined from the HSE relation (equation~\ref{e:hse}) (assuming that $P_{\rm g} = P_{\rm th} + P_{\rm nt}$). The following analytic expression is obtained by fitting to the simulated data using a least squares regression
\begin{equation}
\label{e:nt_eqn0}
P_{\rm nt}(r) = 5.388 \times 10^{13} \left( \frac{r_{200,\mathrm{m}}}{\mathrm{Mpc}} \right)^3 \left( \frac{\rho_{\mathrm{g}}(r)}{\mathrm{g/cm}^3 } \right) \mathrm{erg/cm}^3,
\end{equation}
where $r_{200,\mathrm{m}}$ is the radius at which the average cluster density is $200 \, \times$ the \textit{average} matter density $ \rho_{\rm m}(z) = 3H_{0}^{2}/(8\pi G) \Omega_{\rm M} (1 + z)^3$. Here $H_{0}$ is the Hubble parameter evaluated at $z = 0$. $r_{200,\mathrm{m}}$ can be calculated from $r_{200}$ in a similar way to how $r_{500}$ is (for the NFW case, see equations~\ref{e:r200r5001} through to~\ref{e:r200r5005}, and for Einasto see Section~\ref{s:r500newton}).\\
Expressed in `astronomical' units ($M_{\mathrm{Sun}}$Mpc$^{-1}$s$^{-2}$), equation~\ref{e:nt_eqn0} can be written as
\begin{equation}
\label{e:nt_eqn}
P_{\rm nt}(r) = \beta \left( \frac{r_{200,\mathrm{m}}}{\mathrm{Mpc}} \right)^3 \rho_{\mathrm{g}}(r),
\end{equation} 
where $\beta = 5.658 \times 10^{-36}$~Mpc$^2$s$^{-2}$.


\subsection{Incorporating non-thermal pressure into the physical models}\label{s:nt_calcs}

Redefining $P_{\rm g}(r)$ from equation~\ref{e:pgas_pelec} as 
\begin{equation}\label{e:p_gas_re}
P_{\rm g}(r) \equiv P_{\rm th}(r) + P_{\rm nt}(r),
\end{equation}
where
\begin{equation}\label{e:th_eqn}
P_{\rm th}(r) = \frac{\mu_{\rm e}}{\mu_{\rm g}} \frac{P_{\rm ei}}{\left(\frac{r}{r_{\rm p}}\right)^{c}\left(1+\left(\frac{r}{r_{\rm p}}\right)^{a}\right)^{(b-c)/a}},
\end{equation}
and re-evaluating the HSE relation with the new form of $P_{\rm g}(r)$ gives
\begin{equation} \label{e:hse2}
\frac{\mathrm{d} P_{\rm g}(r)}{\mathrm{d}r} = \frac{\mathrm{d}}{\mathrm{d}r} \left[ P_{\rm th}(r) + P_{\rm nt}(r) \right] = - \frac{G \rho_{\rm{g}}(r) M(r)}{r^2}.
\end{equation}
Equation~\ref{e:hse2} can be rearranged to give
\begin{equation} \label{e:rhog_diff}
\frac{\mathrm{d} \rho_{\rm g}(r)}{\mathrm{d}r} + \frac{G (\mathrm{Mpc})^3 M(r)}{\beta  r_{200,\mathrm{m}}^3} \frac{1}{r^2} \rho_{\rm g}(r) = \frac{\mu_{\rm e}}{\mu_{\rm g}} \left( \frac{\mathrm{Mpc}}{r_{200,\mathrm{m}}} \right)^3  \frac{P_{\rm ei}}{\beta} \left[ \frac{1}{r} \left(\frac{r}{r_{\rm p}}\right)^{-c}\left[1+\left(\frac{r}{r_{\rm p}}\right)^{a}\right]^{-(1+(b-c)/a)}\left[b\left(\frac{r}{r_{\rm p}}\right)^{a}+c\right]  \right],
\end{equation}
which is a inhomogeneous first order linear ODE with dependent variable $\rho_{\rm{g}}(r)$ and independent variable $r$. Since equation~\ref{e:rhog_diff} includes a $M(r)$ term, its final form depends on the dark matter profile considered. Note that when calculating $M(r)$ for either PMN I or PMN II we assume $M(r) \approx M_{\rm dm}(r)$ as we did when profiling PM I and PM II. \\
For PMN I we have the expression for $M(r)$ given by equation~\ref{e:nfw_m_tot_2} and so the differential equation becomes 
\begin{equation} \label{e:rhog_nfw_diff}
\begin{split}
& \frac{\mathrm{d} \rho_{\rm g}(r)}{\mathrm{d}r} + \frac{4\pi G (\mathrm{Mpc})^3 \rho_{\rm s}r_{\rm s}^{3}}{\beta r_{200,\mathrm{m}}^3} \frac{\left[\ln \left(1+\frac{r}{r_{\rm s}}\right)-\left(1+\frac{r_{\rm s}}{r}\right)^{-1}\right]}{r^2} \rho_{\rm g}(r) \\
 & = \frac{\mu_{\rm e}}{\mu_{\rm g}} \left( \frac{\mathrm{Mpc}}{r_{200,\mathrm{m}}} \right)^3  \frac{P_{\rm ei}}{\beta} \left[ \frac{1}{r} \left(\frac{r}{r_{\rm p}}\right)^{-c}\left[1+\left(\frac{r}{r_{\rm p}}\right)^{a}\right]^{-(1+(b-c)/a)}\left[b\left(\frac{r}{r_{\rm p}}\right)^{a}+c\right]  \right].
 \end{split}
\end{equation}
For PMN II we have the expression for $M(r)$ given by equation~\ref{e:ein_mass} and so 
\begin{equation} \label{e:rhog_ein_diff}
\begin{split}
& \frac{\mathrm{d} \rho_{\rm g}(r)}{\mathrm{d}r} +
\frac{4\pi G (\mathrm{Mpc})^3 \rho_{-2}r_{-2}^{3} \exp \left( 2 / \alpha_{\rm Ein} \right) \left( \frac{\alpha_{\rm Ein}}{2} \right) ^{3 / \alpha_{\rm Ein}}}{\beta r_{200,\mathrm{m}}^3 \alpha_{\rm Ein}} \frac{\gamma \left[ \frac{3}{\alpha_{\rm Ein}}, \frac{2}{\alpha_{\rm Ein}} \left( \frac{r}{r_{-2}} \right)^{\alpha_{\rm Ein}} \right]}{r^2} \rho_{\rm g}(r)\\
& = \frac{\mu_{\rm e}}{\mu_{\rm g}} \left( \frac{\mathrm{Mpc}}{r_{200,\mathrm{m}}} \right)^3  \frac{P_{\rm ei}}{\beta} \left[ \frac{1}{r} \left(\frac{r}{r_{\rm p}}\right)^{-c}\left[1+\left(\frac{r}{r_{\rm p}}\right)^{a}\right]^{-(1+(b-c)/a)}\left[b\left(\frac{r}{r_{\rm p}}\right)^{a}+c\right]  \right].
\end{split}
\end{equation}
For brevity we define the following constants
\begin{equation}
\label{e:nt_consts}
\begin{aligned}
&\omega_{\rm NFW} \equiv \frac{4 \pi G \rho_{\rm s} r_{\rm s}^3 (\mathrm{Mpc})^3}{\beta r_{200,\mathrm{m}}^3} \\
&\omega_{\rm Ein} \equiv \frac{4 \pi G \rho_{-2} r_{-2}^3 (\mathrm{Mpc})^3}{\beta r_{200,\mathrm{m}}^3} \frac{1}{\alpha_{\rm Ein}} \left( \frac{\alpha_{\rm Ein}}{2} \right) ^{3 / \alpha_{\rm Ein}} \exp(2 / \alpha_{\rm Ein}) \\
&\sigma \equiv \frac{\mu_{\rm e}}{\mu_{\rm g}} \frac{1}{\beta} \left( \frac{\mathrm{Mpc}}{r_{200,\mathrm{m}}} \right)^3.
\end{aligned}
\end{equation} 
In fact the inhomogeneous ODEs derived above can be transformed into homogeneous ODEs as follows. Consider a general ODE of the form
\begin{equation} \label{e:gen_if_ode}
\frac{\mathrm{d} \rho_{\rm g}(r)}{\mathrm{d}r} + g(r)\rho_{\rm g}(r) = f(r),
\end{equation}
then using an integrating factor defined by
\begin{equation} \label{e:gen_if_if}
I(r) = \exp \left( \int^r g(r') \mathrm{d}r' \right),
\end{equation}
equation~\ref{e:gen_if_ode} can be transformed into a homogeneous separable ODE which gives the result
\begin{equation} \label{e:gen_if_hom}
\rho_{\rm g}(r) I(r) - \rho_{\rm g}(r_{0}) I(r_{0})  =  \int_{r_0}^{r} I(r') f(r') \mathrm{d}r',
\end{equation}
where $r_{0}$ and $\rho_{\rm g}(r_{0})$ are dependent on the input parameters of the problem. \\
For PMN I this gives
\begin{equation} \label{e:rhog_nfw_diff2}
\begin{split}
&\rho_{\rm g}(r) \left(1 + \frac{r}{r_{\rm s}} \right)^{-\omega_{\rm NFW} / r} - \rho_{\rm g}(r_{0}) \left(1 + \frac{r_{0}}{r_{\rm s}} \right)^{-\omega_{\rm NFW} / r_{0}}\\  
&= P_{\rm ei} \sigma  \int_{r_{0}}^{r} \left(1 + \frac{r'}{r_{\rm s}} \right)^{-\omega_{\rm NFW} / r'}  \left[ \frac{1}{r'} \left(\frac{r'}{r_{\rm p}}\right)^{-c}\left[1+\left(\frac{r'}{r_{\rm p}}\right)^{a}\right]^{-(1+(b-c)/a)}\left[b\left(\frac{r'}{r_{\rm p}}\right)^{a}+c\right]  \right] \mathrm{d}r'.
\end{split}
\end{equation}
For PMN II the integrating factor does not have an analytical form, hence the homogeneous form can only be simplified to
\begin{equation} \label{e:rhog_ein_diff2}
\begin{split}
&\rho_{\rm g}(r) \exp \left( \int^{r} \omega_{\rm Ein} \frac{\gamma \left[ \frac{3}{\alpha_{\rm Ein}}, \frac{2}{\alpha_{\rm Ein}} \left( \frac{r'}{r_{-2}} \right)^{\alpha_{\rm Ein}} \right]}{r'^2} \mathrm{d}r' \right) - \rho_{\rm g}(r_{0}) \exp \left( \int^{r_{0}} \omega_{\rm Ein} \frac{\gamma \left[ \frac{3}{\alpha_{\rm Ein}}, \frac{2}{\alpha_{\rm Ein}} \left( \frac{r'}{r_{-2}} \right)^{\alpha_{\rm Ein}} \right]}{r'^2} \mathrm{d}r' \right)\\
&= P_{\rm ei} \sigma \int_{r_{0}}^{r} \left[ \exp \left( \omega_{\rm Ein} \int^{r'}  \frac{\gamma \left[ \frac{3}{\alpha_{\rm Ein}}, \frac{2}{\alpha_{\rm Ein}} \left( \frac{r''}{r_{-2}} \right)^{\alpha_{\rm Ein}} \right]}{r''^2} \mathrm{d}r'' \right) \right] \left[ \frac{1}{r'} \left(\frac{r'}{r_{\rm p}}\right)^{-c}\left[1+\left(\frac{r'}{r_{\rm p}}\right)^{a}\right]^{-(1+(b-c)/a)}\left[b\left(\frac{r'}{r_{\rm p}}\right)^{a}+c\right]  \right] \mathrm{d}r'.
\end{split}
\end{equation}

It is also interesting to see if the non-thermal only pressure term provides a solution to the HSE (I have already verified this is the case for $P_{\rm th}(r)$ in PM~I and PM~II, by deriving the relevant expressions for $\rho_{\rm{g}}(r)$). Putting the expression for $P_{\rm nt}(r)$ into the HSE gives
\begin{equation}\label{e:p_nth_gas}
\int_{r_{0}}^{r_{1}} \frac{1}{\rho_{\rm{g}}(r') } \frac{\mathrm{d} \rho_{\rm{g}}(r')}{\mathrm{d}r'} \mathrm{d}r' = \left( \frac{\mathrm{Mpc}}{ r_{200,\mathrm{m}}} \right)^{3} \frac{G}{\beta} \int_{r_{0}}^{r_{1}} \frac{M(r')}{r'^2} \mathrm{d}r'.
\end{equation}
For the NFW dark matter profile this gives
\begin{equation}\label{e:p_nth_gas_nfw}
\rho_{\rm{g}}(r_{1}) = \rho_{\rm{g}}(r_{0}) \frac{\left(1 + \frac{r_{1}}{r_{\mathrm{s} } } \right) ^(\omega_{\mathrm{NFW}} / r_{1}) } {\left(1 + \frac{r_{0}}{r_{\mathrm{s} } } \right) ^(\omega_{\mathrm{NFW}} / r_{0}) }.
\end{equation}
As was the case with the integrating factor in the full solution for the thermal and non-thermal pressure, the Einasto dark matter profile does not give an analytic solution the non-thermal only case
\begin{equation}\label{e:p_nth_gas_ein}
\rho_{\rm{g}}(r_{1}) = \rho_{\rm{g}}(r_{0}) \exp \left[ \omega_{\mathrm{Ein}} \int_{r_{0}}^{r_{1}} \frac{\gamma \left[ \frac{3}{\alpha_{\rm Ein}}, \frac{2}{\alpha_{\rm Ein}} \left( \frac{r'}{r_{-2}} \right)^{\alpha_{\rm Ein}} \right]}{r'^2} \mathrm{d}r' \right].
\end{equation}
Note that since the HSE is a inhomogeneous differential equation, the solutions associated with $P_{\mathrm{th}}$ and $P_{\mathrm{nt}}$ do not sum to solution asspcoated with $P_{\mathrm{g}} = P_{\mathrm{th}} + P_{\mathrm{nt}}$.
\subsection{Determining $P_{\rm ei}$ for the non-thermal case}\label{s:nt_pei}

Equation~\ref{e:rhog_nfw_diff} (equation~\ref{e:rhog_ein_diff}) has four unknown parameters: $r_{\rm s}$ ($r_{-2}$), $\rho_{\rm s}$ ($\rho_{-2}$), $r_{\rm p}$ and $P_{\rm ei}$. The first three of these can be calculated in the same way as in PM I and PM II. However, as was the case with the full mass modelling in Chapter~\ref{c:sixth}, $P_{\rm ei}$ cannot be calculated trivially from the input parameters and calculations derived above for the PMNs. Hence we consider the methods described in Sections~\ref{s:tot_mass_pars_det1},~\ref{s:tot_mass_pars_det2}, and~\ref{s:tot_mass_pars_det3} which we denote method I, method II and method III respectively.

\begin{figure*}
  \begin{center}
    \begin{tabular}{@{}cc@{}}
     \includegraphics[ width=0.50\linewidth]{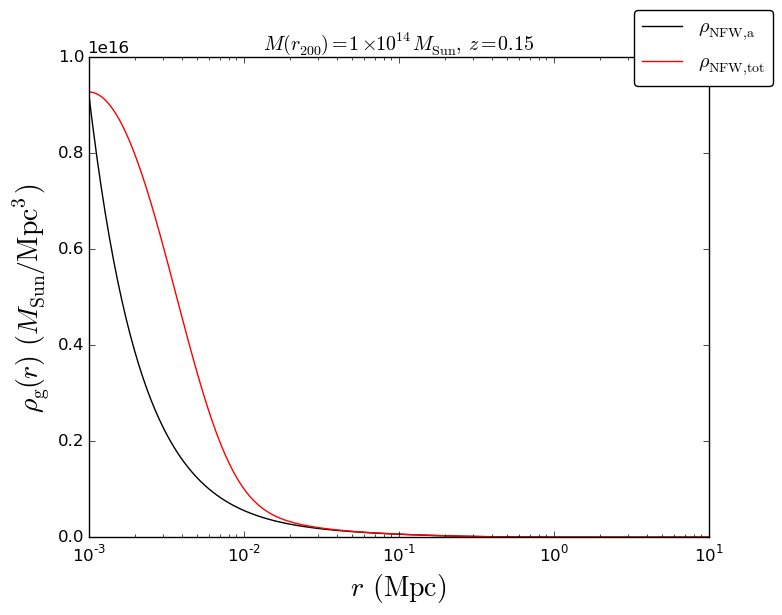} &
     \includegraphics[ width=0.50\linewidth]{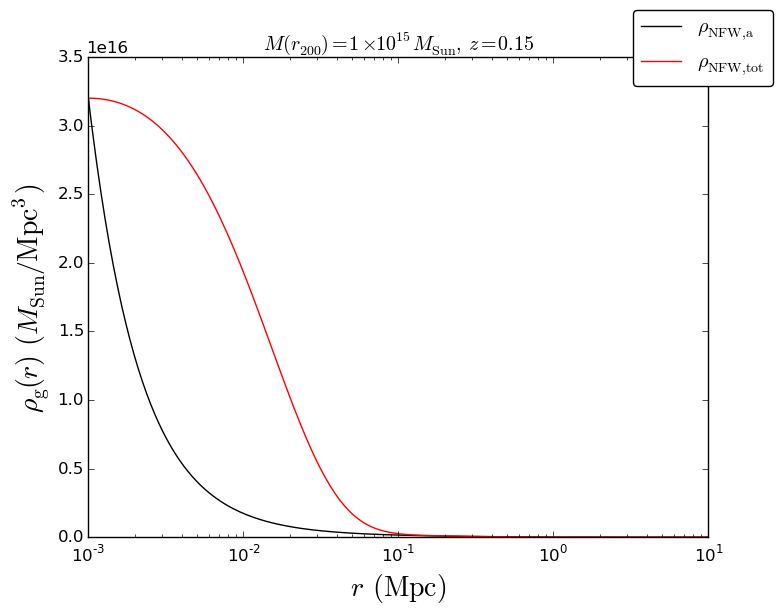} \\
     \includegraphics[ width=0.50\linewidth]{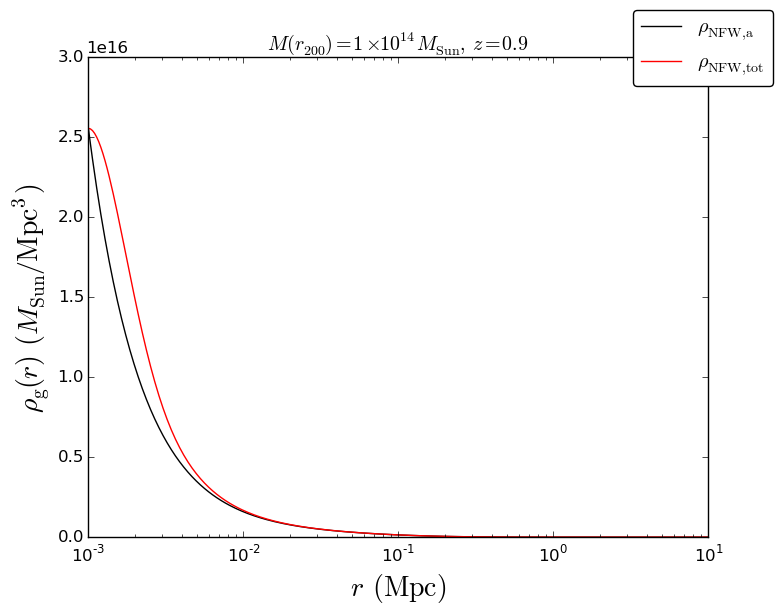} &
     \includegraphics[ width=0.50\linewidth]{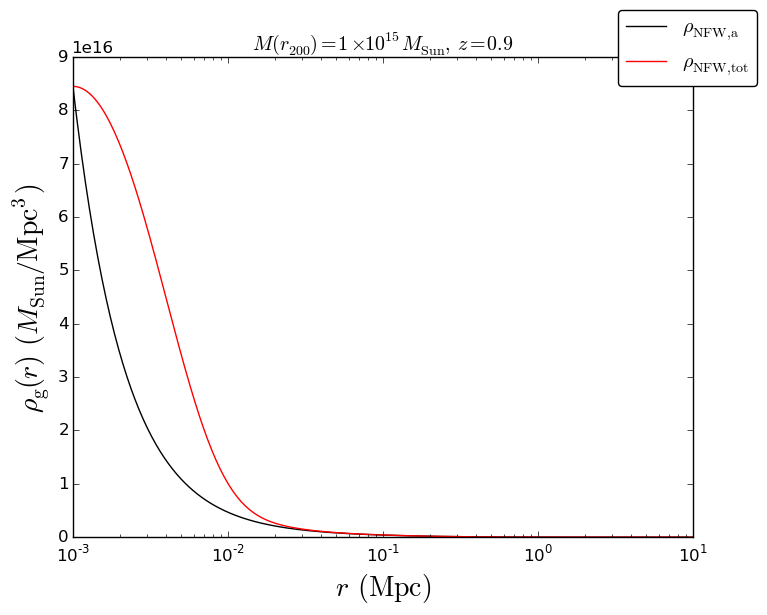} \\
    \end{tabular}

  \caption{$\rho_{\rm g}(r)$ profiles for PM I and PMN I. Each graph features both profiles for one of the four different input parameter sets. Top row has $z = 0.15$, bottom row has $z = 0.9$. Left column has $M(r_{200}) = 1\times 10^{14} M_{\mathrm{Sun}}$, right column has $M(r_{200}) = 1\times 10^{15} M_{\mathrm{Sun}}$}
\label{f:nt_rhog_nfw}
  \end{center}
\end{figure*}

For method I, since there is only one unknown parameter we only need to know the value of $\rho_{\rm g}(r)$ and its derivative at one point. It makes sense to consider the asymptotic case ($r \rightarrow \infty$) in which case $\rho_{\rm g}(r)$ and its derivative tend to zero. However, since there is no constant term in the ODEs, this gives us (using equation~\ref{e:rhog_nfw_diff} or~\ref{e:rhog_ein_diff}) $ 0 = P_{\rm ei} \times 0$ and thus $P_{\rm ei}$ cannot be determined. I have not been able to come up with any physically justified estimates for $\rho_{\rm g}(r)$ and its derivative at finite $r$, and so I do not pursue this method any further. \\
Method II presents the same potential difficulties as in the Chapter~\ref{c:sixth}, and so I do not pursue it here. \\
Method III would require us to get an approximate value for $P_{\rm ei}$ from the calculational steps of PM I and PM II. This requires us to ignore the non-thermal contribution in the HSE equation and determine an analytic form for $\rho_{\rm g}(r)$. As was the case in Chapter~\ref{c:sixth}, this is by far the simplest way of determining $P_{\rm ei}$, I therefore use it to obtain cluster parameter profiles for PMN I \& PMN II and compare them with those from PM I \& PM II for illustrative purposes.

\begin{figure*}
  \begin{center}
    \begin{tabular}{@{}cc@{}}
     \includegraphics[ width=0.50\linewidth]{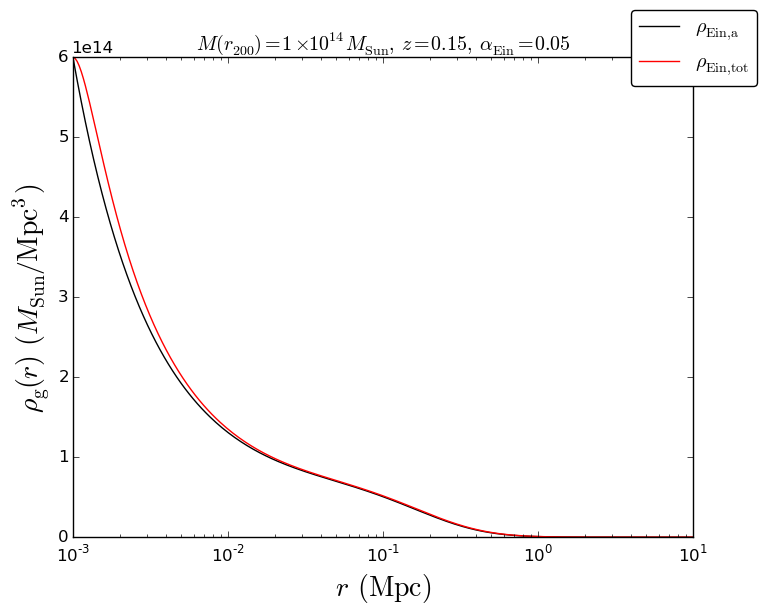} &
     \includegraphics[ width=0.50\linewidth]{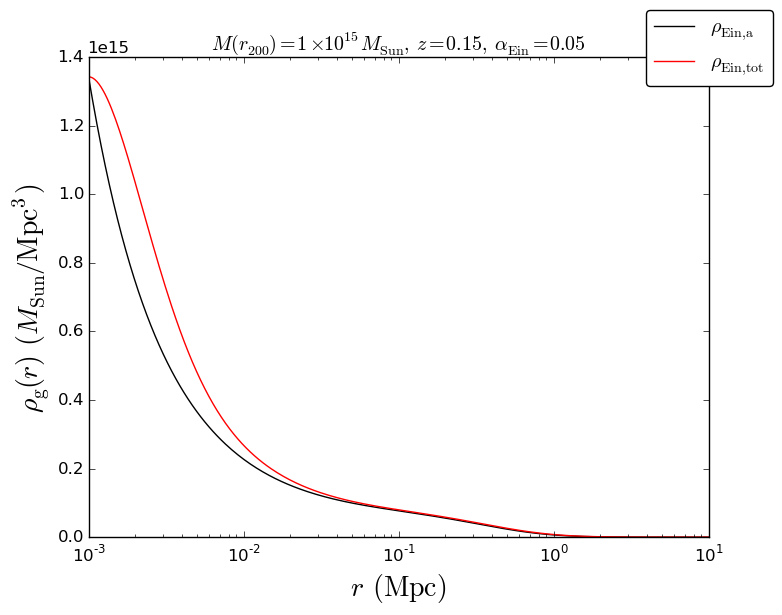} \\
     \includegraphics[ width=0.50\linewidth]{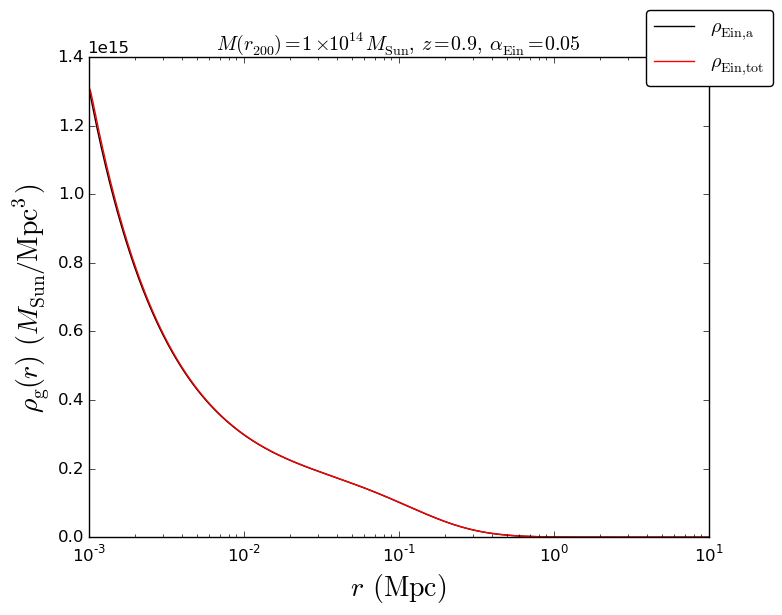} &
     \includegraphics[ width=0.50\linewidth]{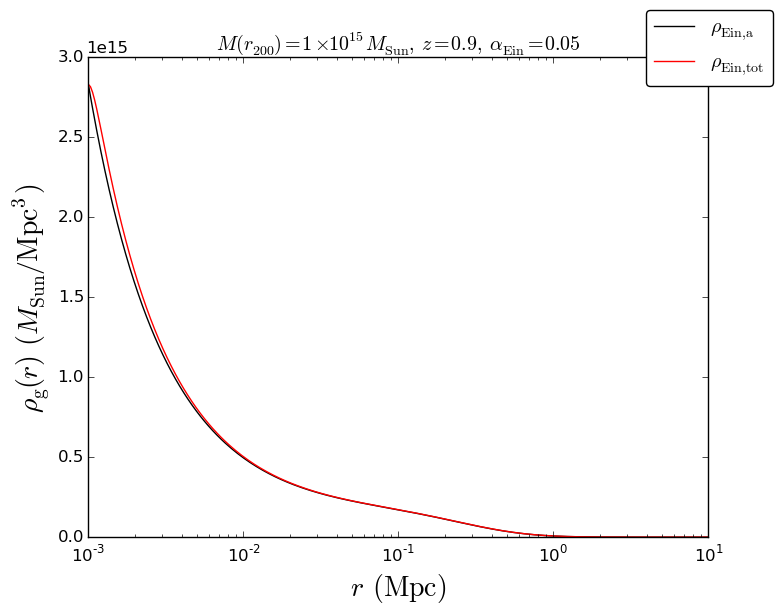} \\
    \end{tabular}

  \caption{$\rho_{\rm g}(r)$ profiles for PM II and PMN II with $\alpha_{\rm Ein} = 0.05$. Graphs are laid out as described in Figure \ref{f:nt_rhog_nfw}.}
\label{f:nt_rhog_einlow}
  \end{center}
\end{figure*}

\begin{figure*}
  \begin{center}
    \begin{tabular}{@{}cc@{}}
     \includegraphics[ width=0.50\linewidth]{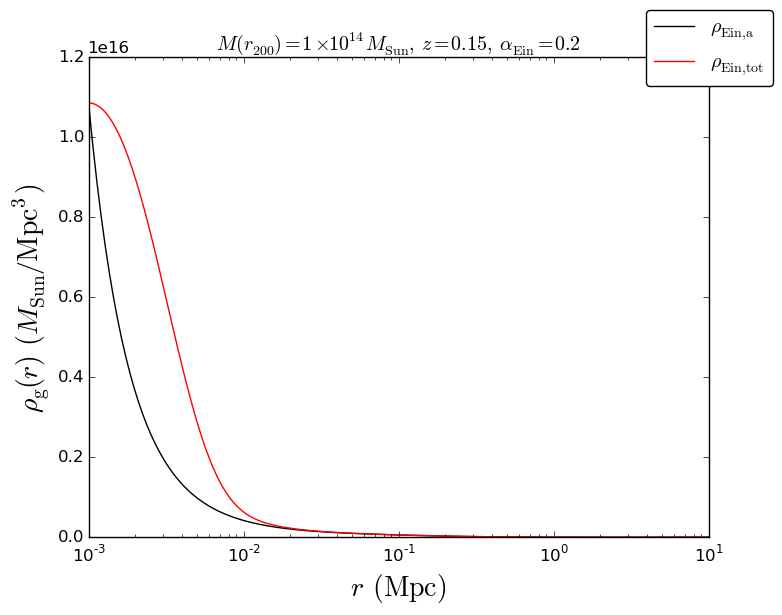} &
     \includegraphics[ width=0.50\linewidth]{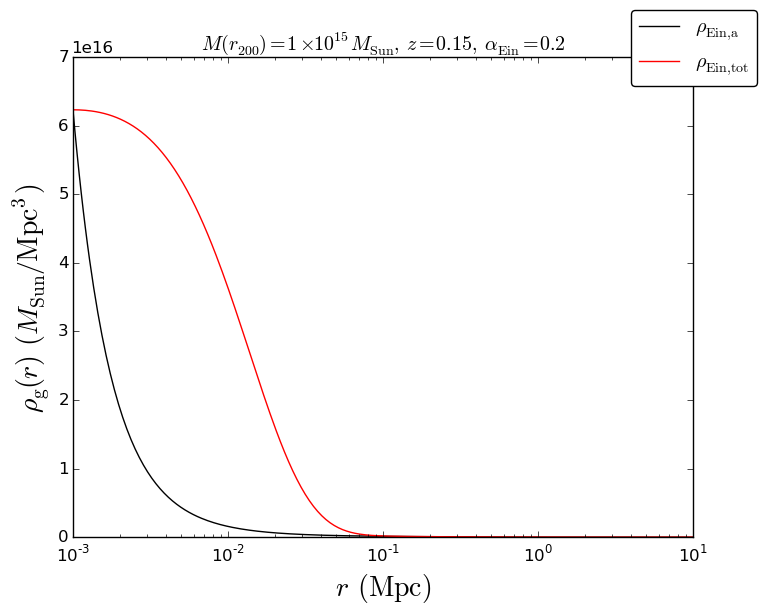} \\
     \includegraphics[ width=0.50\linewidth]{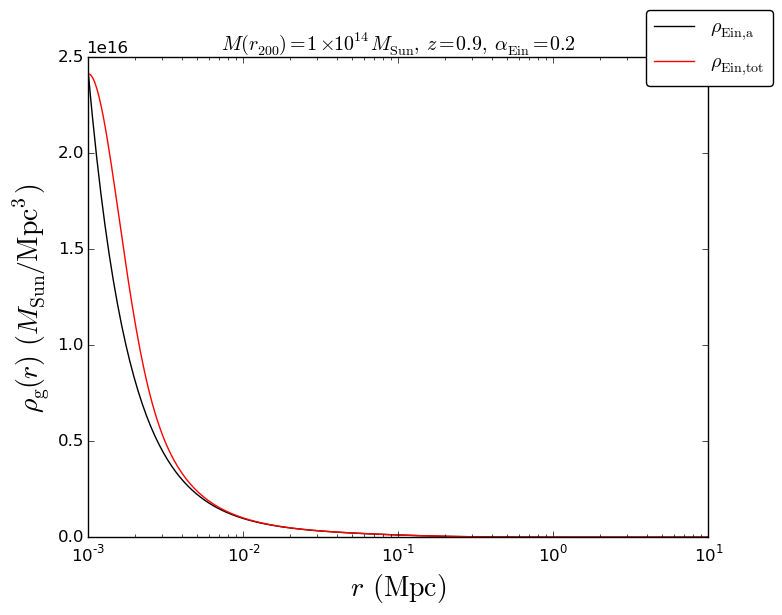} &
     \includegraphics[ width=0.50\linewidth]{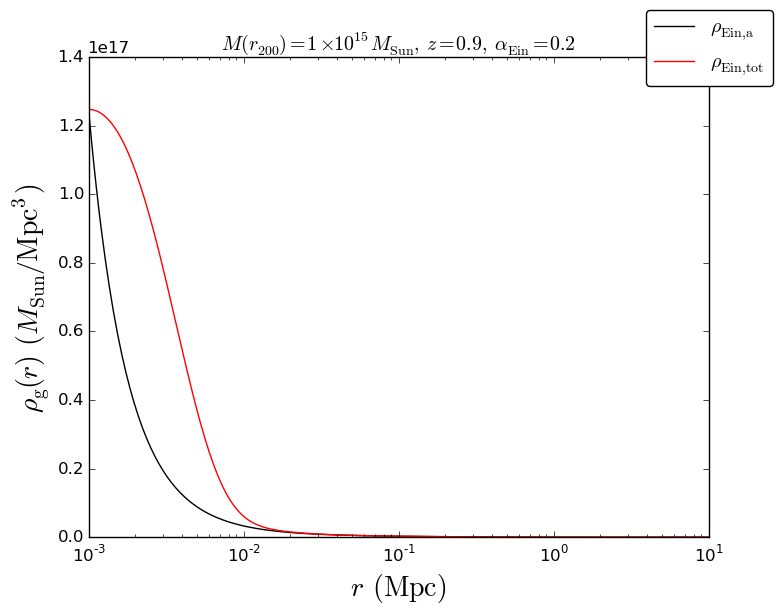} \\
    \end{tabular}

  \caption{$\rho_{\rm g}(r)$ profiles for PM II and PMN II with $\alpha_{\rm Ein} = 0.2$. Graphs are laid out as described in Figure \ref{f:nt_rhog_nfw}.}
\label{f:nt_rhog_einmed}
  \end{center}
\end{figure*}

\begin{figure*}
  \begin{center}
    \begin{tabular}{@{}cc@{}}
     \includegraphics[ width=0.50\linewidth]{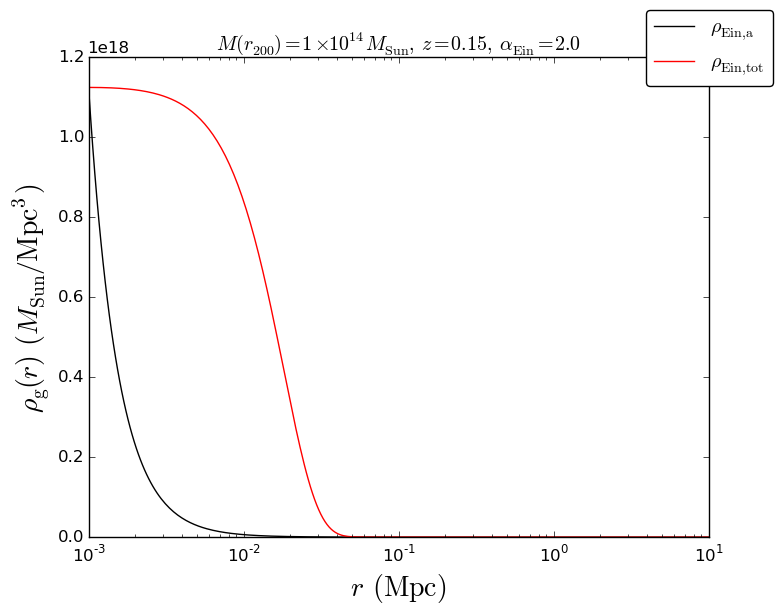} &
     \includegraphics[ width=0.50\linewidth]{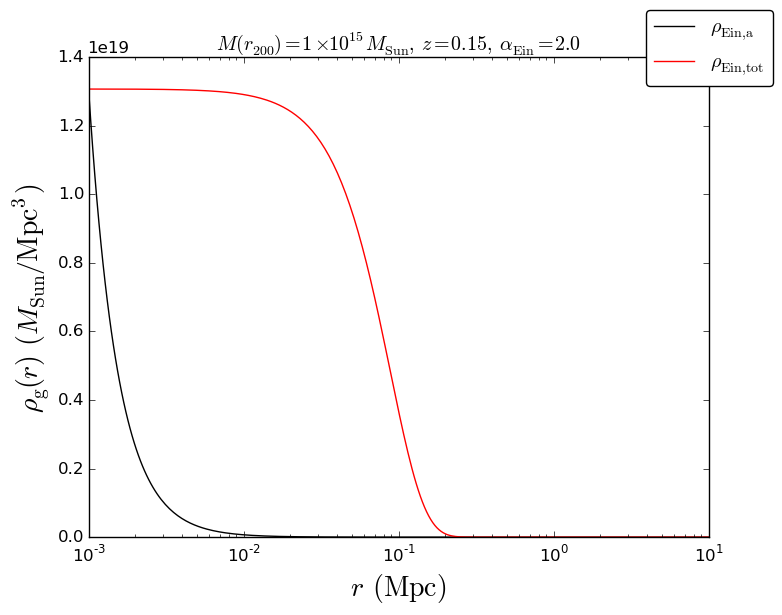} \\
     \includegraphics[ width=0.50\linewidth]{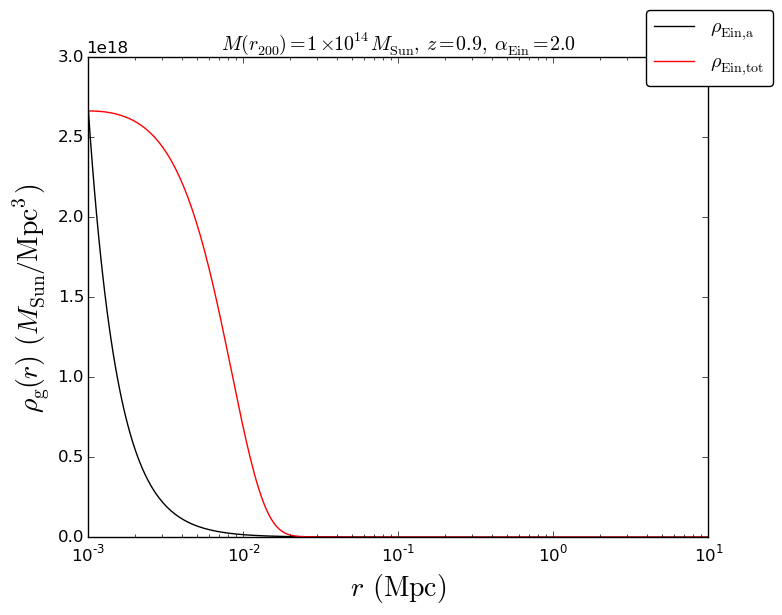} &
     \includegraphics[ width=0.50\linewidth]{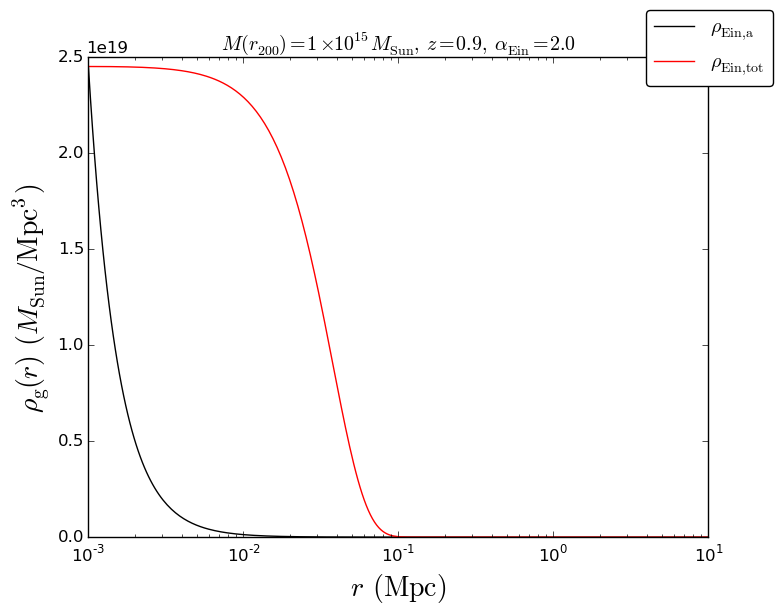} \\
    \end{tabular}

  \caption{$\rho_{\rm g}(r)$ profiles for PM II and PMN II with $\alpha_{\rm Ein} = 2$. Graphs are laid out as described in Figure \ref{f:nt_rhog_nfw}.}
\label{f:nt_rhog_einhigh}
  \end{center}
\end{figure*}

\subsection{Boundary conditions for $\rho_{\rm g}(r)$}\label{s:nt_ic}

I first tried setting $\rho_{\rm g}(r_{\rm max}) = 0$ where $r_{\rm max}$ is the upper limit on $r$ used in the ODE solver. However, this failed to generate a sensible profile for $\rho_{\rm g}(r)$. This is expected, as such an initial condition surely provides `too little' information on the form of $\rho_{\rm g}(r)$ to constrain its profile at low $r$.
I next applied the initial condition $\rho_{\rm g}(r_{\rm min}) = \rho_{\rm g,a}(r_{\rm min})$, where $\rho_{\rm g,a}$ is the value obtained from PM I / PM II. I think this assumption is sensible, given that the non-thermal contributions are generally thought to be less and less significant at smaller radii as pointed out in Section~\ref{s:nt_p}.
I generally found that the latter initial condition produced solution curves for $\rho_{\rm g}(r)$ when solving the ODEs given by equations~\ref{e:rhog_nfw_diff} and~\ref{e:rhog_ein_diff}. 

\section{Non-thermal pressure profiling} \label{s:nt_plots}

As in the previous Sections which focus on cluster profiling, I create plots for clusters with input values of $M(r_{200}) = 1\times 10^{14} M_{\mathrm{Sun}}$ \& $M(r_{200}) = 1\times 10^{15} M_{\mathrm{Sun}}$, and  $z = 0.15$ \& $z = 0.9$. For PM II and PMN II we consider $\alpha_{\rm Ein}$ values of $0.05, \, 0.2$, and $ 2.0$. 

\subsection{Gas density profiles} \label{s:nt_rhog_plots}
Figures~\ref{f:nt_rhog_nfw},~\ref{f:nt_rhog_einlow},~\ref{f:nt_rhog_einmed}, and~\ref{f:nt_rhog_einhigh} compare the PM and PMN profiles for the NFW, $\alpha_{\rm Ein} = 0.05$, $\alpha_{\rm Ein} = 0.2$, and $\alpha_{\rm Ein} = 2$ cases respectively. The most striking feature of these graphs is the fact that the PMN profiles have higher gas densities than their PM equivalent for radii $> r_{0}$, until they decay to $\approx 0$ at high $r$. As was the case with the PMT profiles, changing the mass / $z$ input parameters does not seem to effect the shape of the PMN gas density profiles. However, unlike the comparison between the PM and PMT models, changing the input parameters here does seem to have an effect on the level of disparity between the PM and PMN profiles.

\begin{figure*}
  \begin{center}
    \begin{tabular}{@{}cc@{}}
     \includegraphics[ width=0.50\linewidth]{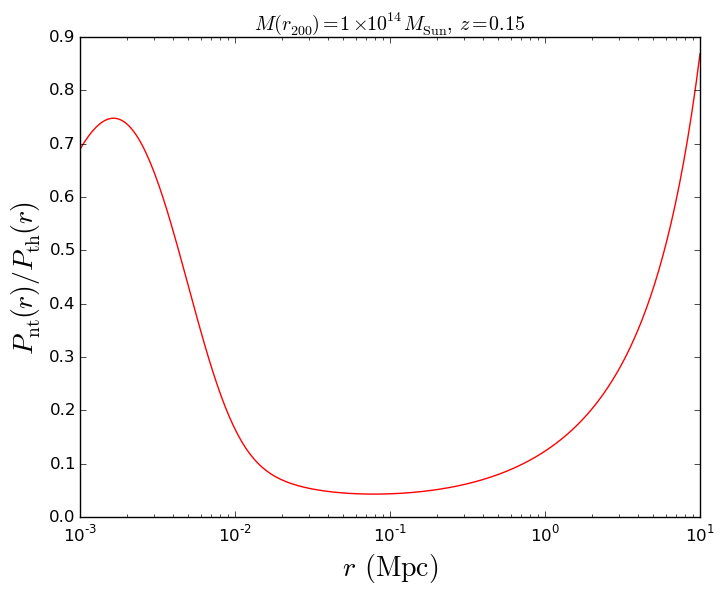} &
     \includegraphics[ width=0.50\linewidth]{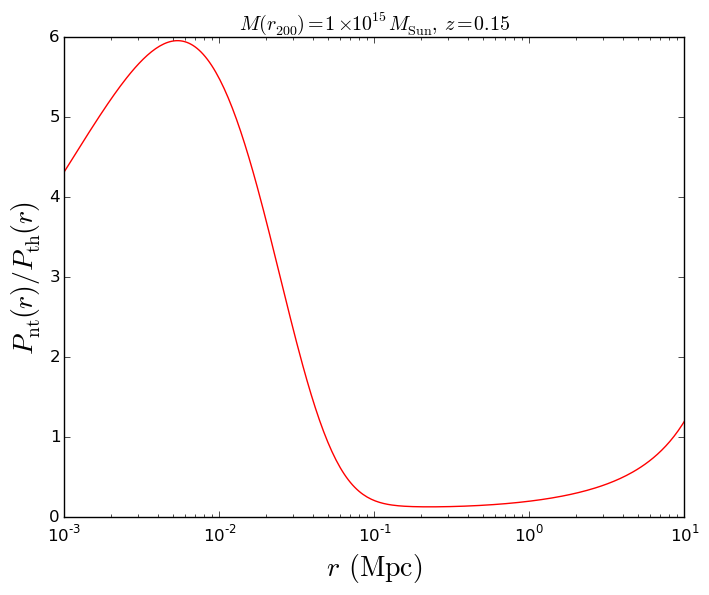} \\
     \includegraphics[ width=0.50\linewidth]{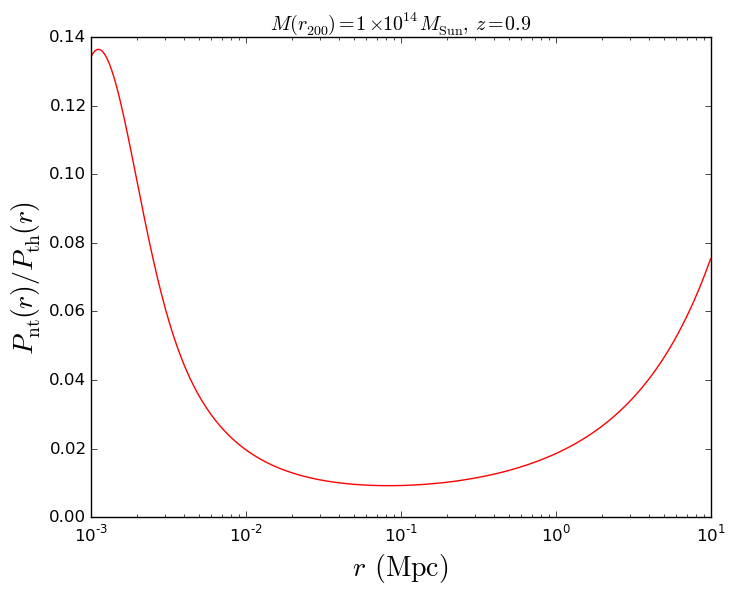} &
     \includegraphics[ width=0.50\linewidth]{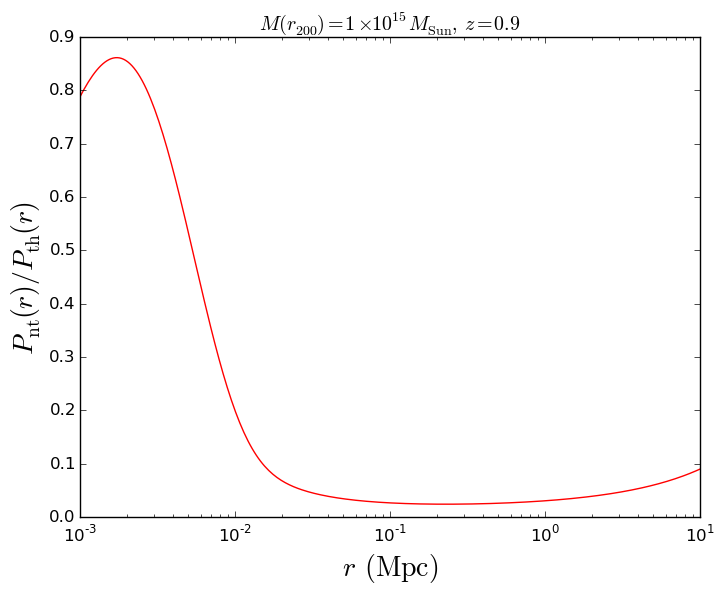} \\
    \end{tabular}

  \caption{Ratio of $P_{\rm nt}$ to $P_{\rm th}$ profiles for PM I and PMN I. Each graph features both profiles for one of the four different input parameter sets. Top row has $z = 0.15$, bottom row has $z = 0.9$. Left column has $M(r_{200}) = 1\times 10^{14} M_{\mathrm{Sun}}$, right column has $M(r_{200}) = 1\times 10^{15} M_{\mathrm{Sun}}$}
\label{f:nt_prat_nfw}
  \end{center}
\end{figure*}

\subsection{Thermal and non-thermal pressure profiles} \label{s:nt_p_plots}
Once $\rho_{\rm g}(r)$ has been determined, $P_{\rm nt}$ can be calculated from equation~\ref{e:nt_eqn}. $P_{\rm th}$ is given by equation~\ref{e:th_eqn}, and so is the same as the profiles of $P_{\rm g}$ calculated for the PMs.
\begin{figure*}
  \begin{center}
    \begin{tabular}{@{}cc@{}}
     \includegraphics[ width=0.50\linewidth]{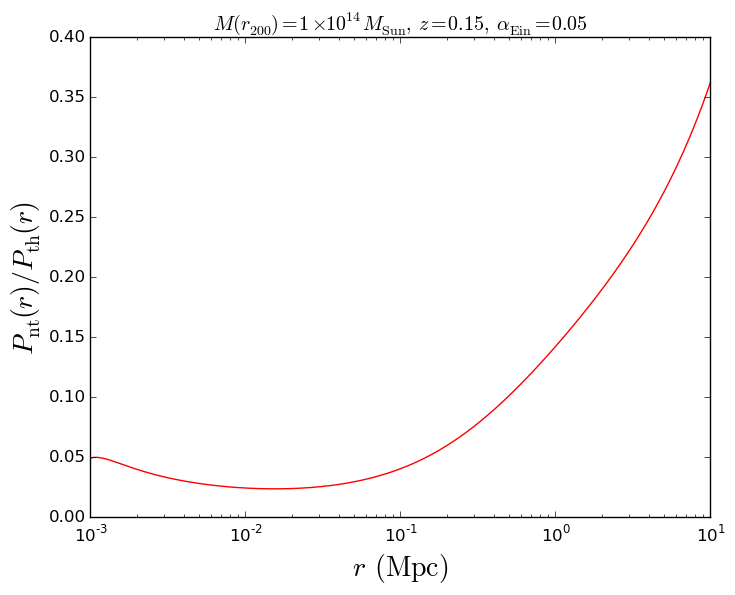} &
     \includegraphics[ width=0.50\linewidth]{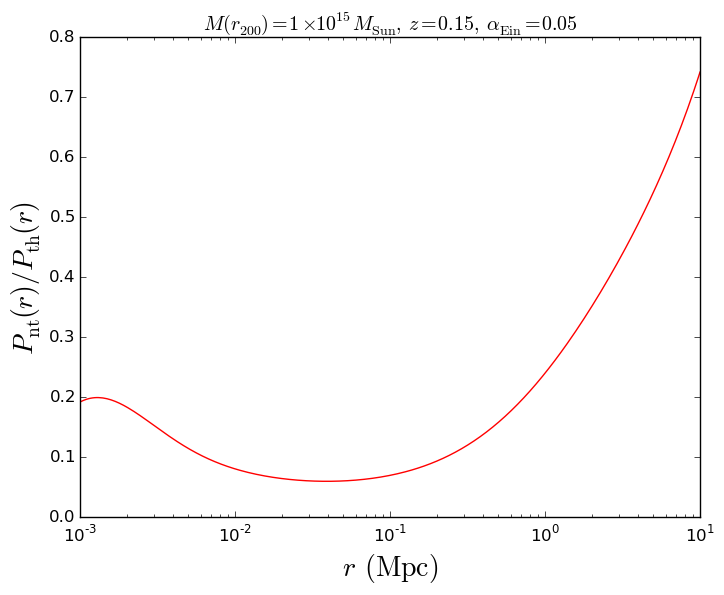} \\
     \includegraphics[ width=0.50\linewidth]{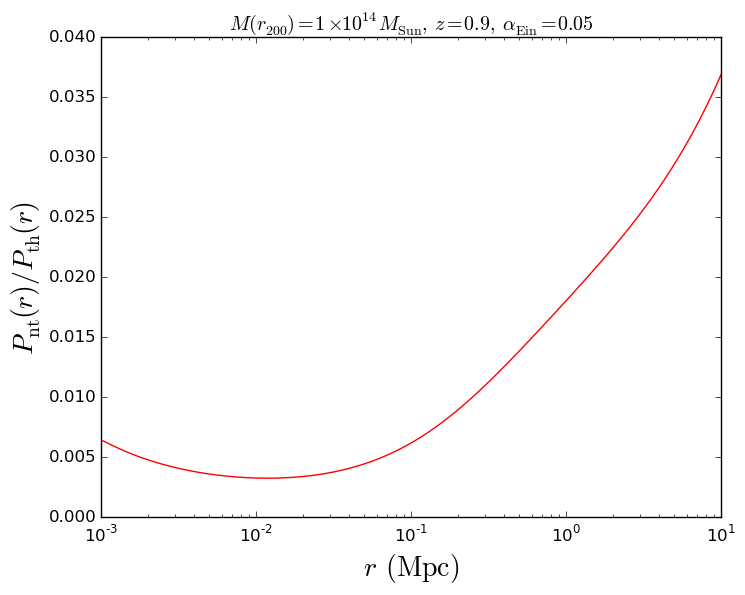} &
     \includegraphics[ width=0.50\linewidth]{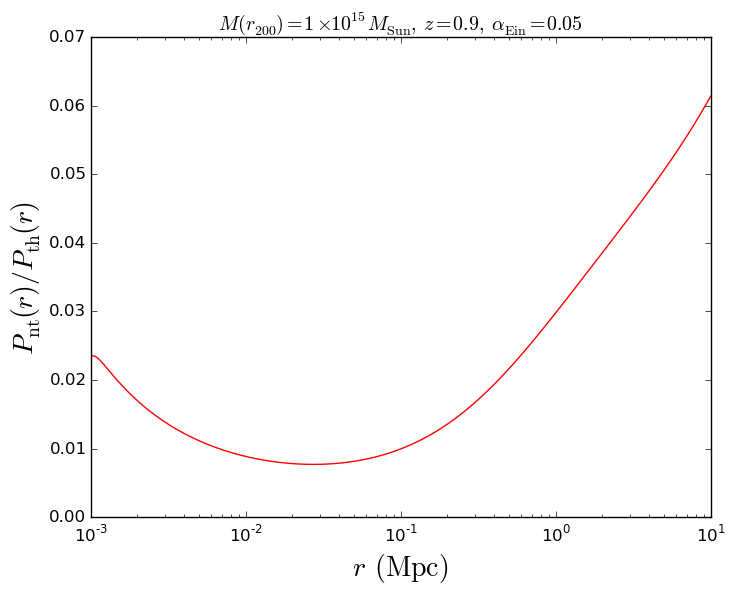} \\
    \end{tabular}

  \caption{Ratio of $P_{\rm nt}$ to $P_{\rm th}$ profiles for PM II and PMN II with $\alpha_{\rm Ein} = 0.05$. Graphs are laid out as described in Figure \ref{f:nt_prat_nfw}.}
\label{f:nt_prat_einlow}
  \end{center}
\end{figure*}
Furthermore the mass is still calculated using the approximation $M \approx M_{\rm dm}$ and so it has identical values between the PMs and PMNs.
Figures~\ref{f:nt_prat_nfw},~\ref{f:nt_prat_einlow},~\ref{f:nt_prat_einmed}, and~\ref{f:nt_prat_einhigh} show the ratio of non-thermal to thermal pressure (calculated from the PMNs) for the NFW, $\alpha_{\rm Ein} = 0.05$, $\alpha_{\rm Ein} = 0.2$, and $\alpha_{\rm Ein} = 2$ cases respectively. Given that in simulations (\citealt{2004MNRAS.351..237R}, \citealt{2009MNRAS.394..479A}, \citealt{2008A&A...491...71P}, and \citealt{2011MNRAS.413..573B}), non-thermal pressure was found to be at a maximum $\approx 15\%$ of the thermal pressure, these Figures show that the PMNs considered here are unphysical, particularly as the value of $\alpha_{\rm Ein}$ increases. The only profiles which give sensible values are the $\alpha_{\rm Ein} = 0.05$ cases. Here, the non-thermal pressure does go above $20\%$, but only at high $r$, where both types of pressure should take negligibly small values. Even though the ratio profiles look sensible for $\alpha_{\rm Ein} = 0.05$, the fact they are off by such a large amount for the other clusters implies the models formulated here are probably unfeasible (including the validity of the method used to determine $P_{\rm ei}$), and that the case of one good result has probably been obtained by chance.
However we do note that in DM16 the ratio approaches unity for five of the ten cluster sample of simulations at $r \approx r_{200, \mathrm{m}}$ (see Figure~5 of DM16). Whilst this doesn't add any validity to the results, it does suggest that non-thermal pressure can contribute greatly (up to $\approx$ the majority) towards the total pressure, and thus further work on incorporating its effect into cluster SZ models is important in improving their performance.

\begin{figure*}
  \begin{center}
    \begin{tabular}{@{}cc@{}}
     \includegraphics[ width=0.50\linewidth]{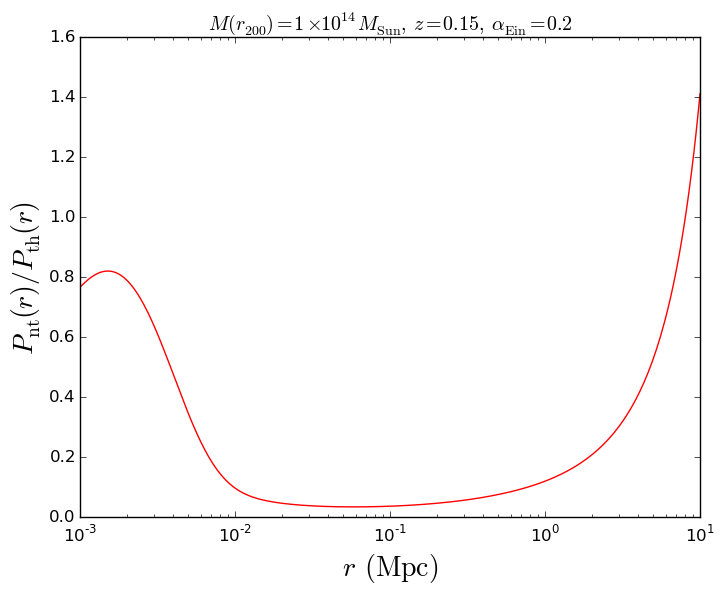} &
     \includegraphics[ width=0.50\linewidth]{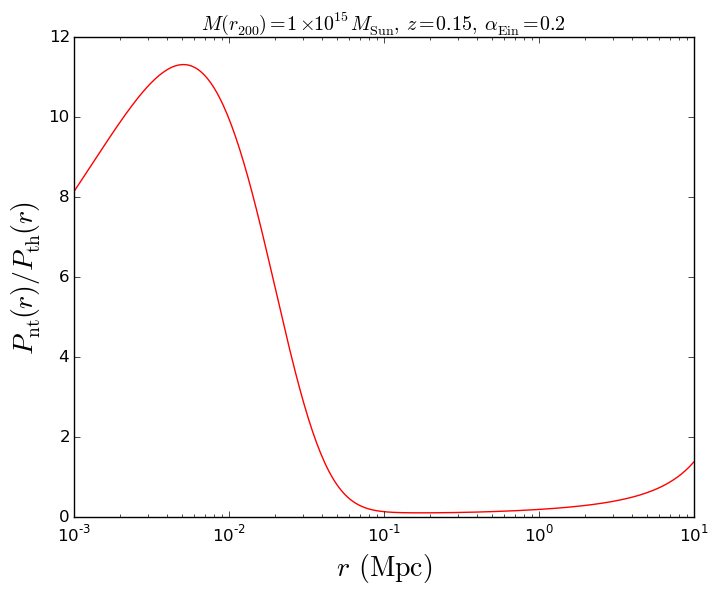} \\
     \includegraphics[ width=0.50\linewidth]{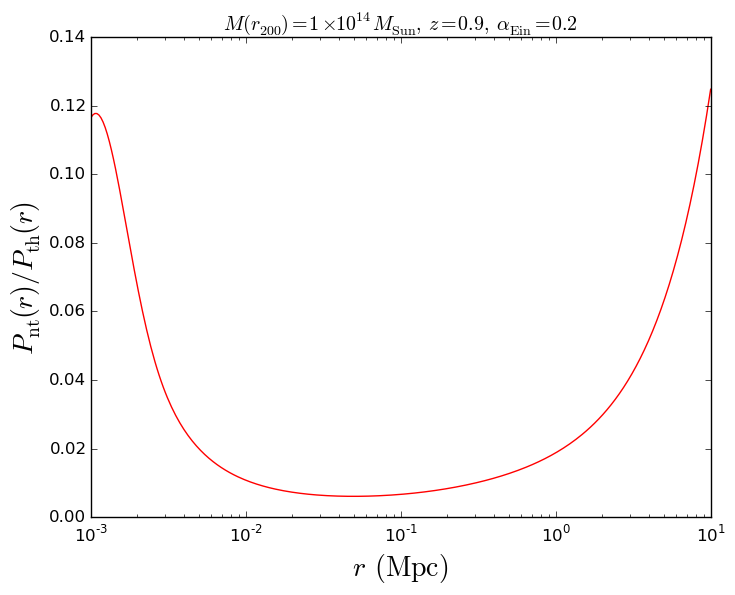} &
     \includegraphics[ width=0.50\linewidth]{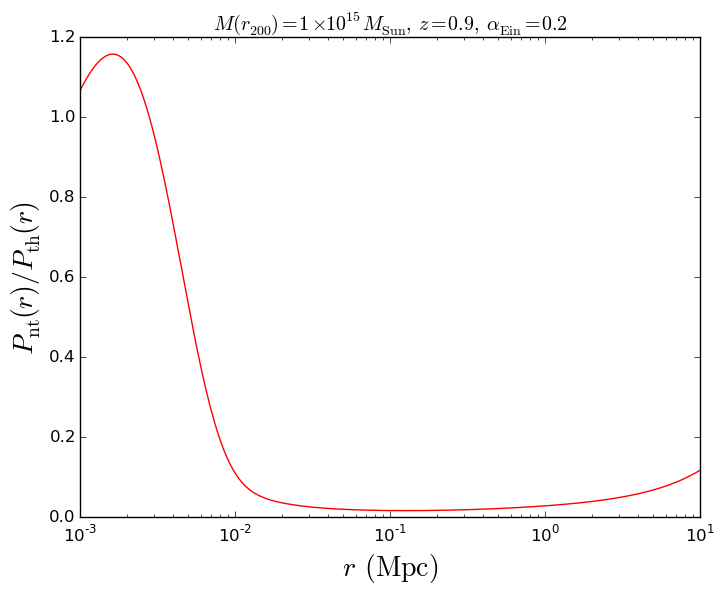} \\
    \end{tabular}

  \caption{Ratio of $P_{\rm nt}$ to $P_{\rm th}$ profiles for PM II and PMN II with $\alpha_{\rm Ein} = 0.2$. Graphs are laid out as described in Figure \ref{f:nt_prat_nfw}.}
\label{f:nt_prat_einmed}
  \end{center}
\end{figure*}

\begin{figure*}
  \begin{center}
    \begin{tabular}{@{}cc@{}}
     \includegraphics[ width=0.50\linewidth]{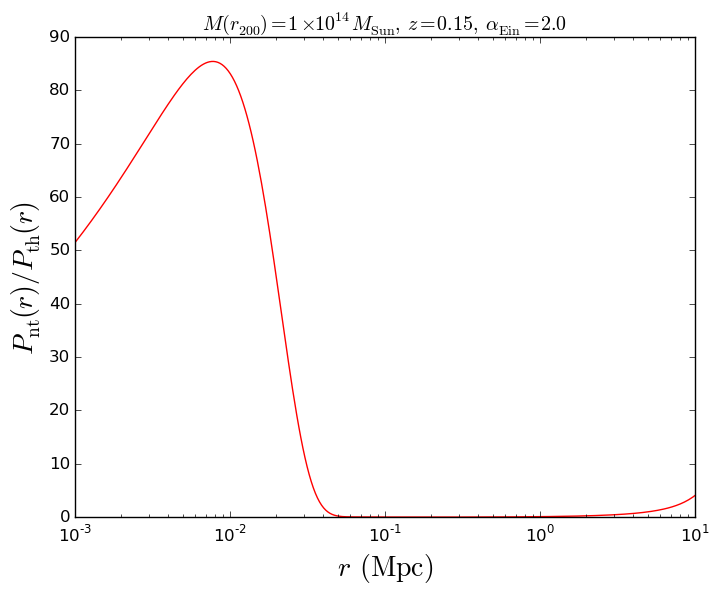} &
     \includegraphics[ width=0.50\linewidth]{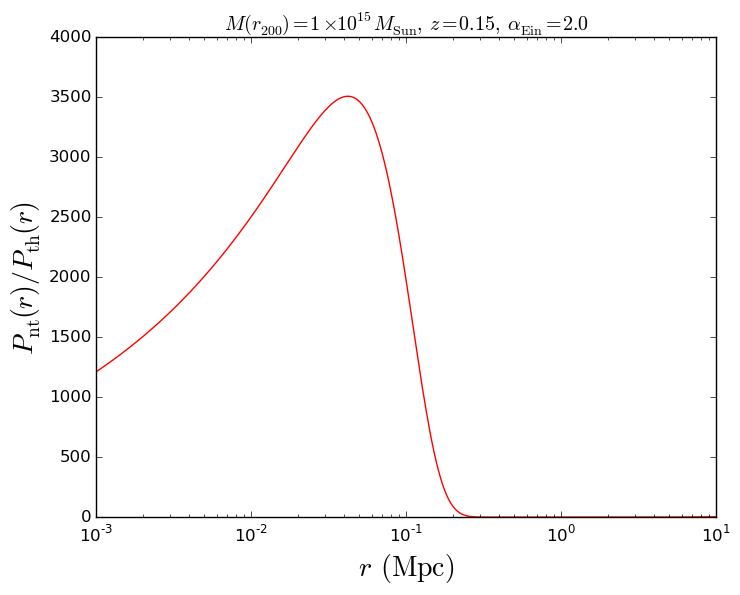} \\
     \includegraphics[ width=0.50\linewidth]{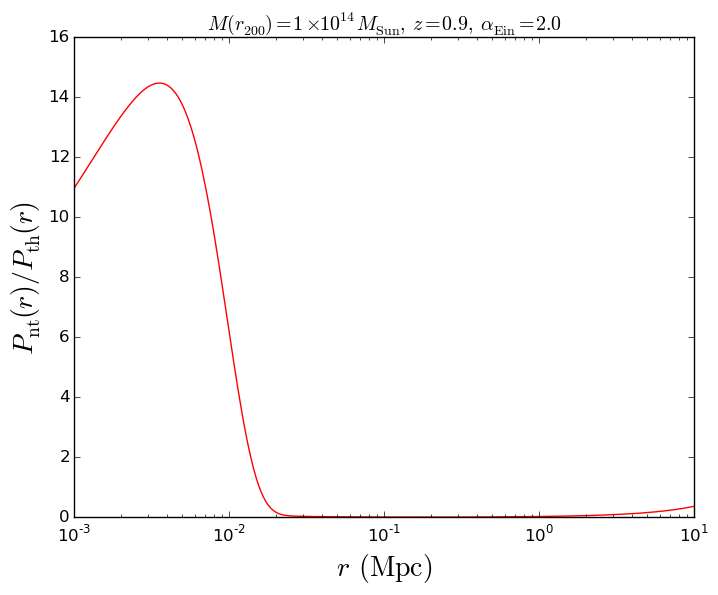} &
     \includegraphics[ width=0.50\linewidth]{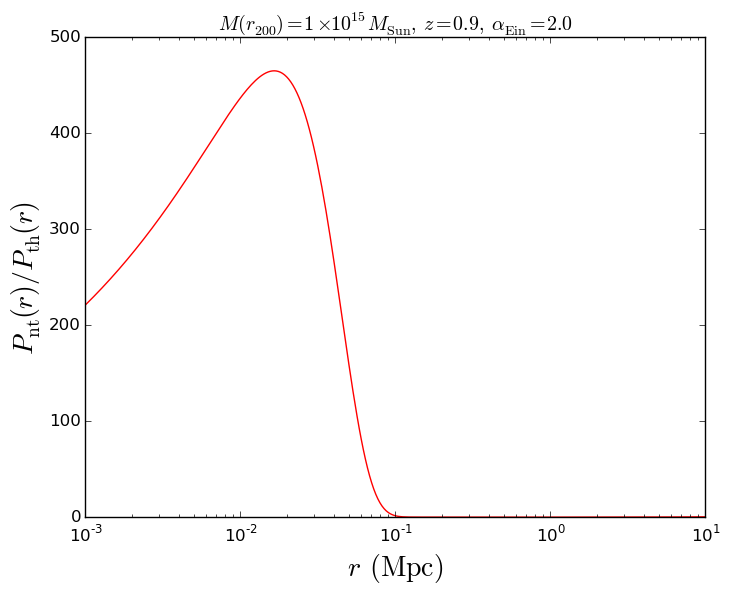} \\
    \end{tabular}

  \caption{Ratio of $P_{\rm nt}$ to $P_{\rm th}$ profiles for PM II and PMN II with $\alpha_{\rm Ein} = 2$. Graphs are laid out as described in Figure \ref{f:nt_prat_nfw}.}
\label{f:nt_prat_einhigh}
  \end{center}
\end{figure*}


\section{Conclusions}
\label{s:nt_conc}

In this Chapter I incorporated non-thermal pressure into the physical models presented in Section~\ref{s:phys_mod} (PM I) and the equivalent Einasto physical model (Section~\ref{s:ein_mod}, PM II) to see if this would produce more physically plausible models for clusters. I derive two new models PMN I and PMN II based on the analytical equation for non-thermal gas pressure in a cluster derived in \citet{2016arXiv160804388M}. Both PMNs require linear inhomogeneous ordinary differential equations in $\rho_{\rm g}(r)$ to be solved. However to do so, values for $P_{\rm ei}$ need to be determined as in Chapter~\ref{c:sixth}. 
Due to its simplicity, I used the method outlined in Section~\ref{s:tot_mass_pars_det3} to determine an approximate value for $P_{\rm ei}$. I then solved the ODEs in $\rho_{\rm g}(r)$ for various cluster input parameters and found the following.
\begin{itemize}
\item The PMN profiles have higher gas densities than their PM equivalent, until they decay to $\approx 0$ at high $r$. 
\item As was the case with the PMT profiles in Chapter~\ref{c:sixth}, changing the mass or $z$ input parameters does not seem to effect the shape of the PMN gas density profiles. However, unlike the comparison between the PM and PMT models, changing the input parameters here does seem to have an effect on the level of disparity between the PM and PMN profiles.
\end{itemize}
I then plotted the ratio of non-thermal to thermal pressure for different cluster inputs, to see how the ratio compared with those obtained from simulations in the literature, and found the following.
\begin{itemize}
\item For all but the $\alpha_{\rm Ein} = 0.05$ clusters, the ratio of non-thermal to thermal pressure was unphysical, as it exceeded values well over $100\%$ (which seems unfeasible in isolation, and even more unreasonable when compared to the values of $\approx 15\%$ obtained in simulations).
\item Even though the ratio profiles looked sensible for $\alpha_{\rm Ein} = 0.05$, the fact they were off by such a large amount for the other clusters implies the models formulated here are probably unfeasible, and that the case of one good result was probably obtained by chance.
\end{itemize}


%% file: CHAP-8/chapter8.tex
\chapter{Joint analysis of AMI and Planck data}\label{c:eighth}

Analysis of data obtained from different telescopes allows one to compare and verify inferences from measurements of different quantities, are subject to different systematic errors, and are obtained from different wavebands and on different angular scales. Simultaneous analysis of multiple datasets can lead to results different from those obtained in the individual cases, and can be used to investigate problems which cannot be resolved by the individual analyses. \\
In the context of galaxy clusters 
\citet{2006ApJ...652..917L} have used joint X-ray--SZ data in an attempt to constrain the dark energy equation of state parameter $w$. Similarly, cosmological constants have been estimated from X-ray analyses (see e.g. \citealt{2009ApJ...692.1060V} and \citealt{2010MNRAS.406.1759M}), SZ measurements (see e.g. \citealt{2011ApJ...732...28M}, \citealt{2011ApJ...738..139W}, and \citealt{2011ApJ...731..100M}) and a joint X-ray--SZ analysis \citep{2012ApJ...748..113H}. \\
Joint analysis of data from galaxy clusters is not restricted to telescopes which measure different quantities. \citet{2015A&A...576A..12A}, \citet{2015ApJ...807..121R}, \citet{2016A&A...586A.122A}, and \citet{2017arXiv170706113R} all use SZ measurements from instruments including the Planck satellite \citep{2006astro.ph..4069T}, Bolocam \citep{2013ApJ...768..177S} and \citep{2015ApJ...806...18C}, Green Bank telescope \citep{2011ApJ...734...10K}, and IRAM 30-metre telescope \citep{2014JLTP..176..787M}, that probe different angular scales and operate over different frequency ranges, to infer profiles of cluster parameters such as pressure, temperature and mass.

In this Chapter 
I carry out joint analysis of SZ data from AMI and from the Planck satellite. Note that I conduct \textit{separate} analyses on these data in Chapter~\ref{c:third}. I apply Bayesian analysis using a joint likelihood for data from both instruments, to simulated cluster data generated with observational and physical models (largely based on the ones introduced in Chapters~\ref{c:second},~\ref{c:fourth}, and~\ref{c:fifth}). 
I analyse the resulting posterior distributions and compare them with results obtained from analysing the likelihoods for AMI and Planck separately. \\
I also apply the joint analysis to real data from the Planck detected cluster PSZ2G063.80+11.42, whose mass estimates derived from AMI and Planck data in Chapter~\ref{c:third} showed discrepancies with one another. Note the work in this Chapter has been published as a paper in MNRAS \citep{2019MNRAS.486.2116P}, which I am a lead author of. The paper includes more information on how the Planck simulations were generated, and presents results of analyses where the simulated data was much better understood (and less prone to bugs).

\section{Joint likelihood analysis}

The key aspects of Bayesian inference have already been highlighted in Chapter~\ref{c:second}. Nevertheless it is useful to highlight how we evaluate the joint likelihood function of datasets which have previously been analysed in isolation and with different analysis pipelines.

\subsection{AMI data analysis}

As previously, \textsc{McAdam} is used to calculate the posterior distribution for AMI data (see Section~\ref{s:ap_AMIlhood}). 

\subsection{Planck detection algorithms}

The $Y$ and $M$ values published in the Planck catalogue PSZ2 are derived from data from one of three detection algorithms: MMF1, MMF3 (\citealt{2009ApJ...701...32S}; \citealt{2011ApJ...737...61M}) and PowellSnakes (PwS, \citealt{2012MNRAS.427.1384C}). The mass estimates presented in Chapter~\ref{c:third} that are based on Planck data were calculated from the outputs of the PwS algorithm. Similarly the joint AMI-Planck analysis here uses PwS to process the data for the Planck part of the analysis (see Section~\ref{s:ap_PwSlhood}).

\subsection{Joint likelihood function}
\label{s:ap_jlhood}

If one has an AMI dataset $\vec{d}_{\mathrm{AMI}}$ and a Planck dataset $\vec{d}_{\mathrm{Pl}}$, then the \textit{joint} likelihood function for the data is given by
\begin{equation}
\label{e:ap_lhood}
\mathcal{L}(\vec{\Theta}) = \mathcal{L}\left(\vec{d}_{\mathrm{AMI}}, \vec{d}_{\mathrm{Pl}} |\vec{\Theta}, \mathcal{M}\right).
\end{equation}
In this analysis we treat $\vec{d}_{\mathrm{AMI}}$ and $\vec{d}_{\mathrm{Pl}}$ as being independent (see Section~\ref{s:ap_sims_cov} for justification), and since the Planck-predicted data only rely on the cluster parameters we can write
\begin{equation}
\label{e:ap_lhood2}
\mathcal{L}(\vec{\Theta}) = \mathcal{L}_{\mathrm{AMI}}\left(\vec{d}_{\mathrm{AMI}} | \vec{\Theta}, \mathcal{M} \right) \mathcal{L}_{\mathrm{Pl}} \left( \vec{d}_{\mathrm{Pl}} |\vec{\Theta}_{\rm cl}, \mathcal{M}\right).
\end{equation}

\subsubsection{AMI likelihood function}
\label{s:ap_AMIlhood}
The form of the AMI likelihood function used here is exactly the same as the one presented in Section~\ref{s:likelihood}. Note also that the AMI covariance matrix $\boldsymbol{\mathsf{C}}_{\mathrm{AMI},\nu,\nu'}$  is comprised of the same components as noted in Section~\ref{s:likelihood} (which are described in Sections~\ref{s:inst_noise},~\ref{s:prim_cmb}, and~\ref{s:conf_noise}), and recognised radio-sources are also treated the same way as previously described (Sections~\ref{s:identified_sources} and~\ref{s:rs_priors}) for \textit{AMI} data. For clarity I note that the predicted AMI data are denoted $\vec{d}_{\mathrm{AMI},\nu'}^{\rm p}(\vec{\Theta})$

\subsubsection{PwS likelihood function}
\label{s:ap_PwSlhood}
For a single source and given observing frequency, PwS treats the data observed by Planck as a superposition of background sky emission (including foreground emission and primordial CMB) $\vec{b}_{\nu}$, instrumental noise $\vec{n}_{\nu}$, and signal from the source $\vec{s}_{\nu}$. The model for the predicted data vector is thus
\begin{equation}
\label{e:ap_PwSd}
\vec{d}_{\mathrm{Pl},\nu}^{\rm p}(\vec{\Theta}_{\rm cl}) = \vec{s}_{\nu}(\vec{\Theta}_{\rm cl}) + \vec{b}_{\nu} + \vec{n}_{\nu}.
\end{equation}
PwS works with patches of sky sufficiently small such that it can be assumed the noise contributions are statistically homogeneous. In this limit it is more convenient to work in Fourier space, as the Fourier modes are uncorrelated assuming the noise contributions are Gaussian. This assumption is fair in the case of instrumental noise, but more questionable for $\vec{b}_{\nu}$. The deviations from Gaussianity of $\vec{b}_{\nu}$ are discussed in Section~4.3 of the second PwS paper \citep{2012MNRAS.427.1384C}. 
Since PwS is a detection algorithm, it calculates the ratio of the likelihood of detecting a cluster parameterised by $\vec{\Theta}_{\rm cl}$ and the likelihood of the data with no cluster signal ($\vec{s}_{\nu}(\vec{\Theta}_{\rm cl, 0}$) = 0). Thus the log-likelihood ratio
 of the Fourier transformed quantities is 
\begin{equation}
\label{e:ap_PwSlhoodr}
\begin{split}
\ln \left[ \frac{\mathcal{L}_{\mathrm{Pl}}\left(\vec{\Theta}_{\rm cl}\right)}{\mathcal{L}_{\mathrm{Pl}}\left(\vec{\Theta}_{\rm cl,0}\right)} \right] = & \sum\limits_{\nu,\nu'} \tilde{\vec{d}}_{\mathrm{Pl},\nu}^{\rm p}(\vec{\Theta}_{\rm cl})^{\rm T} \boldsymbol{\mathsf{C}}_{\mathrm{Pl},\nu,\nu'}^{-1} \tilde{\vec{d}}_{\mathrm{Pl},\nu}(\vec{\Theta}_{\rm cl}) \\
 & - \frac{1}{2} \tilde{\vec{d}}_{\mathrm{Pl},\nu}^{\rm p}(\vec{\Theta}_{\rm cl})^{\rm T} \boldsymbol{\mathsf{C}}_{\mathrm{Pl},\nu,\nu'}^{-1} \tilde{\vec{d}}_{\mathrm{Pl},\nu}^{\rm p}(\vec{\Theta}_{\rm cl}),
\end{split}
\end{equation}
where tildes denote the Fourier transform of a quantity, and $\boldsymbol{\mathsf{C}}_{\mathrm{Pl},\nu,\nu'}$ is the covariance matrix of the data in Fourier space. \\
A full specification of the PwS likelihood ratio is given in \citet{2009MNRAS.393..681C} and \citet{2012MNRAS.427.1384C}.

\section{Joint likelihood analysis hyperparameters}

\citet{2000MNRAS.315L..45L} and \citet{2002MNRAS.335..377H} (MH02 from here on) introduced a Bayesian method for determining the relative weighting of two or more \textit{independent} datasets when analysed simultaneously, while \citet{2014A&C.....5...45M} built on this work to develop a method which works for datasets \textit{correlated} with one another. The basic idea behind the approach is to introduce additional {\em hyperparameters} $\vec{\alpha}$ into the Bayesian inference problem. In other words we extend our parameter space to include not only the parameters of interest ($\vec{\Theta}$), but also the hyperparameters $\vec{\alpha}$. Thus we have
\begin{equation}
\label{e:ap_hp_p}
\mathcal{P}(\vec{\Theta}, \vec{\alpha}) = \frac{\mathcal{L}(\vec{\Theta}, \vec{\alpha}) \pi(\vec{\Theta}, \vec{\alpha})}{\mathcal{Z}}, 
\end{equation}
where $\mathcal{Z}$ is now given by
\begin{equation}
\label{e:ap_hp_z}
\mathcal{Z} = \int \int \mathcal{L}(\vec{\Theta}, \vec{\alpha}) \pi(\vec{\Theta}, \vec{\alpha}) \mathrm{d}\vec{\alpha} \mathrm{d}\vec{\Theta}.
\end{equation}
Equations~\ref{e:ap_hp_p} and~\ref{e:ap_hp_z} tell us that to obtain the quantities of interest ($\mathcal{P}(\vec{\Theta})$ and $\mathcal{Z}$) we have to marginalise over the hyperparameters. \\ 
It is reasonable to assume that the parameters of the original problem and those affecting the weighting of each likelihood are independent of one another, so the priors can be written as 
\begin{equation}
\label{e:ap_hp_pi}
\pi(\vec{\Theta}, \vec{\alpha}) = \pi(\vec{\Theta}) \pi(\vec{\alpha}).
\end{equation}
For more information on the typical priors used for $\vec{\alpha}$ we refer the reader to Section~4.1 of MH02. \\
To see how $\vec{\alpha}$ are incorporated into $\mathcal{L}(\vec{\Theta})$ we consider two independent datasets, so that $\mathcal{L}(\vec{\Theta}, \vec{\alpha})$ can be written as
\begin{equation}
\label{e:ap_hp_l}
\mathcal{L}(\vec{\Theta}, \vec{\alpha}) = \mathcal{L}_1(\vec{\Theta})^{\alpha_1} \mathcal{L}_2(\vec{\Theta})^{\alpha_2}.
\end{equation}
Note we have chosen for the likelihoods to have such dependence on $\vec{\alpha}$ so that if $\mathcal{L}_1$ and $\mathcal{L}_2$ are Gaussian (equation~\ref{e:lhood}) 
\begin{equation}
\label{e:ap_hp_l2}
\mathcal{L}_1(\vec{\Theta})^{\alpha_1} = \frac{1}{Z_{N,1}}e^{-\frac{1}{2}\alpha_1 \chi_1^{2}}
\end{equation}
(and similarly for $\mathcal{L}_2$), then we can write
\begin{equation}
\label{e:ap_hp_chi}
\chi_{\rm joint}^2 = \alpha_1 \chi_1^2 + \alpha_2 \chi_2^2,
\end{equation}
where the $\chi^2$ quantities are defined by equation~\ref{e:chisq}. Thus $\alpha_1$ and $\alpha_2$ control the relative weighting of the goodness-of-fit metrics of the data. 

\subsection{Effects of likelihood hyperparameters}
The effects of including $\vec{\alpha}$ in Bayesian analysis are best illustrated through examples. Here I provide a very brief overview and quote the results of the toy model considered in MH02, to emphasise how the inclusion of hyperparameters affects evidence and posterior estimates of joint analyses compared with not using them (i.e. $\alpha_1 = \alpha_2 = 1$). I refer to the results obtained from including hyperparameters as HP and those from the `vanilla' method as V. The toy problem consists of fitting a straight line through two (independent) sets of data points, and thus is a two-likelihood (one for each set of data), two-parameter problem of inferring the gradient ($m$) and intercept ($c$) of the line. The likelihood thus takes the form given by equation~\ref{e:ap_hp_l} with $\vec{\Theta} = (m,c)$. Several versions of the problem are considered which vary in the standard deviations used for the likelihood functions and how the two datasets are generated.

\subsubsection{Correct likelihood standard deviations and consistent datasets}
\label{s:ap_hp_tm}
The first example considered involves drawing points for each of the two datasets from the same distribution, namely Gaussian distributions with standard deviations $\sigma_{\rm{d},1} = \sigma_{\rm{d},2} = 0.1$, and mean values corresponding to the line with $m_{\rm d} = c_{\rm d} = 1$; the same deviations are used for the likelihood functions: $\sigma_{l,1} = \sigma_{l,2} = 0.1$. The resulting posterior distributions for the HP and V cases are shown in Figure~1 of MH02. The two methods recover the true values of $m$ and $c$ equally well, but the V run leads to a higher evidence estimate. This is to be expected for simple problems (for which the methods provide an equivalent fit to the data), as the added complexity of the HP method decreases the Bayesian evidence according to Occam's razor.

\subsubsection{Incorrect likelihood standard deviations and consistent datasets}
The second example generates the two datasets in the same way, but the standard deviations used in the likelihoods are incorrect: $\sigma_{l,1} = 0.02, \, \sigma_{l,2} = 0.1$. Thus the predicted errors on the first dataset are much smaller than the true values used to generate it. In this case (Figure~2 of MH02) the V posterior underestimates the errors on $m$ and $c$ such that the true value is outside the $99\%$ probability interval; whereas the HP method results in much larger error estimates, leading to the correct value being within the $95\%$ confidence interval. This suggests that $\alpha_1$ on average took relatively small values to accommodate for $\sigma_{l,1}$ being underestimated in the analysis. Furthermore the evidence ratio between the HP and V analyses is greatly in favour of the former, suggesting the data are fit sufficiently better by the HP model to overcome its additional complexity.

\subsubsection{Correct likelihood standard deviations with inconsistent datasets}
The final scenario considered for the toy problem in MH02 involves sampling the two datasets from different distributions i.e. sampling two sets of data which represent different lines. This means that there are two `true' values for $m$ and $c$ corresponding to each dataset, and so a good inference of the data should produce a bimodal distribution with peaks at these values. They first test this out by sampling one dataset from a distribution corresponding to $m_{\rm{d},1} = c_{\rm{d},1} = 1$ and the other from $m_{\rm{d},2} = 0, \, c_{\rm{d},2} = 1.5$. The resultant posterior distributions shown in Figure~3 of MH02 show that the V distribution is unimodal and does not contain either of the true values within its 99\% probability contours, while the HP distribution is bimodal with the peaks occurring close to the true values. \\
They repeat this analysis but sample from distributions corresponding to $m_{\rm{d},1} = c_{\rm{d},1} = 1$ and $m_{\rm{d},2} = 0.7, \, c_{\rm{d},2} = 0.7$ and find again that the V posterior distribution is unimodal and centred far from the true values, while HP results in a bimodal distribution with peaks in the vicinity of the true values (but not as close as in the previous case). The evidence ratio between the V and HP analyses suggests the latter is a more suitable model in both cases.

\subsection{Incorporating the likelihood hyperparameters into AMI-Planck analysis}

\begin{figure*}
  \begin{center}
    \begin{tabular}{@{}c@{}}
  \includegraphics[ width=0.75\linewidth]{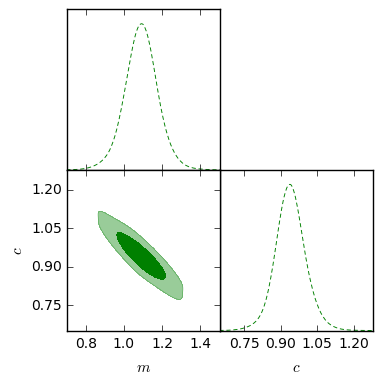} \\
  \includegraphics[ width=0.75\linewidth]{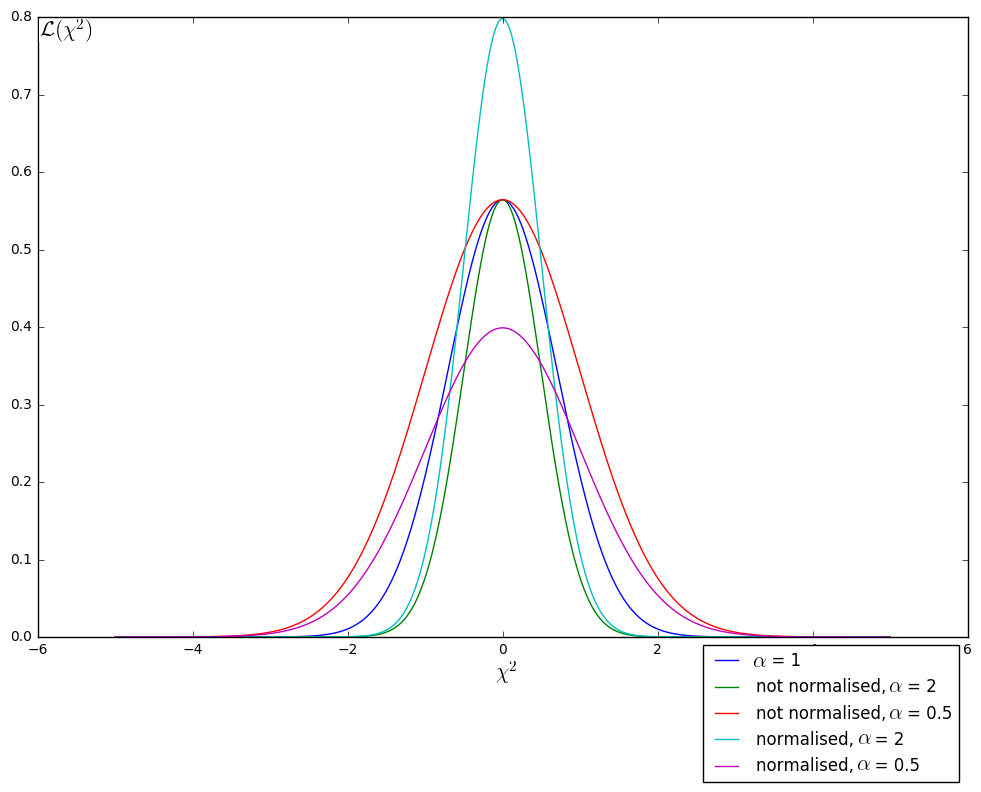}
    \end{tabular}
  \caption{Top: Two-dimensional posterior distribution obtained from application of likelihood hyperparameter method on toy model considered in Section~6.1 of MP02. $m$ and $c$ are the gradient and intercept parameters of the toy model respectively. These results were obtained using the likelihood functions given by equation~\ref{e:ap_hp_l2}. Bottom: Gaussian likelihood for a range of $\alpha$ values, including (normalised) or excluding (un-normalised) $C^{\alpha}$. Note for the $\alpha = 1$ case the normalisation doesn't depend on $\alpha$ since $C(\alpha = 1) = C$). }
\label{f:ap_hp_post}
  \end{center}
\end{figure*}

From the examples reviewed above, it is clear that inclusion of the likelihood hyperparameters leads to inferences more representative of the data in the cases that the errors in the analysis are underestimated or the datasets are systematically different from one another. Thus it makes sense to include them in analyses of data obtained from telescopes operating at different frequencies and angular scales and that are subject to different systematic errors. \\
However the log-ratio given by equation~\ref{e:ap_PwSlhoodr} is not a probability density due to the fact that it is missing a normalisation factor proportional to $\tilde{\vec{d}}_{\mathrm{Pl},\nu}(\vec{\Theta}_{\rm cl})^{\rm T} \boldsymbol{\mathsf{C}}_{\mathrm{Pl},\nu,\nu'}^{-1} \tilde{\vec{d}}_{\mathrm{Pl},\nu}(\vec{\Theta}_{\rm cl}) \equiv C$. Inclusion of the likelihood hyperparameters means that the normalisation factor of a likelihood function is dependent on $\alpha$, since it is marginalised over to obtain $\mathcal{L}(\vec{\Theta})$, so $C \equiv C(\alpha) = C^{\alpha}$. \\
To test whether the inclusion of $C(\alpha)$ was strictly needed for the hyperparameter methodology, I replicated the toy model example considered in Section~6.1 of MP02 (reviewed in Section~\ref{s:ap_hp_tm}), ran the analysis using the `full' hyperparameter likelihood functions (equation~\ref{e:ap_hp_l2}) and also conducted the analysis using hyperparameter likelihood ratios (i.e. using the likelihoods given by equation~\ref{e:ap_hp_l2} but excluding the $C^{\alpha}$ factors present in the $\chi^2$s). The full likelihood analysis produced a a posterior distribution similar to the one obtained in MP02 (left plot of Figure~\ref{f:ap_hp_post}) while the likelihood ratio analysis failed to produce posterior samples. The reason why likelihood ratios are incompatible with the hyperparameter method is shown graphically in the right plot of Figure~\ref{f:ap_hp_post}. From this plot it is clear that $C^{\alpha}$ dictates the shape of the likelihood function as well as its peak. For example around $\chi^2 = 0$ the normalised $\alpha = 2$ curve is above the $\alpha = 1$, while the un-normalised $\alpha = 2$ curve is below it. This inconsistency generalises to all $\chi^2$ and $\alpha$ values and thus one cannot reliably evaluate the effect of $\vec{\alpha}$ on the analysis without knowing $C^{\alpha}$ and hence the hyperparameters cannot be used with likelihood ratios such as the one used by PwS. As a result, the two likelihoods had to be weighted equally (i.e. I set $\alpha_{1} = \alpha_{2} =1$).

\section{Cluster models}

As described in Section~\ref{s:sz_theory}, a radio interferometer measure signal that is the Fourier transform of a quantity proportional to the Comptonisation parameter $y$. Similarly the Planck satellite is also sensitive SZ effect and thus measures a signal $\propto y$. Thus the cluster models introduced previously in this thesis which calculate a map of $y$ can be used to calculate $\vec{d}_{\mathrm{AMI},\nu}^{\rm p}(\vec{\Theta})$ and $\vec{d}_{\mathrm{Pl},\nu}^{\rm p}(\vec{\Theta}_{\rm cl})^{\rm T}$.

\subsection{Observational model}
\label{s:ap_obs}

The observational model used in this Chapter (OM III) is the same as the ones introduced in Chapter~\ref{c:fourth} other than the priors it uses. \\
Here I assign non-informative, independent priors to $Y_{\rm tot}$ and $\theta_{\rm p}$ (see Table~\ref{t:ap_priors}), to get a better idea of how much the joint likelihood function can constrain the parameters. The priors used for $a$ and $b$ vary throughout the analysis (Table~\ref{t:ap_priors}); they are either fixed at some specific value (as was the case in OM I and OM II) or allowed to vary uniformly. 

\subsection{Physical models} 
\label{s:ap_phys}

The physical models used here are the same as the ones presented in Chapters~\ref{c:second} and~\ref{c:fourth} (PM I and PM II) i.e. they model the cluster dark matter content using NFW and Einasto profiles respectively.
The prior distributions the PMS are also given in Table~\ref{t:ap_priors}. \\

All three models can be used to calculate the profile of $P_{\rm e}(r)$ which can be used to produce a $y$ map using equation~\ref{e:sz9}.

\begin{table*}
\begin{center}
\begin{tabular}{{l}{c}{c}}
\hline
Parameter & Model(s) featured in & Prior distribution(s) \\ 
\hline
$x_{\rm c}$ & OM III, PM I, and PM II & $\mathcal{N}(0'', 60'')$ \\
$y_{\rm c}$ & OM III, PM I, and PM II & $\mathcal{N}(0'', 60'')$ \\
$Y_{\rm tot}$ & OM III & $\mathcal{U}[0.00~\mathrm{arcmin}^{2}, 0.02~\mathrm{arcmin}^{2}]$ \\
$\theta_{\rm p}$ & OM III & $\mathcal{U} [1.3', 15']$ \\
$z$ & PM I and PM II & $\delta(z)$ \\
$M(r_{200})$ & PM I and PM II & $\mathcal{U} [ \log (0.5\times 10^{14} M_{\mathrm{Sun}}),\log (50\times 10^{14} M_{\mathrm{Sun}})]$ \\
$f_{\rm gas}(r_{200})$ & PM I and PM II & $\mathcal{N}(0.12, 0.02)$ \\
$\alpha_{\rm Ein}$ & PM II & $\delta(\alpha_{\rm Ein})$ or $\mathcal{U}[0.05, 0.3]$ \\
$a$ & OM III, PM I, and PM II & $\delta(a)$ or $\mathcal{U}[0.3, 3.5]$ \\
$b$ & OM III, PM I, and PM II & $\delta(b)$ or $\mathcal{U}[3.5, 7.5]$ \\
$c$ & OM III, PM I, and PM II & $\delta(c)$ \\
\hline
\end{tabular}
\caption{Cluster parameter prior distributions. $\mathcal{N}$ denotes a normal distribution parameterised by its mean and standard deviation, $\mathcal{U}$ denotes a uniform distribution, and $\delta$ is a Dirac delta function. In the cases where the latter is used, the values used for the function's argument will be stated when the analyses are carried out.}
\label{t:ap_priors}
\end{center}
\end{table*}

\section{Cluster simulations}

The cluster simulations were generated using the in-house package \textsc{Profile} (used in Chapters~\ref{c:third} and~\ref{c:fifth}). For all simulations the $y$ map of a single cluster is generated with either OM III, PM I, or PM II, and primordial CMB noise is sampled from an empirical distribution \citep{2013ApJS..208...19H} and added at random positions to the data. At this point the data are duplicated so that additional noise contributions specific to each telescope can be added. \\
For the AMI simulated data, confusion noise is added as described in Section~\ref{s:pl_phys_results_ii} using the 10C source counts given in \citet{2011MNRAS.415.2708A}. Instrumental noise with an RMS value of $0.379$~Jy per channel per baseline per second is also added. \\
For the Planck simulated data, foreground emission and instrumental noise are added. For more information on the Planck simulations, see \citet{2019MNRAS.486.2116P}. 
Finally, the data are `observed' by AMI and Planck separately to generate $\vec{d}_{\mathrm{AMI},\nu}$ and $\vec{d}_{\mathrm{Pl},\nu}$.

\subsection{Testing the independence of the AMI and Planck datasets}
\label{s:ap_sims_cov}

\begin{figure*}
  \begin{center}
     \includegraphics[width=\textwidth, keepaspectratio]{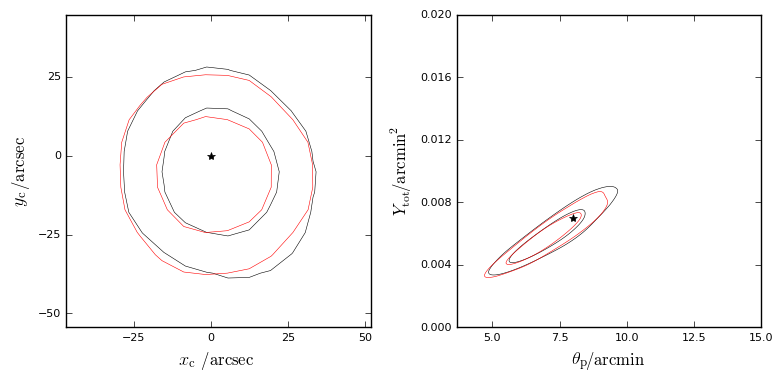}
\caption{Two-dimensional marginalised $x_{\rm c}-y_{\rm c}$ and $Y_{\rm tot}-\theta_{\rm p}$ posterior distributions for a high SNR (see Section~\ref{s:ap_obs_sim}) cluster simulation generated using OM III. The red contours correspond to the posterior distribution associated with the AMI and Planck datasets which had different CMB realisations to each other, while the black ones correspond to datasets generated with the same realisation. The star symbols indicate the values input when generating the simulations.}
\label{f:ap_cov_test}
  \end{center}
\end{figure*}

In Section~\ref{s:ap_jlhood} we made the assumption that $\vec{d}_{\mathrm{AMI},\nu}^{\rm p}$ and $\vec{d}_{\mathrm{Pl},\nu}^{\rm p}$ are not correlated with each other, so that the likelihoods for the two datasets can be separated. The instrumental noises associated with each telescope can safely be assumed to be independent. Due to the telescopes operating at different angular scales and frequencies, the confusion noise present in AMI data and foreground emission present in Planck data are assumed to be independent of one another. A similar argument can be applied for primordial CMB noise, nevertheless I carried out a simple test to see if this is the case. For a given set of cluster parameters, I ran the joint analysis on Planck and AMI datasets which had different CMB realisations to one another. I found that the resultant parameter constraints were not affected by this when compared with the results obtained using AMI and Planck data which had the same CMB realisations as one another (Figure~\ref{f:ap_cov_test}). I thus concluded that the covariance between the datasets was negligible. 

\section{Cluster simulation results}

In the following analysis I generate cluster simulations for different noise realisations and cluster parameter values (and models). I apply the joint analysis to these simulated clusters, and compare results with analyses which use (the same) AMI or Planck data alone. Note that for all examples considered, the model used to \textit{simulate} the cluster was also used to \textit{analyse} the data.

\subsection{Observational model with `universal' shape parameters}
\label{s:ap_obs_sim}
I generate simulations using OM III, with GNFW shape parameter values $a = 1.0510$, $b=5.4905$, and $c = 0.3081$ (i.e. the same ones used in Chapter~\ref{c:fifth}). 
As shown in Table~\ref{t:ap_obs_sim} I consider a `low' and a `high' signal-to-noise ratio (SNR) cluster, which correspond to input values of $Y_{\rm tot} =0.001$~arcmin$^{2}$ and $\theta_{\rm p} = 2$~arcmin and $Y_{\rm tot} =0.007$~arcmin$^{2}$ and $\theta_{\rm p} = 8$~arcmin respectively. I generate 10 simulations for each of these clusters, each of which has a different noise realisation. I then analyse these simulations using the priors given in Table~\ref{t:ap_priors}, with delta priors on $a$, $b$ and $c$ centred on their `true' values (the ones used as inputs to the simulations), and plot the resulting posterior distributions using \textsc{GetDist}.

\begin{table}
\begin{center}
\begin{tabular}{{l}{c}{c}}
\hline
 &  \multicolumn{2}{c}{Simulation input} \\
\hline
Parameter & low SNR & high SNR \\
$x_{\rm c}$ &  \multicolumn{2}{c}{$0$~arcsec} \\
$y_{\rm c}$ &  \multicolumn{2}{c}{$0$~arcsec} \\
$Y_{\rm tot}$ & $0.001$~arcmin$^{2}$ & $0.007$~arcmin$^{2}$ \\
$\theta_{\rm p}$ & $2$~arcmin & $8$~arcmin \\
$a$ & \multicolumn{2}{c}{$1.051$} \\
$b$ & \multicolumn{2}{c}{$5.4905$} \\
$c$ & \multicolumn{2}{c}{$0.3081$} \\
$z$ & \multicolumn{2}{c}{$0.17$} \\
\hline
\end{tabular}
\caption{Cluster simulation inputs using an observational model and the `universal' GNFW shape parameters calculated in \citet{2010A&A...517A..92A}. Although $z$ isn't an input parameter for observational models, it is still required to generate simulations of clusters.}
\label{t:ap_obs_sim}
\end{center}
\end{table}

\subsubsection{Low SNR simulation analyses}

\begin{figure*}
  \begin{center}
    \begin{tabular}{@{}c@{}}
     \includegraphics[width=\textwidth, height = 7.5cm, keepaspectratio]{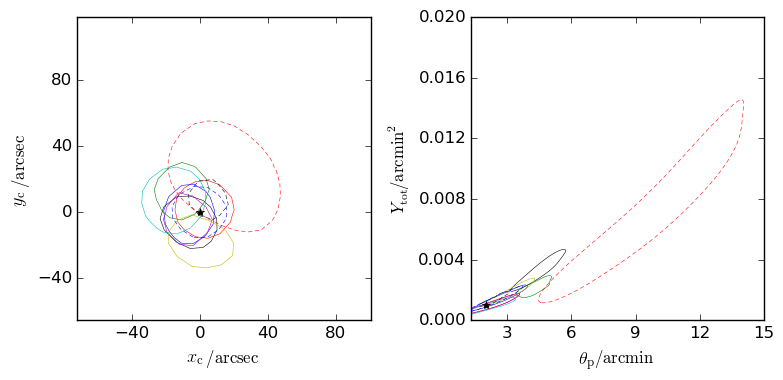} \\
     \includegraphics[width=\textwidth, height = 7.5cm, keepaspectratio]{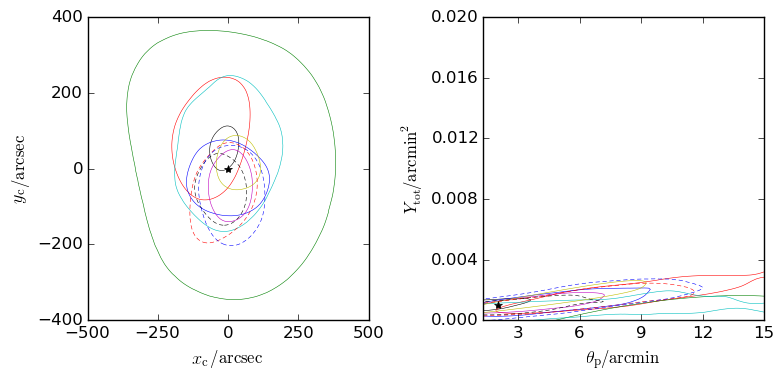} \\
     \includegraphics[width=\textwidth, height = 7.5cm, keepaspectratio]{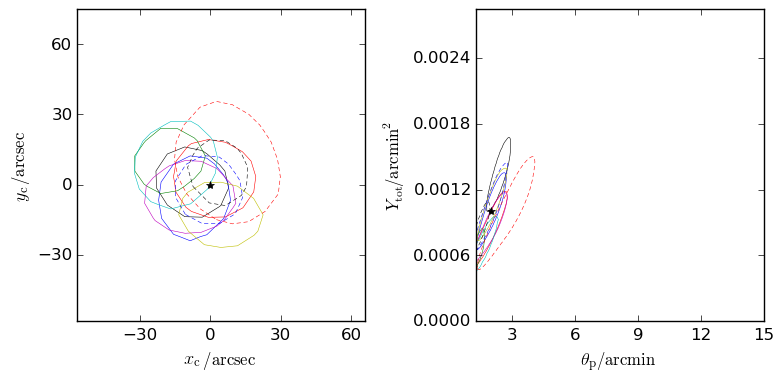} \\
    \end{tabular}
\caption{Two-dimensional marginalised $x_{\rm c}-y_{\rm c}$ and $Y_{\rm tot}-\theta_{\rm p}$ posterior distributions for the 10 OM III low SNR cluster simulations obtained from: AMI data (top row), Planck data (middle row), and AMI and Planck data combined (bottom row). The contours in each plot represent the $68\%$ confidence intervals of the separate posterior distributions obtained from each of the 10 simulations. The star symbols indicate the values input when generating the simulations.}
\label{f:ap_obs_lowsnr_sims}
  \end{center}
\end{figure*}

Figure~\ref{f:ap_obs_lowsnr_sims} shows the two-dimensional marginalised $x_{\rm c} - y_{\rm c}$ and $Y_{\rm tot}- \theta_{\rm p}$ posterior distributions of the joint, AMI-only, and Planck-only analyses of the low SNR cluster. Note that each plot contains the posteriors of the 10 different simulations, each of which is represented by a contour (68\% confidence interval). Looking at the AMI data only analyses, in two of the simulations the correct values for $x_{\rm c}$ and $y_{\rm c}$ are not recovered within a 68\% confidence interval. The plot shows that the constraints in $Y_{\rm tot}-\theta_{\rm p}$ are generally tight, but three contours do not encompass the input value. \\
The Planck-only analyses generally recover the correct values for $x_{\rm c}$ and $y_{\rm c}$ but the contours are much wider. There is a large degeneracy in $\theta_{\rm p}$, suggesting that in this case Planck cannot constrain the geometric size of the clusters very well. \\
The joint analysis shows similar results to the AMI-only analyses for the $x_{\rm c}-y_{\rm c}$ distributions, but the constraints on $Y_{\rm tot}$ and $\theta_{\rm p}$ are very tight (sharper distributions than in the case of AMI-only), which suggests that even though the Planck data in isolation was degenerate, when combined with AMI it can help infer the correct size of a cluster.

\subsubsection{High SNR simulation analyses}

\begin{figure*}
  \begin{center}
    \begin{tabular}{@{}c@{}}
     \includegraphics[width=\textwidth, height = 7.5cm, keepaspectratio]{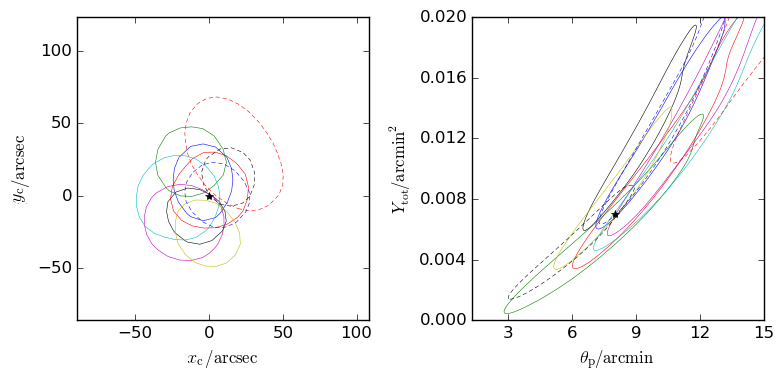} \\
     \includegraphics[width=\textwidth, height = 7.5cm, keepaspectratio]{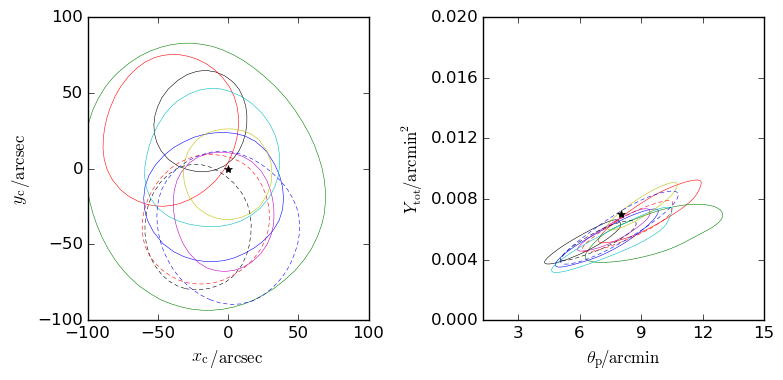} \\
     \includegraphics[width=\textwidth, height = 7.5cm, keepaspectratio]{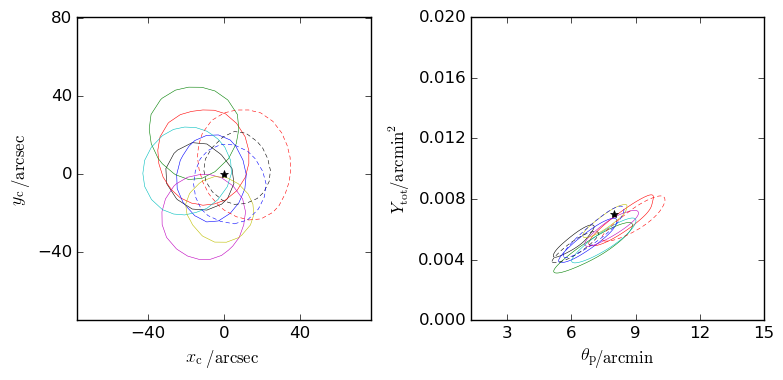} \\
    \end{tabular}
\caption{Two-dimensional marginalised $x_{\rm c}-y_{\rm c}$ and $Y_{\rm tot}-\theta_{\rm p}$ posterior distributions for the 10 OM III high SNR cluster simulations. The Figure layout is as described in Figure~\ref{f:ap_obs_lowsnr_sims}.}
\label{f:ap_obs_highsnr_sims}
  \end{center}
\end{figure*}

Figure~\ref{f:ap_obs_highsnr_sims} shows the contours of the posterior distributions obtained from the high SNR simulations.\\ 
The AMI-only $x_{\rm c}-y_{\rm c}$ posterior contours are similar to the low SNR case, but are generally more offset from the correct value in this instance. The $Y_{\rm tot}-\theta_{\rm p}$ posteriors show large degeneracies along the line of changing $Y_{\rm tot}$ and $\theta_{\rm p}$ (i.e. a large positive covariance between the two parameters). 
The Planck-only data results show tighter constraints on $x_{\rm c}$ and $y_{\rm c}$ relative to the low SNR simulations, but still wider than the other two analysis methods. The $Y_{\rm tot}-\theta_{\rm p}$ posteriors show that Planck arguably does a better job than AMI in recovering the true values, as the contours are generally tighter in the former case, and both analyses give a similar number of distributions where the correct value lies in the proximity of the contours. \\
The joint analysis shows that the cluster offset inferences are driven almost entirely by the AMI data, as they strongly resemble the results of the AMI runs. In contrast the $Y_{\rm tot}-\theta_{\rm p}$ posteriors suggest Planck data is dominating the inferences, and that the joint data distributions provide the tightest constraints on $Y_{\rm tot}-\theta_{\rm p}$ estimates. However, five of these distributions fail to recover the true values within their 68\% confidence intervals. 

\subsubsection{Variable shape parameter analysis}
\label{s:ap_obs_highsnrab_sims}
\begin{sidewaysfigure}

    \begin{tabular}{c}
     \includegraphics[width = 23cm, height = 7.5cm, keepaspectratio]{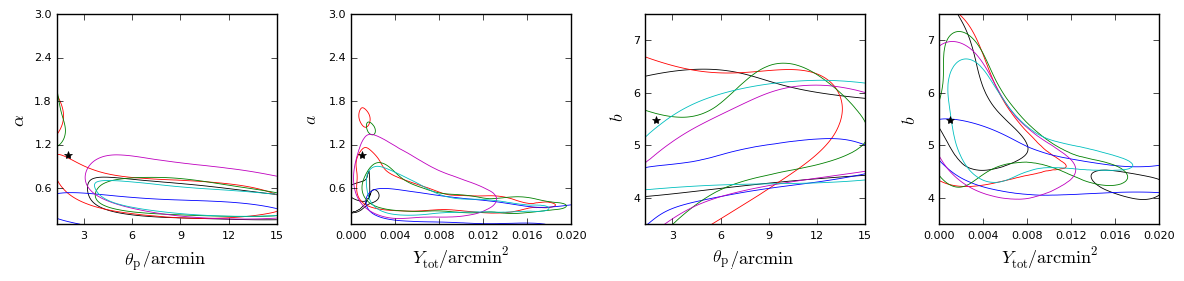} \\
     \includegraphics[width = 23cm, height = 7.5cm, keepaspectratio]{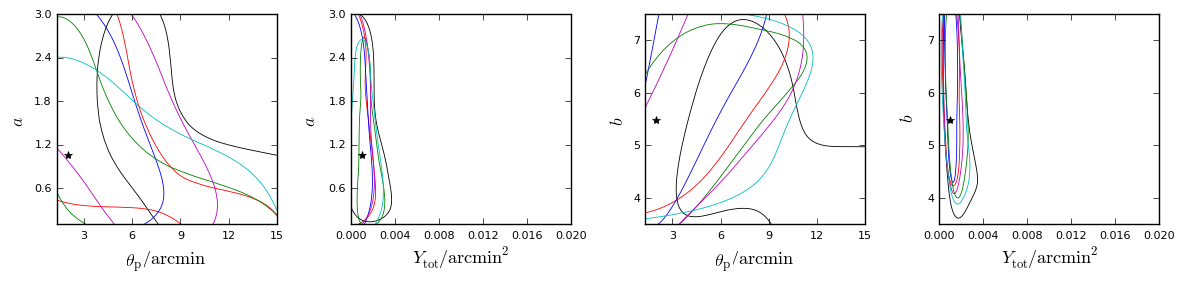} \\
     \includegraphics[width = 23cm, height = 7.5cm, keepaspectratio]{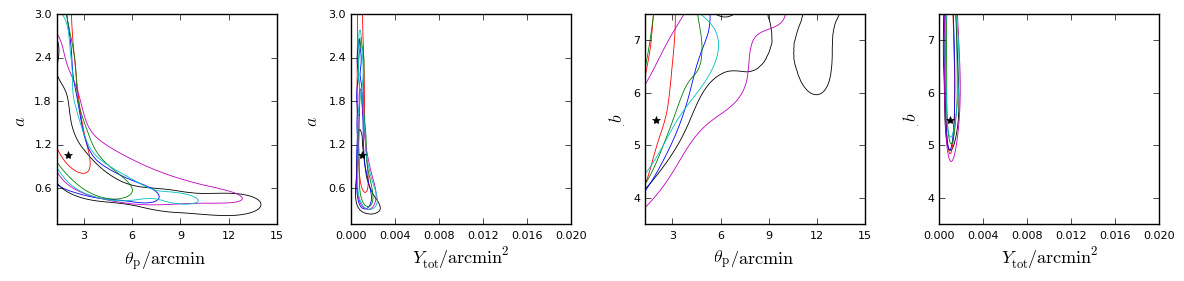} \\ 
     
    \end{tabular}
 \caption{Two-dimensional marginalised posterior distributions for six OM III low SNR cluster simulations obtained from: AMI data (top row), Planck data (middle row), and AMI and Planck data combined (bottom row). In the Bayesian analysis of the data, the values of the GNFW shape parameters $a$ and $b$ were allowed to vary (had uniform priors). The black stars indicate the values input when generating the simulations.}
 \label{f:ap_obs_lowsnrab_sims}
\end{sidewaysfigure}

I next consider the same simulations described in Section~\ref{s:ap_obs_sim}, but allowing the GNFW shape parameters $a$ and $b$ to vary in the \textit{analysis}. I thus assign the uniform priors stated in Table~\ref{t:ap_obs_sim} to $a$ and $b$. I note that throughout the analysis I found that the cluster model used to analyse the data did not affect the posterior constraints on $x_{\rm c}$ and $y_{\rm c}$, and so I do not discuss them in the subsequent analyses. \\
Figure~\ref{f:ap_obs_lowsnrab_sims} shows two-dimensional posterior distributions of pairs of the parameters: $Y_{\rm tot}$, $\theta_{\rm p}$, $a$, and $b$, resultant from six low SNR simulations. The $\theta_{\rm p}$ -- $a$ posteriors show that the AMI-only and Planck-only analyses fail to produce good constraints, as the former has a large degeneracy in $\theta_{\rm p}$ which misses the simulation input while the latter is almost completely uninformative (resembles the prior). The joint analysis leads to results that encompass the true value within the 68\% contour, albeit with large degeneracies in $a$ at low $\theta_{\rm p}$ (where the true $\theta_{\rm p}$ value lies) and in $\theta_{\rm p}$ at low $a$. The $Y_{\rm tot}$ -- $a$ posterior plots shows similar results for AMI, Planck recovers $Y_{\rm tot}$ well but has a large degeneracy in $a$. The joint analysis gives similar results to Planck-only, but with a tighter constraint on $Y_{\rm tot}$ (as was the case in the fixed $a$ and $b$ low SNR analyses). Posteriors in the $Y_{\rm tot}$ -- $b$ plane show similar results, but in this case the joint analysis produces contours which are less degenerate in $b$ than the Planck-only results. The $\theta_{\rm p}$ -- $b$ plots show that all three analyses fail to produce informative (well constrained) posteriors. \\ While the joint analysis tends to show degeneracy in $a$ and $b$, it does produce superior constraints on $Y_{\rm tot}$ and $\theta_{\rm p}$ relative to the single data analyses for marginalised posteriors considered here.

\begin{sidewaysfigure} 
  \begin{center}
    \begin{tabular}{c}
     \includegraphics[width = 23cm, keepaspectratio]{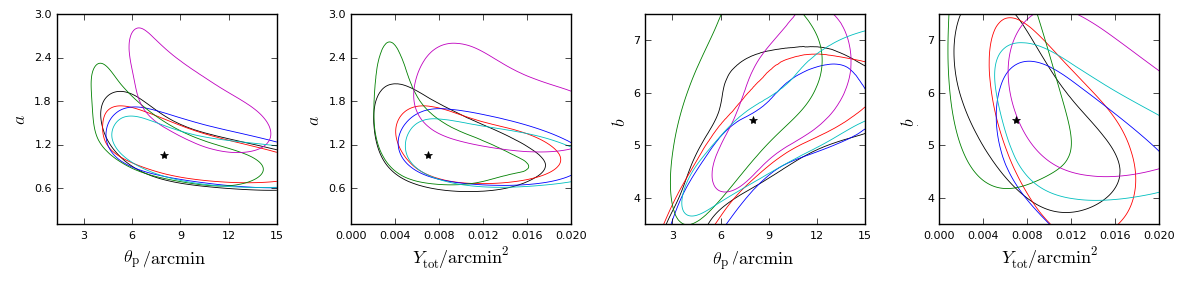} \\
     \includegraphics[width = 23cm, keepaspectratio]{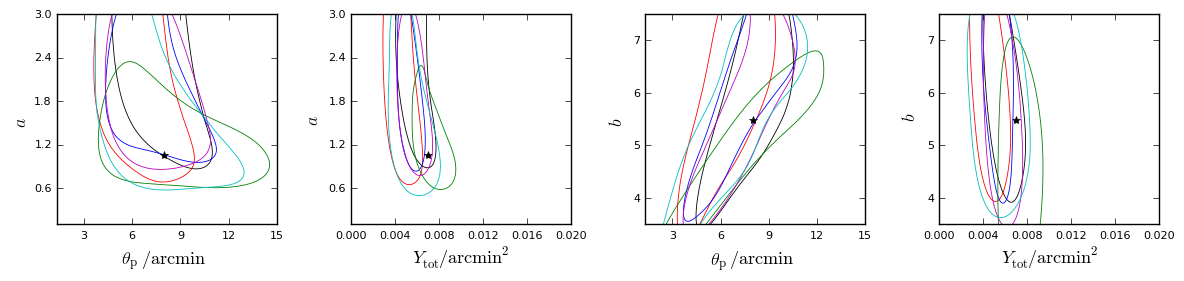} \\
     \includegraphics[width = 23cm, keepaspectratio]{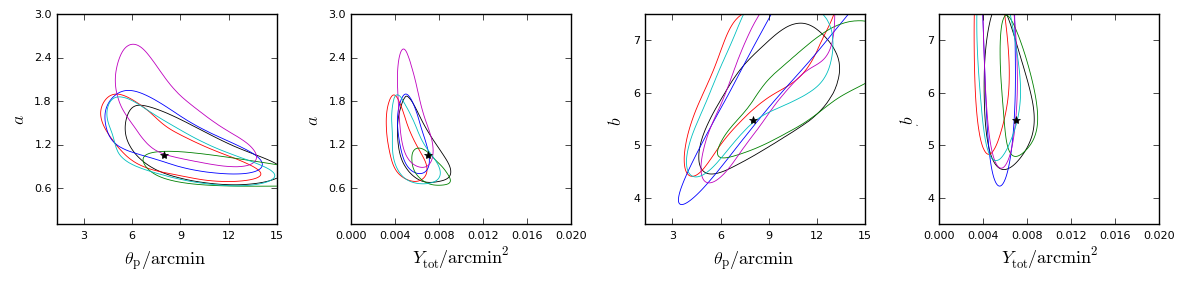} \\
     
    \end{tabular}
\caption{Two-dimensional marginalised posterior distributions for six OM III high SNR cluster simulations where $a$ and $b$ were allowed to vary in the analysis. The plot layout is as described in Figure~\ref{f:ap_obs_lowsnrab_sims}.}
\label{f:ap_obs_highsnrab_sims}
  \end{center}
\end{sidewaysfigure}

%
%
Figure~\ref{f:ap_obs_highsnrab_sims} shows the two-dimensional posterior distributions for the high SNR simulations, in which case the AMI posterior distributions recover $a$ relatively well (with the one clear exception). The AMI posteriors for $b$ are quite wide but generally peak around the input value of $b$. \\
The Planck-only distributions also show some  improvement over the low SNR case. \\
The joint analysis gives slightly worse results for $a$ than the AMI-only case (though the exceptionally bad AMI distribution improves), while the posteriors for $b$ arguably improve in the joint case for five of the six simulations.


     
\subsection{Cluster simulations using physical models}
\label{s:ap_phys_sim}

I repeat the simulation procedure described in Section~\ref{s:ap_obs_sim}, but this time using PM I and PM II in the cluster simulation and analysis. Table~\ref{t:ap_phys_sim} shows the input parameters used for PM simulations; the low SNR simulations have $M(r_{200})$ = $5\times 10^{14}~M_{\mathrm{Sun}}$ while the high SNR use $10\times 10^{14}~M_{\mathrm{Sun}}$.

\begin{table*}
\begin{center}
\begin{tabular}{{l}{c}{c}{c}{c}}
\hline
 &  \multicolumn{4}{c}{Simulation input} \\
\hline
Parameter & PM I low SNR & PM I high SNR & PM II low SNR & PM II high SNR \\
$x_{\rm c}$ &  \multicolumn{4}{c}{$0$~arcsec} \\
$y_{\rm c}$ &  \multicolumn{4}{c}{$0$~arcsec} \\
$M(r_{200})$ & $5\times 10^{14}~M_{\mathrm{Sun}}$ & $10\times 10^{14}~M_{\mathrm{Sun}}$ & $5\times 10^{14}~M_{\mathrm{Sun}}$ & $10\times 10^{14}~M_{\mathrm{Sun}}$ \\
$f_{\rm gas}(r_{200})$ & \multicolumn{4}{c}{$0.12$} \\
$\alpha_{\rm Ein}$ & \multicolumn{2}{c}{--} & \multicolumn{2}{c}{$0.2$} \\
$a$ & \multicolumn{4}{c}{$1.051$} \\
$b$ & \multicolumn{4}{c}{$5.4905$} \\
$c$ & \multicolumn{4}{c}{$0.3081$} \\
$z$ & \multicolumn{4}{c}{$0.17$} \\
\hline
\end{tabular}
\caption{Cluster simulation inputs for PM I and PM II. The cluster centre, GNFW shape parameters, and redshift inputs are the same for all four models. The Einasto shape parameter is only an input for PM II.}
\label{t:ap_phys_sim}
\end{center}
\end{table*}

\subsubsection{PM I low SNR posteriors}
\begin{figure*} 
  \begin{center}
    \begin{tabular}{c}
     \includegraphics[width = \textwidth]{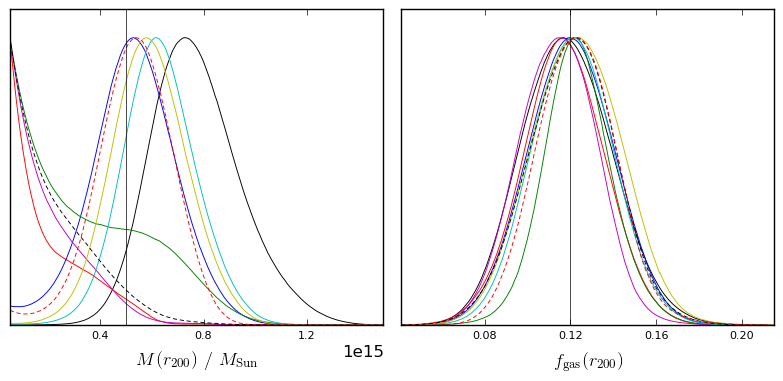} \\
     \includegraphics[width = \textwidth]{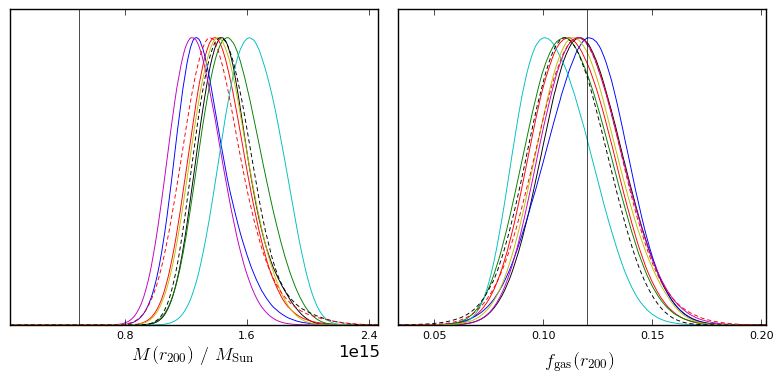} \\
     \includegraphics[width = \textwidth]{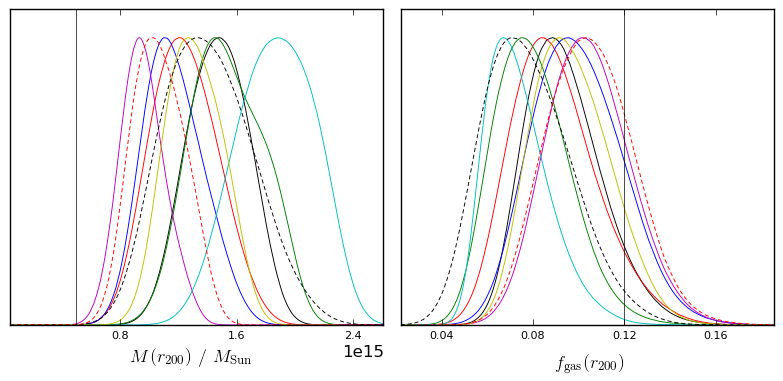} \\
    
    \end{tabular}
\caption{One-dimensional marginalised posterior distributions for the 10 PM I low SNR cluster simulations obtained from: AMI data (top row), Planck data (middle row), and AMI and Planck data combined (bottom row). The black vertical lines indicate the values input when generating the simulations.}
\label{f:ap_phys_lowsnr_sims}
  \end{center}
\end{figure*}

The one-dimensional posterior distributions for $x_{\rm c}$, $y_{\rm c}$, $M(r_{200})$, and $f_{\rm gas}(r_{200})$ for the 10 low SNR simulations are shown in Figure~\ref{f:ap_phys_lowsnr_sims}. 
Four of the AMI mass posteriors replicate the shape of the prior distribution (which has a $1/M(r_{200})$ dependence in linear space), indicating that the likelihood is negligible for these analyses.
$f_{\rm gas}(r_{200})$ is recovered very well by AMI for all ten simulations (and also takes the same shape as the prior). \\
In the case of the Planck mass estimates, the modes of the posteriors overestimate the input value by a factor of at least two. The same statistic slightly underestimates $f_{\rm gas}(r_{200})$ in some cases, but not to the same degree as the $M(r_{200})$ values are overestimated. \\
The combined data also overestimates $M(r_{200})$, with the modes ranging between $\approx 1.75$ -- $4$ times the true values. What is also striking is the values of the modes of the $f_{\rm gas}(r_{200})$ posteriors, which in some cases (which correspond to the larger mass estimates) occur around $f_{\rm gas}(r_{200}) \approx 0.8$. The overestimation of mass and underestimation of $f_{\rm gas}(r_{200})$ suggests that in the joint analysis, it is the composition of the clusters which have been incorrectly inferred, whilst in the Planck-only case it appears that the physical size of the clusters is overestimated.  

    

\subsubsection{PM I high SNR posteriors}

\begin{figure*} 
  \begin{center}
    \begin{tabular}{c}
     \includegraphics[width = \textwidth]{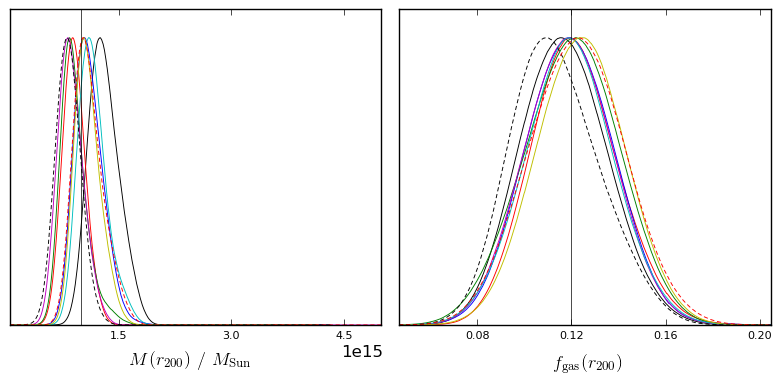} \\
     \includegraphics[width = \textwidth]{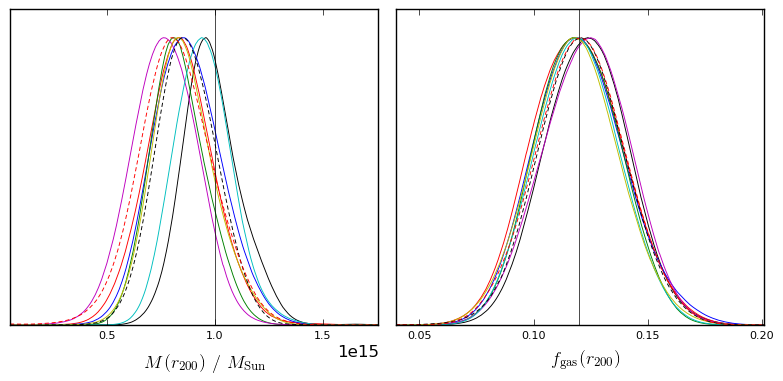} \\
     \includegraphics[width = \textwidth]{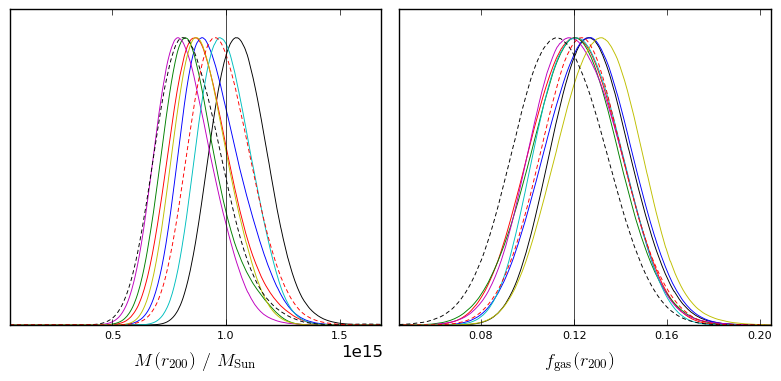} \\
     
    \end{tabular}

\caption{One-dimensional marginalised posterior distributions for the 10 PM I high SNR cluster simulations. The plots are laid out as described in Figure~\ref{f:ap_phys_lowsnr_sims}.}
\label{f:ap_phys_highsnr_sims}
  \end{center}
\end{figure*}

For the high SNR cluster simulations (Figure~\ref{f:ap_phys_highsnr_sims}) the AMI mass estimates on average peak on the true mass value. The Planck mass modal values generally underestimate the input mass, which is in stark contrast to the low SNR case where they massively overestimated it. The Planck estimates of $f_{\rm gas}(r_{200})$ are extremely accurate, which again suggests that it is the size rather than the composition of the cluster that Planck has difficulty with. The joint estimates perform similarly well to the separate analyses. 

     

\subsubsection{PM II cluster simulations}

\begin{figure*} 
  \begin{center}
    \begin{tabular}{c}
     \includegraphics[width = \textwidth]{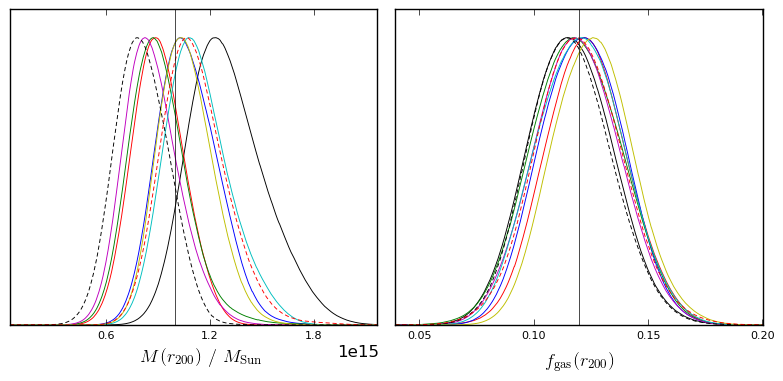} \\
     \includegraphics[width = \textwidth]{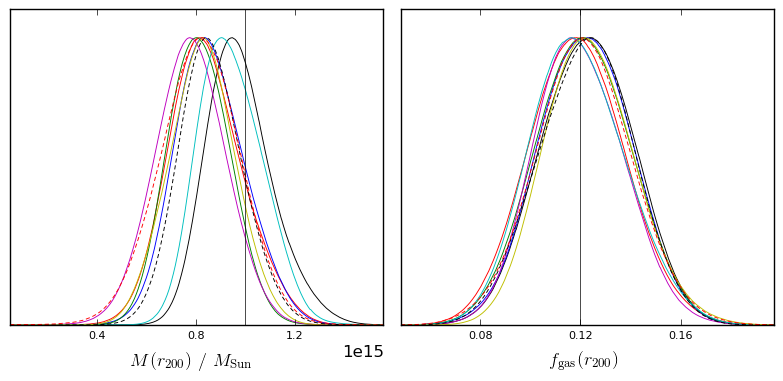} \\
     \includegraphics[width = \textwidth]{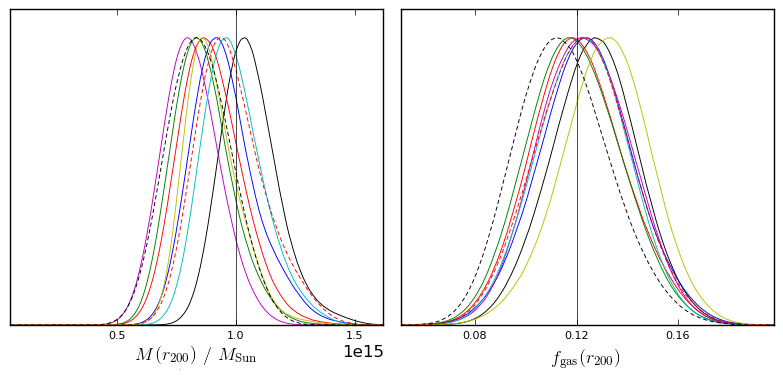} \\
     
    \end{tabular}

\caption{One-dimensional marginalised posterior distributions for the 10 PM II high SNR cluster simulations obtained by marginalising over $\alpha_{\rm Ein}$. The plots are laid out as described in Figure~\ref{f:ap_phys_lowsnr_sims}.}
\label{f:ap_phys2_highsnr_sims}
  \end{center}
\end{figure*}

Cluster simulations were generated with the PM II setting $\alpha_{\rm Ein} = 0.2$. Note this value for $\alpha_{\rm Ein}$ corresponds to a profile similar to that given by the NFW profile (as discussed in \citealt{2014MNRAS.441.3359D} and Chapter~\ref{c:fifth}). The clusters were analysed with a uniform prior on $\alpha_{\rm Ein}$ (given in Table~\ref{t:ap_priors}), but \textsc{GetDist} failed to plot distributions from the resultant posterior samples. This suggests that the marginalised $\alpha_{\rm Ein}$ posterior distributions are not `compatible' with Gaussian kernel density estimation techniques used in the program. Nevertheless \textsc{GetDist} still produced posterior distributions of other parameters (by marginalising over $\alpha_{\rm Ein}$), and gave results similar to the PM I simulations. The posteriors obtained from analysis of the high SNR PM II clusters are shown in Figure~\ref{f:ap_phys2_highsnr_sims}. Likewise Bayesian analysis of the cluster simulations with a delta prior on $\alpha_{\rm Ein}$ resulted in posterior distributions similar to those obtained from PM I simulations and analysis.

Note that the overestimation of cluster parameters has been resolved in \citet{2019MNRAS.486.2116P} by understanding the Planck simulations better (and correcting a couple of associated bugs), but the paper focuses on observational models rather than physical.

     


\section{Application of joint analysis to real cluster data}

\subsection{Mass estimates of cluster PSZ2G063.80+11.42}
\begin{figure}
  \begin{center}
  \includegraphics[ width=0.90\linewidth]{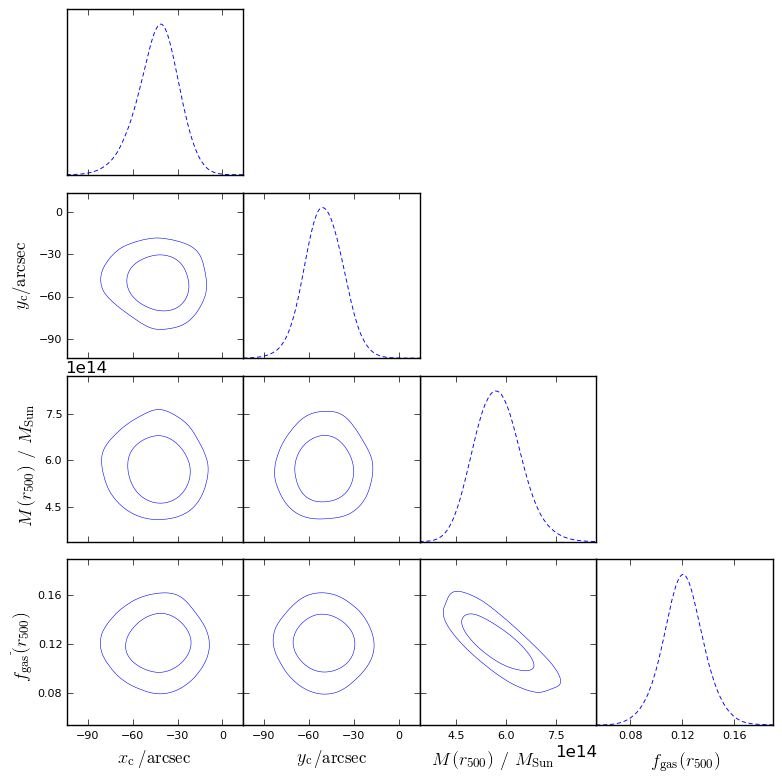}
  \caption{Marginalised posterior distributions obtained from joint AMI-Planck analysis of cluster PSZ2G063.80+11.42, using PM I. The dashed line plots are fully marginalised posterior distributions, while the contour plots are two-dimensional marginalised distributions. The inner contours correspond to the region of $68\%$ confidence, while the outer contours corresponds to $95\%$.  }
\label{f:ap_phys_real}
  \end{center}
\end{figure}

I apply the joint analysis to a cluster featured in PSZ2 (PSZ2G063.80+11.42) and the 54 cluster sample analysed in Chapters~\ref{c:third} and~\ref{c:fourth}. Note that in these Chapters slightly different values for $a$, $b$, and $c$ were used, which were derived in \citet{2010A&A...517A..92A} for the standard self-similar case (Appendix~B of Arnaud et al.). It was shown in \citet{2013MNRAS.430.1344O} that PM I is not affected by which of these two sets of parameters is used. 
In Chapter~\ref{c:third} I calculated the AMI mass estimate to be $M_{\rm AMI}(r_{500}) = (3.37 \pm 0.76) \times 10 ^{14}~M_{\rm Sun}$ and the PwS mass estimate (using the slicing function methodology introduced in PSZ2 and detailed in Section~\ref{s:psz2_m}) to be $M_{\rm Pl, \, slice}(r_{500}) = (6.41 \pm _{0.58}^{0.57} ) \times 10 ^{14}~M_{\rm Sun}$. I chose to run the joint analysis on this cluster due to the fact that its AMI and Planck masses were quite discrepant, despite the AMI radio-source environment not appearing to be problematic on the map of the observation. The cluster redshift is taken from PSZ2 as $z = 0.426$, and the coordinates of the Planck patch centre are within $0.01$~arcmin of the AMI SA pointing centre of the observation. \\
I run the joint analysis with PM I using the priors given in Table~\ref{t:ap_priors} (assigning delta priors to the GNFW shape parameters). The marginalised posterior distribution for $M_{\rm Joint, \, PM I}(r_{500})$ (Figure~\ref{f:ap_phys_real}) gives a mean estimate of $M_{\rm Joint, \, PM I}(r_{500}) = (5.74 \pm 0.70) \times 10^{14} M_{\rm Sun}$. Hence the joint analysis gives a value within one combined standard deviation of the value obtained from Planck data using the PSZ2 slicing function methodology, and within three combined standard deviations of the value obtained from AMI data alone. \\
For further comparison I run the Planck-only analysis for the same cluster using the same model, and find that $M_{\rm Pl, \, PM I}(r_{500}) = (6.98 \pm 1.02) \times 10^{14} M_{\rm Sun}$. For clarity I note that $M_{\rm Pl, \, slice}(r_{500})$ and $M_{\rm Pl, \, PM I}(r_{500})$ are obtained from the same data using the same PwS algorithm, but the former uses the scaling relations and slicing function methodology to obtain a mass estimate, whereas the latter uses PM I in the Bayesian analysis to directly infer mass posterior distributions. 

In Chapter~\ref{c:third} I found that PSZ2 mass estimates were generally higher than those obtained by AMI. In this Chapter the low SNR PM I simulations show similar results, as Planck data analyses gives large overestimates of the true values, whereas AMI underestimates it on average. The real data analysed here suggest the same -- although we do not know the `true' mass value in this case. The fact that both estimates from Planck data are relatively high suggests the data themselves are causing this, not the model being applied. I note however that this is based on just one real cluster, and that the Planck-only analysis of high SNR simulations did not produce mass overestimates.  

\subsection{Variable $a$ and $b$ analysis}
\begin{sidewaysfigure} 
\refstepcounter{figure}
  \begin{center}
    \begin{tabular}{c}
     \includegraphics[width = 23cm, keepaspectratio]{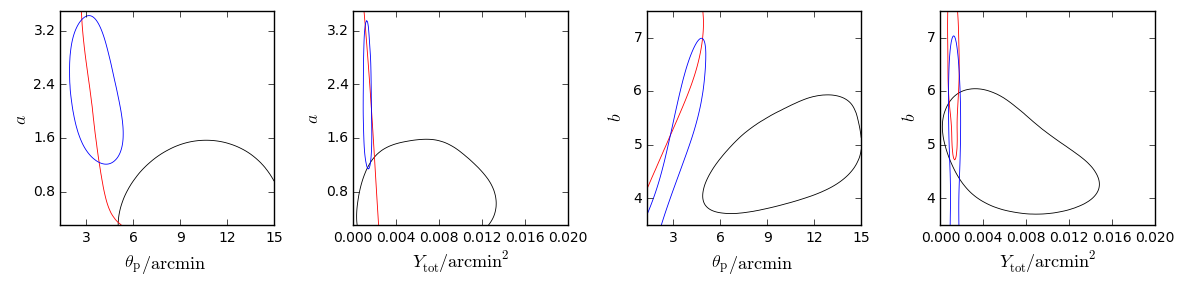} \\
    \end{tabular}
\caption{Two-dimensional posterior distributions for parameter pairs: $(a, \theta_{\rm p})$, $(a, Y_{\rm tot})$, $(b, \theta_{\rm p})$, and $(b, Y_{\rm tot})$. The black, red (in the first three plots, the region of higher probability is to the left of the curve), and blue contours correspond to the 68\% confidence intervals of the AMI, Planck, and AMI-Planck posterior distributions respectively.}
\label{f:ap_obs_ab_real}
  \end{center}
\end{sidewaysfigure}

For comparison with the results obtained from simulated data in Section~\ref{s:ap_obs_highsnrab_sims}, I analysed the PSZ2G063.80+11.42 data using OM III while allowing $a$ and $b$ to vary. Figure~\ref{f:ap_obs_ab_real} shows the resulting posterior distributions for the three analysis methods. As was the case in the simulations, the joint analysis gives a tighter constraint on the $Y_{\rm tot}$ and $\theta_{\rm p}$ parameters, but does show a degeneracy in $a$ and $b$. 

\section{Conclusions}
\label{s:ap_conc}

I have introduced a joint likelihood function for data obtained from Planck and AMI in order to compare inferences obtained using it with those from the individual likelihood functions. The Bayesian analysis of Planck data was carried out using PowellSnakes (PwS, \citealt{2012MNRAS.427.1384C}) and AMI data were analysed in a way similar to the method outlined in \citet{2009MNRAS.398.2049F} (and used in the preceeding Chapters); the joint analysis ran both of these simultaneously. \\
I tried implementing the likelihood hyperparameter method introduced in \citet{2000MNRAS.315L..45L} and \citet{2002MNRAS.335..377H}. I showed that likelihood ratios cannot be used with the hyperparameter method by implementing the toy model considered in \citet{2002MNRAS.335..377H}. Therefore since PwS evaluates a likelihood ratio it is not compatible with this method.  \\
I generated simulations of clusters using an observational model (OM III, similar to the ones used in Chapter~\ref{c:fourth}) for 10 different noise realisations, and analysed the data using the same model. From looking at the resulting posterior distributions I found the following.
\begin{itemize}
\item For low signal-to-noise ratio (SNR) clusters, AMI data alone could be used to constrain values for the integrated Comptonisation parameter $Y$ and angular radius $\theta$ rather well, but Planck data showed large degeneracies in $\theta$. The joint analysis however showed the tightest constraints in $Y-\theta$ space (generally centred around the simulation input values).
\item For high SNR clusters, the Planck-only analyses gave moderate constraints on $\theta$ and good results for $Y$, while the AMI-only analyses showed large covariance between $Y$ and $\theta$. The joint analysis results gave similar results to the former, but with tighter constraints. For all three analyses, the true value was often in the proximity of the $68\%$ confidence interval contours, rather than close to their centres.   
\item When allowing the shape parameters $a$ and $b$ of the generalised NFW model (\citealt{2007ApJ...668....1N}, used to parameterise the electron pressure) to vary in the Bayesian analysis, it was found that the joint analysis could generally constrain the $Y$ and $\theta$ parameters better than the individual analyses, but showed degeneracies in $a$ and $b$.
\end{itemize}
Using physical models derived in Chapters~\ref{c:second} and~\ref{c:fifth} I generated cluster simulations and analysed them with the three likelihood functions to infer cluster mass estimates. From this I found the following.
\begin{itemize}
\item For low SNR clusters I found that AMI underestimated cluster masses on average, but did recover the true value for some noise realisations. Planck systematically overestimated the masses by factors of at least two, while the joint analysis also led to overestimations (but generally to a smaller extent), suggesting it was the Planck likelihood dominating the joint posterior inferences. 
\item The gas fraction estimates from the joint analysis for low SNR clusters are consistently lower than the simulation input values, which suggests that the joint analysis is struggling to correctly infer the composition of the cluster, which is probably the cause of the mass overestimates.
\item Analysis of high SNR clusters with AMI data gave accurate estimates of the input mass, while Planck data led to slight underestimates. Application of the joint analysis gave results similar to the individual analyses.
\end{itemize}
Finally, I applied the joint analysis to real data for the cluster PSZ2G063.80+11.42 which is part of the sample of 54 clusters considered in Chapters~\ref{c:third} and~\ref{c:fourth}. I compared the mass estimates obtained with those obtained from AMI and Planck data and found the following.
\begin{itemize}
\item The AMI estimates and joint analysis mass estimates are $M_{\rm AMI}(r_{500}) = (3.37 \pm 0.76) \times 10 ^{14}~M_{\rm Sun}$ (obtained in Chapter~\ref{c:fourth}) and $M_{\rm Joint}(r_{500}) = (5.74 \pm 0.70) \times 10^{14} M_{\rm Sun}$ respectively. The two estimates derived from Planck data are $M_{\rm Pl, \, PM I}(r_{500}) = (6.98 \pm 1.02) \times 10^{14} M_{\rm Sun}$ and $M_{\rm Pl, \, slice}(r_{500}) = (6.41 \pm _{0.58}^{0.57} ) \times 10 ^{14}~M_{\rm Sun}$. The former of these was inferred directly from the PM I posterior distributions. The latter was obtained from the slicing function method introduced in \citet{2016A&A...594A..27P} and detailed in Section~\ref{s:psz2_m}.
\item The joint analysis estimate is sandwiched in between the other three values, but is closer to $M_{\rm Pl, \, PM I}(r_{500})$ than it is $M_{\rm AMI}(r_{500})$, suggesting that the Planck likelihood has a large effect on the joint analysis posterior distribution.
\item The fact that both Planck data-only mass estimates are higher than the AMI value suggests that it is the data which are causing the relatively high estimates, at least for the real example considered here.
\item When allowing the GNFW shape parameters $a$ and $b$ to vary, the joint analysis generally provides much tighter parameter constraints than the individual analyses.
\end{itemize}  


%% file: CHAP-9/chapter9.tex
\chapter{Monte Carlo sampling methods}\label{c:ninth}

For most astrophysical problems, calculating the Bayesian evidence numerically is unfeasible, especially for high dimensional problems. Likewise, attempting to calculate parameter probability distributions exactly is computationally impossible. Thus one usually resorts to statistical sampling to make estimates of these quantities. \\
Monte Carlo sampling methods are a broad class of computational algorithms that rely on repeated random sampling of some distribution to obtain a numerical approximation of the true results. In the context of Bayesian inference, this amounts to representing a posterior distribution via a set of $n_{\rm s}$ weighted samples
\begin{equation}
\mathcal{S} = \{(\vec{\Theta}_1,\mathcal{P}_1),~..., (\vec{\Theta}_{n_{\rm s}},\mathcal{P}_{n_{\rm s}})\}, 
\end{equation}
where $\mathcal{P}_i$ is the weight of each sample and $\sum_{i = 1}^{n_{\rm s}} \mathcal{P}_i = 1$. In this Chapter I give a brief review of how these samples can be obtained and used to plot approximations of the true posterior distribution. It serves as a reference to astrophysicists who are new to sampling, and refers to methods which are well known in the field of statistics.

\section{Inverse transform sampling}

\begin{figure*}
  \begin{center}
    \begin{tabular}{@{}cc@{}}
  \includegraphics[ width=0.45\linewidth]{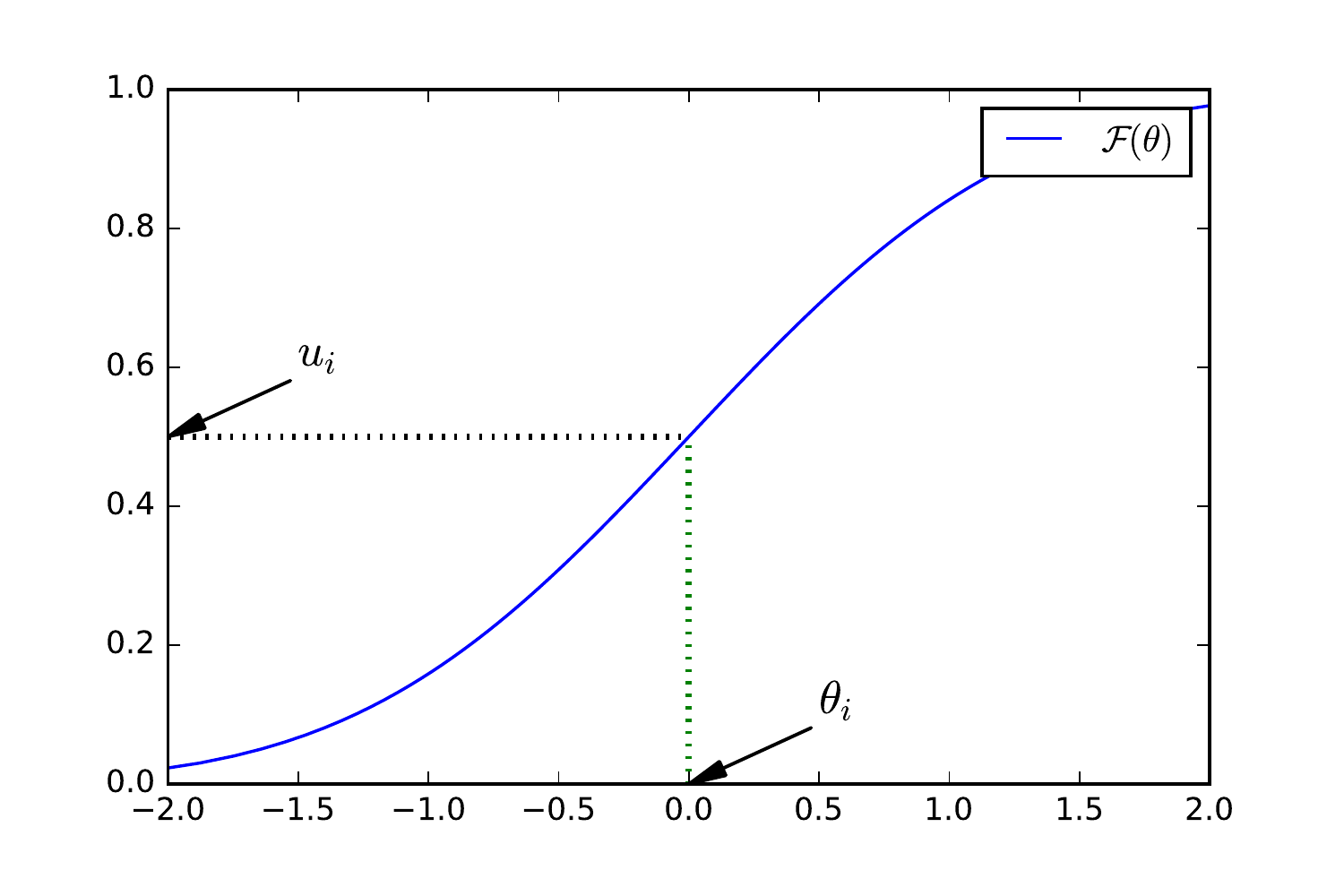} &
  \includegraphics[ width=0.45\linewidth]{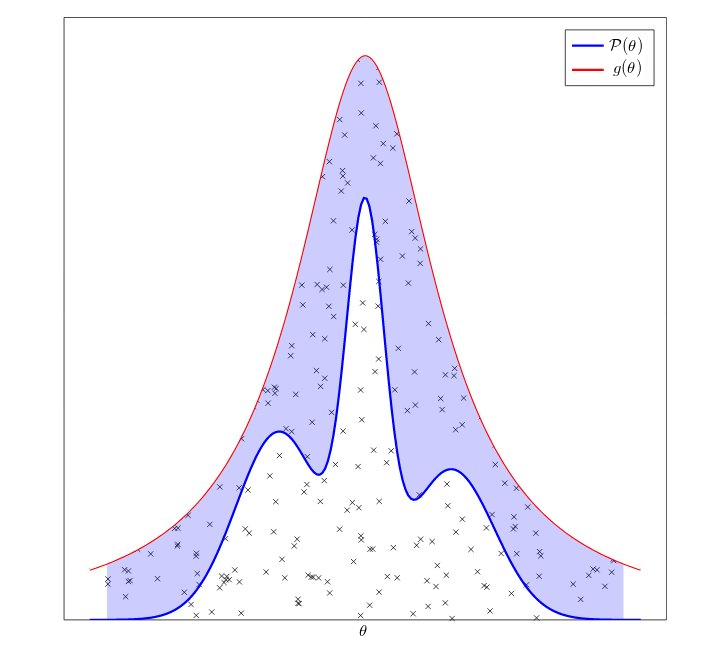}
    \end{tabular}
  \caption{Left: Illustrating inverse transform sampling for a one-dimensional distribution $\mathcal{P}(\theta)$. Once a value for $u_i$ is obtained, one draws a horizontal line from $(0, u)$ until it intersects with $\mathcal{F}(\theta)$ (black dotted line). The value of $\theta$ at the point of intersection is the point sampled from the distribution ($\theta_i$, green dotted line). Clearly the steeper $\mathcal{F}(\theta)$ is over an interval $\delta \theta$, the higher the chance of it intersecting with the horizontal lines corresponding to the uniform $[0,1]$ samples. Right: Illustrating rejection sampling, taken from \citet{handley}. The crosses correspond to samples from the distribution $g(\theta)$. Provided $g(\theta) > \mathcal{P}(\theta)$ for the domain of interest, then the samples beneath the blue curve (white area) can be regarded as samples from $\mathcal{P}(\theta)$. }
\label{f:mc_samp}
  \end{center}
\end{figure*}

Assuming we can draw independent, identically distributed random variables $u$ that are uniformly distributed on $[0,1]$, and provided we can calculate the inverse of the cumulative distribution function of the posterior $\mathcal{F}^{-1}$, then we can draw random samples from $\mathcal{P}(\vec{\Theta})$. We can interpret $u$ as being a probability, and thus by evaluating $\mathcal{F}^{-1}(u)$ we are finding the value of $\vec{\Theta}$ which satisfies
\begin{equation}
u = \mathcal{F}(\vec{\Theta}) = \int_{\vec{\Theta}_{m}}^{\vec{\Theta}} \mathcal{P}(\vec{\Theta'}) \mathrm{d}\vec{\Theta'},
\end{equation}  
where $\vec{\Theta}_{m}$ is the component-wise minimum value of $\vec{\Theta}$ over which $\mathcal{P}(\vec{\Theta})$ is defined. The steepness of $\mathcal{F}$ at a given point is proportional to the value of $\mathcal{P}(\vec{\Theta})$ and thus regions of higher probability density will be sampled from more often as shown in the left plot of Figure~\ref{f:mc_samp}. Consequently the weights of the samples are proportional to the number of times a value of $\vec{\Theta}$ is sampled. The difficulty in inverse transform sampling arises when $\mathcal{F}^{-1}$ is hard to evaluate.

\section{Rejection sampling}

Rejection sampling involves sampling from a proposal distribution $g(\theta)$ to ultimately draw samples from the distribution of interest $\mathcal{P}(\theta)$. The only requirement on $g(\theta)$ is that $g(\theta) > \mathcal{P}(\theta)$ for the domain of interest. The method works as follows.
\begin{enumerate}
\item Sample a value of $\theta$ ($\equiv \theta_g$) from $g(\theta)$ by using e.g. inverse transform sampling.
\item Sample a random variable $u_g$ uniformly from the range $[0, g(\theta_g)]$.
\item If $\mathcal{P}(\theta_g) > u_g$, \textit{accept} the point $\theta_g$ as a sample from $\mathcal{P}$ and \textit{reject} it otherwise. The sample weights are thus once again proportional to the number of times a value is sampled. 
\end{enumerate}
Rejection sampling is demonstrated graphically in the right plot of Figure~\ref{f:mc_samp}. Note that rejection sampling can be inefficient (reject a lot of samples) when $g$ and $\mathcal{P}$ are dissimilar. The similarity between the two can be quantified by some distribution distance metric such as the Earth Mover's Distance used in Chapter~\ref{c:fourth}.

\section{Markov Chain Monte Carlo sampling}
\label{s:mcmc}
Before talking about Markov Chain Monte Carlo (MCMC) sampling methods I give a primer on Markov chains and state some of their key properties relevant to MCMC.

\subsection{Markov chains}
\subsubsection{Types of Markov chain}
A Markov chain is a sequence of random variables for which the probability of outcomes for a particular element of the chain depends only on the state attained in the previous step of the chain. A Markov chain $X$ can be continuous in time i.e. $X \equiv X(t)$ for $t \geq 0$ or discrete, $X \equiv \{X_0,X_1,...,X_{n-1}\}$. In the case of the former we are saying that the chain can be measured at any time $t$, while for the latter we are saying $X$ can only be measured at discrete times defined by the index $n$. The possible values that $X$ can take (often referred to as the \textit{state space}, $\Theta$) can also be continuous or discrete. A continuous state space refers to one in which $X$ can take any of the (uncountably infinite) values defined on the space. A discrete state space can include a finite or a countably infinite number of states. \\
\subsubsection{Discrete time discrete state space Markov chains}
For the properties considered here we will consider discrete time discrete state space Markov chains only, but note that these ideas generalise to the continuous cases. For more information on continuous Markov chains we refer the reader to \citet{mackay2002}, \citet{robert_casella2004}, and \citet{johansen}.
A Markov chain $X$ with discrete time domain and discrete state space $\Theta$ can be stated mathematically as
\begin{equation}
\label{e:mc1}  
P(X_{n} = \theta_i | X_{n-1} = \theta_{j},...,X_{0} = \theta_k) = P(X_{n} = \theta_i | X_{n-1} = \theta_{j}) \equiv T_{n,ji},
\end{equation}
where the $\theta_{l} \in \Theta$.
$T_{n,ji}$ is the transition probability from $\theta_{j}$ to $\theta_i$ between steps $n-1$ and $n$. For a \textbf{homogeneous} Markov chain the transition probability between two states is independent of time, thus we can write $T_{n,ji} \equiv T_{ji}$. 
\subsubsection{State properties}
We will now 
focus on \textit{homogeneous} Markov chains and introduce some of their properties relevant to Monte Carlo sampling. \\
A state $\theta_i$ is said to be \textbf{accessible} from state $\theta_j$ (denoted $\theta_j \rightarrow \theta_i$) if
\begin{equation}
\label{e:mc2}  
\inf\{ n: P(X_{n} = \theta_i | X_{0} = \theta_{j}) > 0\} < \infty,
\end{equation}
or equivalently $\inf \{ n: T_{ji}^n > 0 \} < \infty$. $\inf$ refers to the infimum of the set (greatest lower bound of the set). If this condition is satisfied it means that there is a finite probability of moving from state $\theta_j$ to state $\theta_i$ after a finite number of steps $n$. The definition of \textbf{communication} follows from accessibility: two states
$\theta_{j}$ and $\theta_i$ are said to \textit{communicate} with each other ($\theta_j \leftrightarrow \theta_i$) if they are \textit{accessible} from one another
\begin{equation}
\label{e:mc3}
\theta_j \leftrightarrow \theta_i \quad \Leftrightarrow \quad \theta_j \rightarrow \theta_i \, \mathrm{and} \, \theta_i \rightarrow \theta_j.
\end{equation}
A Markov chain is said to be \textbf{irreducible} if all states \textit{communicate} with each other, that is $\theta_j \leftrightarrow \theta_i$ for all $\theta_i, \theta_j \in \Theta$. This is important in the context of MCMC as a chain with this property can explore the entire state space without being confined to some portion of it (which could be determined by the chain's initial state). The chain is said to be \textbf{strongly irreducible} if any state can be reached from any other state in a single step i.e. if $T_{ji} > 0$ for all $i$ and $j$.  \\
It is also important to consider the number of paths can take from a state $\theta_i$ before the chain returns to $\theta_i$, as this will tell us something about the presence of long-range correlation between the states of the chain. A state $\theta_i$ has period $d(\theta_i)$ which is given by
\begin{equation}
\label{e:mc4}
d(\theta_i) = \gcd \{ n \geq 1 : T_{i,i}^{n} > 0 \},
\end{equation}
where $\gcd$ denotes the greatest common denominator of the set. It can be shown that all states which communicate have the same period, hence for an \textit{irreducible} Markov chain all states have the same period. An irreducible Markov chain with $d(\theta_i) = 1$ is said to be \textbf{aperiodic}. This essentially means that the Markov chain can transition back into the same state that it was in at the previous step. In the context of MCMC this means that the same value can be sampled consecutively. \\
Another quantity relevant to MCMC is the number of times a state is visited, $n_{\theta_i}$, in the asymptotic limit $n \rightarrow \infty$. We define this as
\begin{equation}
\label{e:mc5}
n_{\theta_i} = \sum_{j = 0}^{\infty} \mathcal{I}(X_{j} - \theta_i),
\end{equation}
where $\mathcal{I}(Y)$ equals one for $Y = 0$ and zero otherwise. The introduction of $n_{\theta_i}$ allows us to introduce two more properties of Markov chains: \textbf{transience} and \textbf{recurrence}. A state is said to be \textit{transient} if
\begin{equation}
\label{e:mc6}
\mathbb{E}(n_{\theta_i}) < \infty,
\end{equation}
while it is said to be \textit{recurrent} if
\begin{equation}
\label{e:mc7}
\mathbb{E}(n_{\theta_i}) = \infty,
\end{equation}
where the expectations are taken in the asymptotic limit. In the case of irreducible chains, transience and recurrence are properties of the chain itself rather than its individual states, so we can say that for such a chain all states are either transient or they are all recurrent. If the Markov chain is recurrent then the samples from MCMC can take any value in $\Theta$ an infinite number of times.
Another notion of recurrence can be defined with respect to time rather than frequency of transitions to a state: if the `time' (number of steps) between a chain moving to state $\theta_i$ and revisiting the state, $\tau_{\theta_i, \theta_i}$ has a finite first moment, then the state is said to be \textbf{positive recurrent}. Note that positive recurrence is also a property of the whole Markov chain in the case that it is irreducible.
\subsubsection{Stationarity and reversibility of Markov chains}
A distribution $\mu$ defined on $\Theta$ is said be \textbf{stationary} if
\begin{equation}
\label{e:mc8}
\vec{\mu} \boldsymbol{\mathsf{T}} = \vec{\mu},
\end{equation} 
where $\vec{\mu}$ is a row vector of the values of $\mu(\theta_i) \equiv \mu_i$ for all $\theta_i \in \Theta$ and $\boldsymbol{\mathsf{T}}$ is a matrix of transition probabilities $T_{ij}$ for all valid $i$ and $j$. If at any step along the Markov chain its marginal distribution $P(X_i)$ is distributed according to its stationary distribution $\mu$, then it stays distributed according to $\mu$ since $\vec{\mu} \boldsymbol{\mathsf{T}}^{n} = \vec{\mu}$ for arbitrary $n$. \\
A \textit{stationary} stochastic process is said to be \textbf{reversible} if the statistics of the time-reversed version of the process match those obtained in the original. An alternative way of interpreting this is that the distribution of any collection of future states given the past states must match the conditional distribution of the past states given the future states. This means that we require
\begin{equation}
\label{e:mc9}
P(X_{0} = \theta_i | X_{-1} = \theta_{j}) = P(X_{0} = \theta_i | X_{1} = \theta_{j}).
\end{equation}
It can be shown that if a Markov chain satisfies the \textbf{detailed balance} relation given by
\begin{equation}
\label{e:mc10}
T_{ij} \mu_i = T_{ji} \mu_j,
\end{equation}
then the chain is \textit{reversible}. Note that satisfying equation~\ref{e:mc9} is a \textit{sufficient} condition for a Markov chain to converge to its stationary distribution ($\vec{\mu}$). The reversibility property can be shown by substituting equation~\ref{e:mc10} into~\ref{e:mc9}
\begin{equation}
\begin{split}
\label{e:mc11}
P(X_{0} = \theta_i | X_{-1} = \theta_{j}) & = T_{ji} \\
& = \frac{T_{ij}\mu_i}{\mu_j} \\
& = \frac{P(X_{1} = \theta_j| X_{0} = \theta_{i}) P(X_{0} = \theta_i)}{P(X_{1} = \theta_j)} \\
& = P(X_{0} = \theta_i | X_{1} = \theta_{j}).
\end{split}
\end{equation}
Note that the \textit{necessary} conditions for a Markov chain (with a discrete state space) to converge on the target distribution are for it to be irreducible, aperiodic, and for the stationary distribution to be the target distribution.

\subsection{Examples of MCMC algorithms}
To use MCMC to sample from continuous probability distributions, we must assume that our Markov chain has a continuous state space for $\theta$, but we still work in discrete time. In this case the detailed balance relation between steps $k$ and $k+1$ along the chain is given by 
\begin{equation}
\label{e:mc12}
T(\theta_{k+1}, \theta_{k}) \mu(\theta_k) = T(\theta_{k}, \theta_{k+1}) \mu(\theta_{k+1}),
\end{equation}
where $T(\theta_{k+1}, \theta_{k}) \mu(\theta_k) = P(X_{k+1} \in \Theta | X_{k} \in \Theta)$ and $\mu(\theta_{k}) = \mu(X_k \in \Theta)$. In the context of Bayesian inference, the posterior distribution should be the target distribution of the Markov chain and so we want $\mu(\theta_k) \equiv \mathcal{P}(\theta_k)$. All that is left is to find a form for the transition distribution that satisfies equation~\ref{e:mc12} (a \textit{sufficient} condition for the Markov chain to converge to $\mathcal{P}(\theta)$). 

\subsubsection{Metropolis-Hastings algorithm}
The Metropolis-Hastings algorithm (MH, \citealt{1970Bimka..57...97H}) generates samples from $\mathcal{P}(\theta)$ using a relatively simple trial distribution $q$. For a step along the Markov chain from $k$ to $k+1$ the algorithm operates as follows.
\begin{enumerate}
	\item Sample a trial point $\theta'$ from the trial distribution $q(\theta' | \theta_k)$.
	\item Calculate the acceptance probability $\alpha(\theta', \theta_k) = \min \left( \frac{q(\theta_{k} | \theta') \mathcal{P}(\theta')}{  q(\theta' | \theta_{k}) \mathcal{P}(\theta_{k})}, 1\right)$.
	\item Draw a uniform random variable $u$ from $[0,1]$. If $u < \alpha(\theta', \theta_k)$ set $\theta_{k+1} \rightarrow \theta'$. Otherwise $\theta_{k+1} \rightarrow \theta_k$.
\end{enumerate}
In Section~\ref{s:ns_mh_db} we show that the MH algorithm satisfies detailed balance, that the MH acceptance probability can be derived from the detailed balance relation, and that the MH acceptance probability is optimal in the sense that it permits the most steps along the chain without violating detailed balance. The Appendix also gives the relation between $T$ and $\alpha$. \\ 
Like the previous sampling techniques considered, MH produces posterior samples with weights proportional to the number of times each state is visited.

\subsubsection{Metropolis algorithm}
When the trial distribution $q(\theta'|\theta_k)$ is symmetric in its arguments, i.e. $q(\theta'|\theta_k) = q(\theta_k|\theta')$, then the trial acceptance probability simplifies to \citep{1953JChPh..21.1087M}
\begin{equation}
\label{e:mc13}
\alpha(\theta', \theta_k) = \min \left( \frac{\mathcal{P}(\theta')}{\mathcal{P}(\theta_{k})} \right).
\end{equation}
This form for $\alpha$ still satisfies detailed balance (for suitable $\mathcal{P}$ and $q$) and can be useful when calculating the trial distribution (not necessarily sampling from it) is difficult, as is the case in Chapter~\ref{c:tenth}.

\section{Nested sampling}
\label{s:ns}
\citet{2004AIPC..735..395S} introduced a novel sampling method referred to as nested sampling. This algorithm focuses on calculating the evidence, but also generates samples from the \textit{posterior} probability distribution. The key computational expense associated with nested sampling is the constraint that newly generated samples must be above a certain likelihood value which increases at each iteration. \\
Initially, \citet{Sivia2006} suggested satisfying this constraint by evolving a Markov chain starting at one of the pre-existing samples and evaluating an acceptance ratio based on the one used by the Metropolis algorithm \citep{1953JChPh..21.1087M} used in Markov Chain Monte Carlo (MCMC) sampling (see e.g. \citealt{mackay2002} for a review). A variant of the nested sampling algorithm which focused on sampling from ellipsoids which approximate the region in which the likelihood constraint is satisfied was also developed \citep{2006ApJ...638L..51M}. A major breakthrough in the applicability of nested sampling to highly multi-modal distributions came with the invention of clustering nested sampling algorithms (\citealt{2007MNRAS.378.1365S}, \citealt{2008MNRAS.384..449F}, and \textsc{MultiNest}. The latter of these was used extensively in the preceeding Chapters to carry out Bayesian inference). These algorithms effectively sample from multiple ellipsoids determined by some clustering algorithm, with the aim of approximating the likelihood constraint for each mode of the distribution.
More recently, the slice sampling algorithm \textsc{POLYCHORD} (\citealt{2015MNRAS.450L..61H}, \citealt{2015MNRAS.453.4384H}) has been introduced and is effective at navigating high dimensional spaces, due to the fact that it is not a rejection sampling algorithm. Section~4.1 of \citet{2015MNRAS.453.4384H} gives further examples of nested sampling algorithms which have different ways of satisfying the likelihood constraint.

\subsection{Overview of the nested sampling algorithm} 

\label{s:gns_ns}
\begin{figure}
  \begin{center}
  \includegraphics[ width=0.90\linewidth]{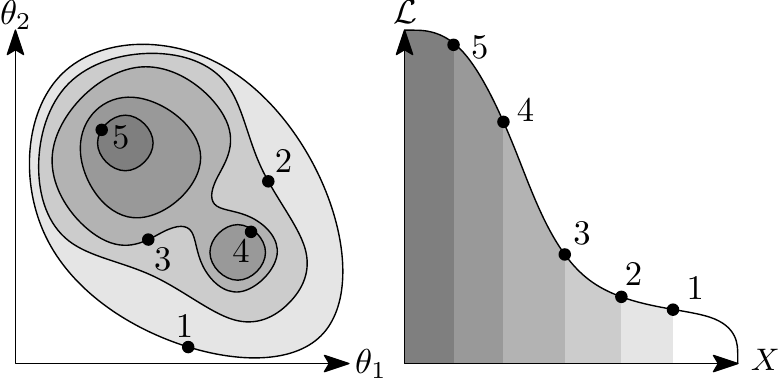}
  \caption{Left plot: Five iso-likelihood contours of a two-dimensional, multi-modal likelihood $\mathcal{L}(\theta_1, \theta_2)$. Each contour encloses some fraction of the prior $X$, with the colourscale indicating the value of $X$ (darkest: smallest $X$). Right: Corresponding $\mathcal{L}$ as a function of $X$ plot (not to scale). The area under the curve is equal to $\mathcal{Z}$.} 
  \label{f:gns_ns_plots}
  \end{center}
\end{figure}

Nested sampling exploits the relation between the likelihood and `prior volume' to transform the N-dimensional integral given by equation~\ref{e:evidence} into a one-dimensional integral. The prior volume $X$ is defined by $\mathrm{d}X = \pi\left(\vec{\Theta}\right) \mathrm{d}\vec{\Theta}$ for parameter space $\vec{\Theta}$, thus $X$ is defined on $[0,1]$ and we can set
\begin{equation}
\label{e:gns_ns}
X(\lambda) = \int_{\mathcal{L}\left(\vec{\Theta}\right) > \lambda} \pi\left(\vec{\Theta}\right) \mathrm{d}\vec{\Theta}.
\end{equation}
The integral extends over the region(s) of the parameter space contained within the iso-likelihood contour $\mathcal{L}\left(\vec{\Theta}\right) = \lambda$ (see Figure~\ref{f:gns_ns_plots}). Assuming that the inverse of equation~\ref{e:gns_ns} ($\lambda(X) = X^{-1}(\lambda) \equiv \mathcal{L}(X)$) exists which is the case when $\pi$ is strictly positive, then the evidence integral can be written as (see Section~\ref{s:ns_Z_proof})
\begin{equation}\label{e:gns_nsz}
\mathcal{Z} = \int_{0}^{1} \mathcal{L}(X) \mathrm{d}X.
\end{equation}
Thus, if one can evaluate $\mathcal{L}(X)$ at $n_{\rm s}$ values of $X$, the integral given by equation~\ref{e:gns_nsz} can be approximated by standard quadrature methods
\begin{equation}\label{e:gns_nsz_sum}
\mathcal{Z} \approx \sum_{i = 1}^{n_{\rm s}} \mathcal{L}_{i} (X_{i-1} - X_{i}),
\end{equation}
where 
\begin{equation}\label{e:gns_X_vals}
0 < X_{n_{\rm s}} < ... < X_{1} < X_{0} = 1.
\end{equation}
Note that one can use more accurate approximations to the integral~\ref{e:gns_nsz} such as the trapezium rule (which has numerical error $\mathcal{O} \left( \frac{1}{n_{\rm s}^2} \right)$, compared with $\mathcal{O} \left( \frac{1}{n_{\rm s}} \right)$ for the sum given above) 
\begin{equation}\label{e:gns_nsz_sum_trap}
\mathcal{Z} \approx \sum_{i = 1}^{n_{\rm s}} \frac{1}{2}(\mathcal{L}_{i-1} + \mathcal{L}_{i}) (X_{i-1} - X_{i}).
\end{equation}
However, I use the method given by equation~\ref{e:gns_nsz_sum} in our implementation of the geometric nested sampler (Chapter~\ref{c:tenth}) for simplicity.
Note further that the first inequality in equation~\ref{e:gns_X_vals} follows from the fact that there could always remain some tiny prior volume containing a larger likelihood value than $\mathcal{L}_{n_{\rm s}}$, unless that can be ruled out by some \textit{a-priori} knowledge of the maximum value of $\mathcal{L}$.

\subsubsection{Determining the nested sampling sum} 
\label{s:gns_ns_sum}
The nested sampling algorithm performs the summation~\ref{e:gns_nsz_sum} as follows. At initiation $n_{l}$ `livepoints' are sampled from the prior $\pi\left(\vec{\Theta}\right)$ which are uniformly distributed in the region $X_{i-1}$ ($= 1$ upon initiation). 
Note also that $\mathcal{L}_{0} = 0$ (relevant when the trapezium rule is used). 
$\mathcal{L}$ is calculated for each of these points, and the livepoint corresponding to the lowest likelihood value $\mathcal{L}_i$ is removed from the livepoint set. This `deadpoint' is replaced by a point drawn from $\pi\left(\vec{\Theta}\right)$, say $\vec{\Theta}_{\rm t}$, subject to the constraint that $\mathcal{L}_{\rm t} > \mathcal{L}_i$. Once this constraint has been satisfied, $\vec{\Theta}_{\rm t}$ is added to the livepoint set. As noted in \citet{2004AIPC..735..395S}, it is intuitive to assume that the shrinkage in $X$ associated with each iso-likelihood contour is geometric. Hence we can write
\begin{equation}\label{e:gns_X_shrinks}
X_1 = t_1 X_0,~ X_2 = t_2 X_1,~ ... ,~X_{n_{\mathrm{s} - 1}} = t_{n_{\mathrm{s} - 1}} X_{n_{\mathrm{s} - 2}}, ~ X_{n_{\rm s}} = t_{n_{\mathrm{s}}} X_{n_{\mathrm{s} - 1}},
\end{equation}
where each $t_i$ lies between zero and one, and can be thought of as the shrinkage factor between successive shells of the prior volume. In practice it is difficult to determine the exact values of $t_i$, as the amount of prior volume shrinkage between iso-likelihood contours $\mathcal{L}_i$ and $\mathcal{L}_{i-1}$ is in general, non-trivial to calculate. Nevertheless, we can estimate $t_i$ statistically as follows. 
Since at each iteration of shrinking the prior volume, there are $n_{l}$ livepoints uniformly distributed in $X_{i-1}$, then we can take $t_i$ to be the largest of $n_{l}$ uniformly distributed numbers between zero and one, since the lowest likelihood should be attributed with the smallest volume shrinkage. This gives the following distribution for the shrinkage factor (derived in Section~\ref{s:ns_tdist})
\begin{equation}\label{e:gns_t_prob}
P(t_i) = n_{l} t_i^{n_{l} -1}. 
\end{equation}
This statistical treatment of the $t_i$ can be used to calculate the expected value of $\mathcal{Z}$ as well as its error, as detailed in \citet{2011MNRAS.414.1418K}.
Once $t_i$ has been calculated, $X_{i}$ can be determined and one is left with $n_l$ livepoints uniformly distributed in the range $[0, X_i]$. For the next iteration of the algorithm the process is repeated from the step of determining the livepoint with the lowest likelihood. 

As explained in \citet{2004AIPC..735..395S}, the geometric uncertainty associated with the $X_i$ leads to the idea that $\log(\mathcal{Z})$ rather than $\mathcal{Z}$ is a normally distributed variable. Assuming the latter to be normally distributed can result in distributions of $\mathcal{Z}$ with variances that suggest $\mathcal{Z}$ can take negative values, which is unphysical. This is the case with the likelihood describing gravitational wave detection used in Section~\ref{s:gns_gw_like}.
The mean and variance of a log-normally distributed random variable, $\mathbb{E}\left[ \log(\mathcal{Z}) \right]$ and $\mathrm{var}\left[ \log(\mathcal{Z}) \right]$, can be calculated from the moments of the non-logarithmic variables as
\begin{gather} 
\label{e:gns_lognorm_mean}
\mathbb{E}\left[ \log(\mathcal{Z}) \right] = 2 \log \left( \mathbb{E}[\mathcal{Z}] \right) - \frac{1}{2} \log \left( \mathbb{E}\left[\mathcal{Z}^2\right] \right), \\
\label{e:gns_lognorm_var}
\mathrm{var}\left[ \log(\mathcal{Z}) \right] = \log \left( \mathbb{E}\left[ \mathcal{Z}^2 \right] \right) - 2 \log \left( \mathbb{E}[\mathcal{Z}] \right).
\end{gather}
Hence our geometric nested sampling algorithm calculates the moments of the linear variables following \citet{2011MNRAS.414.1418K} (in log-space to avoid numerical difficulties, see Section~\ref{s:keet_log}) but the final evidence estimate and its associated error are calculated using equations~\ref{e:gns_lognorm_mean} and~\ref{e:gns_lognorm_var}.  

\subsubsection{Stopping criterion}
\label{s:gns_stop_crit}
\begin{figure}
  \begin{center}
  \includegraphics[ width=0.90\linewidth]{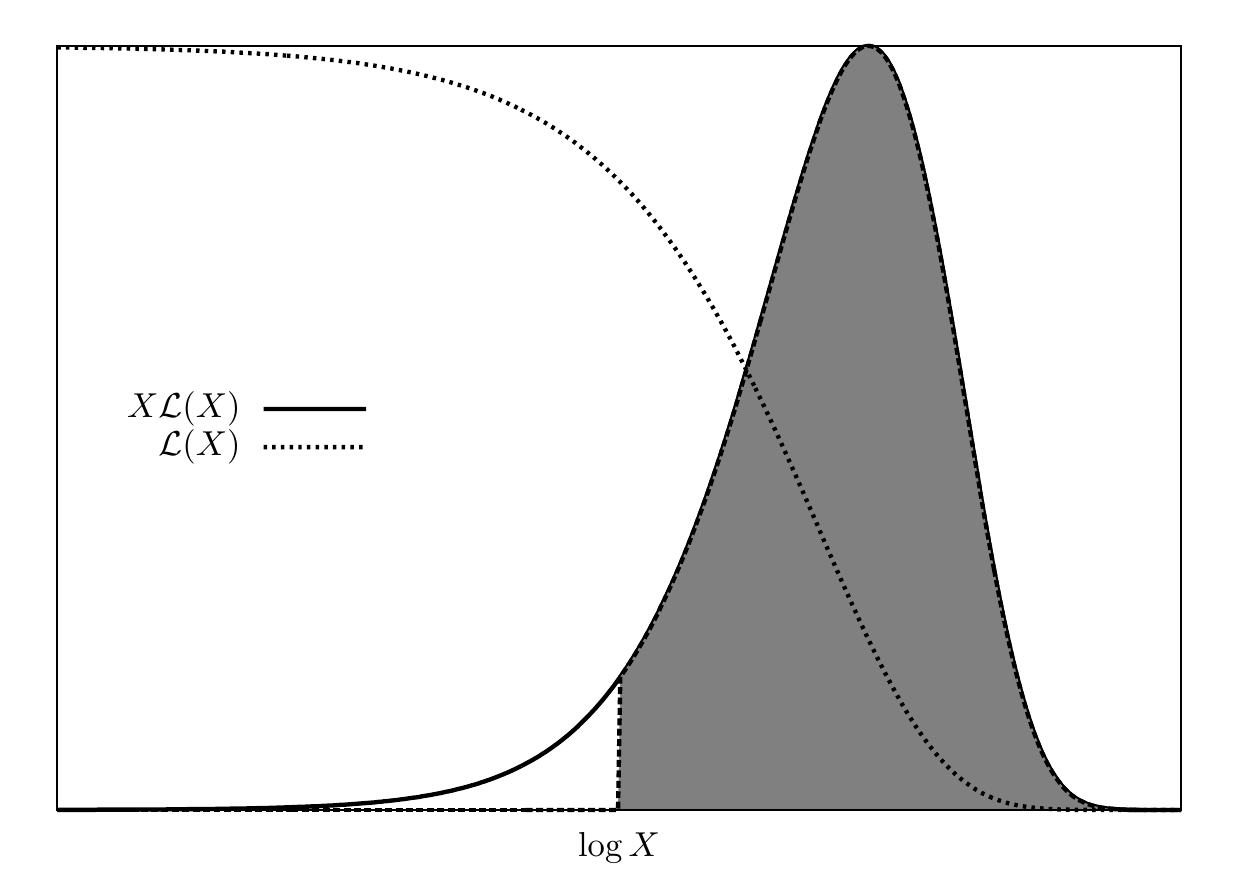}
  \caption{Plots of $\mathcal{L}(X)$ and $\mathcal{L}(X)X$ for typical likelihood functions. The area under the $\mathcal{L}(X)$ curve corresponds to $\mathcal{Z}$. The height of the curve $\mathcal{L}(X)X$, gives an indication of the contribution to $\mathcal{Z}$ for a small fractional change in $X$. After a number of nested sampling iterations, this contribution becomes negligible.} 
  \label{f:gns_LX}
  \end{center}
\end{figure}

The nested sampling algorithm can be terminated based on an estimate of how precisely the evidence value has been calculated up to the current iteration. One measure of this is to look at the ratio of the current estimate of $\mathcal{Z}$ to its value plus an estimate of the `remaining' evidence associated with the current livepoints. Since after iteration $n_{\rm s}$ the livepoints are uniformly distributed in the range $[0, X_{n_{\rm s}}]$, we can approximate their final contribution to the evidence as
\begin{equation}\label{e:ngs_final}
\mathcal{Z}_{\rm f} \approx \frac{X_{n_{\rm s}}}{n_{l}} \sum_{i = 1}^{n_{l}} \mathcal{L}_i,
\end{equation}
where $\mathcal{L}_i$ is the likelihood value of the $i^{\mathrm{th}}$ remaining livepoint. The stopping criterion can then be quantified as
\begin{equation}\label{e:ngs_stop_crit}
\frac{\mathcal{Z}_{\rm f}}{\mathcal{Z}_{\rm f} + \mathcal{Z}} < \epsilon. 
\end{equation} 
$\epsilon$ is a user defined parameter, which I set to $0.01$ in the nested sampling implementations used in Chapter~\ref{c:tenth}.
The final estimate of $\mathcal{Z}$ is then updated to be $\mathcal{Z} \rightarrow \mathcal{Z} + \mathcal{Z}_{\rm f}$.
Note that after a large number of iterations of the nested sampling algorithm, we can be fairly confident that the remaining contribution to the evidence is small. Referring back to equation~\ref{e:gns_nsz_sum}, as the sampling progresses the value of $(X_{i-1} - X_{i})$ gets smaller and there will be a point part way through the process, where its value decreases at a rate faster than $\mathcal{L}_{i}$ increases. Thus after this point, the contribution to the evidence at each iteration becomes smaller, until at some point it becomes negligible (see Figure~\ref{f:gns_LX}). 

\subsubsection{Posterior inferences}
\label{s:gns_post_points}
Once $\mathcal{Z}$ has been determined, posterior inferences can easily be generated using the deadpoints and final livepoints from the nested sampling process to give a total of $n_{\rm s} + n_{l}$ samples (and we set $n_{\rm s} \rightarrow n_{\rm s} + n_{l}$). Each such point is assigned the weight
\begin{equation}\label{e:ngs_post_weights}
\mathcal{P}_{i} = \frac{\mathcal{L}_{i} \left(X_{i-1} - X_{i}\right)}{\mathcal{Z}}.
\end{equation}
Note that for the $n_{l}$ samples obtained from the final set of livepoints $X_{i-1} - X_{i} = \frac{X_{n_{\rm s}}}{n_{l}}$. The weights (along with the corresponding values of $\vec{\Theta}$) can be used to calculate statistics of the posterior distribution, or plot it using software such as \textsc{getdist} or \textsc{corner}\footnote{\url{https://pypi.python.org/pypi/corner}.}.

\section{Plotting posterior samples}
The set of discrete samples $\mathcal{S}$ can be used to determine functional approximations to $\mathcal{P}$. Histograms and kernel density estimation (KDE) are two popular methods deployed to obtain distribution approximations from samples.

\subsection{Histograms}

\begin{figure}
  \begin{center}
  \includegraphics[ width=0.90\linewidth]{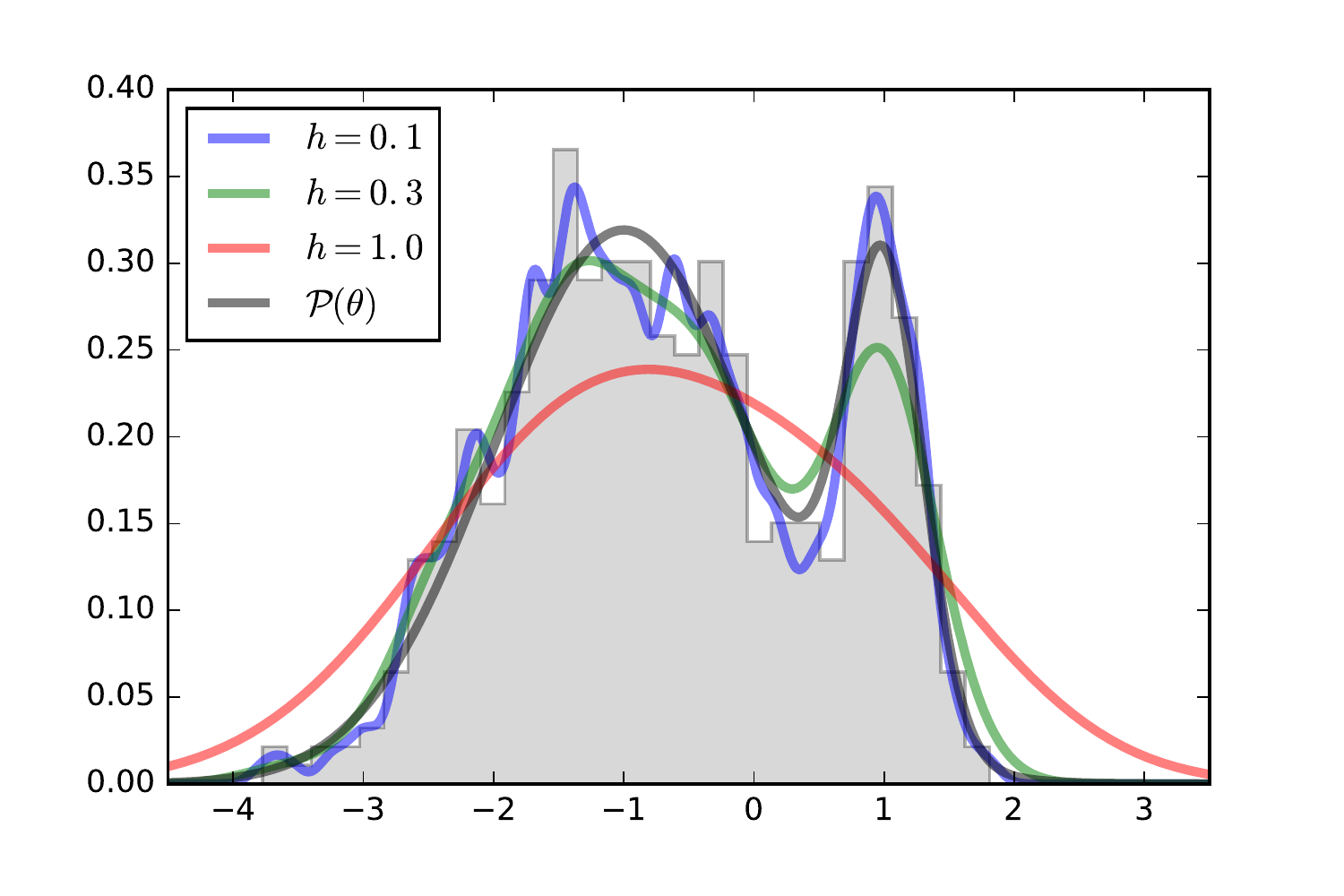}
  \caption{Illustration of approximationg a one-dimensional posterior function $\mathcal{P}(\theta)$ using a histogram or KDE. $\mathcal{P}(\theta)$ is a Gaussian mixture model (parameterised in terms of means and standard deviations): $\mathcal{P}(\theta) = 0.8 \times \mathcal{N}(-1,1) + 0.2 \times \mathcal{N}(1,0.3)$. The samples $\mathcal{S} = \{(\theta_1,\mathcal{P}_1),~..., (\theta_{n_{\rm s}},\mathcal{P}_{n_{\rm s}})\}$ are generated by drawing $100$ samples from $\mathcal{N}(1,0.3)$ \& $400$ samples from $\mathcal{N}(-1,1)$, and are assigned to the $\theta_{n_{i}}$ while the weights are set to unity. The histogram is generated by binning the samples into $30$ bins of uniform width over the range $[-4.5,3.5]$. The KDE estimates are generated using a Gaussian function for $K_{h}$, with either $h = 0.1$, $h = 0.3$, or $h = 1$. The black curve represents the `true' form of the function $\mathcal{P}(\theta)$, the grey region represents the (discrete) histogram approximation, while the blue, green and red curves correspond to the $\hat{\mathcal{P}}(\theta)$ obtained for the different values of $h$. } 
  \label{f:hist_kde}
  \end{center}
\end{figure}

Histograms provide a quick way to generate a piecewise discontinuous approximation of $\mathcal{P}(\vec{\Theta})$. The sample weights are `binned' into a series of intervals separating $\vec{\Theta}$. The new sample weight for each bin $b$, $\mathcal{P}(\vec{\Theta}_b)$, is simply the sum of the $\mathcal{P}(\vec{\Theta}_i)$ associated with that bin, and the value of $\vec{\Theta}_b$ is defined as some function of the corresponding $\vec{\Theta}_i$ (e.g. their average). For each bin $\mathcal{P}$ is constant over the corresponding interval on $\vec{\Theta}$ and so the function approximation is discontinuous. using a small number of bins reduces the noise associated with the sampling process, but can lead to key features of the true $\mathcal{P}(\vec{\Theta})$ being missed, while a large number of bins will tend to overfit to the samples \& produce a very `peaky' approximation. Figure~\ref{f:hist_kde} shows an example of a histogram with a moderate number of bins, which catches the main features of $\mathcal{P}$ but also includes a noticable amount of sampling noise.

\subsection{Kernel density estimation}
\label{s:kde}
KDE is a non-parametric method for estimating probability densities from samples, which `improves' on simple histograms by smoothing the resulting curve. A continuous function approximation for the posterior, $\hat{\mathcal{P}}(\vec{\Theta})$ is given by
\begin{equation}
\label{e:kde1}
\hat{\mathcal{P}}(\vec{\Theta})= \sum_{i = 1}^{n_{\rm s}} \mathcal{P}(\vec{\Theta}_i) K_{h}(\vec{\Theta} - \vec{\Theta}_i), 
\end{equation}
where $K_{h}$ is a smoothing kernel with width parameter $h$. $K_{h}$ must integrate over its domain to one (i.e. be a probability density function) to ensure that $\hat{\mathcal{P}}(\vec{\Theta})$ is also normalised. $h$ determines the variance of the smoothing kernel and thus how smooth $\hat{\mathcal{P}}(\vec{\Theta})$ is. Figure~\ref{f:hist_kde} illustrates the use of KDE with a Gaussian smoothing kernel and either $h = 0.1$, $h = 0.3$, or $h = 1$\footnote{Example inspired by \url{https://jakevdp.github.io/blog/2013/12/01/kernel-density-estimation/}.}. The latter value corresponds to a $\hat{\mathcal{P}}(\theta)$ which is a poor estimation of $\mathcal{P}(\theta)$ (due to `oversmoothing' $\hat{\mathcal{P}}(\theta)$ does not reveal the bimodality of $\mathcal{P}(\theta)$). The $\hat{\mathcal{P}}(\theta)$ corresponding to $h = 0.1$ and $h = 0.3$ capture the bimodality of $\mathcal{P}(\theta)$, but include a lot of small peaks not present in the true distribution (`undersmoothing'). \textsc{GetDist} uses a truncated Gaussian for $K_{h}$ with the determination of $h$ based on minimisation of the mean integrated square error \footnote{For more information on the specifics of the implementation of KDE used in GetDist, see \url{https://cosmologist.info/notes/GetDist.pdf}.}
\begin{equation}
\label{e:kde2}
\min\displaylimits_{h} \left[ \int \mathbb{E} \left[ (\mathcal{P}(\vec{\Theta}) - \hat{\mathcal{P}}(\vec{\Theta}))^2 \right] \mathrm{d}\vec{\Theta} \right].
\end{equation}


%% file: CHAP-10/chapter10.tex
\chapter{Geometric nested sampling}\label{c:tenth}

Here I present a nested sampling algorithm 
which provides a new method for satisfying the nested sampling likelihood constraint (see Section~\ref{s:ns}) based on the Markov method used in \citet{Sivia2006} (and also applied in \citealt{2008MNRAS.384..449F}). Certain parameters relevant to astrophysics exhibit special properties which mean they naturally parameterise points on geometric objects such as circles, tori and spheres. The algorithm we introduce here which we refer to as the geometric nested sampler, exploits these properties to generate samples efficiently and enables mobile exploration of distributions which are defined on such geometries. My implementation of the algorithm can be found at \url{https://github.com/SuperKam91/nested_sampling} \citep{javid2020geometric}. A paper corresponding to the work carried out in this Chapter is going to be submitted to MNRAS \citep{2019arXiv190509110J}, and contains several more motivating toy examples for the geometric nested sampler.

\section{Nested sampling prior distributions}
\label{s:gns_priors}
Bayesian inference has been reviewed in Section~\ref{s:bayes_inf} and nested sampling in Section~\ref{s:ns}. Here we make a note about the form of the prior distribution $\pi$ of the parameter set $\vec{\Theta}$ used throughout this Chapter. \\
In general for nested sampling, $\pi\left(\vec{\Theta}\right)$ can take any form as long as the distribution integrates to one and has a connected support (\citealt{2008arXiv0801.3887C};  this roughly means that the parts of the domain at which $\pi\left(\vec{\Theta}\right) \neq 0$ is not `separated' by the parts at which $\pi\left(\vec{\Theta}\right) = 0$). For simplicity, in all examples considered here I assume that each component of the $N$-dimensional vector $\vec{\Theta} = (\theta_1,...,\theta_N)$ is independent of one another, and that each $\pi(\theta_i)$ is a uniform probability distribution, so that
\begin{equation}\label{e:gns_priors}
\pi\left(\vec{\Theta}\right) = \prod_{i=1}^{N} \pi_{i}(\theta_{i}) = \prod_{i=1}^{N} \frac{1}{\theta_{\mathrm{max},i} - \theta_{\mathrm{min},i}},
\end{equation}
where $\theta_{\mathrm{max},i}$ and $\theta_{\mathrm{min},i}$ are respectively the upper and lower bounds on $\theta_i$. Values for $\theta_{\mathrm{max},i}$ and $\theta_{\mathrm{min},i}$ used in the examples presented here will be stated in the following Sections.

\section{Satisfying the likelihood constraint} 
\label{s:gns_lhood_const}
At each step of the nested sampling iteration, one needs to sample a new point which satisfies $\mathcal{L}_{\rm t} > \mathcal{L}_{i}$. As mentioned in Section~\ref{s:ns}, considerable work has been put into increasing the efficiency of this process, as it is by far the most computationally expensive step of the nested sampling algorithm. I now give a review of the Metropolis nested sampling method used by \citet{Sivia2006} and \citet{2008MNRAS.384..449F}, which forms the basis of the method used in geometric nested sampling.

\subsection{Metropolis nested sampling}
\label{s:gns_m_lhood_const}
The Metropolis nested sampling method is an adaption of the Metropolis algorithm used in MCMC sampling of a posterior distribution (see Sections~\ref{s:mcmc} and~\ref{s:ns_mh_db}). 
The acceptance ratio for the Metropolis nested sampling algorithm takes the form
\begin{equation}
\label{e:gns_ns_accept}
\alpha = \begin{cases}
\mathrm{min}\left[\pi\left(\vec{\Theta}_{\rm t}\right) / \pi\left(\vec{\Theta}_{l}\right), 1\right] \quad \mathrm{ if } \quad \mathcal{L}_{\rm t} > \mathcal{L}_{i}, \\
 0 \quad \mathrm{ otherwise.}
\end{cases}
\end{equation}
Here $\vec{\Theta}_{l}$ is obtained by picking one of the current livepoints at random, and using its value of $\vec{\Theta}$. The value for $\vec{\Theta}_{\rm t}$ is sampled from a trial distribution $q\left(\vec{\Theta}_{\rm t} | \vec{\Theta}_{l}\right)$. Sivia \& Skilling and Feroz et al. use symmetric Gaussian distributions centred on $\vec{\Theta}_{l}$ for $q\left(\vec{\Theta}_{\rm t} | \vec{\Theta}_{l}\right)$. The trial point is accepted to be a new livepoint (replacing the deadpoint associated with $\mathcal{L}_{i}$) with probability $\alpha$. Note that equation~\ref{e:gns_ns_accept} implicitly assumes that the proposal distribution is symmetric in its arguments, that is $q\left(\vec{\Theta}_{\rm t} | \vec{\Theta}_{l}\right) = q\left(\vec{\Theta}_{l} | \vec{\Theta}_{\rm t}\right)$. In the case that the proposal distribution is asymmetric, the acceptance ratio includes an additional factor $q\left(\vec{\Theta}_{l} | \vec{\Theta}_{\rm t}\right)q\left(\vec{\Theta}_{\rm t} | \vec{\Theta}_{l}\right)$ (in which case the algorithm is referred to as the Metropolis-Hastings algorithm, see Section~\ref{s:mcmc}).
The fact that the Metropolis nested sampling method uses the current livepoints as a `starting point' for selecting $\vec{\Theta}_{\rm t}$, means that the autocorrelation between the livepoints is high, which in turn leads to biased sampling. This can be prevented by increasing the variance of the trial distribution used, or by requiring that multiple trial points must be accepted before the final one is accepted as a livepoint, i.e. after the first accepted trial point is found, set $\vec{\Theta}_{l} \rightarrow \vec{\Theta}_{\rm t}$ and use this to sample a new $\vec{\Theta}_{\rm t}$ from $q\left(\vec{\Theta}_{\rm t} | \vec{\Theta}_{l}\right)$. This can be repeated an arbitrary number of times, but in general more iterations leads to a lower correlation between the livepoint used at the beginning of the chain and the final accepted trial point which is added to the livepoint set. Sivia \& Skilling suggest that at each nested sampling iteration, the number of trial points generated $n_{\rm t}$ to get a new livepoint should be $\approx 20$. In my implementation I set this number to $20 \times N$ where $N$ is the dimensionality of the parameter estimation problem. Note that $n_{\rm t}$ includes both accepted and rejected trial points. Sivia and Skilling also suggest that the acceptance rate for the trial points at each nested sampling iteration should be $\approx 50\%$. This is because a high acceptance rate usually suggests high auto-correlation between the successive trial points, whilst a low acceptance rate can suggest high correlation between the final accepted trial point and the one used to initialise the chain, as too few steps have been made between the two. In the extreme case that the acceptance rate is zero, the process of picking a new livepoint has failed, as one cannot have two livepoints corresponding to the same $\vec{\Theta}$. The acceptance rate is affected by the variance of the trial distribution, a large variance usually results in more trial points being rejected (especially near the peaks of the posterior). Sivia \& Skilling suggest updating the trial standard deviation as
\begin{equation}
\label{e:gns_trialsig}
\sigma_{\rm t} \rightarrow \begin{cases}
 \sigma_{\rm t} \exp(1/N_{\rm a}) \quad \mathrm{ if } \quad N_{\rm a} > N_{\rm r}, \\
 \sigma_{\rm t} \exp(-1/N_{\rm r}) \quad \mathrm{ if } \quad N_{\rm a} \leq N_{\rm r},
\end{cases}
\end{equation}
where $N_{\rm a}$ and $N_{\rm r}$ are the number of accepted and rejected trial points in the current nested sampling iteration respectively. Note however that I determine the variance using different methods (see Sections~\ref{s:gns_trialvar} and~\ref{s:gns_spherevar}).

Feroz et al. incorporate the Metropolis likelihood sampling into their clustering nested sampling algorithm rather than use it in isolation. The geometric likelihood sampling I introduce in the next Section is a modified version of the Metropolis algorithm used in isolation. 

\section{Geometric nested sampling}
\label{s:gns_geom_ns}

\begin{figure}
  \begin{center}
  \includegraphics[ width=0.90\linewidth]{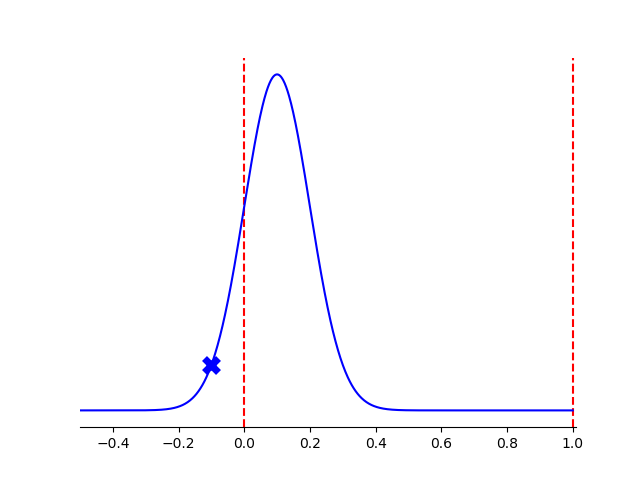}
  \caption{`Vanilla' non-wrapped trial distribution. The blue curve represents a Gaussian `vanilla' trial distribution $q(\theta'|\theta)$ with starting point $\theta = 0.1$, and sampled trial point $\theta' = -0.1$ shown by the blue cross.
	The support of $\pi$ is indicated by the red dashed lines ($[0,1]$). Since $\theta'$ lies outside the support of $\pi$,
	it would automatically be rejected by the Metropolis algorithm.} 
  \label{f:gns_nonwrap}
  \end{center}
\end{figure}

One key issue with Metropolis nested sampling is that at each nested sampling iteration, if too many trial points are rejected, then the livepoints will be highly correlated with each other after a number of nested sampling iterations. To prevent this one must sample a large number of trial points in order to increase the number of acceptances and decrease the auto-correlation of the trial point chain. This solution can be problematic if computing the likelihood is computationally expensive. One particular case in which the sampled point is guaranteed to be rejected, is if the point lies outside of the domain of $\mathcal{P}$ (support of $\pi$). Such a case is illustrated in Figure~\ref{f:gns_nonwrap} for parameter $\theta$. Of course, this can be avoided by adapting $q\left(\vec{\Theta}_{\rm t} | \vec{\Theta}_{l}\right)$ so that it is truncated to fit the support of $\pi$, but in high dimensions this can be tedious, and inefficient in itself. Hence one desires an algorithm which does not sample outside the support of $\pi$, without having to truncate $q$.

Another issue which most sampling algorithms are subject to occurs when the modes of the posterior distribution are far away from each other in $\vec{\Theta}$ space, e.g. when they are at `opposite ends' of the domain of $\pi$. In the context of nested sampling this can result in one or more of the modes not being sampled accurately, particularly in the case of low livepoint runs.
Thus a sampling algorithm should be able to efficiently manoeuvre between well separated modes which lie at the `edges' of $\pi$'s support. 

Geometric nested sampling attempts to solve these two issues by interpreting parameter values as points on geometric objects, namely on circles, tori and spheres.

\subsection{Wrapping the trial distribution}
\label{s:gns_wrap}
\begin{figure}
  \begin{center}
  \includegraphics[ width=0.90\linewidth]{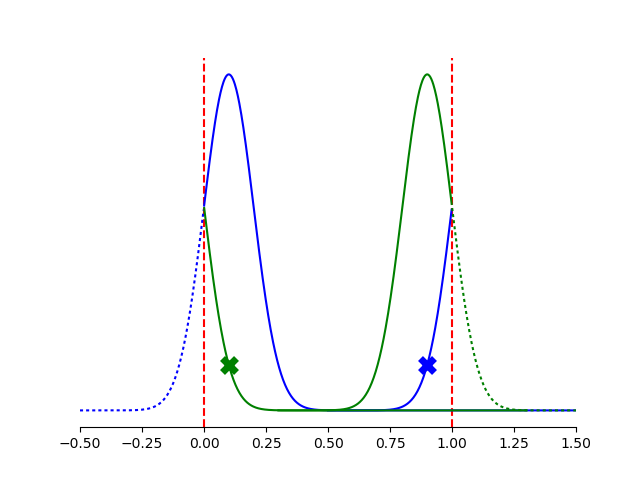}
  \caption{Wrapped trial distributions. The solid blue curve represents a Gaussian trial distribution $q(\theta'|\theta)$ as in previous Figure, but now incorporating the wrapping methodology. As a result of the wrapping, $\theta'$ (blue cross)
	is at $0.9$, and so won't be automatically rejected by the Metropolis algorithm.
	The green curve shows the same trial distribution $q(\theta|\theta')$ centred on $0.9$.
	The fact that $\theta =0.1$ (green cross) is sampled from $q(\theta|\theta')$ with the same probability as $\theta'$ is from $q(\theta'|\theta)$ shows that the wrapped trial distribution is still symmetric with respect to its arguments (provided $q(a|b)$ is a symmetric function about the point $b$).} 
  \label{f:gns_wrap}
  \end{center}
\end{figure}

A relatively straightforward way of ensuring that the trial points sampled from $q$ are in the support of $\pi$ is to `wrap' $q$. This is illustrated in Figure~\ref{f:gns_wrap}, where we consider a one-dimensional uniform prior on $[0, 1]$. 
For any point $\theta$, there will be a non-zero probability of sampling a value of $\theta'$ from the trial distribution $q(\theta' | \theta)$ that lies outside $[0, 1]$. If the point sampled has a value of say $\theta' = -0.1$, then if we consider $q$ to be wrapped around the support this can be interpreted as sampling a point at value $\theta' = 0.9$.
More generally, if $\theta'$ is outside the support of $\pi$ defined by upper and lower bounds $\theta_{\mathrm{max}}$ and $\theta_{\mathrm{min}}$ it will be transformed as
\begin{equation}
\label{e:gns_wrap}
\theta' = \begin{cases}
\theta_{\mathrm{max}} - W(\theta') \quad \mathrm{ if } \quad \theta' > \theta_{\mathrm{max}}, \\
\theta_{\mathrm{min}} + W(\theta') \quad \mathrm{ if } \quad \theta' < \theta_{\mathrm{min}},
\end{cases} 
\end{equation}
where
\begin{equation}
\label{e:gns_wrapmod}
W(\theta) = \begin{cases}
(\theta - \theta_{\mathrm{max}}) \mod (\theta_{\mathrm{max}} - \theta_{\mathrm{min}}) \quad \mathrm{ if } \quad \theta > \theta_{\mathrm{max}}, \\
(\theta_{\mathrm{min}} - \theta) \mod (\theta_{\mathrm{max}} - \theta_{\mathrm{min}}) \quad \mathrm{ if } \quad \theta < \theta_{\mathrm{min}}.
\end{cases}
\end{equation}
Assuming the support of $\pi$ is connected (a requirement of nested sampling, as stated in Section~\ref{s:gns_priors}), then this operation will be well defined for all $\pi$ with bounded supports, of arbitrary dimension. Using this transformation does not affect the argument symmetry of $q$, thus the value of $\alpha$ given by equation~\ref{e:gns_ns_accept} still holds. Furthermore, this symmetry ensures that the detailed balance relation given by equation~\ref{e:mc12} is still satisfied.

\subsection{Circular parameters}
\label{s:gns_circular}

\begin{figure}
  \begin{center}
  \includegraphics[ width=0.90\linewidth]{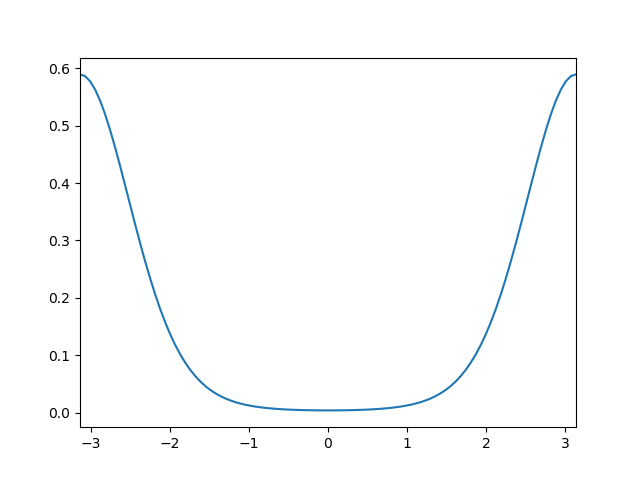}
  \caption{von Mises distribution with domain $[-\pi, \pi]$, centred on $\pi$. The peak wraps around at edges of domain, so that it appears as two half peaks on a linear space.} 
  \label{f:gns_vm}
  \end{center}
\end{figure}

As well as ensuring that none of the sampled trial points lie outside the support of $\pi$, the wrapped trial distribution can also improve the manoeuvrability of the sampling process, since the trial point chain can always `move in either direction' without stepping outside of the support of $\pi$. This proves to be particularly useful for `circular parameters'. Here I define circular parameters to be those whose value at $\theta_{\mathrm{max}}$ and $\theta_{\mathrm{min}}$ correspond physically to the same point. Examples of circular parameters include angles (which are circular at e.g. zero and $2\pi$) and time periods (e.g. $00$:$00$ and $24$:$00$). Often, circular parameters have probability distributions associated with them which are also circular. An example of a circular distribution is the von Mises distribution, an example of which is shown in Figure~\ref{f:gns_vm} (and defined in Section~\ref{s:gns_tmI}). This particular example shows that the function's peak(s) may be split by the wrapping, so that when plotted linearly, they appear to have to `half peaks' about $\theta_{\mathrm{max}}$ and $\theta_{\mathrm{min}}$.
Such half peaks would be classified as two separate peaks by clustering nested sampling algorithms. Thus in general, the number of livepoints would need to be increased to accommodate for the higher number of modes, to ensure both half peaks are sampled adequately without one cluster `dying out'. Furthermore, the two half peaks occur at opposite ends of the domain of a linear space, making it more difficult for a sampler to explore the regions of higher probability efficiently.
The wrapped trial distribution resolves both of these issues, as the two half peaks in linear space are treated as one full peak as far as the sampling (and allocation of livepoints) is concerned. Consequently, the second issue of the half peaks being far away from each other is automatically eradicated. 
The wrapped trial distribution methodology can thus be applied to problems which involve sampling on non-Euclidean spaces. I apply the method to toy models with distributions defined on circles and tori in Sections~\ref{s:gns_tmI} and~\ref{s:gns_tmII} respectively. Furthermore, I apply the methodology to a practical example in Section~\ref{s:gns_ligo}.

\subsection{Variance of the trial distribution}
\label{s:gns_trialvar}
As with any sampling procedure which relies on a trial distribution, picking a variance for the distribution is difficult without a-priori knowledge of the posterior distribution you are sampling from. A low variance results in a lot of trial points being accepted, but a high auto correlation between these points. A high variance gives a lot of trial rejections, but when these points are accepted, their correlation with the starting point is often low.
Since picking the trial variance can in itself be a mammoth task, I use a simplistic approach and take it to be 
\begin{equation}
\label{e:gns_trialvar}
0.1 \times \left\lvert \max\displaylimits_{\mathrm{livepoints}}\left(\theta_{i}\right) - \min\displaylimits_{\mathrm{livepoints}}\left(\theta_{i}\right) \right\rvert ,
\end{equation}
for each component $i$ of $\vec{\Theta}$. I use this approach to avoid the sampler from taking large steps when the livepoints are close together. However, I acknowledge that this method is far from optimal when the livepoints are compactly located at the edges of the domain of $\mathcal{P}\left(\vec{\Theta}\right)$. 

\subsection{Non-Euclidean sampling via coordinate transformations}
\label{s:gns_coordtrans}
The wrapped trial distribution introduced in Section~\ref{s:gns_wrap} can in theory be used in Metropolis nested sampling to sample effectively from circular and toroidal spaces parameterised in terms of circular variables. However, it is not particularly effective at sampling from spherical spaces, since wrapping around the zenith angle (usually defined on $[0, \pi]$) would result in discontinuous jumps between the poles of the sphere. One could of course just wrap the trial distribution in the dimension representing the azimuthal angle (usually defined on $[0, 2\pi]$), rather than in both angles. However, this would re-introduce the issues stated in Section~\ref{s:gns_geom_ns}, i.e. wasting samples and inefficient exploration of the parameter space.
I therefore propose an alternative method for exploring spherical spaces which I incorporate in the geometric nested sampling algorithm.

\subsection{Spherical coordinate transformations}
\label{s:gns_spheretrans}
\begin{figure*}
  \begin{center}
  \includegraphics[ width=0.65\linewidth]{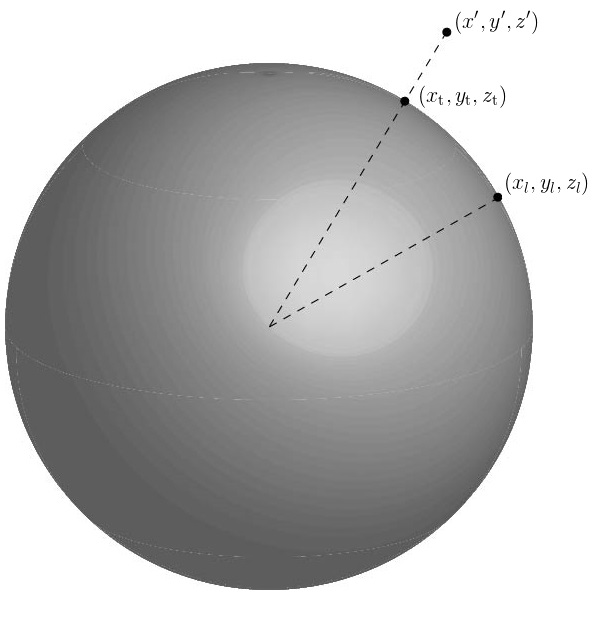}
  \caption{Sampling points on the surface of a sphere in Cartesian coordinates. The three-dimensional trial distribution is centred at the point $(x_l,y_l,z_l)$, which corresponds to $(\phi_l,\theta_l)$. The point $(x',y',z')$ sampled from $q$ in general will not lie on the surface of the sphere, however the point is implicitly projected onto the sphere at $(x_{\rm t},y_{\rm t},z_{\rm t})$ when calculating $(\phi',\theta')$ [ $\equiv (\phi_{\rm t},\theta_{\rm t})$].} 
  \label{f:gns_sphere}
  \end{center}
\end{figure*}

Assuming the surface of a unit sphere is parameterised by azimuthal angle $\phi$ on $[0, 2\pi]$ and zenith angle $\theta$ on $[0, \pi]$, then the corresponding Cartesian coordinates are
\begin{equation}
\label{e:gns_spherecoords}
\begin{split}
&x = r \cos(\phi)\sin(\theta), \\
&y = r \sin(\phi)\sin(\theta), \\
&z = r \cos(\theta),
\end{split}
\end{equation}
with $r = 1$. Note that $\phi$ is the angle measured anti-clockwise from the positive $x$-axis in the $x$--$y$ plane and $\theta$ is the angle measured from the positive $z$-axis. Thus a trial point $\phi_{\rm t}, \theta_{\rm t}$ can be sampled as follows. 
Starting from a point $\phi_{l}, \theta_l$, calculate $x_l, y_l, z_l$, from which a trial point $x', y', z'$ can be sampled from $q(x', y', z' | x_l, y_l, z_l)$. We use a three-dimensional spherically symmetric Gaussian distribution for $q(x', y', z' | x_l, y_l, z_l)$. In general, the point $x', y', z'$ will not lie on the unit sphere. Nevertheless the point is implicitly projected onto it by solving the equations given by~\ref{e:gns_spherecoords} simultaneously for $\phi$ and $\theta$, where we set $x = x'$, $y = y'$, $z = z'$, and $r = r'$ (see Figure~\ref{f:gns_sphere}). 
The resulting values are $\phi_{\rm t}$ and $\theta_{\rm t}$, from which the acceptance ratio given by equation~\ref{e:gns_ns_accept} can be evaluated as normal. There are a few things to note about sampling the trial point in the Cartesian space. Firstly, for equation~\ref{e:gns_ns_accept} to hold we must have $q(\phi_{\rm t}, \theta_{\rm t} | \phi_l, \theta_l) = q(\phi_l, \theta_l | \phi_{\rm t}, \theta_{\rm t})$, which is equivalent to 
\begin{equation}
\label{e:gns_carttrials}
\int \displaylimits_{\vec{x}' \in \{\vec{x}_{\mathrm{t}, \phi,\theta}\}} q(\vec{x}'|\vec{x}) \mathrm{d}\vec{x}' = 
\int\displaylimits_{\vec{x} \in \{\vec{x}_{l,\phi,\theta}\}} q(\vec{x}|\vec{x}') \mathrm{d}\vec{x},
\end{equation}
where $\vec{x}' = (x',y',z')$ and $\vec{x} = (x,y,z)$. $\{\vec{x}_{\mathrm{t}, \phi,\theta}\}$ are the set of Cartesian coordinates which satisfy~\ref{e:gns_spherecoords} for $\phi = \phi_{\rm t}$, $\theta = \theta_{\rm t}$, \& all $r \neq 0$. Similarly $\{\vec{x}_{l,\phi,\theta}\}$ are the $\vec{x}$ which satisfy~\ref{e:gns_spherecoords} for $\phi = \phi_l$ \& $\theta = \theta_l$ (see Figure~\ref{f:gns_sphere}). 
Due to the symmetry of the spherical coordinate system, these sets of vectors lie along the lines given by $(\phi_{\rm t}, \theta_{\rm t})$ and $(\phi_l, \theta_l)$ respectively. The only additional requirement for equation~\ref{e:gns_carttrials} to hold is that $q(x',y',z'|x,y,z)$ is symmetric in its arguments, which it is provided that $q(a|b)$ is a symmetric function about the point $b$.
As in Section~\ref{s:gns_wrap}, the symmetry of the trial distribution ensures that the detailed balance relation given by equation~\ref{e:mc12} is still satisfied.

Sampling in Cartesian coordinates eliminates the risk of sampling points which are automatically rejected (due to being outside the support of $\pi(\phi,\theta)$) to a negligible level, since the only points in Cartesian coordinates which are ill-defined in spherical coordinates are $x=y=0$ for all $z$.
How the coordinate transformation improves the manoeuvrability of the sampler relative to sampling in the original parameter space is less clear-cut. For the latter, when the variance is fixed the step sizes taken by the sampler along the surface of the sphere depend on where you start from. 
 For example, at $\theta \approx 0$, large moves in $\phi$ will result in relatively small steps along the sphere whereas at $\theta \approx \pi/2$ such moves in $\phi$ would result in large steps along the sphere. However when sampling in a Cartesian coordinate system, for a constant variance (see below), the trial points sampled will have the same average step size in Euclidean space regardless of the starting point. Furthermore due to the symmetry of a sphere, when the sampled point ($x',y',z'$) is projected back onto the sphere (implicitly when determining $\phi_{\rm t}$ and $\theta_{\rm t}$), the variance of the steps along the sphere is still independent of the starting point. In either the original parameter space or the transformed space, the variance of the trial distribution can be tweaked to adjust the average step size of the sampler. Nevertheless, it seems more intuitive to me to perform the sampling in the space in which adjusting the variance has an effect which is independent of where you are sampling from. 

A spherical distribution is used in the toy model presented in Section~\ref{s:gns_tmII}, and also features in the gravitational wave detection likelihood function in Section~\ref{s:gns_ligo}.

\subsection{Variance of the Cartesian trial distribution}
\label{s:gns_spherevar}

For given variances of $\phi$ and $\theta$: $\sigma_{\phi}^2$ \& $\sigma_{\theta}^2$, the variance corresponding to a function of these two variables is given by 
\begin{equation}
\label{e:gns_funcvar}
\sigma_{f}^2 = \left( \frac{\partial f}{\partial \phi} \right)^2 \sigma_{\phi}^2 + \left( \frac{\partial f}{\partial \theta} \right)^2 \sigma_{\theta}^2 + 2 \frac{\partial f}{\partial \phi} \frac{\partial f}{\partial \theta} \sigma_{\phi,\theta},
\end{equation}
where $\sigma_{\phi,\theta}$ is the covariance between $\phi$ and $\theta$. Hence one can calculate the corresponding variance in Cartesian coordinates, $\sigma_{x}^2$, $\sigma_{y}^2$, and $\sigma_{z}^2$ by substituting the equations given by~\ref{e:gns_spherecoords} into equation~\ref{e:gns_funcvar}. Using these values for $q(x',y',z'|x,y,z)$ however, leads to an asymmetric trial distribution in its arguments, since the variance is now a function of $\theta$ and $\phi$. Our entire formulation of the geometric nested sampling algorithm requires $q$ to be symmetric in order for equations~\ref{e:gns_ns_accept} and~\ref{e:mc12} to hold. Thus we set $\sigma_{x}^2 = \sigma_{y}^2 = \sigma_{z}^2 = 4 / 100$ to ensure $q$ is symmetric.  

\subsection{Non-spherical coordinate transformations}
\label{s:gns_nonspheretrans}

The transformation of the trial sampling problem introduced in the previous Section need not be unique to the case of a sphere. Indeed, our implementation of geometric nested sampling includes the option to transform to Cartesian coordinates from circular or toroidal parameters. This is done in the same way as described for the spherical case, but with the relations given by~\ref{e:gns_spherecoords} replaced with the equivalent transformations for a circle or torus. 

\subsubsection{Circular coordinate transformations}
\label{s:gns_circtrans}
\begin{figure*}
  \begin{center}
  \includegraphics[ width=0.6\linewidth]{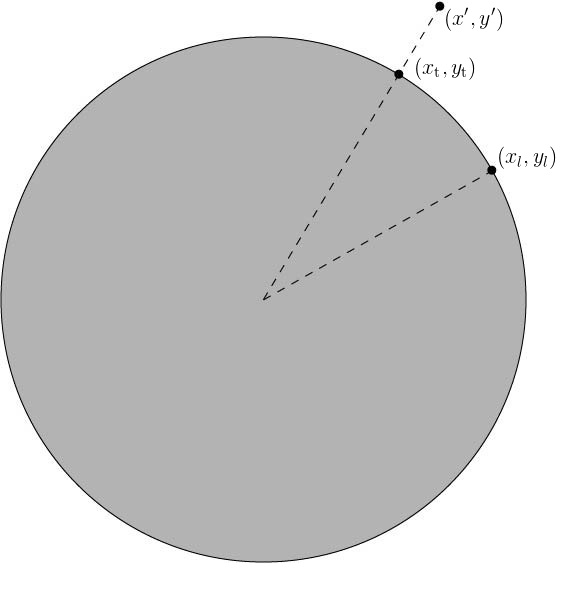}
  \caption{Sampling points on the perimeter of a circle in Cartesian coordinates. The two-dimensional trial distribution is centred at the point $(x_l,y_l)$, which corresponds to $\phi_l$. The point $(x',y')$ sampled from $q$ in general will not lie on the perimeter of the circle, however the point is implicitly projected onto it at $(x_{\rm t},y_{\rm t})$ when calculating $\phi'$ [ $\equiv \phi_{\rm t}$].} 
  \label{f:gns_circle}
  \end{center}
\end{figure*}

For a parameter which can be interpreted as representing points on a circle e.g. $\phi \in [0, 2\pi]$, we can transform $\phi$ into the Cartesian coordinates of a unit circle, 
\begin{equation}
\label{e:gns_circcoords}
\begin{split}
&x = r \cos(\phi), \\
&y = r \sin(\phi), \\
\end{split}
\end{equation}
with $r = 1$. A trial point can be sampled as described for the spherical case but working in two dimensions instead. The symmetry of a circle ensures that the trial distribution $q(\phi_{\rm t} | \phi_l)$ is symmetric in its arguments as long as the Cartesian trial distribution $q(x',y'|x,y)$ adheres to the same symmetry. This is indeed true when a circularly symmetric Gaussian distribution is used for $q(x',y'|x,y)$. The circular transformation and sampling process is illustrated in Figure~\ref{f:gns_circle}.

\subsubsection{Toroidal coordinate transformations}
\label{s:gns_torustrans}
\begin{figure*}
  \begin{center}
  \includegraphics[ width=0.75\linewidth]{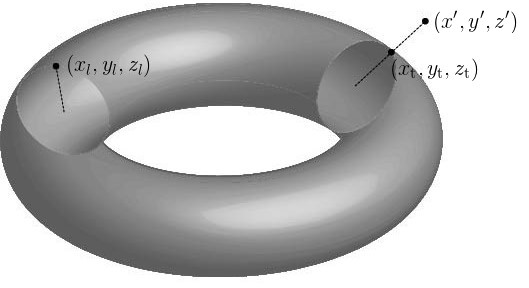}
  \caption{Sampling points on the surface of a torus with major and minor radii $R$ and $r$ in Cartesian coordinates. $R$ corresponds to the distance from the centre of the torus (centre of the whitespace in the middle of the grey tube) to the centre of the cross section (depicted as grey circles) of the torus, while $r$ is the radius of the torus' cross section. The three-dimensional trial distribution is centred at the point $(x_l,y_l,z_l)$, which corresponds to $(\phi_l,\theta_l)$. The point $(x',y',z')$ sampled from $q$ in general will not lay on the surface of the torus, however the point is implicitly projected onto it at $(x_{\rm t},y_{\rm t},z_{\rm t})$ when calculating $(\phi',\theta')$ [ $\equiv (\phi_{\rm t},\theta_{\rm t})$]. The projection of a general point $(x,y,z)$ onto a torus will be such that the distance between the point and the centre of the torus cross section (corresponding to the point it is projected to) is minimised.} 
  \label{f:gns_torus}
  \end{center}
\end{figure*}

\begin{figure}
  \begin{center}
  \includegraphics[ width=0.90\linewidth]{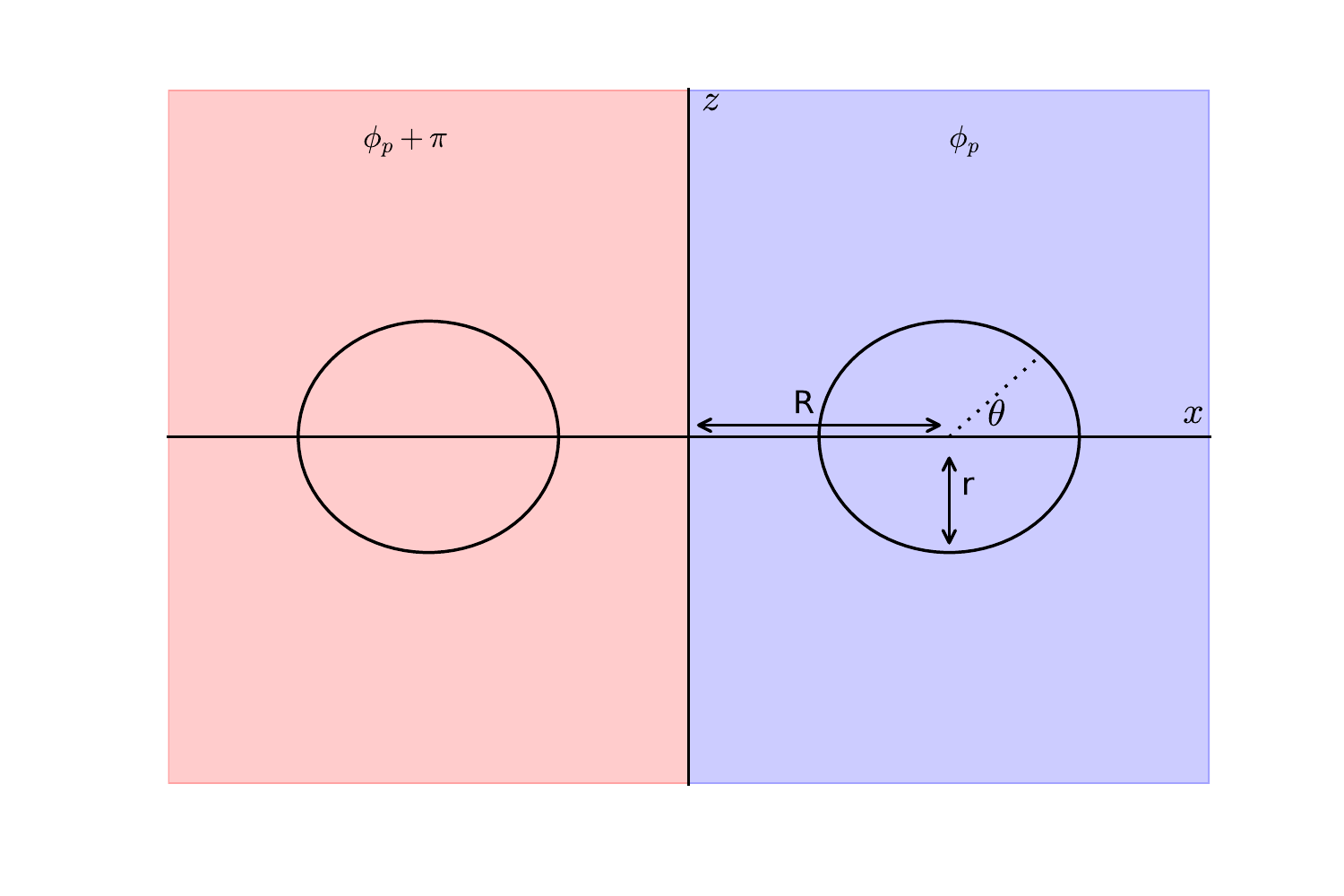}
  \caption{Torus cross section at $\phi = \phi_p$ and $\phi = \phi_p + \pi$ in $x$--$z$ plane. If the sampled point lies in the half-plane (shaded blue) defined by $\phi = \phi_p$, it will be projected onto the circle in this half-plane, otherwise it will be projected onto the circle in the $\phi = \phi_p + \pi$ half-plane (shaded pink).} 
  \label{f:gns_torus_cross}
  \end{center}
\end{figure}

In the case of two parameters representing points on a circle e.g. $\phi \in [0, 2\pi]$ and $\theta \in [0, 2\pi]$, either we can apply separate circular coordinate transformations to each parameter, or we can say that together they parameterise points on the surface of a torus (Figure~\ref{f:gns_torus}). In the latter case $\phi$ and $\theta$ can be expressed in terms of Cartesian coordinates through 
\begin{equation}
\label{e:gns_toruscoords}
\begin{split}
&x = (R + r \cos(\theta))\cos(\phi), \\
&y = (R + r \cos(\theta))\sin(\phi), \\
&z = r \sin(\theta),
\end{split}
\end{equation}
where: $R$ is the distance from the centre of the tube to the centre of the torus and $r$ is the radius of the tube; $\phi$ is the angle between the positive $x$-axis and the line from the centre of the torus to the point $(x,y)$, measured anti-clockwise; and $\theta$ is the angle between (a) the line in the $x-y$ plane pointing `outwards relative to the centre of the torus' from the centre of the tube, and (b) the line from the centre of the tube to point $(x,y,z)$ (also measured anti-clockwise). \\
In the case of a torus the Cartesian sampling has an additional complication compared with the circular and spherical cases with regards to $q(\phi_{\rm t}, \theta_{\rm t} | \phi_l, \theta_l)$ being symmetric in its arguments.\\
If we first restrict our thinking to the two-dimensional half-plane defined by $\phi = \phi_p$ for arbitrary $\phi_p$, the torus maps out a circle with radius $r$ at a distance $R$ from the origin (note that this is just the cross-section of the torus at $\phi = \phi_p$, see Figure~\ref{f:gns_torus_cross}). If we consider sampling (in two dimensions) from a point on this circle, if the sampled point is at $\theta' = \pi$ and the distance between this point and the centre of the circle is $> R$, then the sampled point is not on the half-plane corresponding to $\phi_p$ but is instead on the one defined by $\phi = \phi_p + \pi$. Consequently when the trial point is projected back onto the torus, it is projected onto a point corresponding to $\phi = \phi_p + \pi$. This implies that there is an asymmetry in the probability of sampling a point which is projected onto the part of the torus corresponding to $\pi / 2 < \theta \leq 3 \pi / 2$ relative to sampling a point which projects onto the part corresponding to $0 \geq \theta \leq \pi / 2$ plus $3 \pi / 2 < \theta \leq 2 \pi$; the probability of picking a point in the region given by the latter is higher for an unrestricted trial distribution since the half-plane extends out to infinity. This can be avoided by restricting the range in which $(x',y',z')$ is sampled from such that the shortest distance between the point $(x',y',z')$ and the centre of the tube of the torus is $\leq R$. This ensures that for a symmetric $q$ the probability of sampling a point from the range $\pi / 2 < \theta \leq 3 \pi / 2$ is the same as from $0 \geq \theta \leq \pi / 2$ plus $3 \pi / 2 < \theta \leq 2 \pi$, and thus $q(x',y',z'|x,y,z)$ is symmetric in its arguments for fixed $\phi$. \\
A similar thought experiment can be applied to the case when $\theta$ is fixed and $\phi$ is allowed to vary. For arbitrary $\theta$ this maps out two-dimensional surfaces in the three-dimensional sampling space, for which the restricted sampling stated above results in $q$ being symmetric in its arguments as long as $\theta$ remains fixed. \\
When varying $\phi$ and $\theta$ simultaneously during (three-dimensional) sampling (as you would in the real implementation of the algorithm) there is no trivial way to truncate the trial distribution to ensure $q(x',y',z'|x,y,z)$ is symmetric in its arguments. Thus one is required to evaluate the set of integrals given by~\ref{e:gns_carttrials} (but over integration domains which satisfy~\ref{e:gns_toruscoords} for given $\phi_l, \theta_l$ and $\phi_{\rm t}, \theta_{\rm t}$) to determine $q(\phi_{\rm t}, \theta_{\rm t} | \phi_l, \theta_l)$ and $q(\phi_l, \theta_l | \phi_{\rm t}, \theta_{\rm t} )$. Using the truncated trial distribution (introduced when considering fixed $\phi$) with the variance stated in Section~\ref{s:gns_spherevar} I found that $q(\phi_{\rm t}, \theta_{\rm t} | \phi_l, \theta_l)$ and $q(\phi_l, \theta_l | \phi_{\rm t}, \theta_{\rm t} )$ vary by no more than $\mathcal{O}(10^{-6})$ and on average by $\mathcal{O}(10^{-8})$. 

Given the circular nature of the variables parameterising the points on a circle / torus, I do not think that performing coordinate transformations for these objects will give any advantages over using the wrapped trial distributions in the original parameter spaces. Hence in the applications considered in this thesis, parameters which exhibit circular or toroidal properties will be sampled using the wrapped trial distribution, whilst those of a spherical nature will be sampled using the coordinate transformation methodology.
The coordinate transformation methodology can be applied to arbitrary geometries. However geometries which lack symmetry will in general be much more difficult to sample from without breaking the trial distribution symmetry requirement of the Metropolis acceptance ratio. In this case the Metropolis-Hastings acceptance ratio for nested sampling must be used 
\begin{equation}
\label{e:gns_ns_accept2}
\alpha_{\mathrm{MH}} = \begin{cases}
\mathrm{min}\left[\frac{\pi\left(\vec{\Theta}_{\rm t}\right)q(\vec{\Theta}_{l}| \vec{\Theta}_{\rm t})}{\pi\left(\vec{\Theta}_{l}\right)q(\vec{\Theta}_{\rm t}| \vec{\Theta}_{l})}, 1\right] \quad \mathrm{ if } \quad \mathcal{L}_{\rm t} > \mathcal{L}_{i}, \\
 0 \quad \mathrm{ otherwise.}
\end{cases}
\end{equation}
One can assume that such unsymmetrical geometries mean the integrals associated with calculating the trial distributions distributions in Euclidean space become non-trivial to evaluate. Failure to evaluate equation~\ref{e:gns_ns_accept2} correctly would likely lead to violation of detailed balance which is a \textit{sufficient} condition for a Markov chain to asymptotically converge to the target distribution.

\section{Applications of geometric nested sampling}
\label{s:gns_applications}

I now apply the geometric nested sampling algorithm to models which include circular, toroidal and spherical parameters. I evaluate the algorithm's performance by plotting the posterior samples using \textsc{corner}. I also conduct the analysis with the `vanilla' Metropolis nested sampling algorithm. For circular and toroidal parameters, the vanilla algorithm doesn't use a wrapped trial distribution. In the case of spherical parameters, the vanilla algorithm does not transform to Cartesian coordinates before sampling from the trial distribution.
For further comparison, I calculate posterior samples using \textsc{MultiNest} \citep{2009MNRAS.398.1601F} (i.e. the algorithm I have used for all Bayesian inferences done in the preceeding Chapters), a state of the art clustering nested sampling algorithm, effective in low dimensional problems. \\
I refer to the samples / distributions obtained from the geometric nested sampler as MG (Metropolis geometric nested sampling), those obtained from the vanilla Metropolis nested sampler as M, and those obtained from \textsc{MultiNest} as MN. \\
For all applications I run the algorithms twice, once with a low number of livepoints ($50$), and once with a high number of livepoints ($500$). 

\subsection{Toy model I: circular distribution}
\label{s:gns_tmI}
I first consider the problem of a one-dimensional circular distribution from which we would like to sample from. The model is parameterised by one variable $\phi$, which is defined on $[0, 2\pi]$. Referring back to Section~\ref{s:gns_priors} I take $\pi(\phi)$ to be uniform on $[0, 2\pi]$. For the likelihood function, I use the von Mises distribution introduced in Section~\ref{s:gns_circular} and defined by 
\begin{equation}
\label{e:gns_circleL}
\mathcal{L}\left(\phi | \mu, \sigma^2\right) = \frac{\exp(\cos(\phi - \pi - \mu)/ \sigma^2)}{2\pi I_{0}\left(\frac{1}{\sigma^2}\right)},
\end{equation}
where $\mu$ and $\sigma$ are the mean and standard deviation of the distribution, and $I_{0}(x)$ is the zeroth order modified Bessel function. Here I set $\mu = 0$ so that the peak of the posterior distribution is wrapped around $[0, 2\pi]$, and appears as two half peaks. I set the variance equal to $0.25$. \\
Since the problem involves the circular parameter $\phi$, the geometric nested sampling algorithm uses a wrapped trial distribution.

\subsubsection{Low livepoint runs}
\label{s:gns_tmI_l}

\begin{figure}
  \begin{center}
  \includegraphics[ width=0.90\linewidth]{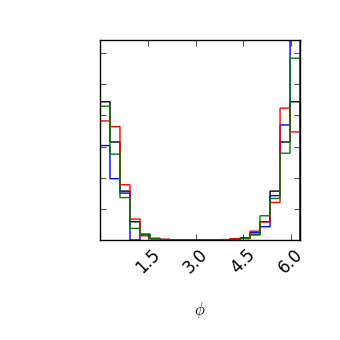}
  \caption{Posterior distributions of the circular toy model defined in Section~\ref{s:gns_tmI}. The black curve corresponds to samples obtained from the analytical expression for $\mathcal{P}(\phi)$ evaluated over a range of $\phi$ values. The blue, red and green curves correspond to the samples obtained from the M, MG and MN algorithms respectively. All three samplers were run with $50$ livepoints.}\label{f:gns_l_c_post}
  \end{center}
\end{figure}

Figure~\ref{f:gns_l_c_post} shows the posterior distribution obtained for toy model I from the three samplers using a low number of livepoints. Note that the Figure also includes a curve plotted from samples which were obtained by evaluating the posterior distribution analytically over a uniform range of $\phi$ values. I refer to this curve as the theoretical (T) result.
The three samplers obtain similar results in the central bins where the probability density is low. However the distributions become asymmetric towards the edges of the domain when compared with the T curve. Overall the MG and MN samplers marginally outperform the M sampler, given the latter has a large asymmetry between the first ($\phi \approx 0$) and final ($\phi \approx 2\pi$) bins.

\subsubsection{High livepoint runs}
\label{s:gns_tmI_h}

\begin{figure}
  \begin{center}
  \includegraphics[ width=0.90\linewidth]{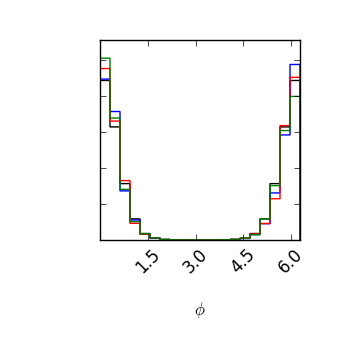}
  \caption{Posterior distributions of the circular toy model with the number of livepoints set to $500$. The colour coding of the plot is as described in Figure~\ref{f:gns_l_c_post}.}\label{f:gns_h_c_post}
  \end{center}
\end{figure}

Figure~\ref{f:gns_h_c_post} shows the results when a high number of livepoints is used for the nested sampling algorithms. The plot shows that all three algorithms do a much better job of replicating the T curve than when they were used with a low number of livepoints, with the MG samples giving the curve most similar to the T result.

\subsection{Toy model II: toroidal distribution}
\label{s:gns_tmII}
I next consider a two-dimensional problem where each parameter is circular. I refer to this as a toroidal model, as it is equivalent to sampling from the surface of a torus parameterised by two angles $\phi$ and $\theta$. We take both $\pi(\phi)$ and $\pi(\theta)$ to be uniform on $[0, 2\pi]$. For the likelihood function, I again use the von Mises distribution, and take the likelihood functions for $\phi$ and $\theta$ to be independent so that 
\begin{equation}
\label{e:gns_torusL}
\mathcal{L}\left(\phi, \theta | \mu_{\phi}, \sigma_{\phi}^2, \mu_{\theta}, \sigma_{\theta}^2\right) = \mathcal{L}\left(\phi | \mu_{\phi}, \sigma_{\phi}^2\right) \mathcal{L}\left(\theta | \mu_{\theta}, \sigma_{\theta}^2\right),
\end{equation}
where the likelihood for each individual parameter takes the form of equation~\ref{e:gns_circleL}. I set $\mu_{\phi} = \mu_{\theta} = 0$ so that the two-dimensional posterior contains four `quarter peaks' at the corners of its domain. I also take $\sigma_{\phi}^2 = \sigma_{\theta}^2 = 0.25$. \\
Since this model represents a toroidal distribution (or two circular distributions), the geometric nested sampling algorithm uses wrapped trial distributions to sample $\phi$ and $\theta$.

\subsubsection{Low livepoint runs}
\label{s:gns_tmII_l}

\begin{figure}
  \begin{center}
  \includegraphics[ width=0.90\linewidth]{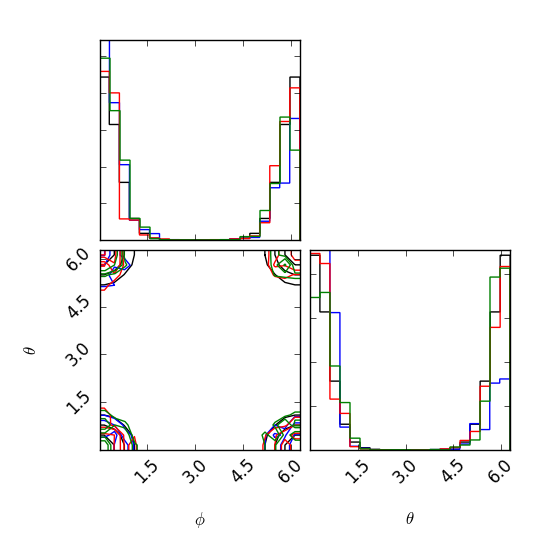}
  \caption{Posterior distributions of the toroidal toy model defined in Section~\ref{s:gns_tmII}, with the number of livepoints set to $50$. The colour coding is as described in Figure~\ref{f:gns_l_c_post}. The plots along the diagonal show the one-dimensional marginalised posteriors for $\phi$ and $\theta$. The centre plot shows the joint two-dimensional posterior.}\label{f:gns_l_t_post}
  \end{center}
\end{figure}

Figure~\ref{f:gns_l_t_post} shows the posterior distributions obtained for toy model II from the three samplers using a low number of livepoints. As in Section~\ref{s:gns_tmI}, samples of the analytical posterior are included for comparison. Looking at the one-dimensional marginalised posteriors for $\phi$ and $\theta$, the M algorithm does a poor job at recovering the true distribution, overestimating the half peaks at low values of $\phi,\theta$ and overestimating them at high $\phi,\theta$. The MG algorithm does a relatively good job of replicating the T distribution, and looking at the marginalised posteriors, outperforms MN at three of the four half peaks (MN does better at the $\theta \approx 0$ peak). One may expect MN to struggle with such a distribution, using a low number of livepoints. Since the four quarter peaks will appear to a clustering algorithm as four separate peaks, MN will on average assign $12.5$ livepoints to each of these peaks, which may not be enough to sample each peak adequately. The MG algorithm on the other hand treats these four quarter peaks as one, and so you would expect it to be able to use all $50$ livepoints to sample this peak relatively well.

\subsubsection{High livepoint runs}
\label{s:gns_tmII_h}

\begin{figure}
  \begin{center}
  \includegraphics[ width=0.90\linewidth]{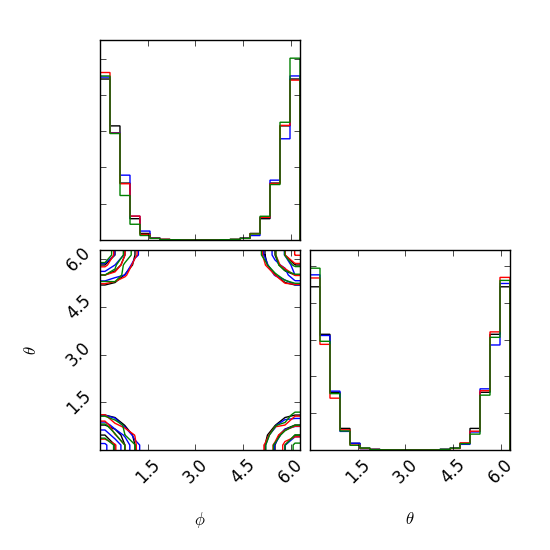}
  \caption{Posterior distributions of toroidal toy model with the number of livepoints set to $500$. The colour coding and layout of the plots is as explained in Figure~\ref{f:gns_l_t_post}.}\label{f:gns_h_t_post}
  \end{center}
\end{figure}

The high livepoint run results for the toroidal distribution are shown in Figure~\ref{f:gns_h_t_post}. All three samplers recover the true distribution well, with the M and MG giving marginally better results than MN. This is perhaps surprising since one would expect MN to easily be able to cope with four modes using 500 livepoints. It does however, highlight the possibility that it is not the number of peaks that MN is struggling with, it is their shape that is causing it to underperform relative to the other two samplers.

\subsection{Toy model III: spherical distribution}
\label{s:gns_tmIII}
For the final toy model I consider the posterior distribution of two angles which parameterise the surface of a sphere. As in Section~\ref{s:gns_spheretrans}, $\phi$ and $\theta$ represent the azimuthal and zenith angles respectively. I take $\pi(\phi)$ to be uniform on $[0, 2\pi]$, and $\pi(\theta)$ to be uniform on $[0, \pi]$. I use a von Mises distribution for $\mathcal{L}\left(\phi | \mu_{\phi}, \sigma_{\phi}^2\right)$ with $\mu_{\phi} = 0$ and $\sigma_{\phi}^2 = 0.25$. For $\mathcal{L}\left(\theta | \mu_{\theta}, \sigma_{\theta}^2\right)$ I use a truncated Gaussian (defined on $[0, \pi]$) with $\mu_{\theta} = \pi /2$ and $\sigma_{\theta}^2 = 0.25$. \\
For this model the geometric nested sampling algorithm uses the spherical transformation sampling procedure detailed in Section~\ref{s:gns_spheretrans} to sample $\phi_{\rm t}$ and $\theta_{\rm t}$.

\subsubsection{Low livepoint runs}
\label{s:gns_tmIII_l}

\begin{figure}
  \begin{center}
  \includegraphics[ width=0.90\linewidth]{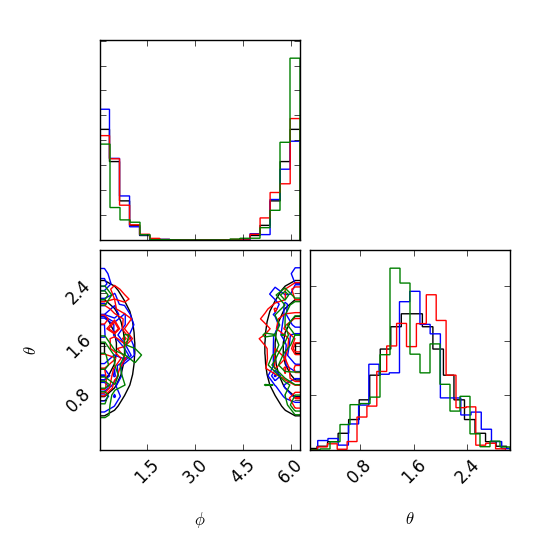}
  \caption{Posterior distributions of the spherical toy model with the number of livepoints set to $50$. The colour coding and layout of the plots is as explained in Figure~\ref{f:gns_l_t_post}.}\label{f:gns_l_s_post}
  \end{center}
\end{figure}

Figure~\ref{f:gns_l_s_post} shows the posterior distributions obtained for toy model III from the three samplers using a low number of livepoints, plus the T samples. The circular distribution of $\phi$ is well recovered by the M and MG algorithms, but less so by MN. All three samplers do a relatively poor job of recovering the truncated Gaussian distribution of $\theta$, with M probably giving the best results due to the symmetry of its distribution.

\subsubsection{High livepoint runs}
\label{s:gns_tmIII_h}

\begin{figure}
  \begin{center}
  \includegraphics[ width=0.90\linewidth]{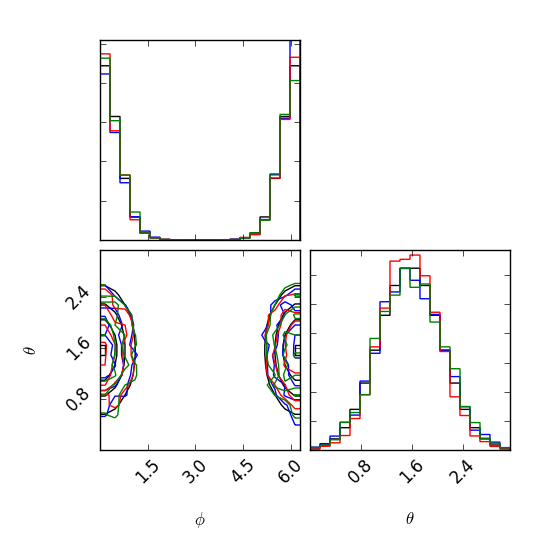}
  \caption{Posterior distributions of the spherical toy model with the number of livepoints set to $500$. The colour coding and layout of the plots is as explained in Figure~\ref{f:gns_l_t_post}.}\label{f:gns_h_s_post}
  \end{center}
\end{figure}

When $500$ livepoints are used for the samplers (Figure~\ref{f:gns_h_s_post}), the MG and MN algorithms recover the $\phi$ profile similarly well. 
However, the MG sampler seems to slightly overestimate $\mathcal{P}(\theta)$ at high probability densities, and underestimate it to a similar extent at low densities. 

\subsection{Practical example: gravitational wave emission from binary black hole mergers}
\label{s:gns_ligo}

I now consider a likelihood function which corresponds to detecting gravitational waves from (binary) black hole mergers. The data for the likelihood are obtained from the LIGO\footnote{\url{https://www.ligo.caltech.edu/page/ligo-gw-interferometer}.} and Virgo\footnote{\url{http://www.virgo-gw.eu/}.} interferometers (see e.g. \citealt{2016PhRvL.116f1102A} and \citealt{2018AAS...23132501L}). 
I now give a brief overview on gravitational waves and how they are detected, but for more thorough analysis see e.g. \citet{Hobson2006}, \citet{2008RvMA...20..140K}, or \citet{2014LRR....17....2B}.

\subsubsection{Origin of gravitational waves}
\label{s:gns_gw_orig}
For an observer lying in a region of spacetime satisfying the Minkowski metric $\eta_{\mu, \nu}$, fluctuations in the metric can be described by a linear perturbation
\begin{equation}
\label{e:gns_metric}
g_{\mu \nu} = \eta_{\mu \nu} + h_{\mu \nu}, 
\end{equation}
where the perturbations are assumed to be small ($|h_{\mu \nu} \ll 1|$). By solving Einstein's field equations using the metric given by equation~\ref{e:gns_metric}, it can be shown that the tensor which represents the gravitational field
\begin{equation}
\label{e:gns_tensor}
\tilde{h}_{\mu \nu} = h_{\mu \nu} - \frac{1}{2} \eta_{\mu \nu} h_{\alpha}^{\alpha}, 
\end{equation}
satisfies the wave equation for a vacuum, and hence has a solution
\begin{equation}
\label{e:gns_wave}
\tilde{h}_{\mu \nu} = A_{\mu \nu} \exp(ik^{\alpha}x_{\alpha}), 
\end{equation}
where $A_{\mu \nu}$ describe the wave's polarisation and amplitude.
It can be shown that by setting an appropriate gauge (the Transverse-Traceless gauge) that $A_{\mu \nu}$ can be defined in terms of two polarisation states $h_{+}$ and $h_{\times}$. For a gravitational wave travelling in the $z$ direction, the tensor $h_{+}$ causes simultaneous expansion (contraction) in the $x$ direction and contraction (expansion) in the $y$ direction. $h_{\times}$ acts similarly at an angle $\pi/4$ to the $x$--$y$ axes.  

Exact solutions of Einstein's field equations have not yet been found, leading to the development of analytic approximations such as the Post-Newtonian (PN) approximation (see e.g. \citealt{1997PThPS.128..123A}) to determine $h_{+}$ and $h_{\times}$. Here we consider the PN approximation up to second order for inspiralling black hole binary systems as described in \citet{1996CQGra..13..575B}.

\subsubsection{Detection of gravitational waves}
\label{s:gns_gw_det}
Laser beam interferometers such as LIGO and Virgo detect gravitational waves by measuring the differential arm length between perpendicular arms of the interferometers. The differential measured is proportional to the gravitational strain $h$, which describes the fractional change in proper space caused by the gravitational perturbation. $h$ can be written as a linear combination of the two polarisation states $h_{+}$ and $h_{\times}$
\begin{equation} 
\label{e:gns_h_lin}
h(t)=F_{+} h_{+}(t) + F_{\times} h_{\times}(t),
\end{equation} 
where $t$ denotes the time at which the strain is measured, and $F_{+}$ \& $F_{\times}$ are functions dependent on the geometry of the detector. Here we consider three detectors: LIGO Hanford, LIGO Livingston and Virgo. The geometries used in this analysis for these detectors can be found at \url{https://www.ligo.org/scientists/GW100916/GW100916-geometry.html}.

\subsubsection{Likelihood function for gravitational wave detection}
\label{s:gns_gw_like}
Assuming we have $n_{d}$ data points $\{x_{i,j}\}$ recorded at times $\{t_i\}$ for each detector $j$, then the likelihood function is given by 
\begin{equation}
\label{e:gns_h_like}
\mathcal{L}\left(\vec{\Theta}\right) = \prod_{i=1}^{n_{d}} \prod_{j=1}^{3} \frac{1}{\sqrt{2\pi\sigma}}\exp \left( - \frac{\left(x_{i,j} - h_{j}\left(t_i, \vec{\Theta}\right)\right)^2}{2\sigma^2}  \right),
\end{equation}
where $h_{i,j}\left(t_i, \vec{\Theta}\right)$ is the theoretical strain and is dependent on the model parameters $\vec{\Theta}$ (defined below). In the analysis presented here we consider data which are simulated by evaluating $h_{j}\left(t_i, \vec{\Theta}\right)$ for fixed model parameters (say $\vec{\theta_{\rm{r}}}$), i.e. we set  
\begin{equation}
\label{e:gns_h_sim}
x_{i,j} \equiv h_{j,\vec{\theta_{\rm{r}}}}(t_i).
\end{equation}
Furthermore, we set $\sigma = 1 \times 10^{-21}$ and $n_{d} = 1000$. 

\subsubsection{Model parameters}
\label{s:gns_h_pars}
$\vec{\Theta}$ is a nine-dimensional vector with components
\begin{equation}
\label{e:gns_h_pars}
\vec{\Theta} = (m_1, m_2, r, t_{\rm c}, \phi_{\rm c}, \phi, \theta, p, i)
\end{equation}
here $m_{1}$ and $m_{2}$ are the masses of the individual black holes, $r$ is the luminosity distance to the centre of the binary system, and $t_{\rm c}$ is the time of coalescence of the two black holes (i.e. the time at which they merge). $\phi_{\rm c}$ is the orbital phase of the binary system at time $t_{\rm c}$ (and is defined on $[0, 2\pi]$), and $\phi$ \& $\theta$ are the angular location of the merger system in the sky (as observed from a detector). The inclination angle $i$ is the angle between the line of sight from the binary system to a detector, and the normal to the orbital plane. The normal is chosen to be right-handed with respect to the sense of motion so that $i$ is defined on $[0, \pi]$. $p$ is the corresponding azimuthal angle as observed from the binary system.
Table~\ref{t:gns_h_params} gives the values of these parameters used in the simulated data, and how they are sampled using the geometric nested sampler. Notice that I only vary the angular parameters ($\phi_{\rm c}, \phi, \theta, p, i $) in the Bayesian analysis, making it a five-dimensional parameter estimation problem. All five parameters are assigned uniform priors over the ranges they are defined on.

Referring back to equation~\ref{e:gns_h_like}, the time values $t_i$ are spaced uniformly between $t_{\rm c} - t$ and $t_{\rm c} + t$, where
\begin{equation}
\label{e:gns_ligo_t}
t = \frac{1000 G (m_1 + m_2)}{c^3}.
\end{equation}
Here $G$ is Newton's gravitational constant and $c$ is the speed of light in a vacuum.
\begin{table*}
\begin{center}
\begin{tabular}{{l}{c}{c}}
\hline
Parameter & Simulation input value & Sampling procedure \\ 
\hline
$m_1$ & $35~M_{\mathrm{Sun}}$ & fixed \\
$m_2$ & $25~M_{\mathrm{Sun}}$ & fixed \\
$r$ & $390~\mathrm{Mpc}$ & fixed \\
$t_{\rm c}$ & $0$ & fixed \\
$\phi_{\rm c}$ & $0$ & circular (wrapped trial distribution) \\
$\phi$ & $0$ & spherical coordinate transformation (azimuthal angle) \\
$\theta$ & $\pi/2$ & spherical coordinate transformation (zenith angle) \\
$p$ & $0$ & spherical coordinate transformation (azimuthal angle) \\
$i$ & $\pi/2$ & spherical coordinate transformation (zenith angle) \\
\hline
\end{tabular}
\caption{Gravitational wave detection model parameters, their simulation input values, and how the parameters are sampled by the geometric nested sampler. The parameters `fixed' sampling procedures were not sampled from, instead their true (simulation input) value was used in each evaluation of $h_{j}(t_i, \vec{\Theta})$. $\phi_{\rm c}$ is interpreted as a circular quantity by the geometric nested sampler, and the pairs of angles $(\phi,\theta)$, \& $(p,i)$ are treated as two independent sets of spherical coordinates (and thus are transformed independently).}
\label{t:gns_h_params}
\end{center}
\end{table*}

\subsubsection{Posterior sampling}
\label{s:gns_h_post}
For the toy models I calculated the posterior distributions analytically over uniform grids so that I could benchmark the sampling algorithms' performance with the `true' distributions. However, since we are sampling from a five-dimensional parameter space in this example, obtaining samples analytically is no longer feasible. We thus run the MN algorithm with a very high number of livepoints ($2000$) and refer to this as the mega MultiNest run (MMN). We use the MMN result as a reference distribution for our low and high livepoint runs of the MG and MN algorithms (we do not include the M algorithm in our comparison here). I note however, in the toy model applications I found evidence to suggest that MN struggles recovering quarter / half peaks even with $500$ livepoints, and when sampling from low dimensional \& low number of mode models. Thus I can make no guarantees that the MMN distribution is the `true' posterior distribution.

\subsubsection{Low livepoint run}
\label{s:gns_h_l_lp}

\begin{figure*}
  \begin{center}
  \includegraphics[ width=\textwidth]{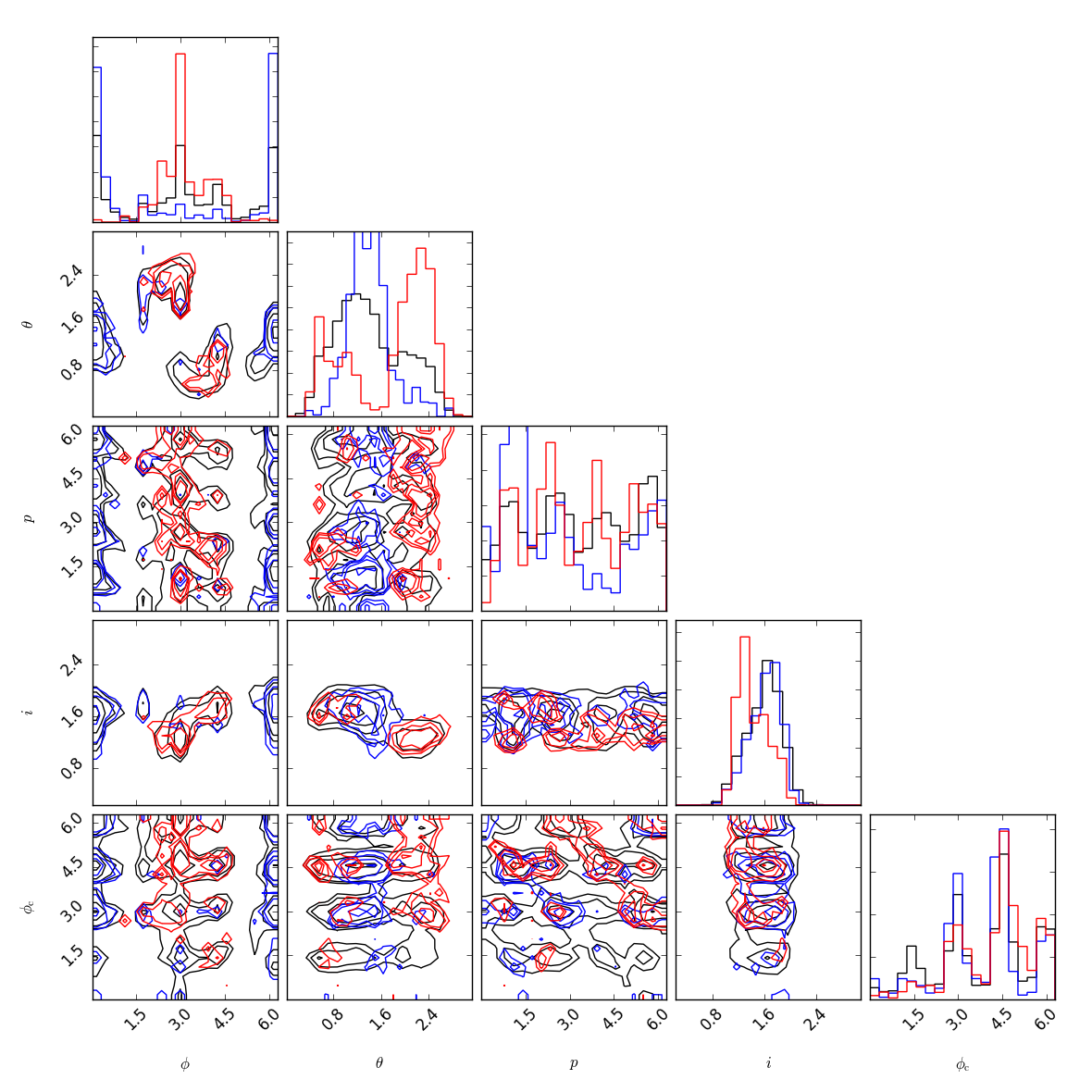}
  \caption{Marginalised one- and two-dimensional posterior distributions for the five angular parameters, $\phi_{\rm c},\phi, \theta, p$, and $i$. The black curves are the results from the $2000$ livepoint MN run. The blue and red curves are plotted using the samples of the MG and MN algorithms respectively, which are obtained from runs with $50$ livepoints.}\label{f:gns_ligo_l_angles}
  \end{center}
\end{figure*}

Figure \ref{f:gns_ligo_l_angles} shows the posterior distributions for the angular parameters obtained from the low livepoint run. Looking at the one-dimensional posterior for $\phi$, it is clear that MG picks up on the two half peaks at $0$ and $2\pi$, but overestimates them compared to the values obtained with MMN. It also underestimates the middle peak ($\phi \approx \pi$) compared to MMN. In fact, one could argue that it doesn't really infer this peak at all. The MN run does the opposite, it overestimates the middle peak, but completely misses the half peaks.
Looking at $\theta$, MG finds a peak around $\theta = \pi/3$, whereas MMN puts the peak at slightly lower $\theta$. The MMN curve shows a flat, high probability density region around $\theta = 2\pi/3$, but MG misses this. The MN run puts the biggest peak at $\theta \approx 2\pi/3$, and a smaller one at $\theta \approx 0.75$.
Both MG and MN do a relatively poor job at constraining $p$ correctly, as the former misses the fourth peak present on the MMN curve at $p \approx 3\pi/2$ (and instead overestimates the first peak at $\approx \pi/2$). The MN algorithm more or less gets the correct number of peaks when compared with MMN, but systematically gets their shape wrong.
MG does a better job than MN in recovering the distribution of $i$ relative to MMN.
MG and MN recover similar profiles for $\phi_{\mathrm{c}}$, and roughly get the shape of the distribution correct when comparing with the MMN result.

\subsubsection{High livepoint run}
\label{s:gns_h_h_lp}

\begin{figure*}
  \begin{center}
  \includegraphics[ width=\textwidth]{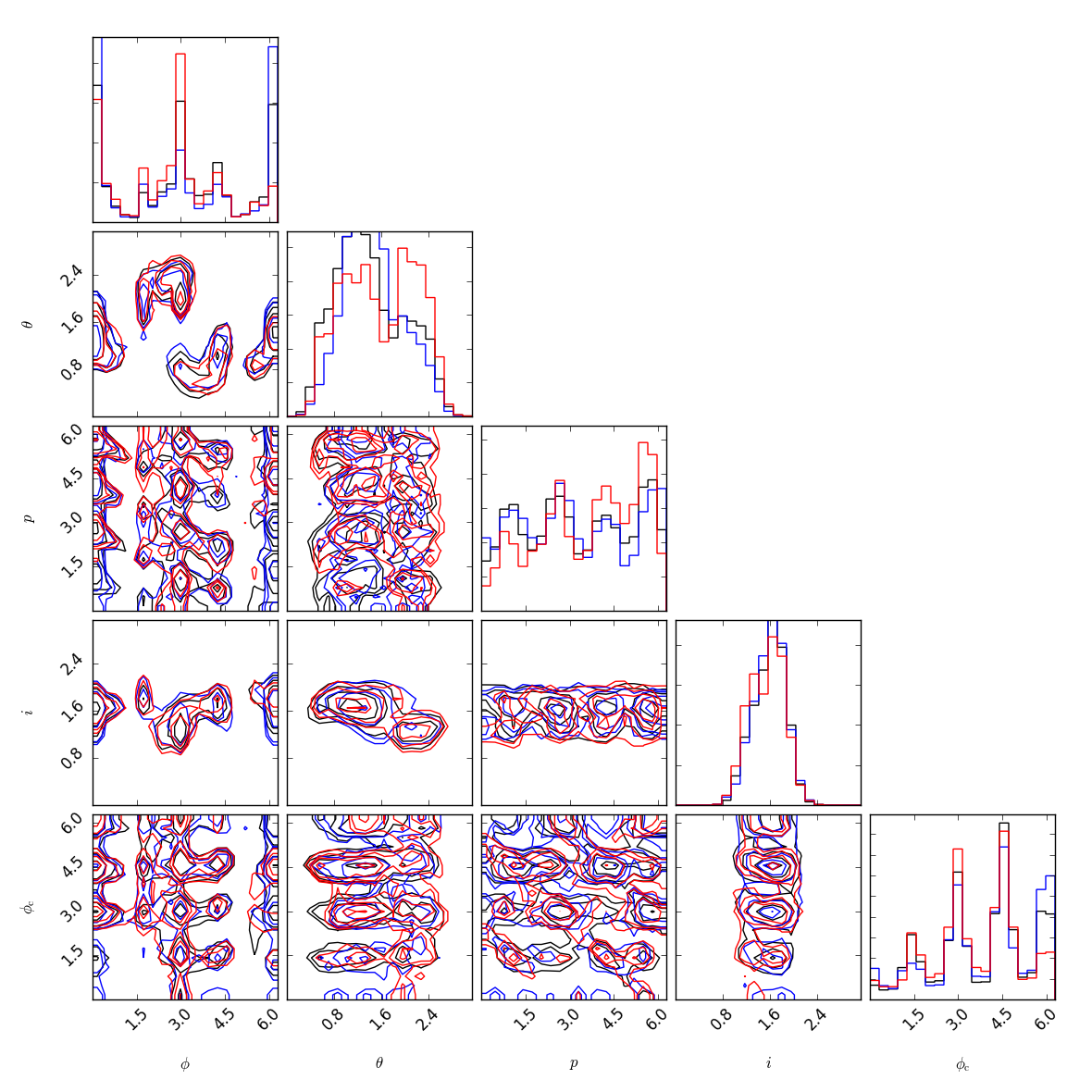}
  \caption{Same plot as Figure~\ref{f:gns_ligo_l_angles} but the blue and red curves show the MG and MN runs with $500$ livepoints.}\label{f:gns_ligo_h_angles}
  \end{center}
\end{figure*}

Figure \ref{f:gns_ligo_h_angles} shows the posterior distributions for the angular parameters obtained from the $500$ livepoint run. In this case MG and MN do a reasonable job of recovering the MMN profile for $\phi$, but still underestimate / overestimate in the same way they did in the low livepoint case. For $\theta$ MG does a good job at replicating the MMN result. MG and MN have similar levels of success in recovering the MMN profiles of $p$, $i$ and $\phi_{\rm c}$.

Overall the MG algorithm performs well relative to MN for the example considered here, given the relative simplicity of the algorithm. To make a statement on which algorithm obtained more accurate inferences of $\mathcal{P}$ for this multi-modal ($\mathcal{O}(10)$ modes), five-dimensional distribution, I believe that a more thorough comparison than the visual inspection conducted here is required. One possible solution to this would be to calculate a distance metric between the MG and MMN posteriors (e.g. Earth mover's distance or the Kullback-Leibler divergence) and compare it with the corresponding value between the MN and MMN distributions. However, I do not make this comparison here. It would also be interesting to see if anyone else has acquired results for this set of simulations, I do not consider this here, however.

\section{Geometric nested sampling implementation}
\label{s:gns_imp}
The implementation of the geometric nested sampler (and the vanilla Metropolis nested sampler) used in this paper, along with the toy models and the gravitational wave likelihood function can be found at \url{https://github.com/SuperKam91/nested_sampling} \citet{javid2020geometric}. The algorithm is written in \textsc{Python 2.7}, hence our \textit{implementation} of the algorithm cannot match that of the state of the art nested sampling algorithms such as \textsc{MultiNest} or \textsc{POLYCHORD} \citep{2015MNRAS.453.4384H}. These algorithms are implemented in \textsc{FORTRAN 90}, and parallelised using a master-slave paradigm (see Section~5.4 of Handley, Hobson, \& Lasenby). Nevertheless there is no reason why geometric nested sampling cannot be implemented more efficiently and parallelised using this method. Furthermore as already mentioned in Section~\ref{s:gns_m_lhood_const}, \citet{2008MNRAS.384..449F} incorporate the Metropolis likelihood sampler into a livepoint clustering algorithm. This same idea could be applied to the geometric nested sampling algorithm. However, in the case of circular parameters, the clustering would also need to be wrapped around the domain of $\mathcal{P}$ along with the trial distribution. This could be avoided by instead performing coordinate transformations (Section~\ref{s:gns_nonspheretrans}) for circular and toroidal parameters before sampling from the ellipsoids resultant from the livepoint clustering. The clustering could be performed in either the original parameter or the transformed Euclidean space, but it is important to note that in either case samples could still be automatically rejected if they lie outside the ellipsoid. Nevertheless the algorithm would still provide the benefit of sampling in the `natural' topology of the problem as discussed in Section~\ref{s:gns_spheretrans}. 

\section{Conclusions}
\label{s:gns_conc}

I have presented a new nested sampling algorithm based on the Metropolis nested sampler proposed in \citet{Sivia2006} and applied in \citet{2008MNRAS.384..449F}. The algorithm exploits the geometric properties of certain kinds of parameters which describe points on circles, tori and spheres, to sample the parameters more efficiently in the context of nested sampling. The algorithm should be more mobile in sampling distributions defined on such geometries. \\
The algorithm consists of two key sampling modes which can be summarised as follows.
\begin{itemize}
\item For circular and toroidal problems, the trial distribution used in the sampling process is wrapped around the support of the prior distribution $\pi$ (domain of the posterior distribution $\mathcal{P}$).
\item This wrapping ensures that no trial points are automatically rejected when evaluating the Metropolis acceptance ratio as a consequence of the point being outside the sampling space of the model.
\item The wrapped trial distribution also makes the sampling more mobile at the edges of the domain of $\mathcal{P}$, meaning that circular and toroidal distributions should be easier to sample, particularly in the case of posteriors with high probability densities at these edges.
\item For spherical problems, parameters specifying the coordinates on a sphere are transformed to Cartesian coordinates and sampled from the corresponding Euclidean space.
\item This again ensures that no trial points are automatically rejected because they are outside the domain of $\mathcal{P}$.
\item It also enhances the mobility of the sampler, whose average step size along the surface of the sphere is not dependent on the location at which the trial distribution is centred. 
\end{itemize}
I applied the geometric nested sampling algorithm (MG) to three toy models, which respectively represented models on a circle, torus and a sphere. I compared the posterior plots with those obtained from a `vanilla' Metropolis nested sampler (M) based on the one used in \citet{2008MNRAS.384..449F}, and with the distributions obtained with the livepoint clustering nested sampling algorithm \textsc{MultiNest} (MN, \citealt{2009MNRAS.398.1601F}). For each model, all three samplers were run twice, once with a low number of livepoints ($50$), and once with a high number of livepoints ($500$). I included the distributions obtained from evaluating $\mathcal{P}$ analytically as means of reference to the `correct' distribution (T). The results can be summarised as follows.
\begin{itemize}
\item For the low livepoint run on the circular toy model (von Mises distribution centred on the origin), the MG and MN samplers marginally outperform the M sampler.
\item For the high livepoint run on the circular toy model, all samplers perform similarly, with the MG algorithm giving slightly superior results with respect to the T distribution.
\item The low livepoint run for the toroidal model (two-dimensional von Mises distribution centred on the origin) the MG outperforms both M and MN. One would maybe expect MN to struggle on a four-mode problem with only $50$ livepoints, whereas the MG effectively treats these four modes as one given their location in the domain of $\mathcal{P}$.
\item The high livepoint run gives better results for all three samplers, but the MN distribution seems the least accurate. This highlights the potential issues which clustering algorithms face with modes which occur at the edges of $\mathcal{P}$, independent of the number of livepoints used. 
\item The spherical toy model which consists of a von Mises distribution on the azimuthal angle $\phi$ and a truncated Gaussian on the zenith angle $\theta$ shows that in the case of low livepoint runs, the M algorithm surprisingly performs the best, as it does a better job at recovering the profile of $\theta$ than the MG algorithm.
\item For the $500$ livepoint run the MG and MN algorithms recover $\phi$ similarly well, but the former systematically overestimates the probability density in $\theta$ around its peak, and underestimates it at low densities. 
\end{itemize}
I then applied the MG and MN sampling algorithms to a model representing the detection of gravitational waves generated by binary black hole mergers and detected with the LIGO and Virgo instruments \citep{2018AAS...23132501L}. Using simulated datasets, we obtained inferences of a five-dimensional (all circular / spherical parameters), multi-modal ($\mathcal{O}(10)$ modes) posterior distribution. For this example my `correct' reference distribution was a MN run with $2000$ livepoints. I found the following. 
\begin{itemize}
\item Overall for the low livepoint run, both algorithms struggle to correctly infer all the peaks of the distribution (of the $2000$ livepoint MN run). 
\item However, this is to be expected for MN since it can only attribute $\approx$ a few livepoints to each mode. Furthermore, the locations of the modes, which occur not just at the edges of $\mathcal{P}$, mean that the MG algorithm must also allocate its livepoints separately to different modes, a task which it is not designed to cope well with.
\item With $500$ livepoints the MG algorithm recovers all the modes inferred from the $2000$ livepoint MN run. MN performs similarly well, but slightly overestimates the number of modes; further quantitative work is needed to home in on this. 
\end{itemize}


%% file: APP-A/appendixa.tex

\chapter{Appendix A: Results of physical modelling of Planck clusters}
\label{c:appendixa}
\newgeometry{margin=1cm} 
\begin{landscape}
\section{Results table}\label{s:pl_phys_results_Table}
\begin{center}
\begin{small}
\begin{longtable}{llllllllllll}
\caption{Summary of values for final sample of 54 clusters. The redshift types correspond to S: spectroscopically measured and P: photometrically measured. $z$, $M(r_{200})$, $x_{\rm c}$, $y_{\rm c}$ and $f_{\rm gas}(r_{200})$ are the physical model sampling parameters. 
$M_{\rm AMI}(r_{500})$,  $M_{\rm Pl, marg} (r_{500})$ and $M_{\rm Pl, slice} (r_{500})$ are the $M(r_{500})$ estimates obtained from the AMI and Planck data respectively. All masses are given in units of $\times 10^{14}~M_{\rm{Sun}}$ and all cluster centre coordinates are measured in arcseconds. }\label{t:pl_phys_results} \\

\hline \multicolumn{1}{c}{Row} & \multicolumn{1}{c}{Planck ID} & \multicolumn{1}{c}{Alias} & \multicolumn{1}{c}{$z$} & \multicolumn{1}{c}{$z$ type} & \multicolumn{1}{c}{$M_{\rm AMI}(r_{200})$} & \multicolumn{1}{c}{$x_{\rm c}$} & \multicolumn{1}{c}{$y_{\rm c}$} & \multicolumn{1}{c}{$f_{\rm gas}(r_{200})$} & \multicolumn{1}{c}{$M_{\rm AMI}(r_{500})$} & \multicolumn{1}{c}{$M_{\rm Pl, marg}(r_{500})$} & \multicolumn{1}{c}{$M_{\rm Pl, slice} (r_{500})$} \\ \hline 
\endfirsthead

\multicolumn{12}{c}%
{{\tablename\ \thetable{} -- continued from previous page}} \\
\hline \multicolumn{1}{c}{Row} & \multicolumn{1}{c}{Planck ID} & \multicolumn{1}{c}{Alias} & \multicolumn{1}{c}{$z$} & \multicolumn{1}{c}{$z$ type} & \multicolumn{1}{c}{$M_{\rm AMI}(r_{200})$} & \multicolumn{1}{c}{$x_{\rm c}$} & \multicolumn{1}{c}{$y_{\rm c}$} & \multicolumn{1}{c}{$f_{\rm gas}(r_{200})$} & \multicolumn{1}{c}{$M_{\rm AMI}(r_{500})$} & \multicolumn{1}{c}{$M_{\rm Pl, marg}(r_{500})$} & \multicolumn{1}{c}{$M_{\rm Pl, slice} (r_{500})$} \\ \hline 
\endhead

\tabulinesep=_1mm
\extrarowsep=1mm
\LTcapwidth=\textwidth

1	&	PSZ2G044.20+48.66	&	ACO2142	&	$	0.0894	$	&	S 	&	$	13.49	\pm	2.35	$	&	$	9.14	\pm	18.20	$	&	$	8.80	\pm	15.08	$	&	$	0.13	\pm	0.02	$	&	$	9.25	\pm	1.58	$	&	$	10.81	\pm	0.42	$	&	$	8.76	\pm	^{	0.19	}	_{	0.21	}	$	\\
2	&	PSZ2G053.53+59.52	&	ACO2034	&	$	0.113	$	&	S 	&	$	8.51	\pm	1.28	$	&	$	-1.80	\pm	13.10	$	&	$	19.39	\pm	9.86	$	&	$	0.13	\pm	0.02	$	&	$	5.87	\pm	0.86	$	&	$	5.38	\pm	0.39	$	&	$	5.48	\pm	^{	0.24	}	_{	0.24	}	$	\\
3	&	PSZ2G151.90+11.63	&	CIZAJ0515.3+5845	&	$	0.12	$	&	S 	&	$	5.74	\pm	1.24	$	&	$	67.58	\pm	27.09	$	&	$	68.01	\pm	18.58	$	&	$	0.13	\pm	0.02	$	&	$	3.99	\pm	0.84	$	&	$	4.23	\pm	1.03	$	&	$	3.65	\pm	^{	0.50	}	_{	0.47	}	$	\\
4	&	PSZ2G218.59+71.31	&	ACO1272	&	$	0.137	$	&	S 	&	$	2.70	\pm	0.99	$	&	$	2.82	\pm	25.21	$	&	$	-16.62	\pm	25.98	$	&	$	0.13	\pm	0.02	$	&	$	1.90	\pm	0.68	$	&	$	4.79	\pm	0.80	$	&	$	3.62	\pm	^{	0.30	}	_{	0.30	}	$	\\
5	&	PSZ2G226.18+76.79	&	ACO1413	&	$	0.1427	$	&	S 	&	$	8.19	\pm	1.23	$	&	$	-35.33	\pm	10.98	$	&	$	-1.13	\pm	13.44	$	&	$	0.13	\pm	0.02	$	&	$	5.62	\pm	0.82	$	&	$	6.14	\pm	0.55	$	&	$	5.98	\pm	^{	0.25	}	_{	0.25	}	$	\\
6	&	PSZ2G165.06+54.13	&	ACO990 	&	$	0.144	$	&	S 	&	$	7.80	\pm	1.35	$	&	$	32.43	\pm	13.21	$	&	$	-27.57	\pm	15.52	$	&	$	0.14	\pm	0.02	$	&	$	5.36	\pm	0.90	$	&	$	5.13	\pm	0.51	$	&	$	4.83	\pm	^{	0.28	}	_{	0.29	}	$	\\
7	&	PSZ2G077.90-26.63	&	ACO2409 	&	$	0.147	$	&	S 	&	$	9.09	\pm	1.32	$	&	$	-26.87	\pm	10.89	$	&	$	18.00	\pm	11.85	$	&	$	0.14	\pm	0.02	$	&	$	6.22	\pm	0.88	$	&	$	5.92	\pm	0.58	$	&	$	5.08	\pm	^{	0.27	}	_{	0.27	}	$	\\
8	&	PSZ2G050.40+31.17	&	ACO2259	&	$	0.164	$	&	S 	&	$	5.52	\pm	1.19	$	&	$	35.72	\pm	21.77	$	&	$	9.31	\pm	19.56	$	&	$	0.13	\pm	0.02	$	&	$	3.80	\pm	0.80	$	&	$	4.53	\pm	0.62	$	&	$	4.36	\pm	^{	0.35	}	_{	0.36	}	$	\\
9	&	PSZ2G097.72+38.12	&	ACO2218 	&	$	0.1709	$	&	S 	&	$	10.65	\pm	1.68	$	&	$	31.99	\pm	15.25	$	&	$	-0.95	\pm	13.52	$	&	$	0.13	\pm	0.02	$	&	$	7.23	\pm	1.11	$	&	$	7.44	\pm	0.40	$	&	$	6.64	\pm	^{	0.17	}	_{	0.17	}	$	\\
10	&	PSZ2G099.30+20.92	&	MCXCJ1935.3+6734	&	$	0.171	$	&	S 	&	$	5.57	\pm	1.24	$	&	$	-37.19	\pm	19.92	$	&	$	-24.50	\pm	21.16	$	&	$	0.13	\pm	0.02	$	&	$	3.83	\pm	0.83	$	&	$	5.88	\pm	0.93	$	&	$	3.91	\pm	^{	0.23	}	_{	0.25	}	$	\\
11	&	PSZ2G067.17+67.46	&	ACO1914	&	$	0.1712	$	&	S 	&	$	10.45	\pm	1.49	$	&	$	31.39	\pm	12.81	$	&	$	-33.15	\pm	11.99	$	&	$	0.13	\pm	0.02	$	&	$	7.09	\pm	0.99	$	&	$	7.14	\pm	0.47	$	&	$	7.04	\pm	^{	0.26	}	_{	0.27	}	$	\\
12	&	PSZ2G167.67+17.63	&	RXJ0638.1+4747	&	$	0.174	$	&	S 	&	$	4.78	\pm	1.36	$	&	$	-28.70	\pm	31.24	$	&	$	10.76	\pm	28.64	$	&	$	0.13	\pm	0.02	$	&	$	3.30	\pm	0.92	$	&	$	7.72	\pm	0.81	$	&	$	6.31	\pm	^{	0.33	}	_{	0.34	}	$	\\
13	&	PSZ2G066.68+68.44	&	ACO1902	&	$	0.181	$	&	S 	&	$	4.95	\pm	1.43	$	&	$	56.07	\pm	25.47	$	&	$	8.14	\pm	33.23	$	&	$	0.13	\pm	0.02	$	&	$	3.41	\pm	0.97	$	&	$	5.27	\pm	0.84	$	&	$	3.98	\pm	^{	0.33	}	_{	0.37	}	$	\\
14	&	PSZ2G065.28+44.53	&	ACO2187	&	$	0.183	$	&	S 	&	$	5.24	\pm	1.28	$	&	$	-16.66	\pm	22.61	$	&	$	-16.54	\pm	21.65	$	&	$	0.13	\pm	0.02	$	&	$	3.60	\pm	0.86	$	&	$	3.89	\pm	0.98	$	&	$	3.56	\pm	^{	0.47	}	_{	0.51	}	$	\\
15	&	PSZ2G084.47+12.63	&	MCXCJ1948.3+5113	&	$	0.185	$	&	S 	&	$	4.79	\pm	1.22	$	&	$	-73.73	\pm	31.17	$	&	$	-16.97	\pm	20.93	$	&	$	0.13	\pm	0.02	$	&	$	3.30	\pm	0.82	$	&	$	5.98	\pm	0.65	$	&	$	4.94	\pm	^{	0.33	}	_{	0.34	}	$	\\
16	&	PSZ2G100.04+23.73	&	ACO2317 	&	$	0.21	$	&	S 	&	$	5.44	\pm	1.13	$	&	$	20.24	\pm	19.02	$	&	$	-22.73	\pm	20.90	$	&	$	0.13	\pm	0.02	$	&	$	3.72	\pm	0.75	$	&	$	4.10	\pm	0.80	$	&	$	3.73	\pm	^{	0.29	}	_{	0.31	}	$	\\
17	&	PSZ2G180.60+76.65	&	SDSSCGB26344.3	&	$	0.2138	$	&	S 	&	$	5.38	\pm	1.21	$	&	$	37.81	\pm	15.59	$	&	$	-66.98	\pm	19.41	$	&	$	0.13	\pm	0.02	$	&	$	3.68	\pm	0.81	$	&	$	6.76	\pm	0.75	$	&	$	6.00	\pm	^{	0.35	}	_{	0.34	}	$	\\
18	&	PSZ2G166.09+43.38	&	ACO773N	&	$	0.2172	$	&	S 	&	$	9.84	\pm	1.39	$	&	$	-5.35	\pm	10.66	$	&	$	-3.98	\pm	9.70	$	&	$	0.13	\pm	0.02	$	&	$	6.63	\pm	0.92	$	&	$	7.76	\pm	0.73	$	&	$	6.87	\pm	^{	0.34	}	_{	0.32	}	$	\\
19	&	PSZ2G125.30-27.99	&	N/A	&	$	0.223	$	&	P	&	$	4.51	\pm	1.31	$	&	$	-8.08	\pm	26.99	$	&	$	8.82	\pm	30.24	$	&	$	0.13	\pm	0.02	$	&	$	3.09	\pm	0.87	$	&	$	5.54	\pm	0.98	$	&	$	4.70	\pm	^{	0.56	}	_{	0.55	}	$	\\
20	&	PSZ2G060.13+11.44	&	N/A	&	$	0.224	$	&	S 	&	$	7.47	\pm	1.22	$	&	$	-64.79	\pm	12.50	$	&	$	-49.27	\pm	14.16	$	&	$	0.13	\pm	0.02	$	&	$	5.06	\pm	0.80	$	&	$	7.55	\pm	1.09	$	&	$	5.34	\pm	^{	0.49	}	_{	0.50	}	$	\\
21	&	PSZ2G166.62+42.13	&	ACO746	&	$	0.232	$	&	P 	&	$	3.56	\pm	1.07	$	&	$	-38.98	\pm	29.87	$	&	$	-38.09	\pm	37.84	$	&	$	0.13	\pm	0.02	$	&	$	2.44	\pm	0.72	$	&	$	5.60	\pm	0.71	$	&	$	5.36	\pm	^{	0.39	}	_{	0.41	}	$	\\
22	&	PSZ2G097.94+19.43	&	4C 65.28	&	$	0.25	$	&	S 	&	$	5.01	\pm	1.31	$	&	$	-114.76	\pm	22.50	$	&	$	-13.64	\pm	34.07	$	&	$	0.13	\pm	0.02	$	&	$	3.40	\pm	0.87	$	&	$	5.69	\pm	0.85	$	&	$	4.04	\pm	^{	0.30	}	_{	0.33	}	$	\\
23	&	PSZ2G164.29+08.94	&	N/A	&	$	0.251	$	&	P 	&	$	5.97	\pm	1.06	$	&	$	-62.17	\pm	14.03	$	&	$	18.12	\pm	17.06	$	&	$	0.13	\pm	0.02	$	&	$	4.04	\pm	0.70	$	&	$	7.91	\pm	1.36	$	&	$	6.24	\pm	^{	0.62	}	_{	0.64	}	$	\\
24	&	PSZ2G133.60+69.04	&	RXJ1229.0+4737	&	$	0.254	$	&	S 	&	$	5.26	\pm	1.60	$	&	$	5.87	\pm	25.04	$	&	$	59.40	\pm	37.35	$	&	$	0.13	\pm	0.02	$	&	$	3.57	\pm	1.06	$	&	$	7.04	\pm	0.97	$	&	$	5.42	\pm	^{	0.38	}	_{	0.43	}	$	\\
25	&	PSZ2G086.47+15.31	&	MCXCJ1938.3+5409	&	$	0.26	$	&	S 	&	$	10.89	\pm	1.87	$	&	$	-39.65	\pm	13.24	$	&	$	19.83	\pm	12.61	$	&	$	0.13	\pm	0.02	$	&	$	7.25	\pm	1.21	$	&	$	9.54	\pm	0.63	$	&	$	7.76	\pm	^{	0.29	}	_{	0.28	}	$	\\
26	&	PSZ2G139.62+24.18	&	N/A	&	$	0.2671	$	&	S 	&	$	8.13	\pm	1.28	$	&	$	36.66	\pm	11.64	$	&	$	-12.58	\pm	10.80	$	&	$	0.13	\pm	0.02	$	&	$	5.45	\pm	0.84	$	&	$	8.34	\pm	1.06	$	&	$	7.11	\pm	^{	0.48	}	_{	0.47	}	$	\\
27	&	PSZ2G184.68+28.91	&	ACO611	&	$	0.288	$	&	S 	&	$	7.90	\pm	1.02	$	&	$	22.61	\pm	10.45	$	&	$	13.48	\pm	9.97	$	&	$	0.13	\pm	0.02	$	&	$	5.28	\pm	0.67	$	&	$	11.44	\pm	2.30	$	&	$	5.61	\pm	^{	0.52	}	_{	0.53	}	$	\\
28	&	PSZ2G154.13+40.19	&	ACO747	&	$	0.29	$	&	P	&	$	6.46	\pm	1.13	$	&	$	70.99	\pm	14.72	$	&	$	-42.86	\pm	13.25	$	&	$	0.13	\pm	0.02	$	&	$	4.33	\pm	0.74	$	&	$	6.09	\pm	1.10	$	&	$	5.48	\pm	^{	0.45	}	_{	0.46	}	$	\\
29	&	PSZ2G095.49+16.41	&	N/A	&	$	0.3	$	&	S 	&	$	5.43	\pm	1.12	$	&	$	-24.47	\pm	19.10	$	&	$	-102.18	\pm	18.33	$	&	$	0.13	\pm	0.02	$	&	$	3.65	\pm	0.74	$	&	$	4.91	\pm	0.99	$	&	$	4.38	\pm	^{	0.48	}	_{	0.49	}	$	\\
30	&	PSZ2G109.52-19.16	&	N/A	&	$	0.3092	$	&	P 	&	$	8.53	\pm	1.40	$	&	$	-30.38	\pm	13.77	$	&	$	-15.21	\pm	15.15	$	&	$	0.13	\pm	0.02	$	&	$	5.66	\pm	0.91	$	&	$	8.34	\pm	1.79	$	&	$	5.78	\pm	^{	0.48	}	_{	0.52	}	$	\\
31	&	PSZ2G198.90+18.16	&	[SPD2011] 298	&	$	0.3184	$	&	P 	&	$	7.61	\pm	1.18	$	&	$	26.76	\pm	14.62	$	&	$	-58.07	\pm	11.95	$	&	$	0.13	\pm	0.02	$	&	$	5.06	\pm	0.77	$	&	$	7.99	\pm	1.47	$	&	$	5.87	\pm	^{	0.55	}	_{	0.57	}	$	\\
32	&	PSZ2G152.33+81.28	&	MCXCJ1230.7+3439 	&	$	0.333	$	&	S 	&	$	6.27	\pm	1.12	$	&	$	-52.81	\pm	20.78	$	&	$	44.11	\pm	14.62	$	&	$	0.13	\pm	0.02	$	&	$	4.17	\pm	0.73	$	&	$	5.08	\pm	0.96	$	&	$	5.05	\pm	^{	0.53	}	_{	0.57	}	$	\\
33	&	PSZ2G108.17-11.56	&	N/A	&	$	0.336	$	&	S 	&	$	8.00	\pm	1.23	$	&	$	35.19	\pm	13.14	$	&	$	-70.15	\pm	19.09	$	&	$	0.13	\pm	0.02	$	&	$	5.29	\pm	0.80	$	&	$	9.82	\pm	1.29	$	&	$	7.42	\pm	^{	0.57	}	_{	0.60	}	$	\\
34	&	PSZ2G132.47-17.27	&	MCXCJ0142.9+4438 	&	$	0.341	$	&	S 	&	$	12.43	\pm	1.85	$	&	$	31.87	\pm	10.19	$	&	$	15.27	\pm	12.93	$	&	$	0.13	\pm	0.02	$	&	$	8.13	\pm	1.18	$	&	$	8.27	\pm	1.12	$	&	$	8.07	\pm	^{	0.61	}	_{	0.65	}	$	\\
35	&	PSZ2G207.88+81.31	&	ACO1489	&	$	0.353	$	&	S 	&	$	11.26	\pm	1.61	$	&	$	68.55	\pm	8.44	$	&	$	62.56	\pm	11.55	$	&	$	0.13	\pm	0.02	$	&	$	7.36	\pm	1.02	$	&	$	8.01	\pm	0.95	$	&	$	7.54	\pm	^{	0.45	}	_{	0.45	}	$	\\
36	&	PSZ2G157.32-26.77	&	MCSJ0308.9+2645 	&	$	0.356	$	&	S 	&	$	14.28	\pm	2.12	$	&	$	0.33	\pm	8.12	$	&	$	17.65	\pm	11.53	$	&	$	0.13	\pm	0.02	$	&	$	9.27	\pm	1.34	$	&	$	10.95	\pm	1.12	$	&	$	10.67	\pm	^{	0.64	}	_{	0.65	}	$	\\
37	&	PSZ2G071.21+28.86	&	RXSJ175201.5+444046 	&	$	0.366	$	&	S 	&	$	9.26	\pm	1.51	$	&	$	-29.82	\pm	9.95	$	&	$	-12.58	\pm	13.26	$	&	$	0.13	\pm	0.02	$	&	$	6.07	\pm	0.96	$	&	$	6.15	\pm	0.80	$	&	$	6.70	\pm	^{	0.44	}	_{	0.46	}	$	\\
38	&	PSZ2G194.98+54.12	&	MCSJ1006.9+3200	&	$	0.375	$	&	P	&	$	8.90	\pm	1.56	$	&	$	32.58	\pm	12.17	$	&	$	-0.22	\pm	19.18	$	&	$	0.13	\pm	0.02	$	&	$	5.83	\pm	1.00	$	&	$	6.31	\pm	1.38	$	&	$	5.30	\pm	^{	0.65	}	_{	0.68	}	$	\\
39	&	PSZ2G109.86+27.94	&	N/A	&	$	0.4	$	&	S 	&	$	4.57	\pm	1.28	$	&	$	3.98	\pm	22.50	$	&	$	7.39	\pm	18.70	$	&	$	0.13	\pm	0.02	$	&	$	3.03	\pm	0.83	$	&	$	5.23	\pm	0.91	$	&	$	5.23	\pm	^{	0.45	}	_{	0.48	}	$	\\
40	&	PSZ2G083.29-31.03	&	MCXCJ2228.6+2036	&	$	0.412	$	&	S 	&	$	11.85	\pm	1.73	$	&	$	81.05	\pm	13.29	$	&	$	-3.42	\pm	12.73	$	&	$	0.13	\pm	0.02	$	&	$	7.65	\pm	1.09	$	&	$	9.21	\pm	0.95	$	&	$	8.31	\pm	^{	0.44	}	_{	0.45	}	$	\\
41	&	PSZ2G063.38+53.44	&	NSCJ1537+392702 	&	$	0.422	$	&	S 	&	$	12.17	\pm	1.94	$	&	$	46.13	\pm	12.01	$	&	$	46.02	\pm	9.37	$	&	$	0.13	\pm	0.02	$	&	$	7.84	\pm	1.22	$	&	$	7.78	\pm	1.54	$	&	$	6.17	\pm	^{	0.58	}	_{	0.62	}	$	\\
42	&	PSZ2G063.80+11.42	&	N/A	&	$	0.426	$	&	S 	&	$	5.13	\pm	1.19	$	&	$	-36.41	\pm	22.22	$	&	$	-47.14	\pm	19.79	$	&	$	0.13	\pm	0.02	$	&	$	3.37	\pm	0.76	$	&	$	5.53	\pm	0.63	$	&	$	6.41	\pm	^{	0.57	}	_{	0.58	}	$	\\
43	&	PSZ2G157.43+30.34	&	RXJ0748.6+5940	&	$	0.45	$	&	P 	&	$	11.64	\pm	1.56	$	&	$	-61.32	\pm	7.38	$	&	$	4.53	\pm	8.27	$	&	$	0.13	\pm	0.02	$	&	$	7.47	\pm	0.98	$	&	$	6.71	\pm	0.44	$	&	$	8.16	\pm	^{	0.54	}	_{	0.54	}	$	\\
44	&	PSZ2G150.56+58.32	&	CLGJ1115+5319	&	$	0.47	$	&	S 	&	$	12.77	\pm	2.40	$	&	$	10.18	\pm	13.31	$	&	$	34.06	\pm	18.57	$	&	$	0.13	\pm	0.02	$	&	$	8.14	\pm	1.49	$	&	$	10.04	\pm	1.61	$	&	$	7.44	\pm	^{	0.50	}	_{	0.53	}	$	\\
45	&	PSZ2G170.98+39.45	&	[SPD2011] 16774	&	$	0.5131	$	&	S 	&	$	10.11	\pm	1.38	$	&	$	31.48	\pm	10.20	$	&	$	-30.87	\pm	12.67	$	&	$	0.12	\pm	0.02	$	&	$	6.43	\pm	0.86	$	&	$	8.24	\pm	1.30	$	&	$	7.55	\pm	^{	0.65	}	_{	0.71	}	$	\\
46	&	PSZ2G094.56+51.03	&	N/A	&	$	0.5392	$	&	S 	&	$	10.83	\pm	1.43	$	&	$	81.61	\pm	8.09	$	&	$	52.86	\pm	8.80	$	&	$	0.13	\pm	0.02	$	&	$	6.85	\pm	0.88	$	&	$	6.46	\pm	0.93	$	&	$	5.90	\pm	^{	0.45	}	_{	0.44	}	$	\\
47	&	PSZ2G228.16+75.20	&	CLGJ1149+2223 	&	$	0.545	$	&	S 	&	$	15.63	\pm	1.66	$	&	$	-15.49	\pm	5.32	$	&	$	17.11	\pm	4.75	$	&	$	0.13	\pm	0.01	$	&	$	9.78	\pm	1.01	$	&	$	9.64	\pm	0.94	$	&	$	9.69	\pm	^{	0.53	}	_{	0.55	}	$	\\
48	&	PSZ2G213.39+80.59	&	SDSSCGB41791	&	$	0.5586	$	&	S 	&	$	9.31	\pm	1.32	$	&	$	-9.73	\pm	11.90	$	&	$	69.37	\pm	12.14	$	&	$	0.13	\pm	0.02	$	&	$	5.89	\pm	0.81	$	&	$	8.03	\pm	1.39	$	&	$	6.77	\pm	^{	0.63	}	_{	0.65	}	$	\\
49	&	PSZ2G066.41+27.03	&	N/A	&	$	0.5699	$	&	S 	&	$	13.23	\pm	2.05	$	&	$	-33.18	\pm	11.12	$	&	$	97.03	\pm	11.32	$	&	$	0.13	\pm	0.02	$	&	$	8.27	\pm	1.25	$	&	$	7.33	\pm	0.82	$	&	$	7.72	\pm	^{	0.52	}	_{	0.54	}	$	\\
50	&	PSZ2G144.83+25.11	&	CLGJ0647+7015 	&	$	0.584	$	&	S 	&	$	11.69	\pm	1.46	$	&	$	4.15	\pm	7.87	$	&	$	-1.21	\pm	8.54	$	&	$	0.13	\pm	0.02	$	&	$	7.32	\pm	0.89	$	&	$	8.50	\pm	1.27	$	&	$	7.80	\pm	^{	0.72	}	_{	0.74	}	$	\\
51	&	PSZ2G045.87+57.70	&	N/A	&	$	0.611	$	&	S 	&	$	9.22	\pm	1.97	$	&	$	11.71	\pm	14.87	$	&	$	24.21	\pm	12.21	$	&	$	0.13	\pm	0.02	$	&	$	5.78	\pm	1.20	$	&	$	8.49	\pm	1.61	$	&	$	7.05	\pm	^{	0.66	}	_{	0.71	}	$	\\
52	&	PSZ2G108.27+48.66	&	N/A	&	$	0.674	$	&	S 	&	$	9.31	\pm	1.46	$	&	$	9.99	\pm	11.34	$	&	$	35.79	\pm	11.45	$	&	$	0.13	\pm	0.02	$	&	$	5.77	\pm	0.88	$	&	$	8.44	\pm	1.58	$	&	$	4.96	\pm	^{	0.48	}	_{	0.52	}	$	\\
53	&	PSZ2G086.93+53.18	&	N/A	&	$	0.6752	$	&	P 	&	$	9.85	\pm	1.69	$	&	$	-47.72	\pm	14.38	$	&	$	27.69	\pm	10.67	$	&	$	0.13	\pm	0.02	$	&	$	6.10	\pm	1.01	$	&	$	6.07	\pm	1.09	$	&	$	5.46	\pm	^{	0.51	}	_{	0.52	}	$	\\
54	&	PSZ2G141.77+14.19	&	N/A	&	$	0.83	$	&	P 	&	$	10.99	\pm	1.50	$	&	$	-4.36	\pm	8.54	$	&	$	-19.02	\pm	8.85	$	&	$	0.13	\pm	0.02	$	&	$	6.61	\pm	0.87	$	&	$	9.94	\pm	2.01	$	&	$	7.77	\pm	^{	0.90	}	_{	0.95	}	$	\\

\hline

\end{longtable}
\end{small}
\end{center}

\end{landscape}
\restoregeometry

\clearpage

%% file: APP-B/appendixb.tex
%
%
\ifpdf

\chapter{Appendix B: $Y_{\rm tot}$ derivation and results of physical and observational modelling comparison}
\label{c:appendixb}

\section{GNFW $Y_{\rm tot}$ analytical solution}
\label{s:pl_obs_ytot}

We wish to solve the integral
\begin{equation}
\int_{0}^{\infty} r^2 \left( \frac{r}{r_{\rm p}} \right)^{-c} \left[ 1 + \left( \frac{r}{r_{\rm p}} \right)^{a} \right] ^{\frac{c - b}{a}} \mathrm{d}r.
\end{equation}
First we use the substitution $s = r / r_{\rm p}$ to get
\begin{equation}
r_{\rm p}^3 \int_{0}^{\infty} s^2 s^{-c} \left[ 1 + s^{a} \right] ^{\frac{c - b}{a}} \mathrm{d}s.
\end{equation}
Next we use the substitution $t = s^a$ which gives
\begin{equation}
\begin{aligned}
& \frac{r_{\rm p}^3}{a} \int_{0}^{\infty} s^{3 - a - c} \left[ 1 + t \right] ^{\frac{c - b}{a}} \mathrm{d}t \\
&= \frac{r_{\rm p}^3}{a} \int_{0}^{\infty} \frac{t^{\frac{3 - c}{a} - 1}}{\left( 1 + t \right) ^{\frac{b - c}{a}}} \mathrm{d}t.
\end{aligned}
\end{equation}
Now using the following form for the beta function
\begin{equation}
\beta(x, y) = \int_{0}^{\infty} \frac{t^{x - 1}}{\left( 1 + t \right)^{x + y}} \mathrm{d}t,
\end{equation}
and the relation between the beta and Gamma functions given by $\beta(x, y) = \Gamma(x) \Gamma(y) / \Gamma(x + y)$ then we get the result
\begin{equation}
\frac{r_{\rm p}^3}{a} \frac{\Gamma \left( \frac{3 - c}{a} \right) \Gamma \left( \frac{b - 3}{a} \right)}{\Gamma \left( \frac{b - c}{a} \right)},
\end{equation}
as required. 
\clearpage

\newgeometry{margin=1cm}
\begin{landscape}
\section{Results tables}\label{s:pl_obs_results}
\begin{center}
\begin{longtable}{llllllllll}
\caption{Summary of parameter estimates for final sample of 54 clusters. 
All $Y$ values are given in units of $\times 10^{-3}~\rm (arcmin^{2})$, and all cluster centre coordinates are given in arcseconds. The cluster centre estimates from the physical model are omitted here but can be found in the results Table~\ref{t:pl_phys_results} in Appendix~\ref{c:appendixa}, which is ordered in the same order as this Table. Note the Table in Appendix~\ref{c:appendixa} also gives external names associated with these clusters, as well as the method used to measure the respective redshifts (i.e. spectroscopic or photometric). 
}\label{t:pl_obs_results1} \\

\hline \multicolumn{1}{c}{Row} & \multicolumn{1}{c}{Planck ID}  & \multicolumn{1}{c}{$z$} & \multicolumn{1}{c}{$Y_{\rm PM}(r_{500})$} & \multicolumn{1}{c}{$Y_{\rm OM\, I}(r_{500})$} & \multicolumn{1}{c}{$x_{0, \rm OM I}$} & \multicolumn{1}{c}{$y_{0, \rm OM\, I}$} & \multicolumn{1}{c}{$Y_{\rm OM\, II}(r_{500})$} & \multicolumn{1}{c}{$x_{0, \rm OM\, II}$} & \multicolumn{1}{c}{$y_{0, \rm OM\, II}$} \\ \hline 
\endfirsthead

\multicolumn{10}{c}%
{{\tablename\ \thetable{} -- continued from previous page}} \\
\hline \multicolumn{1}{c}{Row} & \multicolumn{1}{c}{Planck ID}  & \multicolumn{1}{c}{$z$} & \multicolumn{1}{c}{$Y_{\rm PM}(r_{500})$} & \multicolumn{1}{c}{$Y_{\rm OM\, I}(r_{500})$} & \multicolumn{1}{c}{$x_{0, \rm OM I}$} & \multicolumn{1}{c}{$y_{0, \rm OM\, I}$} & \multicolumn{1}{c}{$Y_{\rm OM\, II}(r_{500})$} & \multicolumn{1}{c}{$x_{0, \rm OM\, II}$} & \multicolumn{1}{c}{$y_{0, \rm OM\, II}$} \\ \hline 
\endhead

\tabulinesep=_1mm
\extrarowsep=1mm
\LTcapwidth=\textwidth

1	&	PSZ2G044.20+48.66	&	$	0.0894	$	&	$	11.59	\pm	2.28	$	&	$	6.77	\pm	3.32	$	&	$	6.53	\pm	18.56	$	&	$	8.93	\pm	14.41	$	&	$	20.48	\pm	6.19	$	&	$	10.36	\pm	18.38	$	&	$	8.32	\pm	15.32	$	\\
2	&	PSZ2G053.53+59.52	&	$	0.113	$	&	$	3.81	\pm	0.67	$	&	$	2.02	\pm	0.90	$	&	$	-1.77	\pm	12.69	$	&	$	23.19	\pm	9.38	$	&	$	3.12	\pm	1.74	$	&	$	-1.07	\pm	12.67	$	&	$	20.89	\pm	9.88	$	\\
3	&	PSZ2G151.90+11.63	&	$	0.12	$	&	$	1.76	\pm	0.50	$	&	$	2.55	\pm	1.56	$	&	$	63.93	\pm	28.11	$	&	$	67.61	\pm	18.86	$	&	$	4.09	\pm	1.83	$	&	$	59.05	\pm	27.67	$	&	$	67.19	\pm	19.36	$	\\
4	&	PSZ2G218.59+71.31	&	$	0.137	$	&	$	0.45	\pm	0.25	$	&	$	0.35	\pm	0.15	$	&	$	-8.85	\pm	14.58	$	&	$	-17.72	\pm	14.59	$	&	$	0.43	\pm	0.27	$	&	$	0.04	\pm	23.62	$	&	$	-16.95	\pm	24.66	$	\\
5	&	PSZ2G226.18+76.79	&	$	0.1427	$	&	$	2.45	\pm	0.45	$	&	$	0.91	\pm	0.45	$	&	$	-45.20	\pm	10.61	$	&	$	6.46	\pm	12.25	$	&	$	1.21	\pm	0.51	$	&	$	-42.92	\pm	10.66	$	&	$	3.80	\pm	12.00	$	\\
6	&	PSZ2G165.06+54.13	&	$	0.144	$	&	$	2.26	\pm	0.54	$	&	$	0.70	\pm	0.25	$	&	$	29.82	\pm	10.17	$	&	$	-29.36	\pm	12.22	$	&	$	0.95	\pm	0.27	$	&	$	31.51	\pm	10.76	$	&	$	-29.04	\pm	12.83	$	\\
7	&	PSZ2G077.90-26.63	&	$	0.147	$	&	$	2.80	\pm	0.46	$	&	$	1.35	\pm	0.48	$	&	$	-27.99	\pm	9.91	$	&	$	20.12	\pm	11.23	$	&	$	1.48	\pm	0.49	$	&	$	-28.06	\pm	10.13	$	&	$	19.93	\pm	11.07	$	\\
8	&	PSZ2G050.40+31.17	&	$	0.164	$	&	$	1.01	\pm	0.29	$	&	$	1.07	\pm	0.70	$	&	$	37.21	\pm	20.82	$	&	$	9.59	\pm	19.09	$	&	$	1.18	\pm	0.76	$	&	$	36.11	\pm	22.25	$	&	$	9.30	\pm	19.70	$	\\
9	&	PSZ2G097.72+38.12	&	$	0.1709	$	&	$	2.65	\pm	0.46	$	&	$	2.72	\pm	1.26	$	&	$	29.79	\pm	15.43	$	&	$	-2.59	\pm	13.68	$	&	$	3.97	\pm	1.49	$	&	$	32.13	\pm	15.62	$	&	$	-1.56	\pm	13.81	$	\\
10	&	PSZ2G099.30+20.92	&	$	0.171	$	&	$	0.97	\pm	0.31	$	&	$	0.79	\pm	0.49	$	&	$	-35.09	\pm	19.11	$	&	$	-24.57	\pm	21.53	$	&	$	0.86	\pm	0.51	$	&	$	-36.16	\pm	19.13	$	&	$	-25.55	\pm	21.67	$	\\
11	&	PSZ2G067.17+67.46	&	$	0.1712	$	&	$	2.70	\pm	0.46	$	&	$	1.30	\pm	0.54	$	&	$	34.00	\pm	11.65	$	&	$	-30.54	\pm	10.97	$	&	$	1.48	\pm	0.60	$	&	$	33.18	\pm	11.61	$	&	$	-31.32	\pm	11.16	$	\\
12	&	PSZ2G167.67+17.63	&	$	0.174	$	&	$	0.72	\pm	0.30	$	&	$	1.69	\pm	1.05	$	&	$	-24.86	\pm	32.03	$	&	$	10.55	\pm	28.11	$	&	$	1.33	\pm	0.77	$	&	$	-23.41	\pm	33.17	$	&	$	11.93	\pm	29.04	$	\\
13	&	PSZ2G066.68+68.44	&	$	0.181	$	&	$	0.72	\pm	0.29	$	&	$	1.24	\pm	0.79	$	&	$	55.97	\pm	25.19	$	&	$	9.20	\pm	32.13	$	&	$	1.12	\pm	0.72	$	&	$	56.41	\pm	26.70	$	&	$	7.31	\pm	32.63	$	\\
14	&	PSZ2G065.28+44.53	&	$	0.183	$	&	$	0.79	\pm	0.28	$	&	$	0.65	\pm	0.38	$	&	$	-21.13	\pm	20.72	$	&	$	-15.63	\pm	18.96	$	&	$	0.61	\pm	0.34	$	&	$	-19.57	\pm	22.13	$	&	$	-16.08	\pm	20.64	$	\\
15	&	PSZ2G084.47+12.63	&	$	0.185	$	&	$	0.67	\pm	0.25	$	&	$	0.58	\pm	0.29	$	&	$	-67.12	\pm	29.59	$	&	$	-23.26	\pm	18.01	$	&	$	0.53	\pm	0.28	$	&	$	-69.03	\pm	30.83	$	&	$	-20.78	\pm	20.07	$	\\
16	&	PSZ2G100.04+23.73	&	$	0.21	$	&	$	0.65	\pm	0.18	$	&	$	1.28	\pm	0.75	$	&	$	17.47	\pm	19.11	$	&	$	-22.73	\pm	22.21	$	&	$	1.05	\pm	0.55	$	&	$	17.93	\pm	19.85	$	&	$	-23.27	\pm	22.53	$	\\
17	&	PSZ2G180.60+76.65	&	$	0.2138	$	&	$	0.63	\pm	0.20	$	&	$	1.73	\pm	0.93	$	&	$	36.57	\pm	16.66	$	&	$	-73.38	\pm	20.39	$	&	$	1.11	\pm	0.50	$	&	$	35.90	\pm	17.29	$	&	$	-70.57	\pm	22.18	$	\\
18	&	PSZ2G166.09+43.38	&	$	0.2172	$	&	$	1.67	\pm	0.28	$	&	$	1.10	\pm	0.50	$	&	$	-4.29	\pm	10.57	$	&	$	-6.54	\pm	9.54	$	&	$	1.14	\pm	0.46	$	&	$	-4.73	\pm	10.32	$	&	$	-6.66	\pm	9.63	$	\\
19	&	PSZ2G125.30-27.99	&	$	0.223	$	&	$	0.45	\pm	0.18	$	&	$	0.99	\pm	0.64	$	&	$	-8.12	\pm	26.53	$	&	$	2.49	\pm	30.79	$	&	$	0.60	\pm	0.38	$	&	$	-9.03	\pm	28.36	$	&	$	6.48	\pm	31.71	$	\\
20	&	PSZ2G060.13+11.44	&	$	0.224	$	&	$	1.00	\pm	0.20	$	&	$	1.17	\pm	0.64	$	&	$	-64.93	\pm	12.76	$	&	$	-49.60	\pm	15.02	$	&	$	1.12	\pm	0.56	$	&	$	-64.67	\pm	12.69	$	&	$	-49.56	\pm	14.77	$	\\
21	&	PSZ2G166.62+42.13	&	$	0.232	$	&	$	0.29	\pm	0.13	$	&	$	1.57	\pm	0.92	$	&	$	-36.13	\pm	30.51	$	&	$	-54.22	\pm	32.52	$	&	$	0.53	\pm	0.35	$	&	$	-34.92	\pm	31.92	$	&	$	-40.79	\pm	38.30	$	\\
22	&	PSZ2G097.94+19.43	&	$	0.25	$	&	$	0.45	\pm	0.17	$	&	$	1.24	\pm	0.69	$	&	$	-121.19	\pm	21.52	$	&	$	-2.42	\pm	32.74	$	&	$	0.73	\pm	0.41	$	&	$	-115.20	\pm	27.60	$	&	$	-5.84	\pm	34.15	$	\\
23	&	PSZ2G164.29+08.94	&	$	0.251	$	&	$	0.59	\pm	0.13	$	&	$	0.87	\pm	0.43	$	&	$	-62.15	\pm	13.92	$	&	$	20.46	\pm	17.35	$	&	$	0.73	\pm	0.35	$	&	$	-62.23	\pm	13.90	$	&	$	18.67	\pm	17.99	$	\\
24	&	PSZ2G133.60+69.04	&	$	0.254	$	&	$	0.47	\pm	0.20	$	&	$	1.60	\pm	1.12	$	&	$	0.13	\pm	24.80	$	&	$	66.74	\pm	35.89	$	&	$	0.80	\pm	0.45	$	&	$	3.35	\pm	25.98	$	&	$	63.00	\pm	37.13	$	\\
25	&	PSZ2G086.47+15.31	&	$	0.26	$	&	$	1.48	\pm	0.33	$	&	$	1.70	\pm	0.71	$	&	$	-41.40	\pm	14.66	$	&	$	19.45	\pm	13.73	$	&	$	1.58	\pm	0.60	$	&	$	-40.08	\pm	14.39	$	&	$	20.08	\pm	13.75	$	\\
26	&	PSZ2G139.62+24.18	&	$	0.2671	$	&	$	0.89	\pm	0.16	$	&	$	0.77	\pm	0.34	$	&	$	35.74	\pm	11.80	$	&	$	-13.45	\pm	11.11	$	&	$	0.70	\pm	0.33	$	&	$	35.83	\pm	11.49	$	&	$	-13.78	\pm	10.76	$	\\
27	&	PSZ2G184.68+28.91	&	$	0.288	$	&	$	0.76	\pm	0.12	$	&	$	0.95	\pm	0.38	$	&	$	22.66	\pm	10.55	$	&	$	12.19	\pm	10.37	$	&	$	0.83	\pm	0.31	$	&	$	22.58	\pm	10.48	$	&	$	13.03	\pm	10.41	$	\\
28	&	PSZ2G154.13+40.19	&	$	0.29	$	&	$	0.55	\pm	0.13	$	&	$	0.72	\pm	0.50	$	&	$	71.59	\pm	15.07	$	&	$	-42.78	\pm	13.41	$	&	$	0.46	\pm	0.23	$	&	$	69.88	\pm	14.52	$	&	$	-42.45	\pm	13.20	$	\\
29	&	PSZ2G095.49+16.41	&	$	0.3	$	&	$	0.39	\pm	0.12	$	&	$	0.87	\pm	0.54	$	&	$	-19.80	\pm	21.12	$	&	$	-94.58	\pm	19.43	$	&	$	0.48	\pm	0.26	$	&	$	-22.58	\pm	20.72	$	&	$	-98.75	\pm	20.62	$	\\
30	&	PSZ2G109.52-19.16	&	$	0.3092	$	&	$	0.78	\pm	0.16	$	&	$	1.00	\pm	0.57	$	&	$	-31.66	\pm	14.34	$	&	$	-15.21	\pm	15.68	$	&	$	0.82	\pm	0.39	$	&	$	-31.16	\pm	14.43	$	&	$	-15.23	\pm	15.95	$	\\
31	&	PSZ2G198.90+18.16	&	$	0.3184	$	&	$	0.62	\pm	0.12	$	&	$	0.86	\pm	0.40	$	&	$	27.42	\pm	15.36	$	&	$	-59.55	\pm	12.35	$	&	$	0.69	\pm	0.27	$	&	$	27.03	\pm	15.25	$	&	$	-57.65	\pm	12.64	$	\\
32	&	PSZ2G152.33+81.28	&	$	0.333	$	&	$	0.43	\pm	0.11	$	&	$	0.78	\pm	0.42	$	&	$	-49.96	\pm	20.35	$	&	$	44.73	\pm	15.45	$	&	$	0.48	\pm	0.22	$	&	$	-53.60	\pm	20.93	$	&	$	43.79	\pm	15.28	$	\\
33	&	PSZ2G108.17-11.56	&	$	0.336	$	&	$	0.61	\pm	0.12	$	&	$	2.24	\pm	1.10	$	&	$	27.48	\pm	14.92	$	&	$	-36.56	\pm	20.44	$	&	$	1.12	\pm	0.25	$	&	$	30.62	\pm	13.89	$	&	$	-51.07	\pm	19.77	$	\\
34	&	PSZ2G132.47-17.27	&	$	0.341	$	&	$	1.25	\pm	0.21	$	&	$	1.38	\pm	0.52	$	&	$	32.53	\pm	10.83	$	&	$	16.82	\pm	12.65	$	&	$	1.37	\pm	0.47	$	&	$	32.34	\pm	10.66	$	&	$	16.61	\pm	12.56	$	\\
35	&	PSZ2G207.88+81.31	&	$	0.353	$	&	$	1.05	\pm	0.18	$	&	$	0.90	\pm	0.34	$	&	$	67.45	\pm	8.46	$	&	$	61.30	\pm	11.45	$	&	$	0.82	\pm	0.29	$	&	$	66.90	\pm	8.21	$	&	$	59.84	\pm	11.43	$	\\
36	&	PSZ2G157.32-26.77	&	$	0.356	$	&	$	1.52	\pm	0.27	$	&	$	1.25	\pm	0.42	$	&	$	-0.28	\pm	8.01	$	&	$	19.15	\pm	11.86	$	&	$	1.23	\pm	0.39	$	&	$	-1.07	\pm	7.59	$	&	$	17.73	\pm	11.58	$	\\
37	&	PSZ2G071.21+28.86	&	$	0.366	$	&	$	0.72	\pm	0.15	$	&	$	0.91	\pm	0.34	$	&	$	-29.47	\pm	10.86	$	&	$	-12.29	\pm	14.04	$	&	$	0.75	\pm	0.25	$	&	$	-29.64	\pm	10.48	$	&	$	-12.13	\pm	13.74	$	\\
38	&	PSZ2G194.98+54.12	&	$	0.375	$	&	$	0.65	\pm	0.15	$	&	$	1.28	\pm	0.69	$	&	$	32.85	\pm	12.59	$	&	$	-5.89	\pm	18.85	$	&	$	0.93	\pm	0.32	$	&	$	32.71	\pm	12.45	$	&	$	-3.46	\pm	19.90	$	\\
39	&	PSZ2G109.86+27.94	&	$	0.4	$	&	$	0.21	\pm	0.09	$	&	$	0.30	\pm	0.11	$	&	$	8.03	\pm	16.29	$	&	$	-1.95	\pm	14.87	$	&	$	0.20	\pm	0.07	$	&	$	7.15	\pm	21.69	$	&	$	2.87	\pm	17.97	$	\\
40	&	PSZ2G083.29-31.03	&	$	0.412	$	&	$	0.95	\pm	0.17	$	&	$	0.66	\pm	0.21	$	&	$	75.26	\pm	13.22	$	&	$	-0.29	\pm	12.25	$	&	$	0.60	\pm	0.20	$	&	$	72.16	\pm	13.03	$	&	$	2.13	\pm	11.88	$	\\
41	&	PSZ2G063.38+53.44	&	$	0.422	$	&	$	0.93	\pm	0.19	$	&	$	1.28	\pm	0.45	$	&	$	39.37	\pm	14.20	$	&	$	49.33	\pm	10.77	$	&	$	1.12	\pm	0.29	$	&	$	41.65	\pm	13.30	$	&	$	48.43	\pm	10.17	$	\\
42	&	PSZ2G063.80+11.42	&	$	0.426	$	&	$	0.24	\pm	0.08	$	&	$	0.80	\pm	0.46	$	&	$	-42.04	\pm	23.06	$	&	$	-44.32	\pm	20.40	$	&	$	0.29	\pm	0.14	$	&	$	-36.98	\pm	23.28	$	&	$	-45.28	\pm	20.74	$	\\
43	&	PSZ2G157.43+30.34	&	$	0.45	$	&	$	0.82	\pm	0.13	$	&	$	0.91	\pm	0.26	$	&	$	-61.41	\pm	7.56	$	&	$	4.85	\pm	8.34	$	&	$	0.85	\pm	0.23	$	&	$	-61.63	\pm	7.29	$	&	$	4.79	\pm	8.26	$	\\
44	&	PSZ2G150.56+58.32	&	$	0.47	$	&	$	0.93	\pm	0.25	$	&	$	0.86	\pm	0.38	$	&	$	9.81	\pm	14.03	$	&	$	35.97	\pm	18.29	$	&	$	0.70	\pm	0.25	$	&	$	8.34	\pm	12.93	$	&	$	36.51	\pm	18.01	$	\\
45	&	PSZ2G170.98+39.45	&	$	0.5131	$	&	$	0.54	\pm	0.08	$	&	$	1.62	\pm	0.68	$	&	$	23.91	\pm	12.09	$	&	$	-18.32	\pm	13.31	$	&	$	0.88	\pm	0.17	$	&	$	26.68	\pm	11.52	$	&	$	-22.95	\pm	12.68	$	\\
46	&	PSZ2G094.56+51.03	&	$	0.5392	$	&	$	0.63	\pm	0.10	$	&	$	0.50	\pm	0.09	$	&	$	82.24	\pm	7.64	$	&	$	50.61	\pm	8.76	$	&	$	0.45	\pm	0.08	$	&	$	81.87	\pm	7.67	$	&	$	50.51	\pm	8.62	$	\\
47	&	PSZ2G228.16+75.20	&	$	0.545	$	&	$	1.06	\pm	0.10	$	&	$	1.35	\pm	0.27	$	&	$	-14.53	\pm	5.57	$	&	$	16.35	\pm	5.31	$	&	$	1.25	\pm	0.21	$	&	$	-14.39	\pm	5.59	$	&	$	16.50	\pm	5.08	$	\\
48	&	PSZ2G213.39+80.59	&	$	0.5586	$	&	$	0.45	\pm	0.08	$	&	$	0.89	\pm	0.36	$	&	$	-5.34	\pm	12.49	$	&	$	65.15	\pm	12.29	$	&	$	0.58	\pm	0.18	$	&	$	-8.19	\pm	12.21	$	&	$	68.13	\pm	12.60	$	\\
49	&	PSZ2G066.41+27.03	&	$	0.5699	$	&	$	0.79	\pm	0.16	$	&	$	1.76	\pm	0.73	$	&	$	-37.37	\pm	11.95	$	&	$	100.92	\pm	13.20	$	&	$	1.00	\pm	0.24	$	&	$	-34.28	\pm	11.21	$	&	$	97.77	\pm	11.89	$	\\
50	&	PSZ2G144.83+25.11	&	$	0.584	$	&	$	0.61	\pm	0.07	$	&	$	1.34	\pm	0.45	$	&	$	1.55	\pm	9.00	$	&	$	-3.86	\pm	8.95	$	&	$	0.89	\pm	0.17	$	&	$	3.09	\pm	8.57	$	&	$	-2.97	\pm	8.79	$	\\
51	&	PSZ2G045.87+57.70	&	$	0.611	$	&	$	0.41	\pm	0.12	$	&	$	0.93	\pm	0.46	$	&	$	20.59	\pm	17.97	$	&	$	16.79	\pm	15.76	$	&	$	0.52	\pm	0.16	$	&	$	16.61	\pm	16.65	$	&	$	20.54	\pm	14.20	$	\\
52	&	PSZ2G108.27+48.66	&	$	0.674	$	&	$	0.40	\pm	0.08	$	&	$	0.55	\pm	0.20	$	&	$	8.45	\pm	11.83	$	&	$	35.26	\pm	11.93	$	&	$	0.42	\pm	0.12	$	&	$	9.91	\pm	12.03	$	&	$	35.53	\pm	11.69	$	\\
53	&	PSZ2G086.93+53.18	&	$	0.6752	$	&	$	0.43	\pm	0.10	$	&	$	1.28	\pm	0.57	$	&	$	-40.06	\pm	16.39	$	&	$	30.84	\pm	12.08	$	&	$	0.59	\pm	0.15	$	&	$	-44.92	\pm	15.26	$	&	$	29.36	\pm	11.53	$	\\
54	&	PSZ2G141.77+14.19	&	$	0.83	$	&	$	0.45	\pm	0.06	$	&	$	0.56	\pm	0.17	$	&	$	-3.40	\pm	8.77	$	&	$	-18.18	\pm	9.36	$	&	$	0.47	\pm	0.11	$	&	$	-4.11	\pm	8.78	$	&	$	-18.97	\pm	9.37	$	\\

\hline

\end{longtable}
\end{center}

\begin{center}
\begin{small}
\begin{longtable}{lllllllllll}
\caption{Summary of model comparison statistics for final sample of 54 clusters. The Planck IDs are omitted but are the same as in Table~\ref{t:pl_obs_results1}.}\label{t:pl_obs_results2} \\

\hline  \multicolumn{1}{c}{Row} & \multicolumn{1}{c}{$z$} & \multicolumn{1}{c}{$d_{\rm EMD}(\mathcal{P}_{\rm PM}, \mathcal{P}_{\rm OM\, I})$} & \multicolumn{1}{c}{$d_{\rm EMD}(\mathcal{P}_{\rm OM\, II}, \mathcal{P}_{\rm OM\, I})$} & \multicolumn{1}{c}{$d_{\rm EMD}(\mathcal{P}_{\rm PM}, \mathcal{P}_{\rm OM\, II})$} & \multicolumn{1}{c}{$\ln (\mathcal{Z}_{\rm PM} / \mathcal{Z}_{\rm null})$} & \multicolumn{1}{c}{$\ln (\mathcal{Z}_{\rm OM \, I} / \mathcal{Z}_{\rm null})$} & \multicolumn{1}{c}{$\ln (\mathcal{Z}_{\rm OM \, II} / \mathcal{Z}_{\rm null})$} & \multicolumn{1}{c}{$\ln (\mathcal{Z}_{\rm PM} / \mathcal{Z}_{\rm OM \, I})$} & \multicolumn{1}{c}{$\ln (\mathcal{Z}_{\rm OM \, II} / \mathcal{Z}_{\rm OM \, I})$} & \multicolumn{1}{c}{$\ln (\mathcal{Z}_{\rm PM} / \mathcal{Z}_{\rm OM \, II})$} \\ \hline 
\endfirsthead

\multicolumn{11}{c}%
{{\tablename\ \thetable{} -- continued from previous page}} \\
\hline 
\multicolumn{1}{c}{Row} & \multicolumn{1}{c}{$z$} & \multicolumn{1}{c}{$d_{\rm EMD}(\mathcal{P}_{\rm PM}, \mathcal{P}_{\rm OM\, I})$} & \multicolumn{1}{c}{$d_{\rm EMD}(\mathcal{P}_{\rm OM\, II}, \mathcal{P}_{\rm OM\, I})$} & \multicolumn{1}{c}{$d_{\rm EMD}(\mathcal{P}_{\rm PM}, \mathcal{P}_{\rm OM\, II})$} & \multicolumn{1}{c}{$\ln (\mathcal{Z}_{\rm PM} / \mathcal{Z}_{\rm null})$} & \multicolumn{1}{c}{$\ln (\mathcal{Z}_{\rm OM \, I} / \mathcal{Z}_{\rm null})$} & \multicolumn{1}{c}{$\ln (\mathcal{Z}_{\rm OM \, II} / \mathcal{Z}_{\rm null})$} & \multicolumn{1}{c}{$\ln (\mathcal{Z}_{\rm PM} / \mathcal{Z}_{\rm OM \, I})$} & \multicolumn{1}{c}{$\ln (\mathcal{Z}_{\rm OM \, II} / \mathcal{Z}_{\rm OM \, I})$} & \multicolumn{1}{c}{$\ln (\mathcal{Z}_{\rm PM} / \mathcal{Z}_{\rm OM \, II})$} \\ 
\hline 
\endhead

\tabulinesep=_1mm
\extrarowsep=1mm
\LTcapwidth=\textwidth

1	&	$	0.0894	$	&	$	0.222	$	&	$	0.514	$	&	$	0.297	$	&	$	33.90	\pm	0.16	$	&	$	29.17	\pm	0.16	$	&	$	33.38	\pm	0.16	$	&	$	4.73	\pm	0.23	$	&	$	4.21	\pm	0.23	$	&	$	0.52	\pm	0.22	$	\\
2	&	$	0.113	$	&	$	0.152	$	&	$	0.091	$	&	$	0.093	$	&	$	30.94	\pm	0.17	$	&	$	31.06	\pm	0.17	$	&	$	30.01	\pm	0.17	$	&	$	-0.12	\pm	0.24	$	&	$	-1.05	\pm	0.24	$	&	$	0.93	\pm	0.24	$	\\
3	&	$	0.12	$	&	$	0.083	$	&	$	0.123	$	&	$	0.189	$	&	$	10.40	\pm	0.13	$	&	$	10.54	\pm	0.13	$	&	$	10.00	\pm	0.14	$	&	$	-0.14	\pm	0.19	$	&	$	-0.53	\pm	0.19	$	&	$	0.39	\pm	0.19	$	\\
4	&	$	0.137	$	&	$	0.132	$	&	$	0.115	$	&	$	0.051	$	&	$	1.71	\pm	0.17	$	&	$	3.41	\pm	0.17	$	&	$	1.76	\pm	0.17	$	&	$	-1.70	\pm	0.24	$	&	$	-1.65	\pm	0.24	$	&	$	-0.05	\pm	0.24	$	\\
5	&	$	0.1427	$	&	$	0.170	$	&	$	0.033	$	&	$	0.138	$	&	$	23.01	\pm	0.15	$	&	$	24.85	\pm	0.15	$	&	$	23.50	\pm	0.15	$	&	$	-1.84	\pm	0.21	$	&	$	-1.35	\pm	0.21	$	&	$	-0.49	\pm	0.21	$	\\
6	&	$	0.144	$	&	$	0.210	$	&	$	0.045	$	&	$	0.165	$	&	$	13.68	\pm	0.13	$	&	$	17.82	\pm	0.13	$	&	$	15.56	\pm	0.14	$	&	$	-4.14	\pm	0.18	$	&	$	-2.26	\pm	0.19	$	&	$	-1.88	\pm	0.19	$	\\
7	&	$	0.147	$	&	$	0.140	$	&	$	0.014	$	&	$	0.126	$	&	$	32.94	\pm	0.12	$	&	$	34.76	\pm	0.12	$	&	$	33.50	\pm	0.12	$	&	$	-1.82	\pm	0.17	$	&	$	-1.26	\pm	0.17	$	&	$	-0.56	\pm	0.17	$	\\
8	&	$	0.164	$	&	$	0.065	$	&	$	0.026	$	&	$	0.069	$	&	$	9.61	\pm	0.08	$	&	$	10.32	\pm	0.08	$	&	$	9.10	\pm	0.08	$	&	$	-0.71	\pm	0.11	$	&	$	-1.23	\pm	0.11	$	&	$	0.51	\pm	0.12	$	\\
9	&	$	0.1709	$	&	$	0.049	$	&	$	0.082	$	&	$	0.087	$	&	$	33.10	\pm	0.16	$	&	$	33.00	\pm	0.16	$	&	$	32.62	\pm	0.16	$	&	$	0.10	\pm	0.22	$	&	$	-0.37	\pm	0.22	$	&	$	0.47	\pm	0.23	$	\\
10	&	$	0.171	$	&	$	0.058	$	&	$	0.012	$	&	$	0.058	$	&	$	7.73	\pm	0.15	$	&	$	8.46	\pm	0.15	$	&	$	7.08	\pm	0.15	$	&	$	-0.73	\pm	0.21	$	&	$	-1.38	\pm	0.21	$	&	$	0.65	\pm	0.21	$	\\
11	&	$	0.1712	$	&	$	0.135	$	&	$	0.022	$	&	$	0.114	$	&	$	26.98	\pm	0.10	$	&	$	28.19	\pm	0.10	$	&	$	27.08	\pm	0.11	$	&	$	-1.21	\pm	0.14	$	&	$	-1.11	\pm	0.15	$	&	$	-0.10	\pm	0.15	$	\\
12	&	$	0.174	$	&	$	0.132	$	&	$	0.029	$	&	$	0.107	$	&	$	3.67	\pm	0.11	$	&	$	4.53	\pm	0.11	$	&	$	3.56	\pm	0.11	$	&	$	-0.86	\pm	0.15	$	&	$	-0.97	\pm	0.16	$	&	$	0.11	\pm	0.16	$	\\
13	&	$	0.181	$	&	$	0.084	$	&	$	0.015	$	&	$	0.080	$	&	$	4.42	\pm	0.13	$	&	$	5.00	\pm	0.12	$	&	$	4.06	\pm	0.13	$	&	$	-0.58	\pm	0.18	$	&	$	-0.95	\pm	0.18	$	&	$	0.36	\pm	0.18	$	\\
14	&	$	0.183	$	&	$	0.068	$	&	$	0.010	$	&	$	0.063	$	&	$	5.57	\pm	0.13	$	&	$	6.52	\pm	0.13	$	&	$	5.35	\pm	0.13	$	&	$	-0.94	\pm	0.18	$	&	$	-1.16	\pm	0.18	$	&	$	0.22	\pm	0.19	$	\\
15	&	$	0.185	$	&	$	0.062	$	&	$	0.010	$	&	$	0.056	$	&	$	3.57	\pm	0.18	$	&	$	4.28	\pm	0.18	$	&	$	3.47	\pm	0.18	$	&	$	-0.71	\pm	0.25	$	&	$	-0.80	\pm	0.25	$	&	$	0.09	\pm	0.25	$	\\
16	&	$	0.21	$	&	$	0.094	$	&	$	0.026	$	&	$	0.076	$	&	$	7.98	\pm	0.14	$	&	$	8.67	\pm	0.14	$	&	$	7.51	\pm	0.14	$	&	$	-0.69	\pm	0.20	$	&	$	-1.15	\pm	0.20	$	&	$	0.46	\pm	0.20	$	\\
17	&	$	0.2138	$	&	$	0.143	$	&	$	0.051	$	&	$	0.094	$	&	$	4.68	\pm	0.18	$	&	$	5.67	\pm	0.18	$	&	$	4.38	\pm	0.18	$	&	$	-0.99	\pm	0.25	$	&	$	-1.29	\pm	0.25	$	&	$	0.30	\pm	0.25	$	\\
18	&	$	0.2172	$	&	$	0.072	$	&	$	0.006	$	&	$	0.069	$	&	$	27.82	\pm	0.12	$	&	$	28.93	\pm	0.12	$	&	$	27.64	\pm	0.13	$	&	$	-1.11	\pm	0.17	$	&	$	-1.29	\pm	0.17	$	&	$	0.18	\pm	0.18	$	\\
19	&	$	0.223	$	&	$	0.097	$	&	$	0.054	$	&	$	0.057	$	&	$	4.36	\pm	0.10	$	&	$	4.84	\pm	0.10	$	&	$	3.95	\pm	0.10	$	&	$	-0.48	\pm	0.14	$	&	$	-0.89	\pm	0.14	$	&	$	0.41	\pm	0.14	$	\\
20	&	$	0.224	$	&	$	0.049	$	&	$	0.009	$	&	$	0.051	$	&	$	16.34	\pm	0.13	$	&	$	17.23	\pm	0.13	$	&	$	15.79	\pm	0.13	$	&	$	-0.89	\pm	0.18	$	&	$	-1.44	\pm	0.19	$	&	$	0.55	\pm	0.19	$	\\
21	&	$	0.232	$	&	$	0.225	$	&	$	0.147	$	&	$	0.083	$	&	$	3.02	\pm	0.15	$	&	$	4.37	\pm	0.15	$	&	$	2.54	\pm	0.15	$	&	$	-1.35	\pm	0.21	$	&	$	-1.82	\pm	0.21	$	&	$	0.48	\pm	0.21	$	\\
22	&	$	0.25	$	&	$	0.136	$	&	$	0.071	$	&	$	0.070	$	&	$	3.03	\pm	0.15	$	&	$	3.96	\pm	0.15	$	&	$	2.26	\pm	0.15	$	&	$	-0.93	\pm	0.21	$	&	$	-1.70	\pm	0.21	$	&	$	0.77	\pm	0.21	$	\\
23	&	$	0.251	$	&	$	0.055	$	&	$	0.024	$	&	$	0.045	$	&	$	12.67	\pm	0.16	$	&	$	13.45	\pm	0.16	$	&	$	11.69	\pm	0.17	$	&	$	-0.78	\pm	0.23	$	&	$	-1.76	\pm	0.23	$	&	$	0.99	\pm	0.23	$	\\
24	&	$	0.254	$	&	$	0.180	$	&	$	0.110	$	&	$	0.076	$	&	$	3.80	\pm	0.11	$	&	$	5.27	\pm	0.11	$	&	$	3.86	\pm	0.11	$	&	$	-1.47	\pm	0.15	$	&	$	-1.41	\pm	0.15	$	&	$	-0.06	\pm	0.15	$	\\
25	&	$	0.26	$	&	$	0.041	$	&	$	0.009	$	&	$	0.040	$	&	$	13.18	\pm	0.17	$	&	$	13.79	\pm	0.16	$	&	$	12.32	\pm	0.17	$	&	$	-0.60	\pm	0.23	$	&	$	-1.46	\pm	0.23	$	&	$	0.86	\pm	0.23	$	\\
26	&	$	0.2671	$	&	$	0.043	$	&	$	0.012	$	&	$	0.051	$	&	$	28.23	\pm	0.14	$	&	$	29.05	\pm	0.14	$	&	$	27.67	\pm	0.14	$	&	$	-0.81	\pm	0.20	$	&	$	-1.38	\pm	0.20	$	&	$	0.56	\pm	0.20	$	\\
27	&	$	0.288	$	&	$	0.038	$	&	$	0.018	$	&	$	0.032	$	&	$	22.61	\pm	0.14	$	&	$	23.45	\pm	0.14	$	&	$	21.90	\pm	0.14	$	&	$	-0.85	\pm	0.19	$	&	$	-1.55	\pm	0.19	$	&	$	0.71	\pm	0.20	$	\\
28	&	$	0.29	$	&	$	0.045	$	&	$	0.034	$	&	$	0.046	$	&	$	9.72	\pm	0.18	$	&	$	10.64	\pm	0.18	$	&	$	9.42	\pm	0.18	$	&	$	-0.92	\pm	0.26	$	&	$	-1.23	\pm	0.26	$	&	$	0.31	\pm	0.26	$	\\
29	&	$	0.3	$	&	$	0.138	$	&	$	0.115	$	&	$	0.045	$	&	$	5.26	\pm	0.20	$	&	$	5.94	\pm	0.19	$	&	$	4.44	\pm	0.20	$	&	$	-0.68	\pm	0.28	$	&	$	-1.51	\pm	0.28	$	&	$	0.83	\pm	0.28	$	\\
30	&	$	0.3092	$	&	$	0.047	$	&	$	0.027	$	&	$	0.041	$	&	$	14.83	\pm	0.12	$	&	$	15.62	\pm	0.12	$	&	$	14.13	\pm	0.12	$	&	$	-0.80	\pm	0.17	$	&	$	-1.49	\pm	0.17	$	&	$	0.70	\pm	0.17	$	\\
31	&	$	0.3184	$	&	$	0.042	$	&	$	0.025	$	&	$	0.032	$	&	$	14.64	\pm	0.11	$	&	$	15.36	\pm	0.10	$	&	$	13.88	\pm	0.11	$	&	$	-0.72	\pm	0.15	$	&	$	-1.48	\pm	0.15	$	&	$	0.76	\pm	0.15	$	\\
32	&	$	0.333	$	&	$	0.071	$	&	$	0.058	$	&	$	0.036	$	&	$	9.30	\pm	0.15	$	&	$	9.89	\pm	0.15	$	&	$	8.59	\pm	0.15	$	&	$	-0.58	\pm	0.21	$	&	$	-1.30	\pm	0.21	$	&	$	0.72	\pm	0.21	$	\\
33	&	$	0.336	$	&	$	0.209	$	&	$	0.122	$	&	$	0.088	$	&	$	10.98	\pm	0.20	$	&	$	14.24	\pm	0.19	$	&	$	12.05	\pm	0.20	$	&	$	-3.26	\pm	0.28	$	&	$	-2.19	\pm	0.28	$	&	$	-1.07	\pm	0.28	$	\\
34	&	$	0.341	$	&	$	0.032	$	&	$	0.006	$	&	$	0.031	$	&	$	32.32	\pm	0.14	$	&	$	33.03	\pm	0.14	$	&	$	31.53	\pm	0.14	$	&	$	-0.71	\pm	0.20	$	&	$	-1.51	\pm	0.20	$	&	$	0.80	\pm	0.20	$	\\
35	&	$	0.353	$	&	$	0.036	$	&	$	0.016	$	&	$	0.045	$	&	$	20.74	\pm	0.16	$	&	$	21.70	\pm	0.15	$	&	$	20.26	\pm	0.16	$	&	$	-0.96	\pm	0.22	$	&	$	-1.44	\pm	0.22	$	&	$	0.48	\pm	0.22	$	\\
36	&	$	0.356	$	&	$	0.039	$	&	$	0.007	$	&	$	0.043	$	&	$	25.23	\pm	0.13	$	&	$	25.70	\pm	0.13	$	&	$	24.79	\pm	0.14	$	&	$	-0.47	\pm	0.19	$	&	$	-0.91	\pm	0.19	$	&	$	0.44	\pm	0.19	$	\\
37	&	$	0.366	$	&	$	0.037	$	&	$	0.018	$	&	$	0.027	$	&	$	11.84	\pm	0.13	$	&	$	12.47	\pm	0.13	$	&	$	11.00	\pm	0.13	$	&	$	-0.62	\pm	0.19	$	&	$	-1.47	\pm	0.19	$	&	$	0.84	\pm	0.19	$	\\
38	&	$	0.375	$	&	$	0.093	$	&	$	0.050	$	&	$	0.047	$	&	$	16.17	\pm	0.14	$	&	$	17.58	\pm	0.14	$	&	$	15.83	\pm	0.14	$	&	$	-1.41	\pm	0.20	$	&	$	-1.74	\pm	0.20	$	&	$	0.34	\pm	0.20	$	\\
39	&	$	0.4	$	&	$	0.023	$	&	$	0.013	$	&	$	0.027	$	&	$	3.36	\pm	0.15	$	&	$	2.77	\pm	0.15	$	&	$	2.75	\pm	0.15	$	&	$	0.59	\pm	0.22	$	&	$	-0.02	\pm	0.22	$	&	$	0.61	\pm	0.22	$	\\
40	&	$	0.412	$	&	$	0.040	$	&	$	0.015	$	&	$	0.054	$	&	$	26.82	\pm	0.16	$	&	$	27.58	\pm	0.16	$	&	$	26.56	\pm	0.16	$	&	$	-0.76	\pm	0.23	$	&	$	-1.01	\pm	0.23	$	&	$	0.26	\pm	0.23	$	\\
41	&	$	0.422	$	&	$	0.058	$	&	$	0.027	$	&	$	0.032	$	&	$	14.70	\pm	0.22	$	&	$	15.84	\pm	0.22	$	&	$	14.37	\pm	0.22	$	&	$	-1.14	\pm	0.31	$	&	$	-1.48	\pm	0.31	$	&	$	0.33	\pm	0.31	$	\\
42	&	$	0.426	$	&	$	0.126	$	&	$	0.106	$	&	$	0.030	$	&	$	4.48	\pm	0.15	$	&	$	4.89	\pm	0.14	$	&	$	4.24	\pm	0.15	$	&	$	-0.41	\pm	0.20	$	&	$	-0.66	\pm	0.20	$	&	$	0.24	\pm	0.21	$	\\
43	&	$	0.45	$	&	$	0.025	$	&	$	0.010	$	&	$	0.020	$	&	$	31.61	\pm	0.16	$	&	$	32.30	\pm	0.15	$	&	$	30.87	\pm	0.16	$	&	$	-0.69	\pm	0.22	$	&	$	-1.43	\pm	0.22	$	&	$	0.74	\pm	0.22	$	\\
44	&	$	0.47	$	&	$	0.032	$	&	$	0.023	$	&	$	0.041	$	&	$	8.28	\pm	0.10	$	&	$	8.74	\pm	0.10	$	&	$	8.14	\pm	0.11	$	&	$	-0.46	\pm	0.14	$	&	$	-0.60	\pm	0.15	$	&	$	0.14	\pm	0.15	$	\\
45	&	$	0.5131	$	&	$	0.133	$	&	$	0.078	$	&	$	0.055	$	&	$	23.66	\pm	0.14	$	&	$	27.24	\pm	0.13	$	&	$	24.82	\pm	0.14	$	&	$	-3.58	\pm	0.19	$	&	$	-2.42	\pm	0.19	$	&	$	-1.16	\pm	0.19	$	\\
46	&	$	0.5392	$	&	$	0.036	$	&	$	0.007	$	&	$	0.043	$	&	$	23.74	\pm	0.18	$	&	$	24.69	\pm	0.18	$	&	$	24.49	\pm	0.18	$	&	$	-0.95	\pm	0.25	$	&	$	-0.20	\pm	0.25	$	&	$	-0.75	\pm	0.25	$	\\
47	&	$	0.545	$	&	$	0.028	$	&	$	0.010	$	&	$	0.020	$	&	$	110.33	\pm	0.19	$	&	$	110.78	\pm	0.19	$	&	$	109.81	\pm	0.19	$	&	$	-0.45	\pm	0.26	$	&	$	-0.97	\pm	0.26	$	&	$	0.52	\pm	0.27	$	\\
48	&	$	0.5586	$	&	$	0.064	$	&	$	0.041	$	&	$	0.027	$	&	$	21.75	\pm	0.20	$	&	$	22.86	\pm	0.20	$	&	$	21.54	\pm	0.20	$	&	$	-1.11	\pm	0.28	$	&	$	-1.31	\pm	0.28	$	&	$	0.21	\pm	0.28	$	\\
49	&	$	0.5699	$	&	$	0.101	$	&	$	0.071	$	&	$	0.031	$	&	$	14.90	\pm	0.17	$	&	$	16.67	\pm	0.17	$	&	$	14.44	\pm	0.17	$	&	$	-1.77	\pm	0.24	$	&	$	-2.23	\pm	0.24	$	&	$	0.46	\pm	0.24	$	\\
50	&	$	0.584	$	&	$	0.080	$	&	$	0.041	$	&	$	0.039	$	&	$	43.03	\pm	0.17	$	&	$	45.57	\pm	0.17	$	&	$	43.52	\pm	0.17	$	&	$	-2.54	\pm	0.24	$	&	$	-2.05	\pm	0.24	$	&	$	-0.49	\pm	0.25	$	\\
51	&	$	0.611	$	&	$	0.112	$	&	$	0.079	$	&	$	0.035	$	&	$	8.60	\pm	0.14	$	&	$	10.46	\pm	0.14	$	&	$	8.54	\pm	0.14	$	&	$	-1.86	\pm	0.20	$	&	$	-1.92	\pm	0.20	$	&	$	0.06	\pm	0.20	$	\\
52	&	$	0.674	$	&	$	0.032	$	&	$	0.026	$	&	$	0.015	$	&	$	13.43	\pm	0.16	$	&	$	13.61	\pm	0.16	$	&	$	12.69	\pm	0.16	$	&	$	-0.18	\pm	0.23	$	&	$	-0.92	\pm	0.23	$	&	$	0.74	\pm	0.23	$	\\
53	&	$	0.6752	$	&	$	0.126	$	&	$	0.090	$	&	$	0.037	$	&	$	13.17	\pm	0.13	$	&	$	15.96	\pm	0.13	$	&	$	13.48	\pm	0.14	$	&	$	-2.79	\pm	0.19	$	&	$	-2.48	\pm	0.19	$	&	$	-0.32	\pm	0.19	$	\\
54	&	$	0.83	$	&	$	0.020	$	&	$	0.014	$	&	$	0.013	$	&	$	35.45	\pm	0.12	$	&	$	35.38	\pm	0.11	$	&	$	34.60	\pm	0.12	$	&	$	0.07	\pm	0.16	$	&	$	-0.78	\pm	0.17	$	&	$	0.85	\pm	0.17	$	\\

\hline

\end{longtable}
\end{small}
\end{center}

\end{landscape}
\restoregeometry

\clearpage

%% file: APP-C/appendixc.tex

%
%
\ifpdf

\chapter{Appendix C: Einasto model derivations and results of PM I and PM II comparisons}
\label{c:appendixc}

\section{Einasto mass integral}
\label{s:einastointegral}

From equations~\ref{e:einasto} and~\ref{e:ein_mass} we have that
\begin{equation}
\label{e:appendixmassrho}
\begin{split}
M(r) &= \int_{0}^{r} 4\pi r'^{2} \rho_{\rm dm, PM II}(r') \, \rm{d}r' \\
     &= 4 \pi \rho_{-2} \exp \left( 2 / \alpha_{\rm Ein} \right) \int_{0}^{r} r'^{2} \exp \left[ \frac{-2}{\alpha_{\rm Ein}} \left( \frac{r'}{r_{-2}} \right)^{\alpha_{\rm Ein}} \right] \rm{d} r'.
\end{split}
\end{equation}
Using the substitution 
\begin{equation}
\label{e:appendixeinsub1}
u = \frac{2^{3/ \alpha_{\mathrm{Ein}}} r'^{3}}{\alpha_{\mathrm{Ein}}^{3 / \alpha_{\mathrm{Ein}}} r_{-2}^{3}} \Rightarrow \mathrm{d} u = \frac{3 \times 2^{3 / \alpha_{\mathrm{Ein}}}r'^{2}}{ \alpha_{\mathrm Ein} ^{3 / \alpha_{\mathrm{Ein}}}r_{-2}^{3}} \mathrm{d}r'
\end{equation}
equation~\ref{e:appendixmassrho} becomes
\begin{equation}
\label{e:appendixmassrho2}
\begin{split}
M(r) &= \frac{4 \pi \rho_{-2} \exp \left( 2 / \alpha_{\rm Ein} \right) \alpha_{\rm Ein}^{3 / \alpha_{\rm Ein}} r_{-2}^{3}}{3 \times 2^{3 / \alpha_{\rm Ein}}} \\
       & \quad \times \int_{u = 0}^{u = \frac{2^{3/ \alpha_{\rm Ein}} r^{3}}{\alpha_{\rm Ein}^{3 / \alpha_{\rm Ein}} r_{-2}^{3}}} \exp \left( -u^{\alpha_{\rm Ein} / 3} \right) \rm{d} u.
\end{split}
\end{equation}
Finally, using the substitution $t = u^{\alpha_{\rm Ein} / 3}$ so that $\mathrm{d}t = \frac{\alpha_{\mathrm{Ein}}}{3} u^{\alpha_{\mathrm{Ein}} / 3 -1} \mathrm{d}u$, then the integral in equation~\ref{e:appendixmassrho2} (ignoring the constant factor) becomes
\begin{equation}
\label{e:appendixmassrho3}
\begin{split}
\frac{3}{\alpha_{\rm Ein}} \int_{0}^{u^{\alpha_{\rm Ein}/3}} u^{1 - \alpha_{\rm Ein} / 3} e^{-t} \rm{d}t &= \frac{3}{\alpha_{\rm Ein}} \int_{0}^{\frac{2r^{\alpha_{\rm Ein}}}{\alpha_{\rm Ein}r_{-2}^{\alpha_{\rm Ein}}}} t^{ 3 / \alpha_{\rm Ein} - 1} e^{-t} \rm{d}t \\
                                                                                                         &= \gamma \left[ \frac{3}{\alpha_{\rm Ein}}, \frac{2}{\alpha_{\rm Ein}} \left( \frac{r}{r_{-2}} \right)^{\alpha_{\rm Ein}} \right],
\end{split}
\end{equation} 
where the last equality follows from the definition of the incomplete lower gamma function $\gamma \left[a, x \right] = \int_{0}^{x} t^{a-1} e^{-t} \rm{d} t $. Including the constant factor in equation~\ref{e:appendixmassrho2} leads to the result
\begin{equation}
\label{e:appendixmassrho4}
M(r) = \frac{4 \pi \rho_{-2}}{\alpha_{\rm Ein}} \exp( 2 / \alpha_{\rm Ein}) \left( \frac{\alpha_{\rm Ein}}{2} \right) ^{3 / \alpha_{\rm Ein}} \gamma \left[ \frac{3}{\alpha_{\rm Ein}}, \frac{2}{\alpha_{\rm Ein}} \left( \frac{r}{r_{-2}} \right)^{\alpha_{\rm Ein}} \right].
\end{equation}


\section{Determining $\lowercase{r}_{500}$ iteratively}
\label{s:r500newton}

Evaluating equations~\ref{e:nfw_m_tot_1} and~\ref{e:ein_mass} at $r_{500}$ and equating we get 
\begin{equation}
\label{e:r500iter1}
\begin{split}
\frac{4\pi}{3} 500 \rho_{\rm crit} (z) r_{500}^{3} &= 4 \pi \rho_{-2} 1 / \alpha_{\rm Ein} \exp( 2 / \alpha_{\rm Ein}) \left( \frac{\alpha_{\rm Ein}}{2} \right) ^{3 / \alpha_{\rm Ein}} r_{-2}^{3} \\
                                                     & \quad \times \gamma \left[ \frac{3}{\alpha_{\rm Ein}}, \frac{2}{\alpha_{\rm Ein}} \left( \frac{r_{500}}{r_{-2}} \right)^{\alpha_{\rm Ein}} \right].
\end{split}
\end{equation}
If we let $R = r_{500} / r_{-2}$, then we can determine $r_{500}$ by solving the following for $R$
\begin{equation}
\label{e:r500iter2}
\begin{split}
&\frac{R^{3}}{\gamma \left[ \frac{3}{\alpha_{\rm Ein}}, \frac{2}{\alpha_{\rm Ein}} R^{\alpha_{\rm Ein}} \right]} \\ 
& - \frac{1}{\rho_{\rm crit}(z)} \frac{3 \rho_{-2}}{500} \left(\frac{\alpha_{\rm Ein}}{2}\right)^{3 / \alpha_{\rm Ein}} \frac{\exp \left( 2 / \alpha_{\rm Ein} \right)}{\alpha_{\rm Ein}} = 0
\end{split}
\end{equation}
by some iterative root finding method e.g. Newton-Raphson. We use the starting point $R_{0} = \frac{2r_{200}}{3r_{-2}}$ which usually results in the algorithm converging in $\mathcal{O}(10)$ iterations. \\
We now show that equation~\ref{e:r500iter2} only has one solution for a given $r_{-2}$. We start by considering both sides of equation~\ref{e:r500iter1} as two different functions, and ignore constant terms for simplicity (this doesn't affect the truth of the final result), i.e. we consider the two functions
\begin{equation}
\label{e:r500iter3}
f(r_{500}) = r_{500}^{3}, \, g(r_{500}) = \gamma \left[ \frac{3}{\alpha_{\rm Ein}}, \frac{2}{\alpha_{\rm Ein}} \left( \frac{r_{500}}{r_{-2}} \right)^{\alpha_{\rm Ein}} \right].
\end{equation}
We first note that $f(0) = g(0) = 0$, and differentiate both functions with respect to $r_{500}$
\begin{equation}
\label{e:r500iter4}
\frac{\mathrm{d}f}{\mathrm{d}r_{500}} \propto r_{500}^{2}, \,  \frac{\mathrm{d}g}{\mathrm{d}r_{500}} \propto r_{500}^{2} \exp \left[ -\frac{2}{\alpha_{\rm Ein}} \left( \left(\frac{r_{500}}{r_{-2}}\right)^{\alpha_{\rm Ein}} - 1 \right) \right].
\end{equation}
Setting these two derivatives equal to each other yields one solution at $r_{500}=r_{-2}$ for all $\alpha_{\rm Ein} \neq 0$, meaning the derivatives only intersect once. Furthermore $\frac{\mathrm{d}g}{\mathrm{d}r_{500}}$ tends to zero for large $r_{500}$ whilst $\frac{\mathrm{d}f}{\mathrm{d}r_{500}}$ is a monotonically increasing function, meaning the former must be larger before the two intersect. This coupled with the fact that $f(0) = g(0) = 0$ means that $g(r_{500}) > f(r_{500})$ until some point (which has to be after the derivatives intersect) when the two intersect, after which $f(r_{500}) > g(r_{500})$ as $g(r_{500})$ flattens off.
This proves that equation~\ref{e:r500iter2} only has one root and that equation~\ref{e:r500iter1} only has one solution in $r_{500}$ for fixed $r_{-2}$.


\newpage
\newgeometry{margin=1cm} 
\begin{landscape}
\section{Simulations results table}
\label{s:ein_simrestab}
\begin{center}
\begin{longtable}{llllllll}
\caption{Input and output values of simulations using NFW and Einasto dark matter profiles. The first column is what dark matter profile was used to \textit{simulate} the cluster. Input $M(r_{200})$ and Input $z$ are the input values used to create the simulation for the given model. Ein out $M(r_{200})$ is the mean and standard deviation of the posterior distribution obtained inferred using an Einasto profile to \textit{model} the cluster. Ein $\ln(\mathcal{Z})$ is the log Bayesian evidence corresponding to the inference. NFW... is as before but using an NFW profile in the \textit{modelling}. $\ln(\mathcal{Z}_{\rm Ein} / \mathcal{Z}_{\rm NFW}) $ is the log ratio of the two evidences obtained.}
\label{t:ein_simrestab} \\

\hline \multicolumn{1}{c}{Model} & \multicolumn{1}{c}{Input $M(r_{200})~(\times10^{14}M_{\mathrm{Sun}})$} & \multicolumn{1}{c}{Input $z$} & \multicolumn{1}{c}{Ein out $M(r_{200})~(\times10^{14}M_{\mathrm{Sun}})$} & \multicolumn{1}{c}{NFW out $M(r_{200})~(\times10^{14}M_{\mathrm{Sun}})$} & \multicolumn{1}{c}{Ein $\ln(\mathcal{Z})$} & \multicolumn{1}{c}{NFW $\ln(\mathcal{Z})$} & \multicolumn{1}{c}{$\ln(\mathcal{Z}_{\rm Ein} / \mathcal{Z}_{\rm NFW}) $}  \\ \hline 

\tabulinesep=_1mm
\extrarowsep=1mm
\LTcapwidth=\textwidth

$\alpha_{\rm Ein} = 0.2$	&$	1	$&$	0.15	$&$	1.05 \pm0.01	$&$	1.08\pm0.01	$&$	47104.4\pm0.4	$&$	47104.6\pm0.4	$&$	-0.2	$	\\
$\alpha_{\rm Ein} = 2.0$	&$	1	$&$	0.15	$&$	1.05\pm0.01	$&$	1.48\pm0.01	$&$	47181.4\pm0.4	$&$	47180.7\pm0.5	$&$	0.6	$	\\
$\alpha_{\rm Ein} = 0.05$	&$	1	$&$	0.15	$&$	1.05\pm0.01	$&$	0.97\pm0.01	$&$	47175.0\pm0.4	$&$	47175.3\pm0.4	$&$	-0.3	$	\\
NFW	&$	1	$&$	0.15	$&$	1.03\pm0.01	$&$	1.05\pm0.01	$&$	47064.3\pm0.4	$&$	47063.6\pm0.4	$&$	0.7	$	\\
$\alpha_{\rm Ein} = 0.2$	&$	1	$&$	0.9	$&$	1.10\pm0.01	$&$	1.21\pm0.01	$&$	47173.2\pm0.5	$&$	47174.5\pm0.5	$&$	-1.3	$	\\
$\alpha_{\rm Ein} = 2.0$	&$	1	$&$	0.9	$&$	1.16\pm0.01	$&$	1.57\pm0.01	$&$	47100.8\pm0.5	$&$	47095.1\pm0.5	$&$	5.7	$	\\
$\alpha_{\rm Ein} = 0.05$	&$	1	$&$	0.9	$&$	1.02\pm0.01	$&$	1.18\pm0.01	$&$	47094.2\pm0.4	$&$	47094.8\pm0.5	$&$	-0.6	$	\\
NFW	&$	1	$&$	0.9	$&$	0.95\pm0.01	$&$	1.05\pm0.01	$&$	47105.2\pm0.4	$&$	47106.7\pm0.4	$&$	-1.5	$	\\
$\alpha_{\rm Ein} = 0.2$	&$	10	$&$	0.15	$&$	10.23\pm0.02	$&$	10.33\pm0.01	$&$	46814.8\pm0.5	$&$	46815.0\pm0.5	$&$	-0.2	$	\\
$\alpha_{\rm Ein} = 2.0$	&$	10	$&$	0.15	$&$	10.18\pm0.01	$&$	15.06\pm0.01	$&$	46640.7\pm0.5	$&$	46638.3\pm0.6	$&$	2.4	$	\\
$\alpha_{\rm Ein} = 0.05$	&$	10	$&$	0.15	$&$	10.21\pm0.02	$&$	9.61\pm0.01	$&$	46844.5\pm0.5	$&$	46844.9\pm0.5	$&$	-0.4	$	\\
NFW	&$	10	$&$	0.15	$&$	10.13\pm0.01	$&$	10.23\pm0.01	$&$	46873.9\pm0.5	$&$	46873.0\pm0.5	$&$	0.9	$	\\
$\alpha_{\rm Ein} = 0.2$	&$	10	$&$	0.9	$&$	11.47\pm0.02	$&$	12.12\pm0.01	$&$	46837.2\pm0.5	$&$	46829.7\pm0.6	$&$	7.5	$	\\
$\alpha_{\rm Ein} = 2.0$	&$	10	$&$	0.9	$&$	11.16\pm0.01	$&$	13.65\pm0.01	$&$	46835.0\pm0.5	$&$	46829.9\pm0.7	$&$	5.1	$	\\
$\alpha_{\rm Ein} = 0.05$	&$	10	$&$	0.9	$&$	11.48\pm0.01	$&$	12.26\pm0.02	$&$	46833.8\pm0.6	$&$	46834.5\pm0.5	$&$	-0.7	$	\\
NFW	&$	10	$&$	0.9	$&$	10.48\pm0.02	$&$	11.57\pm0.02	$&$	46926.4\pm0.6	$&$	46926.6\pm0.5	$&$	-0.2	$	\\
\hline

\end{longtable}
\end{center}

\end{landscape}
\restoregeometry

\clearpage

%% file: APP-D/appendixd.tex

\chapter{Appendix D: Supplementary statistical derivations and results}
\label{c:appendixd}

\section{Metropolis-Hastings and detailed balance}
\label{s:ns_mh_db}

\subsection{Metropolis-Hastings satisfies detailed balance proof}

Following \citet{hauser} we show that for a Markov chain whose values are sampled from target distribution $P$ using proposal distribution $q$ \& the MH acceptance ratio, satisfies detailed balance (a \textit{sufficient} condition for the chain to asymptotically converge to the target distribution).

Consider an arbitrary point along the Markov chain, $\theta_k$, then the proceeding step $\theta_{k+1}$ can lead to one of two possible scenarios which we denote scenario I and scenario II
\begin{equation}
\label{e:ns_mh_db1}
\alpha(\theta_{k+1}, \theta_{k}) \Rightarrow \begin{cases}
 \theta_{k+1} \neq \theta_{k} \quad \mathrm{ scenario \, I,}\\
 \theta_{k+1} = \theta_{k} \quad \mathrm{ scenario \, II.}
\end{cases}
\end{equation}
Scenario I only occurs when an MH step is accepted, which occurs with probability
\begin{equation}
\label{e:ns_mh_db2}
T(\theta_{k+1} | \theta_{k}) = \alpha(\theta_{k+1}, \theta_{k}) q(\theta_{k+1} | \theta_{k}).
\end{equation}
Substituting $\alpha$ for the MH acceptance ratio gives
\begin{equation}
\begin{split}
\label{e:ns_mh_db3}
T(\theta_{k+1} | \theta_{k}) &= \min \left(1, \frac{p(\theta_{k+1})q(\theta_k | \theta_{k+1})
}{p(\theta_k)q(\theta_{k+1} | \theta_k)} \right )q(\theta_{k+1} | \theta_k) \\
&=\frac{1}{p(\theta_k)} \min \left(p(\theta_k) q(\theta_{k+1} | \theta_k), p(\theta_{k+1}) q(\theta_{k} | \theta_{k+1}) \right ),
\end{split}
\end{equation} 
where the second equality follows from taking a factor of $1/p(\theta_k)$ out of both terms in the minimisation and multipling the factor $q(\theta_{k+1} | \theta_k)$ into the function (both actions are allowed if $P$ and $q$ are strictly positive). \\
Observe that the arguments of the the minimisation function are invariant to the relabelling $\theta_{k+1} \rightarrow \theta_{k}$ \& $\theta_{k} \rightarrow \theta_{k+1}$ (except their ordering is switched). Thus $T(\theta_{k} | \theta_{k+1})$ can be written as
\begin{equation}
\label{e:ns_mh_db4}
T(\theta_{k} | \theta_{k+1})= \frac{1}{p(\theta_{k+1})} \min \left(p(\theta_k) q(\theta_{k+1} | \theta_k), p(\theta_{k+1}) q(\theta_{k} | \theta_{k+1}) \right ).
\end{equation}
Substituting these into the detailed balance equation gives 
\begin{equation}
\begin{split}
\label{e:ns_mh_db5}
T(\theta_{k+1} | \theta_{k}) p(\theta_{k}) &= \frac{1}{p(\theta_k)} \min \left(p(\theta_k) q(\theta_{k+1} | \theta_k), p(\theta_{k+1}) q(\theta_{k} | \theta_{k+1}) \right ) p(\theta_{k}) \\
&=\frac{1}{p(\theta_{k+1})} \min \left(p(\theta_k) q(\theta_{k+1} | \theta_k), p(\theta_{k+1}) q(\theta_{k} | \theta_{k+1}) \right ) p(\theta_{k+1})\\
&=T(\theta_{k} | \theta_{k+1})p(\theta_{k+1}),
\end{split}
\end{equation}
and thus the relation is satisfied for scenario I.

Scenario II occurs when either the MH step is accepted and the sampled point happens to be $\theta_{k}$, or when the MH step is rejected
\begin{equation}
\label{e:ns_mh_db6}
T(\theta_{k} | \theta_{k}) = \alpha(\theta_{k}, \theta_{k}) q(\theta_{k} | \theta_{k}) + \int\displaylimits_{\hat{\theta}} q(\theta' | \theta_{k})(1 - \alpha(\theta',\theta_k) )\mathrm{d}\theta',
\end{equation}
where $\hat{\theta}$ is the domain of $p$. Note that substituting $T(\theta_{k} | \theta_{k})$ into the detailed balance equation gives $T(\theta_{k} | \theta_{k}) p(\theta_{k})= T(\theta_{k} | \theta_{k}) p(\theta_{k})$ and so scenario II trivially satisfies the relation.   

\subsection{Deriving the Metropolis-Hastings acceptance ratio from the detailed balance relation}
We can use the same scenario analysis to derive the MH acceptance ratio from the detailed balance relation. For scenario I 
\begin{equation}
\label{e:ns_mh_db7}
\begin{split}
\alpha(\theta_{k+1}, \theta_k) q(\theta_{k+1} | \theta_{k}) p(\theta_{k}) & = T(\theta_{k+1} | \theta_{k}) p(\theta_{k}) \\
&= T(\theta_{k} | \theta_{k+1}) p(\theta_{k+1}) \\
&= \alpha(\theta_{k}, \theta_{k+1}) q(\theta_{k} | \theta_{k+1}) p(\theta_{k+1}),
\end{split}
\end{equation}
which can be rearranged to give
\begin{equation}
\label{e:ns_mh_db8}
\frac{\alpha(\theta_{k+1}, \theta_k)}{\alpha(\theta_{k}, \theta_{k+1})} = \frac{q(\theta_{k} | \theta_{k+1}) p(\theta_{k+1})}{q(\theta_{k+1} | \theta_{k}) p(\theta_{k})}.
\end{equation}
For $\alpha$ to be a probability it must be bounded by $[0,1]$, thus we need a form for $\alpha$ which satisfies this constraint and equation~\ref{e:ns_mh_db8}. If we consider the case where $q(\theta_{k} | \theta_{k+1}) p(\theta_{k+1}) > q(\theta_{k+1} | \theta_{k}) p(\theta_{k})$ then we are saying if $\alpha(\theta_{k+1}, \theta_k) = \alpha(\theta_{k}, \theta_{k+1})$ then the system moves from $\theta_{k+1}$ to $\theta_{k}$ more often than the reverse process happens, and thus detailed balance is violated. This asymmetry suggests that we should maximise $\alpha(\theta_{k+1}, \theta_k)$ i.e. set it equal to one, in which case we can say that
\begin{equation}
\label{e:ns_mh_db9}
\frac{1}{\alpha(\theta_{k}, \theta_{k+1})} = \frac{q(\theta_{k} | \theta_{k+1}) p(\theta_{k+1})}{q(\theta_{k+1} | \theta_{k}) p(\theta_{k})} > 1,
\end{equation}
which is satisfied by the equality
\begin{equation}
\label{e:ns_mh_db10}
\alpha(\theta_{k}, \theta_{k+1}) = \min \left( \frac{q(\theta_{k+1} | \theta_{k}) p(\theta_{k})}{  q(\theta_{k} | \theta_{k+1}) p(\theta_{k+1})} \right).
\end{equation}
Similarly in the case that $q(\theta_{k} | \theta_{k+1}) p(\theta_{k+1}) < q(\theta_{k+1} | \theta_{k}) p(\theta_{k})$ we can set $\alpha(\theta_{k}, \theta_{k+1}) = 1$ and deduce the value of $\alpha(\theta_{k+1}, \theta_k)$. Substituting both of these expressions back in equation~\ref{e:ns_mh_db8} satisfies the equality, and thus we have verified the MH acceptance ratio satisfies detailed balance for scenario I. \\
Scenario II satisfies detailed balance trivially for any $\alpha$ and so does not place any additional constraints on its form. Thus the relation derived from equation~\ref{e:ns_mh_db8} is valid for both scenarios.

Note that the way we have derived the MH acceptance probability implies that there is no alternative acceptance probability $\beta$ that satisfies $\beta(\theta', \theta_k) > \alpha(\theta', \theta_k)$, that does not violate either $\beta \in [0,1]$ or the detailed balance relation.

\section{Evidence integral transformation}
\label{s:ns_Z_proof}

We now show that equations~\ref{e:evidence} and~\ref{e:gns_nsz} are equivalent for the case of a one-dimensional parameter problem: $\vec{\Theta} \equiv \theta$, but note that this proof holds for $\vec{\Theta}$ of arbitrary dimension. \\
The expression for $X$ can be re-written as
\begin{equation}
\label{e:ns_z_proof1}
X(\lambda) = \int_{\hat{\theta}} \pi(\theta) H(\mathcal{L}(\theta) - \lambda) \mathrm{d}\theta,
\end{equation}
where $\hat{\theta}$ is the support of $\pi(\theta)$ and $H(x)$ is the Heaviside step function which satisfies $\frac{\mathrm{d}}{\mathrm{d}x} H(x) = \delta(x)$. Taking the derivative of equation~\ref{e:ns_z_proof1} with respect to $\lambda$ gives
\begin{equation}
\label{e:ns_z_proof2}
\frac{\mathrm{d}X}{\mathrm{d} \lambda} = -\int_{\hat{\theta}} \pi(\theta) \delta(\mathcal{L}(\theta) - \lambda) \mathrm{d}\theta.  
\end{equation}
Equation~\ref{e:gns_nsz} can be re-expressed as an integral over $\mathcal{L}$ by making a change of variables from $X$ to $\mathcal{L}$. Recalling that $\lambda(X) \equiv \mathcal{L}(X)$,
\begin{equation}
\label{e:ns_z_proof3}
\begin{split}
\mathcal{Z} &= \int_{0}^{1} \mathcal{L}(X) dX \\
&= \int \lambda \cdot  \frac{\mathrm{d}X}{\mathrm{d}\lambda}  \mathrm{d}\lambda \\
&= -\int \mathcal{L} \int_{\hat{\theta}} \pi(\theta) \delta(\mathcal{L}(\theta) - \lambda) \mathrm{d}\theta \mathrm{d}\lambda.
\end{split}
\end{equation}
Since the integrals are over all possible values of $\theta$ and $\lambda$ and the limits are not a function of the other variable of integration (i.e. limits$(\theta) \neq g(\lambda)$ and vice versa), the order of integration can be switched according to Fubini's theorem
\begin{equation}
\begin{split}
\label{e:ns_z_proof4}
\mathcal{Z} &= \int_{\hat \theta} \pi(\theta) \left( -\int \lambda \delta(\mathcal{L}(\theta) - \lambda) \mathrm{d} \lambda \right) \mathrm{d}\theta \\
&= \int_{\hat \theta}  \mathcal{L}(\theta) \pi(\theta) \mathrm{d}\theta,
\end{split}
\end{equation}
which is the form for $\mathcal{Z}$ given by equation~\ref{e:evidence}.

\section{Probability density function of the largest of $n$ numbers}
\label{s:ns_tdist}
Let $u_1 , ... , u_n$ be $n$ random variables from the uniform distribution on $[0, 1]$ and let $u_m$ be their maximum. 
We can derive $P(t)$ by considering the cumulative distribution function of $u_m$, $F(u_m)$. 
$u_m$ is less than some value $t$ if and only if all $u_1 , ... , u_n$ are less than $t$. Therefore
\begin{equation}
\label{e:ns_t1}
 F(u_m = t) = F(u_1 = t ,..., u_n = t).
\end{equation}
Since the $u_j$'s are independent this is the same as
\begin{equation}
\label{e:ns_t2}
 F(u_1 = t) ...  F(u_n = t).
\end{equation}
If $t$ is also defined on $[0,1]$, then since $u_j$s are uniformly distributed we get
\begin{equation}
\label{e:ns_t3}
F(u_m = t) = t...t = t^n.
\end{equation}
Hence the probability density function for $u_m$ is
\begin{equation}
\label{e:ns_t4}
F'(u_m = t) = P(t) = n t^{n-1}.
\end{equation}

\section{Determining $\log \left( \mathbb{E}[\mathcal{Z}] \right)$ and $\log \left( \mathbb{E}\left[\mathcal{Z}^2\right] \right)$ from $\log(\mathcal{L})$ and $\log(X)$.}
\label{s:keet_log}

For the standard quadrature approximation of $\mathcal{Z}$ (equation~\ref{e:gns_nsz_sum}) and statistical treatment of $t$ given by equation~\ref{e:gns_t_prob}, \citet{2011MNRAS.414.1418K} derives expressions for $\mathbb{E}[\mathcal{Z}]$ and $\mathrm{var}\left[\mathcal{Z}\right]$. The form of $\mathrm{var}\left[\mathcal{Z}\right]$ derived incorporates the covariance between the value of $\mathcal{Z}$ obtained from the main nested sampling algorithm loop and that obtained from the final contribution to the evidence after the main loop has terminated ($\mathcal{Z}_{\rm f}$, see Section~\ref{s:gns_stop_crit}). When deriving these equations Keeton works in linear space, which is valid as long as $\sqrt{\mathrm{var}\left[\mathcal{Z}\right]} / \mathbb{E}[\mathcal{Z}] \ll 1$, as in this limit $\mathcal{Z}$ is normally distributed. As stated in Section~\ref{s:gns_ns_sum} $\mathcal{Z}$ is log-normally distributed in general and thus we should quote the statistics given by equations~\ref{e:gns_lognorm_mean} and~\ref{e:gns_lognorm_var}. Furthermore, working in linear space can lead to numerical difficulties if $\mathcal{L}$ and $X$ are sufficiently small / large, as is the case in the nested sampling example considered in Section~\ref{s:gns_ligo}. We can adapt the equations derived by Keeton to calculate $\log \left( \mathbb{E}[\mathcal{Z}] \right)$ \& $\log \left( \mathbb{E}\left[\mathcal{Z}^2\right] \right)$ from $\log(\mathcal{L})$ \& $\log(X)$ to obtain estimates of $\mathbb{E}\left[ \log(\mathcal{Z}) \right]$ \& $\mathrm{var}\left[ \log(\mathcal{Z}) \right]$ (which hopefully avoid numerical under / overflow issues), as follows.

We first define a function $L$ which takes a vector $\vec{x}$ as an input, exponentiates this vector component-wise, adds together the resultant values and then takes the logarithm of this sum (known as the LogSumExp function)\footnote{Underflow and / or overflow issues can be avoided to some extent using the trick given in \url{https://hips.seas.harvard.edu/blog/2013/01/09/computing-log-sum-exp/}.}
\begin{equation}
\label{e:keet_log_1}
L(\vec{x}) = \log \left( \sum_{i = 1}^{i = n} \exp(x_i) \right).
\end{equation}
We also define $\log(\vec{x}) \equiv (\log(x_1),~...,\log(x_n))$. $\log \left( \mathbb{E}[\mathcal{Z}] \right)$ (c.f. equation~17 of Keeton) can then be calculated as
\begin{equation}
\label{e:keet_log_2}
\log \left( \mathbb{E}[\mathcal{Z}] \right) = L\left( \log (\vec{\mathcal{L}}) + \log (\vec{\delta X}) \right) - \log(n_l),
\end{equation}
where $\log (\vec{\mathcal{L}})$ is the vector of $\log \mathcal{L}_i$ values obtained in the main nested sampling loop and $\log (\vec{\delta X}) = \left(\log\left(\mathbb{E}[t] \right),...,n_{\rm s} \log\left(\mathbb{E}[t] \right) \right)$. Note that $\log\vec{(\mathcal{L}})$ and $\log (\vec{\delta X})$ are both vectors of length $n_{\rm s}$. \\
$\log \left( \mathbb{E}\left[\mathcal{Z}^2\right] \right)$ (c.f. equation~22 of Keeton) is given by
\begin{equation}
\label{e:keet_log_3}
\log \left( \mathbb{E}\left[\mathcal{Z}^2\right] \right) = \log \left( \frac{2}{(n_l (n_l + 1))} \right) + L\left( \log (\vec{\mathcal{L}}) + \log (\vec{\delta X}) + \log (\vec{I}) \right),
\end{equation}
where $\log (\vec{I)}) = (\log (I_1),...,\log (I_k),...,\log (I_{\rm ns}))$ and
\begin{equation}
\label{e:keet_log_4}
\log (I_k) = L \left( \log(\vec{\mathcal{L}_k}) + \log \left( \vec{\mathbb{E}[t^2]_k} \right) - \log (\vec{\delta X_k}) \right).
\end{equation}
Here the vector quantities denoted $\vec{x_k}$ each have length $k$, and $\log \left( \vec{\mathbb{E}[t^2]_k} \right) = \left(\log \left( \mathbb{E}[t^2] \right),..., k \log \left( \mathbb{E}[t^2] \right) \right)$. \\
The expected contribution to the evidence after the nested sampling algorithm loop terminates, $\log \left(\mathbb{E}[\mathcal{Z}_{\rm f}] \right)$ can be determined from the log-likelihood values of the final set of livepoints $\log(\vec{\mathcal{L}_{\rm f}}) = (\mathcal{L}_1,..., \mathcal{L}_{l})$, and the remaining prior volume $X_{n_{\rm s}}$ through
\begin{equation}
\label{e:keet_log_5}
\log \left( \mathbb{E}[\mathcal{Z}_{\rm f}] \right) = \log(X_{n_{\rm s}}) - \log(n_l) + L(\log(\vec{\mathcal{L}_{\rm f}})).
\end{equation}
Similarly the log of the second moment of $\mathcal{Z}_{\rm f}$ (equation~28 of Keeton) is given by
\begin{equation}
\label{e:keet_log_6}
\log \left( \mathbb{E}[\mathcal{Z}_{\rm f}^2] \right) = L(\log(\vec{\mathcal{L}_{\rm f}})) - 2 \log(n_l) + n_{\rm s} \log \left( \mathbb{E}[t^2] \right) .
\end{equation}
Finally, the log of the cross term $\mathbb{E}[\mathcal{Z} \mathcal{Z}_{\rm f}]$ (Keeton equation~32) can be calculated as
\begin{equation}
\label{e:keet_log_7}
\log \left( \mathbb{E}[\mathcal{Z} \mathcal{Z}_{\rm f}] \right) = L(\log(\vec{\mathcal{L}_{\rm f}})) + \log (\vec{\delta X}_{n_{\rm s}}) - \log (n_l (n_l + 1)) + \log (I_{\rm ns}).
\end{equation}
$\log \left( \mathbb{E}[\mathcal{Z}] \right)$ \& $\log \left( \mathbb{E}\left[\mathcal{Z}^2\right] \right)$ are then updated as
\begin{gather}
\label{e:keet_log_8}
\log \left( \mathbb{E}[\mathcal{Z}] \right) \rightarrow L \left( \left(\log \left( \mathbb{E}[\mathcal{Z}] \right), \log \left( \mathbb{E}[\mathcal{Z}_{\rm f}] \right) \right) \right), \\
\label{e:keet_log_9}
\log \left( \mathbb{E}\left[\mathcal{Z}^2\right] \right) \rightarrow L \left( \left(\log \left( \mathbb{E}\left[\mathcal{Z}^2\right] \right), \log \left( \mathbb{E}[\mathcal{Z}_{\rm f}^2] \right), \log(2) +  \log \left( \mathbb{E}[\mathcal{Z} \mathcal{Z}_{\rm f}] \right) \right)  \right),
\end{gather}
and used in equations~\ref{e:gns_lognorm_mean} \&~\ref{e:gns_lognorm_var} to calculate $\mathbb{E}\left[ \log(\mathcal{Z}) \right]$ \& $\mathrm{var}\left[ \log(\mathcal{Z}) \right]$.


%% file: OTHER/references.tex
\setlength{\bibsep}{0pt}            
\renewcommand{\bibname}{References} 

%
%
%
%
%
%